\documentclass[useAMS,usenatbib]{mn2e}
\usepackage{graphicx}
\usepackage{amsfonts}
\usepackage{pifont}

\newcommand{\bpl}{BolshoiP}
\newcommand{\smdpl}{SMDPL}
\newcommand{\mdpl}{MDPL}

\usepackage{color} 
\definecolor{purple}{RGB}{160,32,240}
\usepackage {comment}
\usepackage{url}

\def\msun{\mbox{M$_{\odot}$}}

\def\sfr{\mbox{SFR}}
\def\gsmf{\mbox{GSMF}}
\def\shmr{\mbox{SHMR}}
\def\phig{\mbox{$\phi_{g_{\rm I}}$}}

\def\phigobs{\mbox{$\phi_{g_{\rm obs}}$}}

\def\phigal{\mbox{$\phi_{\rm *}$}}

\def\phih{\mbox{$\phi_{\rm vir}$}}
\def\phisub{\mbox{$\phi_{\rm sub}$}}
\def\phiDM{\mbox{$\phi_{\rm halo}$}}
\def\phiuv{\mbox{$\phi_{\rm UV}$}}
\def\phinormuv{\mbox{$\phi_{\rm UV}^*$}}

\def\mdm{\mbox{$M_{\rm halo}$}}
\def\reff{\mbox{$R_{\rm eff}$}}

\def\rhom{\mbox{$\rho_{\rm m}$}}

\def\redsfhit{\mbox{$z$}}
\def\ms{\mbox{$M_*$}}
\def\xuv{\mbox{$x_{\rm UV}$}}
\def\muv{\mbox{$M_{\rm UV}$}}
\def\mcuv{\mbox{$M_{\rm UV}^*$}}
\def\mvir{\mbox{$M_{\rm vir}$}}
\def\mpeak{\mbox{$M_{\rm peak}$}}

\def\msmerger{\mbox{$M_{\rm merger}$}}
\def\msSF{\mbox{$M_{\rm SFR}$}}

\def\fmerger{\mbox{$f_{\rm in\, situ}$}}

\defcitealias{RDA12}{RDA12}
\def\ltsima{$\; \buildrel < \over \sim \;$}    
\def\lesssim{\lower.5ex\hbox{\ltsima}}           
\def\gtsima{$\; \buildrel > \over \sim \;$}    
\def\grtsim{\lower.5ex\hbox{\gtsima}}           

\title[The Galaxy-Halo Connection Over The Last 13.3 Gyrs]
 {Constraining the Galaxy-Halo Connection Over The Last 13.3 Gyrs: Star Formation Histories, Galaxy Mergers and Structural Properties}
\author[]{Aldo Rodr\'iguez-Puebla$^{1,2}$\thanks{rodriguez.puebla@gmail.com}, Joel R. Primack$^3$, Vladimir Avila-Reese$^2$, 
\newauthor and S. M. Faber$^4$ 
\\
$^1$ Department of Astronomy \&\ Astrophysics, University of California at Santa Cruz, Santa Cruz, CA 95064, USA\\
$^2$ Instituto de Astronom\'ia, Universidad Nacional Aut\'onoma de M\'exico, A. P. 70-264, 04510, M\'exico, D.F., M\'exico\\
$^3$Physics Department, University of California, Santa Cruz, CA 95064, USA \\
$^4$UCO/Lick Observatory, Department of Astronomy and Astrophysics, University of California, Santa Cruz, CA 95064, USA \\
}

\date{Released 20?? Xxxxx XX}

\pagerange{\pageref{firstpage}--\pageref{lastpage}} \pubyear{20??}

\begin{document}

\label{firstpage}

\maketitle

\begin{abstract}
We present new determinations of the stellar-to-halo mass relation (SHMR) at $z=0-10$ that 
match the evolution of the galaxy stellar mass function, the $\sfr-\ms$ relation,
and the cosmic star formation rate. We utilize a compilation of 
40 observational studies from the literature and correct them for potential biases. 
Using our robust determinations of halo mass assembly and the SHMR, 
we infer star formation histories, merger rates, and structural properties for average galaxies, combining
star-forming and quenched galaxies.
Our main findings: 
(1) The halo mass $M_{50}$ above which 50\% of galaxies are
quenched coincides with sSFR/sMAR$\sim1$, where sMAR is the
specific halo mass accretion rate. 
(2) $M_{50}$ increases with redshift, presumably due to cold streams being more 
efficient at high redshift while virial shocks and AGN feedback become more relevant at lower redshifts. 
(3) The ratio sSFR/sMAR has a peak value, which occurs around $\mvir\sim2\times10^{11}\msun$.
(4) The stellar mass density within 1 kpc, $\Sigma_1$, is a good indicator of the galactic global sSFR.
(5) Galaxies are statistically quenched after they reach a maximum in $\Sigma_1$, consistent with theoretical expectations
of the gas compaction model; this maximum depends on redshift. 
(6) In-situ star formation is responsible for most galactic stellar mass growth, especially for lower-mass galaxies.
(7) Galaxies grow inside out. The marked change in the slope of the size--mass relation when galaxies became quenched, 
from $d\log \reff/d\log\ms\sim0.35$ to $\sim2.5$, could be the result of dry minor mergers. 
\end{abstract}


\section{Introduction}

There has been remarkable recent progress in 
assembling large galaxy samples from multiwavelength sky surveys. Moreover, these advances are not just for observations of
local galaxies but also for very distant galaxies, resulting in reliable 
samples which contain hundreds of thousands of galaxies at $z\sim0.1$, tens of thousands
between $z\sim0.2-4$, hundreds of galaxies between $z\sim6-8$ and a few tens of galaxies confirmed
as distant as $z\sim9-10$.\footnote{These achievements are all the more impressive when one realizes that  
the Universe was only $\sim500$ Myrs old at $z=10$.} Thus, statistical analyses of the 
properties of the galaxies are now possible with unprecedented detail including 
robust determinations of the luminosity functions (LF) 
over a very wide redshift range. 

In parallel, substantial progress has been made on stellar population synthesis (SPS) modelling \citep[for a recent review, see][]{Conroy2013},
allowing the determination of physical parameters that are key to studying galaxy evolution, including
galaxy stellar masses and star formation rates (\sfr s).\footnote{Also empirically motivated diagnostics of galaxy
\sfr\ have been improved in the last few years; for a recent review, see \citet{Kennicutt+2012}.} 
Thus, we are in a era in which robust determinations of the galaxy stellar mass function (\gsmf) and the 
correlation between stellar mass and $\sfr$s can now be robustly explored. 

It is important to have accurate and robust determinations of the $\gsmf$s and $\sfr$s 
because there is a fundamental link between them. 
While the \gsmf\ is a time-integrated quantity (reflecting stellar mass growth 
by in-situ star formation and/or galaxy mergers, and also the mechanisms that have suppressed the formation of stars
especially in massive galaxies), the $\sfr$ is an instantaneous quantity that measures the in-situ transformation
of gas into stars, typically measured over times scales of $\sim10-100$
Myrs. In other words, except for mergers, the galaxy $\sfr$s are proportional 
to the time derivative of the $\gsmf$s. Indeed, previous studies have combined these two
statistical properties to explore the stellar mass growth of galaxies 
\citep[e.g.,][]{Bell+2007,DroryAlvarez2008,Peng+2010,Papovich+2011,Leja+2015} and to understand 
the most relevant mechanisms that shape their present-day properties.   

At the same time, one of the key open questions in modern astronomy is how
galaxy properties relate to those of their host dark matter 
halos  and subhalos. Since galaxies formed and evolved within dark matter halos, 
one expects that the masses and assembly histories of dark matter halos are linked to the 
stellar mass and $\sfr$s of their host galaxies \citep[for recent reviews see][]{Mo+2010,Somerville+2015}. Indeed, a substantial effort 
has focused on establishing robust determinations of the galaxy-halo connection, commonly reported 
in the form of the stellar-to-halo mass relation, \shmr\ 
\citep[][and references therein]{Behroozi+2010,Moster+2010,Yang+2012,Behroozi+2013,Moster+2013}. The galaxy-halo connection can be obtained by
using three different methods, two of which are physically motivated, based on 
hydrodynamical simulations and semi-analytic models of galaxy formation \citep[both discussed in][and references therein]{Somerville+2015}. The third method, which we follow in this paper, relies on semi-empirical modelling of the galaxy-halo connection \citep[see e.g.][]{Mo+2010}.  That is, we use abundance matching to relate galaxies to their host halos, assuming that the average galactic stellar mass growth follows the average halo mass accretion.  Thus we infer the most likely trajectories of the galaxy progenitors as a function of mass and redshift. However,
since we model average growth histories rather than the growth histories of individual galaxies (as is done in semi-analytic models or in hydrodynamic simulations), we do not know ``case by case" how galaxies evolved. 

We use a compilation of data from the literature on galaxy stellar mass functions and star-formation rates, recalibrated to consistent assumptions regarding the initial mass function, photometry, cosmology, etc. (see Tables 1 and 2).  We fit all this data accurately using a galaxy model that consists of eighteen adjustable parameters, fifteen to model the redshift evolution of the \shmr\ and three to model the fraction of stellar mass growth due to in-situ star formation.  Our approach also allows us to model average galaxy radial profiles including surface densities at 1kpc and at the effective radius.  

\subsection{Semi-empirical modelling}

The idea behind semi-empirical modelling of the galaxy-halo connection is to use simple rules to populate dark matter halos and subhalos
with mock galaxies that match the observed distribution of galaxy surveys.  There are three main approaches to doing this.
One of them is subhalo abundance matching, SHAM, which matches the cumulative \gsmf\ to 
the cumulative halo mass function in order to obtain a correlation between galaxy
stellar mass and (sub)halo mass \citep{Kravtsov+2004,ValeOstriker2004,Conroy+2006,Shankar+2006,Behroozi+2010,Moster+2010,
Guo+2010,RDA12,Papastergis+2012,Hearin+2013}. The next two are the halo occupation
distribution (HOD) model \citep{Jing+1998,Seljak+2000,Scoccimarro+2001,Berlind+2002,Cooray+2002,Kravtsov+2004,Zehavi+2005,Zheng+2005,Leauthaud+2011,RAD13,Tinker+2013,Contreras+2017} and the closely 
related conditional stellar mass function model \citep{Yang+2003,Cooray+2006,Yang+2012,RP+2015}.
Both approaches constrain the distribution of central and satellite galaxies by using the
observed two-point correlation function, 2PCF, and/or galaxy group catalogs and galaxy number densities. In this paper we
focus on the first approach, SHAM, described in detail in Section \ref{mod_framework}.  

Semi-empirical modelling of the galaxy-halo connection has received much attention 
recently, in part because of its simplicity but also because of its power in projecting
different observables in terms of the theoretical properties of the halos \citep[for some applications of these ideas see,
e.g.,][]{Shankar+2006,Conroy+2009,Firmani+2010a,Yang+2013,Behroozi+2013,RAD13,Hearin+2013b,Masaki+2013,RP+2015,Li+2016,Micic+2016,Huang+2017,Matthee+2017}. Moreover,   
improvements in building 
more accurate and consistent halo finders for determining halo properties and their progenitors from large $N-$body 
cosmological simulations \citep[e.g.,][and references therein]{Behroozi+2013b} not only make 
possible a more accurate modelling of the galaxy-halo connection but also allow consistent
connections of observables from different redshifts. This is especially true when information on the evolution of the \gsmf\ and
the $\sfr$s is available. Here, we will present a framework that follows the median evolution of halo masses linked to the corresponding stellar masses.

The importance of the galaxy-halo connection, however, not only relies on determining accurate models
that match observations, i.e., having robust determinations of the stellar-to-halo mass relation (\shmr), 
but also on using it as a phenomenological tool to understand the average growth of galaxies by projecting several galaxy 
observables related to galaxy evolution. This does not imply that semi-empirical modelling of the galaxy-halo connection 
should replace more physically motivated studies such as those mentioned previously, but rather that a synergy between 
all these approaches is ideal to have a more complete and accurate picture of galaxy formation and evolution. 

We would like to emphasize that the approach employed here presents a probabilistic description 
of galaxy evolution driven exclusively as a function of mass. That is, the galaxy assembly histories, 
which we also call trajectories, that we will study in this paper refer to {\it average} histories. From this point of view, at least, 
we argue that the results that will be presented on this paper are the typical
assembly histories of galaxies as a function of their stellar or virial masses.  

In this paper we are interested in constraining the galaxy-halo connection and in extending this connection to 
predictions of the evolution of individual (average) galaxies as a function of mass. 
Besides, for the first time, here we introduce the evolution of galaxy {\it radial} structural properties within this framework. 
There is evidence showing a tight correlation between the structural evolution of galaxies and their $\sfr$s.  
For example, \citet{Kauffmann+2003} showed that above a critical effective stellar surface mass density galaxies typically are old and quenched.
Similar conclusions have been found when using other surface mass density definitions \citep{Cheung+2012,Fang+2013,Barro+2013,Barro+2015,Tacchella+2015} 
or S\'ersic index \citep[see e.g.][]{Bell+2012}. Nevertheless, it is not clear whether the galaxy structural transformation is the key driver of quenching, 
since it is also expected that dark matter halos may make SFRs less efficient above $\mvir\sim10^{12}\msun$   \citep[e.g.,][]{Dekel+2006}. 
Using our framework, we are in a good position to investigate how all these quantities relate to each other.  

In short, the present paper formulates a semi-empirical approach based on three 
key assumptions: 
(1) every halo hosts a galaxy; (2) the in-situ/ex-situ stellar mass growth of galaxies is associated to the mass growth of halos 
by accretion/mergers; (3) the average radial stellar mass distribution of galaxies at a given \ms\
is composed of a S\`ersic disc and a de Vacouleours spheroid, where the fractions of these components are associated with
the fractions of star-forming and quenched galaxies at this mass.  
Under these assumptions and once the \shmr s at different redshifts are constrained from observations,
the growth of dark matter halos can be used to determine the stellar in-situ and ex-situ (merger) mass 
growth of the associated galaxies, and thus their star formation histories. In addition, the observed
galaxy size-mass relation at different redshifts can be used to project  
the size evolution of the disc and  spheroid components of galaxies, and probe a possible relation between
their structural and star formation histories. 

Our model, under the assumptions mentioned above, provide a consistent picture of stellar and halo
mass, \sfr, merger rate, and structural evolution of galaxies as  a function of mass across $\sim 13.3$ Gyrs 
of the cosmic time in the context of the $\Lambda$CDM cosmology. 
These results are ideal for constraining semi-analytic models and numerical simulations of
galaxy formation and evolution, not only at the present day but at high redshifts.
On the other hand, our model of galaxy mass assembly and SFR history based on halo mass accretion trajectories
can be used to compare
with those from other empirical approaches and in this way identify possible biases in various approaches. 
For example, by means of the ``fossil record'' method, the global and
radial stellar mass assembly histories of galaxies of different masses have recently been inferred
from spectroscopic surveys of local galaxies under the assumption of no radial displacements of
stars \citep[][]{Perez+2013,Ibarra-Medel+2016}.
The structural and mass assembling histories of galaxies have also been partially inferred
from direct empirical approaches, for instance, from observations of galaxy populations of similar 
masses and comoving number densities at different redshifts \citep[e.g.,][]{vanDokkum+2010,vanDokkum+2013,Leja+2015}.

This paper is organized as follows. In Section \ref{mod_framework}, we describe 
our galaxy-halo-connection model. In Section \ref{Model_Ingredients},
we present the model ingredients, namely, 
the parametric redshift dependence of the \shmr\ and 
the fraction of in-situ and ex-situ stellar mass growth. 
In Section \ref{Obs_inputs}, we present the observations that we utilize to
constrain the galaxy-halo connection. Here we describe in detail the 
potential systematic effects that could affect our derivations, and how we correct for them. 
In Section \ref{mcmc},
we present our best fitting model. In Section \ref{STHMR}
we discuss our resulting \shmr\ and compare to previous determinations
from the literature. Section \ref{growth_SFH} describes the star formation
histories and halo star formation efficiencies for the progenitors of today's galaxies. 
Results from galaxy mergers are presented in Section \ref{Mergers}. 
In Section \ref{Structural_Evolution} we discuss the implications of the structural 
evolution of galaxies. Finally, in Section \ref{Discussion_section} we discuss 
all the results presented in previous sections and in Section \ref{summary}
we summarize and list our main conclusions. 

In this paper we adopt cosmological parameter values from the Planck mission:
$\Omega_\Lambda=0.693,\Omega_{\rm M}=0.307, \Omega_{\rm B}=0.048, h=0.678$.
These are the same parameters used in the Bolshoi-Planck cosmological simulation \citep{Rodriguez-Puebla+2016},
which we use to determine dark matter halo properties and evolution (see the Appendices for details).

\section{Theoretical Framework}
\label{mod_framework}

\subsection{Galaxy-Halo Connection}

Subhalo abundance matching, SHAM, is a simple and yet powerful
statistical approach for connecting galaxies to halos. In its 
simplest form, given some halo property (usually halo mass or maximum circular velocity)
the halo number density and the galaxy number density are matched in order to obtain the 
connection between halos and the galaxies that they host. As a result one obtains
a unique relation between galaxy stellar mass and a halo property. In fact, it is expected that this relation 
has some scatter since the properties of the galaxies might be partly determined by different 
halo properties and/or some environmental factors. For example, in the case of halo mass, 
analysis of large galaxy group catalogs
\citep{Yang+2009b,Reddick+2013}, the kinematics of satellite galaxies  \citep{More+2011},
as well as galaxy clustering \citep{Shankar+2014,RP+2015,Tinker+2016} have found that this dispersion
is of the order of $\sigma_{\rm h} \sim 0.15$ dex.
To take this into account, abundance matching
should be slightly modified to include a physical scatter around the galaxy stellar-to-halo mass relation, 
hereafter \shmr. 

Formally, if we assume that  halo mass \mdm\ is the halo property 
that correlates best with stellar mass, we can model the \gsmf\ of galaxies by defining $\mathcal{H}(\ms|\mdm)$ as 
the probability distribution function that a halo of mass \mdm\ hosts
a galaxy of stellar mass  $\ms$.  
Then the {\it intrinsic} \gsmf, $\phig$, as a function of stellar mass
is given by
\begin{equation}
\phig(\ms) =   \int \mathcal{H}(\ms|\mdm)\phiDM(\mdm)d\log \mdm,
\label{phi_dir_amt}
\end{equation}
where \phiDM\  denotes the total number density of halos  
and subhalos within the mass range $\log\mdm\pm d\log\mdm / 2$.\footnote{Note that observations of the 
\gsmf\ are made over redshift slices. In order to be consistent with
observations, we average \phiDM\ over the observed volumes. For simplicity we
do not show this explicitly in the reminder of this paper.}  
Note that in the above equation the units for $\phig$ and $\phiDM$ are in Mpc$^{-3}$dex$^{-1}$.
Here \mdm\ is interpreted as the virial mass, \mvir, for distinct halos 
and  \mpeak, the peak value of virial mass at or before accretion, for subhalos, thus:
\begin{equation}
	\mdm = \left\{ 
			\begin{array}{c l}
				\mvir & \mbox{Distinct halos}\\
				\mpeak & \mbox{Subhalos}
			\end{array}\right.
\end{equation}
and
\begin{equation}
\phiDM (\mdm) = \phih(\mvir)  + \phisub(\mpeak).
\end{equation}
Subhalos can lose a significant fraction of their mass due to tidal 
striping. Since tidal stripping affects the subhalo more than 
the stars of the galaxy deep inside it, this implies that 
the stellar mass of a satellite galaxy does not correlate in a simple way with its subhalo mass. 
Therefore, in connecting galaxies to dark matter (sub)halos, as in
SHAM, it has been shown that the mass that the halo had when it was
still in a distinct halo correlates better with the stellar mass of the satellite galaxy it hosts. This comes from the 
fact that the observed two-point correlation function is approximately reproduced
when assuming identical stellar-to-halo mass relations for central and satellite galaxies until the satellites are accreted. 
In this paper we use the mass 
$\mpeak$ defined as the maximum mass reached 
along the main progenitor assembly mass. Indeed, 
\citet{Reddick+2013} found that this is the quantity that correlates best 
with galaxy stellar mass and luminosity by reproducing most accurately the observed galaxy 
clustering \citep[see also][]{Moster+2010}. 

We assume that the probability distribution function $\mathcal{H}(\ms|\mdm)$ 
is a log-normal distribution with a scatter $\sigma_{\rm h} = 0.15$ dex \citep{RP+2015}\footnote{Note that \citet{Tinker+2016}
concluded that the upper limit to the intrinsic scatter is 0.16 dex based on galaxy clustering of massive galaxies. This is consistent with the value assumed above.} independent of \mdm:
	\begin{eqnarray}
		\mathcal{H}(\ms|\mdm)=\frac{1}{\sqrt{2\pi\sigma_{\rm h}^2}}\times   & &  \nonumber \\
		\exp\left[-\frac{\left(\log\ms - \langle\log\mathcal{M_*}(\mdm)\rangle\right)^2}{2\sigma_{\rm h}^2}\right],
		\label{H_ms_X}
	\end{eqnarray}
where the mean \shmr\ is denoted by $\langle\log\mathcal{M_*}(\mdm)\rangle$. In the 
absence of scatter around the mean \shmr, Equation (\ref{phi_dir_amt}) reduces to
\begin{equation}
\phig(\ms) = \phiDM(\mdm) \frac{d\log \mdm}{d\log \ms}.
\label{phi_dir}
\end{equation}
The above equation is just simply the traditional SHAM. 

Individual galaxy stellar mass estimates are also subject to 
random errors  \citep{Conroy2013}. Indeed, the \gsmf\ that is inferred from observations 
through the estimation of individual  stellar masses of galaxies (we will
denote this as \phigobs) is the result of the random errors over the
intrinsic \gsmf\  \citep{Behroozi+2010}. Formally, we can represent the observed \phigobs\ as 
the convolution of \phig:
	\begin{equation}
		\phigobs(\ms)=\int  \mathcal{G}(\log\ms/x)\phig(x)d\log x,
		\label{conv_gsmf}
	\end{equation}
where we again assume that random errors follow a lognormal distribution
	\begin{eqnarray}
		 \mathcal{G}(\log\frac{\ms}{x})=\frac{1}{\sqrt{2\pi\sigma_*^2}}\exp\left[-\frac{1}{2\sigma_*^2}\log^2\left(\frac{\ms}{x}\right)\right].		
	\end{eqnarray}
Here $\sigma_*$ will encode all the uncertainties affecting the \gsmf. We use a slightly modified version for $\sigma_*$ as a function
of redshift from \citet{Behroozi+2010}, given by
\begin{equation}
\sigma_* = 0.1 + 0.05z,
\end{equation}
which is more consistent with what is observed in new determinations of $\ms$ at $z\sim0.1$ \citep[e.g.,][]{Mendel+2014}.

Finally, when combining Equations (\ref{phi_dir_amt}) and (\ref{conv_gsmf}) the relation between the observed
\gsmf, \phigobs, and the mass function of dark matter halos \phiDM\ is given by
\begin{equation}
\phigobs(\ms) =   \int P(\ms|\mdm)\phiDM(\mdm)d\log \mdm,
\end{equation}
where $P$ is again assumed to be a log-normal distribution with a scatter $\sigma_{\rm T}$ independent of \mdm
	\begin{eqnarray}
		P(\ms|\mdm)=\frac{1}{\sqrt{2\pi\sigma_{\rm T}^2}}\times   & &  \nonumber \\
		\exp\left[-\frac{\left(\log\ms - \langle\log\mathcal{M_*}(\mdm)\rangle\right)^2}{2\sigma_{\rm T}^2}\right],
		\label{P_ms_X}
	\end{eqnarray}
and $ \sigma_{\rm T}^2 = \sigma_{\rm h}^2 + \sigma_{\rm *}^2$.

\subsection{Star Formation Histories}
\label{SFH_model}

Once we have established the galaxy-halo connection, we can use the growth of dark matter halos, as 
measured from high resolution $N-$body cosmological simulations, to predict the \ms\ of the 
galaxy that they host as a function of $z$, and thus the corresponding galaxy star formation histories.  

Imagine that we want to derive the amount of stars that the galaxies produce between the redshifts $z+\Delta z$ and $z$. 
Now, let us assume that we know the trajectories for the main progenitors of halos at $z_0$
	\begin{equation}
		\mvir(z|z_0) = \mvir( z| M_{\rm vir,0},  z_0),
		\label{HMG}
	\end{equation}
where $M_{\rm vir,0}$ is the final mass of the halo at the redshift of observation $z_0$.
According to Equation (\ref{HMG}), this implies that the mass of the progenitor at
$z+\Delta z$ is $\mvir(z+\Delta z|M_{\rm vir,0},  z_0)$ while at $z$ it is  $\mvir( z| M_{\rm vir,0},z_0)$.
Using the redshift evolution of the \shmr\ described above we can thus infer the amount
of stellar mass the galaxy will grow between $z+\Delta z$ and $z$:
	\begin{eqnarray*}
		\Delta \ms(z | M_{\rm vir,0}, z_0) = \ms \left[ \mvir(z|M_{\rm vir,0}, z_0),z \right] -  & &  \nonumber \\
		\ms \left[  \mvir(z+\Delta z | M_{\rm vir,0},  z_0),z+\Delta z \right] .
		\label{sfr_no_mergers}
	\end{eqnarray*}

Galaxies build their masses via in-situ
star formation and/or through ex-situ process such as galaxy mergers. Following \citet{Behroozi+2013}, we apply corrections
for mergers by assuming that $\Delta\ms $
can be separated by the amount of mass that galaxies grow by mergers
and the amount of mass built up by in-situ star formation during the timescale
	$\Delta t =t(z) - t(z+\Delta z)$:
	\begin{eqnarray*}
		\Delta\ms(z | M_{\rm vir,0}, z_0)=\Delta\msmerger(z | M_{\rm vir,0}, z_0) +  & &  \nonumber \\
		\Delta\msSF(z | M_{\rm vir,0}, z_0),
	\end{eqnarray*}
or equivalently
	\begin{equation}
		\frac{\Delta\msSF}{\Delta t}=\frac{\Delta\ms}{\Delta t}\fmerger,
	\end{equation}
where we have omitted the terms in the parentheses for simplicity and we have defined
	\begin{equation}
		\fmerger= 1 - f_{\rm ex\,  situ} =1-\frac{\Delta\msmerger}{\Delta\ms}.
		\label{fmerger_def}
	\end{equation}
Here, $f_{\rm ex\,  situ}$ is
the fraction of mass acquired through galaxy mergers. 
The fraction $\fmerger$ is a function that depends on \mvir\ and $z$. The 
parametrization that we will use for $\fmerger$ will be described in Section \ref{parameters_insitu}. 
Finally, we calculate the \sfr, taking into account gas recycling, as:
	\begin{equation}
		\sfr(z | M_{\rm vir,0}, z_0) =\frac{\fmerger}{1-R} \left[\frac{\Delta \ms(z | M_{\rm vir,0}, z_0)}{t(z) - t(z+\Delta z)}\right].
		\label{SFR_cen}
	\end{equation}
Here $R$ is the fraction of mass that is returned as gaseous
material into the interstellar medium, ISM, from stellar
winds and short lived stars. In other words, $1-R$ is the fraction
of the change in stellar mass that is kept as long-lived. 
In this paper we make the instantaneous recycling approximation,
we use the fitting form for $R(t)$ given in \citet[][section 2.3]{Behroozi+2013b},
and we assume for simplicity that $t$ in $R(t)$ is the cosmic time since the big bang.

The timescale $\Delta t = t (z) - t (z+\Delta z)$ is a free parameter in our model. 
Star formation indicators are sensitive to probe different time scales. 
The most common estimators based on UV and IR probe star formation
rates on timescales around $\sim100$ Myr, while H$\alpha$ probes the SFR for times less than about
$10$ Myr \citep{Kennicutt+2012}. In this paper we will use observations homogenized to 
timescales of $100$ Myrs as described in \citet{Speagle+2014} and we compute 
SFRs averaged over a timescale of $100$ Myr. That is to say, we fixed 
$\Delta t=100$ Myrs in Equation (\ref{sfr_no_mergers}). 

Before describing our modelling of
the observed distribution of galaxy $\sfr$s, it is worth mentioning that Equation (\ref{SFR_cen})
is only valid for central galaxies. To account for the evolution of satellite galaxies,  we will 
assume, for simplicity, that they evolve as central galaxies, i.e., 
\begin{eqnarray}
\sfr(z | M_{\rm DM,0}, z_0) =   \sfr_{\rm cen} (z | M_{\rm vir,0}, z_0) = & &  \nonumber \\
 \sfr_{\rm sat} (z | M_{\rm peak,0}, z_0). 
\end{eqnarray}
The impact of satellite galaxies has been discussed in detail in previous studies 
\citep[see e.g.,][]{Yang+2009a,Neistein+2011,RDA12,Yang+2012,Wetzel+2012,Zavala+2012,RAD13}.
These studies have shown that assuming similar relations for centrals and satellites
could lead to potential inconsistencies. This problem has been already faced
in previous works \citep[e.g.,][]{Yang+2012,Zavala+2012} by modelling the evolution of satellite galaxies after infall. 
On the other hand, \citet{RDA12} argued that it is still possible to develop a self-consistent model in the light of SHAM 
by relaxing the assumption that the \shmr\ of central and satellite galaxies are identical. In that paper, it was shown
that assuming identical \shmr\ of central and satellite galaxies could lead to underestimate of the spatial clustering
of galaxies from SHAM, see also \citet{Guo+2016} for similar conclusions. A more detailed modelling of
central and satellite galaxies separately would be required to introduce more observational constraints such
as the two point correlation function and/or galaxy groups, which is beyond the scope of 
this paper (see e.g. \citet{RAD13} for details\footnote{\citet{RDA12} and \citet{RAD13} found that
the \shmr\ of central and satellite galaxies are different, an effect that can be explained due to environment effects.
This would imply, that the SFRs of centrals will be on average higher than those of satellite galaxies, specially
at low redshifts.}).

\subsection{The Distribution of Star Formation Rates}
\label{SFRs_model}

In the previous subsection we described the theoretical framework to derive the star formation histories  
in halos of different masses. In this subsection we describe how to compare the above
star formation histories to the observed SFRs from the galaxy population. 

Suppose that we want to measure the mean $\sfr$s for a volume-complete sample 
of galaxies with $\ms$ at some redshift $z$. Then, the {\it intrinsic} probability distribution of having  
galaxies with $\sfr$s in the range $\log\sfr \pm d\log\sfr /2$ is given by
	\begin{eqnarray}
		\mathcal{S}_{*, I}(\sfr|\ms,z) =  \int \mathcal{S}_{\rm h, I}(\sfr | \mdm,z)  \times  & &  \nonumber \\
		\mathcal{H}(\mdm|\ms) d\log\mdm,
		\label{Psfrms}
	\end{eqnarray}
with $\mathcal{S}_{\rm h, I}(\sfr|\ms,z)$ as the  {\it intrinsic}  probability distribution of $\sfr$ in the range $\log\sfr \pm d\log\sfr /2$ in 
(sub)halos of mass \mdm.
Note that $\mathcal{H}(\mdm|\ms)$ is not the same function as $\mathcal{H}(\ms|\mdm)$, but the functions are related to each other
through Bayes' theorem. Similarly to stellar mass, derivation of the SFRs is also subject to random errors. Moreover,
one expects that the observed distribution of SFRs $\mathcal{S}_{*, \rm obs}$, for a given stellar mass,
includes random errors affecting the intrinsic distribution $\mathcal{S}_{*, I}$. In other words, the observed distribution $ \mathcal{S}_{*, \rm obs}$
is the convolution of the distribution of random errors, denoted by  $\mathcal{F}$, with the intrinsic distribution  $\mathcal{S}_{*, \rm I}$:
$ \mathcal{S}_{*, \rm obs} = \mathcal{F} \circ \mathcal{S}_{*, \rm I} $,
with the symbol $\circ$ denoting the convolution operator.  
Note, however, that under the assumption
that random errors have a lognormal distribution with $1\sigma$ statistical fluctuation, in either directions and independent 
on both halo and stellar mass, their effect will only increase the dispersion over the 
relationship \sfr--\ms\ while the mean $\langle \log \sfr\rangle$ is the same. 
As we will show later in this subsection, our framework does not require any other moment 
over the distribution $\mathcal{S}_{*, I}$ (and therefore over $\mathcal{S}_{*, \rm obs}$) than its mean. 

For local galaxies, $z\sim0.1$, the distribution $ \mathcal{S}_{*, \rm obs}$, and consequently $\mathcal{S}_{*, I}$, 
is composed, at a very good approximation, of two lognormal
distributions. One mode is dominated by star-forming galaxies (often referred as the galaxy star-formation main sequence)
while the second mode is dominated by quescient galaxies \citep[e.g.,][]{Salim+2007}. 
This bimodality has also been observed to high redshifts \citep[e.g.,][]{Noeske+2007, Karim+2011}. 
In this paper, however, we are more 
interested in the mean of the above distribution $\langle \log\sfr(\ms,z)\rangle$ rather than
in the full distribution. 

Applying Bayes' theorem to $\mathcal{H}(\mdm|\ms)$ and substituting this in 
Equation (\ref{Psfrms}) we get
	\begin{eqnarray}
		 \mathcal{S}_{*, \rm I}(\sfr|\ms,z)  = \frac{1}{\phig(\ms,z)} \int \mathcal{S}_{\rm h}(\sfr | \mdm,z)  \times  & &  \nonumber \\
		\mathcal{H}(\ms|\mdm,z)\times \phiDM(\mdm,z)d\log\mdm, 
		\label{Psfrms_I}
	\end{eqnarray}
where $\phig(\ms,z)$ is given by Equation (\ref{phi_dir_amt}). By noting that  
	\begin{eqnarray*}
		\langle \log\sfr(\mdm, z)\rangle = \int  \mathcal{S}_{\rm h}(\sfr | \mdm,z) \log\sfr  d\log\sfr,
	\end{eqnarray*}
and using the definition
	\begin{equation}
		\langle \log\sfr(\ms,z)\rangle = \int  \mathcal{S}_{*, \rm I}(\sfr|\ms,z) \log\sfr d\log\sfr,
	\end{equation}
we can conclude that
	\begin{eqnarray}
		\langle \log\sfr(\ms,z)\rangle =\phi_{g_{\rm I}}^{-1} (\ms,z) \int \langle\log\sfr(\mdm, z)\rangle \times  & &  \nonumber \\
		\mathcal{H}(\ms|\mdm,z)\times  \phiDM(\mdm,z)d\log\mdm.
		\label{logsfr_ms_z}
	\end{eqnarray}
We use the above equation to derive SFRs in our model and compare
to observations. Note that this equation does not require any other moment than the mean
of the distribution $\mathcal{S}_{*, \rm I}(\sfr|\ms,z)$. The same argument applies
when using $\mathcal{S}_{*, \rm obs}(\sfr|\ms,z)$ instead.

\subsection{The Cosmic Star Formation Rate}
\label{csfr_model_section}

In this subsection we describe how the Cosmic Star Formation Rate (CSFR) is
calculated in the framework described above. Formally, the CSFR is given by
\begin{equation}
\dot{\rho}_{\rm obs}(z) = \int \sfr\times \phi_{\rm SFR, obs}(\sfr,z)\times d\log\sfr,
\label{csfr_obs}
\end{equation}
with $ \phi_{\rm SFR, obs}(\sfr)$ as the {\it observed} comoving number density of galaxies
with observed $\sfr$s between $\log\sfr\pm d\log\sfr/2$. 

In order to derive a model for the CSFR, we begin by noting that the relation
between the SFRs, denoted by $\phi_{\rm SFR, obs}$, and the observed \gsmf\ is given by
	\begin{eqnarray*}
		\phi_{\rm SFR, obs}(\sfr) = \int  \mathcal{S}_{*, \rm obs}(\sfr|\ms)\phigobs(\ms)d\log\ms,
	\end{eqnarray*}
where we omit the dependence on $z$ for simplicity.  
The next step is to substitute Equation (\ref{Psfrms_I}) to obtain:
	\begin{eqnarray*}
		\phi_{\rm SFR, obs}(\sfr) = & &  \nonumber \\
			\int  \mathcal{S}_{\rm h,obs}(\sfr|\mdm)
			\Theta(\mdm) \phiDM(\mdm)d\log \mdm, & &
	\end{eqnarray*}
with 
	\begin{eqnarray}
		\Theta(\mdm) = \int \frac{\phigobs(\ms)}{\phig(\ms)}\times \mathcal{H}(\ms|\mdm) d\log\ms,
	\end{eqnarray}
and we define $\mathcal{S}_{\rm h,obs} = \mathcal{F}  \circ \mathcal{S}_{\rm h,I}$. Recall that the probability
distribution function $\mathcal{S}_{\rm h,I}$ and the function $\mathcal{F}$  were described in Section \ref{SFRs_model}.
Before moving further into the
description of our model, we briefly discuss the function $\Theta(\mdm)$. Clearly, the function 
$\Theta$ depends strongly on the ratio $\phigobs / \phig$, and in the case that $\phigobs / \phig\sim1$ then
$\Theta\sim1$. Actually, the effect of the ratio on the function $\Theta$ 
is expected to be small for low-mass halos and larger at the high-mass end. This is because
the effect of the convolution in Equation (\ref{conv_gsmf}) depends on the logarithmic slope
of $\phig$ \citep{Cattaneo+2008,Behroozi+2010,Wetzel+2010}. Figure \ref{theta_model} in Appendix \ref{A1} 
illustrates the redshift evolution
of the function $\Theta(\mdm)$ for our best fitting model (see Section \ref{mcmc} below). Note that at high
redshifts this correction becomes very important and it should be taken into account in order 
to have a consistent model and a more accurate calculation of the CSFR. 

When calculating the CSFR, the details behind the 
distribution function $\mathcal{S}_{\rm h,obs}(\sfr|\mdm)$ are relevant. We assume that this distribution
is lognormal with $\sigma_{\rm h,obs} = 0.3$ dex and independent of mass. This assumption  
may sound crude as the observed distribution of $\sfr$s, $\mathcal{S}_{*, \rm obs}$, is actually
bimodal, something that it is also expected for dark matter halos. Note, however, that in this paper we are more interested on the average evolution
of galaxies rather than modelling different populations and therefore a lognormal distribution is an adequate 
approximation for our purposes. Once we have specified the shape of $\mathcal{S}_{\rm h,obs}$ we
can analytically calculate the following integral:
	\begin{equation}
		\langle\sfr(\mdm) \rangle =  \int  \sfr\times \mathcal{S}_{\rm h,obs}(\sfr|\mdm) \times d\log\sfr,
	\end{equation}
which is given by
	\begin{equation}
		\log\langle\sfr(\mdm) \rangle = \langle\log\sfr(\mdm) \rangle + \frac{\sigma_{\rm h,obs}^2}{2}\ln10.
	\end{equation}
Finally, the CSFR can be calculated as:
	\begin{equation}
		\dot{\rho}_{\rm obs}(z) = \int \langle\sfr(\mdm) \rangle \Theta(\mdm) \phiDM(\mdm)d\log \mdm,
		\label{csfr_obs_mod}
	\end{equation}
which is the equation we utilize to calculate the CSFRs and that we will compare with observation. Note that
the value of the scatter $\sigma_{\rm h,obs}$ is relevant only for the calculation of the CSFR but not for the modelling
of $\langle \log\sfr(\ms)\rangle$, see Eq. (\ref{logsfr_ms_z}).

Finally, Appendix \ref{A1} discusses the impact of assuming $\Theta(\mdm)=1$ and that the probability distribution function $\mathcal{S}_{\rm h,obs}$ is a
Dirac$-\delta$ distribution function. In short, we find that the combined effect will result in an underestimation of the CSFR of around 
$\sim30\%$ at low-$z$, but by an order of magnitude at high redshifts. Therefore, {\it taking into account all the above factors
is critical when constraining our model}. We note that most previous authors have failed to recognize the above effects.

\section{Model Ingredients}
\label{Model_Ingredients}

Here we describe the model ingredients, namely, 
the parametric redshift dependence of the \shmr\ and 
the fraction of in-situ and ex-situ stellar mass growth.

\subsection{Dark Matter Halos}

In Appendix \ref{App_DMH}, we present the theoretical ingredients to fully charachterize
the distribution of dark matter halos, namely, the halo and subhalo mass function and halo mass assembly. Briefly, 
we use the updated parameters from \citet{Rodriguez-Puebla+2016} from a set of high resolution $N-$body simulations 
for the \citet{Tinker+2008} halo mass function and the \citet{Behroozi+2013} fitting model to the subhalo population. We also
use the fitting parameters from \citet{Rodriguez-Puebla+2016} for the median halo assembly histories from $N-$body simulations. 

\subsection{Parameterization of the \shmr}

 In order to describe the mean \shmr, we adopt the parametrization 
 proposed in \citet{Behroozi+2013},
\begin{eqnarray}
	 \langle \log \mathcal{M_*} \rangle=\log(\epsilon \times M_{0}) + g(x) - g(0),
 		 \label{msmh}
\end{eqnarray}
where
\begin{equation}
	 g(x) = \delta \frac{(\log(1+e^x))^\gamma}{1+\exp( 10^{-x} ) }-\log(10^{-\alpha x}+1).
		 \label{msmh1}
\end{equation}
and $x=\log(\mvir/M_{0})$.
Previous studies have used simpler functions
with fewer parameters (e.g., \citealp{Yang+2008,Moster+2013});
however, as shown in \citet{Behroozi+2013}, the function as given by Eq. (\ref{msmh1}) is
 necessary in order to map accurately the halo mass function into the observed \gsmf s, which are more 
 complex than a single Schechter function. 
 Recall that at $z\sim0.1$ we
 are using a mass function that is steeper at low masses after 
 correcting for surface brightness incompleteness. 
 
 We assume that the above parameters change with redshift $z$ as
 follows:
	 \begin{eqnarray}
		\log(\epsilon(z)) = \epsilon_0 + \mathcal{P}(\epsilon_1,\epsilon_2,z)\times \mathcal{Q}(z) +  
		\mathcal{P}(\epsilon_3,0,z),
		\label{epsilonz}
	\end{eqnarray}
	 \begin{eqnarray}
		\log(M_{0}(z)) = M_{0,0} + \mathcal{P}(M_{0,1},M_{0,2},z)\times \mathcal{Q}(z),  
	\end{eqnarray}
	 \begin{eqnarray}
		\alpha (z) = \alpha +  \mathcal{P}(\alpha_1,\alpha_2,z)\times \mathcal{Q}(z), 
	\end{eqnarray}
	 \begin{eqnarray}
		\delta(z) = \delta_0 + \mathcal{P}(\delta_1,\delta_2,z)\times \mathcal{Q}(z),  
	\end{eqnarray}
	 \begin{eqnarray}
		\gamma(z) = \gamma_0 + \mathcal{P}(\gamma_1,0,z)\times \mathcal{Q}(z).  
		\label{gammaz}
	\end{eqnarray} 
Here, the functions $\mathcal{P}(x,y,z)$ and $\mathcal{Q}(z)$ are:
	\begin{equation}
		\mathcal{P}(x,y,z) =y\times z - \frac{x\times z}{1+z},
	\end{equation}
and 
	\begin{equation}
		\mathcal{Q}(z)=e^{-4/(1+z)^2}.
	\end{equation}
Similar parameterizations were employed in \citet{Behroozi+2013}.	
	
\subsection{Parameterization of the Fraction of In Situ Stellar Mass Growth}
\label{parameters_insitu}
 
We assume that the function $\fmerger(\mvir,z)$, defined in Equation (\ref{fmerger_def}), which describes the 
fraction of mass assembled via in situ star formation in a period of 100 Myrs at redshift 
$z$, is given by:
	\begin{equation}
		\fmerger(z)=\frac{1}{1+x^{\beta}(z)},
		\label{fmerge}
	\end{equation}
with $x(z)=\mvir/M_{\rm in\, situ}(z)$, 
	\begin{equation}
		\log(M_{\rm in\, situ} (z))=  M_{\rm in\, situ,0} +\mathcal{P}( M_{\rm in\, situ,1},0,z), 
	\end{equation}
 and 
	\begin{equation}
		\beta(z)=\beta + \log(1+z).
	\end{equation}
	
Recall that the complement to $\fmerger$ is $f_{\rm ex\,  situ}$, the fraction of mass
acquired through galaxy mergers. At low redshifts, we expect that at low masses the function described by Equation (\ref{fmerge}) 
asymptotes to $\fmerger(z)\sim1$ given that observations show that 
low-mass galaxies grow mostly by in-situ SFR, while at very 
large masses $\fmerger(z) < 1$ consistent with the fact that
big elliptical galaxies build part of their mass by galaxy mergers.

\section{Observational input: Compilation and Homogenization}
\label{Obs_inputs}

This section describes the observational data that we utilize to constrain our semi-empirical model, namely,
the \gsmf, the SFRs and the CSFR. The main goal of this section is to present a compilation from the literature and 
to calibrate all the observations to the same basis in order to minimize potential systematical effects that can bias our conclusions. 

In our compilation, we do not include observations from dusty submillimeter galaxies, which are the most extreme star formers in the Universe 
\citep[for a recent review, see][]{Casey+2014}.  Submillimeter galaxies are galaxies that emit in the FIR and with high luminosities implying 
SFRs$\geq 300\msun$yr$^{-1}$ at $z\sim1-4$, that is, a factor of $\sim5$ compared to normal star-forming galaxies \citep[see e.g.,][]{Casey+2014}. This means that
the population of submillimeter galaxies is a class of star formers that it is very particular and rare. We do not actually need to consider
individual populations such as this one, given that we assume that the full population of galaxies is described by 
the probability distribution function of SFRs (Section \ref{SFRs_model}) and thus submilimeter 
galaxies will automatically be included as the extreme values of the distribution.  

\subsection{The Galaxy Stellar Mass Function up to $\redsfhit \sim 10$}
\label{MF}

In this Section, we use a compilation of various studies from the literature  
to characterize the redshift evolution of the \gsmf\ from $z\sim0.1$ to $z\sim10$. 
Table \ref{T1}  lists the references we utilize for this section. The reader interested in the details 
is referred to the original sources. 

\begin{table*}
	\caption{Observational data on the galaxy stellar mass function}
	\begin{center}
		\begin{tabular}{l c c c c c c c c c c c c}
			\hline
			\hline
			Author &  Redshift$^a$ & $\Omega$ [deg$^2$]  & Corrections\\
			\hline
			\hline
			\citet{Bell+2003} & $z\sim0.1$  & 462  & I+SP+C\\ 
			\citet{Yang+2009b} & $z\sim0.1$ & 4681  & I+SP+C\\
			\citet{Li+2009} & $z\sim0.1$ & 6437  & I+P+C\\
			\citet{Bernardi+2010} & $z\sim0.1$ & 4681  &  I+SP+C \\
			\citet{Baldry+2012} & $0<z<0.06$ & 143  & C \\
			\citet{Bernardi+2013} & $z\sim0.1$ & 4681 & I+SP+C\\
			Rodriguez-Puebla et al. in prep & $z\sim0.1$ & 7748  & S\\
			\citet{Drory+2009} & $0<z<1$ & 1.73  & SP+C\\
			\citet{Moustakas+2013} & $0<z<1$ & 9  & SP+D+C\\
			\citet{Perez-Gonzalez+2008} & $0.2<z<2.5$ & 0.184  & I+SP+D+C\\
			\citet{Tomczak+2014} & $0.2<z<3$ & 0.0878  & C \\
			\citet{Ilbert+2013} & $0.2<z<4$ & 2  & C \\
			\citet{Muzzin+2013} & $0.2<z<4$ & 1.62  & I+C \\
			\citet{Santini+2012} & $0.6<z<4.5$ & 0.0319  & I+C \\
			\citet{Mortlock+2011} & $1<z<3.5$ & 0.0125 & I+C\\
			\citet{Marchesini+2009} & $1.3<z<4$ & 0.142 & I+C\\
			\citet{Stark+2009} & $z\sim6$ & 0.089 & I \\
			\citet{Lee+2012} & $3<z<7$ & 0.089 & I+SP+C \\
			\citet{Gonzalez+2011} & $4<z<7$ & 0.0778 & I+C \\
			\citet{Duncan+2014} & $4<z<7$ & 0.0778 & C \\
			\citet{Song+2015} & $4<z<8$ & 0.0778 & I \\
			This paper, Appendix D & $4<z<10$ & 0.0778 & --\\
			\hline
		\end{tabular}
		\end{center}
	\label{T1}
	{\bf Notes:} $^a$Indicates the redshifts used in this paper.
	I=IMF; P= photometry corrections; S=Surface Brightness correction;
	D=Dust model; NE= Nebular Emissions: SP = SPS Model: C = Cosmology. 
\end{table*}

The various $\gsmf$s used in this paper were obtained based on different observational campaigns and 
techniques. In order to directly compare the various  published results of the $\gsmf$s, 
and therefore to obtain a consistent evolution of the
\gsmf\ from $z\sim0.1$ to $z\sim10$, 
we apply some calibrations to 
observations. Among the
most important, these calibrations include the initial mass function (IMF), Stellar Population Synthesis models (SPS),
photometry corrections, cosmology, and variations between different search
areas in surveys. Note, however, that in this paper we do not carry out an 
exhaustive analysis for these calibrations. Instead, we apply
standard corrections according to the literature. Additionally, we derive 
the \gsmf\ from $z=4$ to $z=10$ based on the observed
evolution of UV luminosity functions and stellar mass-UV luminosity ratios. Next, we briefly
describe the corrections that we include in our set of $\gsmf$s.

\subsubsection{Systematic Effects on the \gsmf}

One of the most important sources of calibration is the IMF since it determines the 
overall normalization of the stellar mass-to-light ratios. In this paper we will assume that
the IMF is universal, i.e., it is independent of time, galaxy type, and environment. Note, however,
that there is not a consensus on this, since there exist arguments in favour \citep[e.g.][]{Bastian+2010} and against \citep[e.g.][]{Conroy+2013b} 
the universality of the IMF. For the case of a universal IMF, the choice of one or another IMF is a source
of a systematic change in stellar mass, i.e., between two universal IMFs there is a constant offset. The most popular choices for the IMFs
are \citet{Salpeter1955}, diet Salpeter, \citet{Chabrier2003}, and \citet{Kroupa2001}. 
For determining the redshift evolution of the \gsmf, we correct observations to the IMF of 
\citet{Chabrier2003}. We apply the following offsets:
\begin{equation}
M_{\rm C03} = M_{\rm S55} - 0.25 = M_{\rm DS} - 0.1 =  M_{\rm K01} - 0.05,
\end{equation}
The  subscripts refer to  \citet{Salpeter1955}, diet Salpeter, and \citet{Kroupa2001}
respectively. Note that the above offsets are in dex with $M$ referring to the log of the stellar mass derived
using their corresponding IMF. 
We use Table 2 in \citet{Bernardi+2010} for these corrections. 

The second most important source of calibration is the choice of SPS model due
to uncertain stellar evolutionary phases. The treatment of the thermally 
pulsating asymptotic horizontal branch stars (TP-AGB) has received, particularly, much 
attention in recent years. Including the TP-AGB can lower the stellar mass of a single stellar population by $\grtsim 50\%$ for an age of $\sim 1$ Gyr (Chabrier IMF),  but as the population gets older, the mass difference between the models that include or not the TP-AGB tends to disappear \citep{BC07}.  Nonetheless, the SF histories of galaxies are much more complex than the one of a single stellar population.  \citet[][see also, \citealp{Maraston+2005,Maraston+2006,Muzzin+2009, Marchesini+2009}]{Muzzin+2013} have compared the stellar masses of galaxies at different redshifts by using  the SPS models of \citet{Maraston+2005}, which include TP-AGB, with those of  \citet{BC03}, which do not include the TP-AGB. The authors show that there is a systematic effect on the stellar masses of galaxies between the results from  \citet{Maraston+2005}  and  \citet{BC03} at all redshifts, though the effect may change with redshift depending on the galaxy populations.  Here we apply only  conservative corrections.

In this paper we calibrate all observations to 
a  \citet{BC03} SPS model by applying the following systematic offsets:
\begin{eqnarray}
M_{\rm BC03} = M_{\rm BC07} + 0.13 =   M_{\rm P,0.1}  - 0.05 =  & &  \nonumber \\
   =M_{\rm P,z} +0.03 =  M_{\rm M05}  + 0.2=  M_{\rm FSPS} - 0.05.
\end{eqnarray}
The  subscript BC07 refers to  the  \citet{BC07} SPS model, P,0.1 for PEGASE SPS model \citep{Fioc_RoccaVolmerange1997} at $z\sim0.1$
while P,z for PEGASE SPS model at high $z$. The  subscript M05 refers to the \citet{Maraston+2005} SPS model  and FSPS to 
the \citet{Conroy+2009a} SPS model. For the offset between BC03 and BC07 as well
as for the offset between BC03 and M05 we used the analysis from \citet{Muzzin+2009}. For the PEGASE SPS model at $z\sim0.1$ we use the offset noted by \citet{BellJong2001}
in their stellar mass-to-light ratios. For the PEGASE SPS model at high $z$ we use the 0.03 dex offset reported in \citet{Perez-Gonzalez+2008}. Note that
the latter offset is only applied to the $\gsmf$s reported in \citet{Perez-Gonzalez+2008}. 
For the offset between BC03 and FSPS SPS models at $z\sim0.1$ we use the offset estimated in \citet{Moustakas+2013}
based on fitting the SDSS and GALAXES photometry using their {\textsc{iSEDfit}} code for galaxies at $z\sim0.1$, see their figure 19 and 
Figure 6 in \citet{Conroy2013}. 

The choice of the dust attenuation model, star formation history, and metallicity are 
other sources of systematics when deriving stellar masses. 
For the dust attenuation model we calibrate all observations to a \citet{Calzetti+2000} law. Based
on the analysis carried out in \citet{Perez-Gonzalez+2008}, these authors found that  
a \citet{Charlot_Fall2000} dust attenuation model gives, on average, galaxy stellar masses
that are 0.1 dex larger compared to  a \citet{Calzetti+2000} law. In this paper we use this offset. 
Additionally, \citet{Muzzin+2009} found that the median offsets for the Small and Large Magellanic Clouds
and Milky-Way  dust attenuation models are respectively -0.061, +0.036 and +0.05 dex. Finally, we
assume that the choice of star formation histories and metallicity do not have any notable effect on
the \gsmf\ beyond the random uncertainties as demostrated in \citet{Muzzin+2013}. 
Random uncertainties were discussed in Section \ref{mod_framework}. 

To account for differences in cosmologies, 
To account for differences in cosmologies, we scale the GSMFs to our Planck 
?CDM cosmology with $\Omega_\Lambda=0.693,\Omega_{\rm M}=0.307, h=0.678$, and $h=0.678$ using the relation
we scale the $\gsmf$s as:
\begin{equation}
	\phi_{\rm g, us} = \phi_{\rm g, lit} \times \frac{V_{\rm lit}}{V_{\rm us}}, 
\end{equation}
where $V$ is the comoving volume observed for each galaxy redshift survey
\begin{equation}
	V=\int_{\Omega}\int_{z_i}^{z_f}\frac{d^2V_c}{dzd\Omega}dzd\Omega.
\end{equation}
Here, $z_f$ and $z_i$ are the maximum and minimum redshift where each \gsmf\ has been observed,
$\Omega$ is the solid angle of the survey, see Table \ref{T1}, and $V$ is the comoving volume in a  
$\Lambda$CDM universe \citep{Hogg1999}. 
As for stellar masses we correct cosmology by simply $M_{*,\rm us} = M_{*,\rm lit} \times h^2_{\rm lit} / h^2_{\rm us}$, 
where $h$ is the Hubble parameter. Note, however, that
the impact of accounting for different cosmologies is very small. 

Finally, we do not account for any systematic effect due to cosmic variance.

\subsubsection{Other Effects on the \gsmf}

Here we discuss more specific calibrations that are known to affect some observations
of the \gsmf.  

We do not include aperture corrections. Previous studies 
\citep{Bernardi+2010, Bernardi+2013, Bernardi+2016} have found that 
the measurements of the light profiles based on the standard  
SDSS pipeline photometry could be underestimated due to sky subtraction issues.
This could result in a underestimation of the abundance of massive galaxies 
up to a factor of five. While new algorithms have been developed for obtaining more precise 
measurements of the sky subtraction and thus to improve the photometry
\citep{Blanton+2011,Simard+2011,Meert+2015} there is not yet a consensus. For this paper,
we decided to ignore this correction that we may study in more detail in future works. 
Nevertheless, we apply photometric corrections to the \gsmf\ reported in 
\citet{Li+2009}. These authors used stellar masses estimations based on 
the SDSS $r-$band Petrosian magnitudes. It is well known
that using Petrosian magnitudes could result in a underestimation of the total light 
by an amount that could depend on the surface brightness profile of the galaxy and
thus results in the underestimation of the total stellar mass. This will result in   
an artificial shift of the \gsmf\ towards lower masses. In order to account for this shift for the \citet{Li+2009} \gsmf, 
we apply a constant correction of 0.04 dex to all masses. 
As reported by \citet{Guo+2010}, this correction gives an accurate representation of 
the \gsmf\ when the total light is considered, instead. 

At $z\sim0.1$ we use the \gsmf\ derived in Rodriguez-Puebla et al. (in prep.) that has been corrected for the fraction of missing 
galaxies due to surface brightness limits by combining the SDSS NYU-VAGC low-redshift sample and 
the SDSS DR7 based on the methodology described in \citet{Blanton+2005}. Following \citet{Baldry+2012},
we correct the \gsmf\ for the distances based on \citet{Tonry+2000}. We found that including missing galaxies due to 
surface brightness incompleteness could increase the number of galaxies up to a factor of $\sim2-3$ at the lowest 
masses, see Figure \ref{f3}, and therefore have a direct impact on the \shmr. 
As for the new discoveries of the population of ultra-diffuse galaxies
\citep[UDGs][]{vanDokkum+2015c}, their impact on the \gsmf\ is not yet clear, nor is how to incorporate corrections due to surface brightness incompleteness. Note, however, that most of the UDGs have been discovered in massive clusters 
\citep[particularly in the Coma Cluster][]{vanDokkum+2015c,Yagi+2016}, with very few 
examples in the field \citep[e.g.,][]{Martinez-Delgado+2016}. 
If the UDG population is only common in Coma-like clusters, then we could conclude that the impact of UDGs on the 
\gsmf\ would be minimum given the low frequency of such massive objects. 

 Finally, unlike for the local SDSS survey, at higher redshifts we do not expect significant surface brightness incompleteness 
at the low-mass end because (1) most of the high-redshift surveys employed by us are deep 
(e.g., COSMOS, UltraVISTA, CANDELS; these surveys have been shown to be complete in surface brightness 
down to their magnitude (mass) limits); and (2) at higher redshifts the galaxies tend to be more compact in the UV-optical 
bands than the local ones \citep[see e.g.][]{vanderWel+2014}, in such a way that they are expected to have higher surface brightnesses. 

\begin{table*}
	\caption{Observational data on the star formation rates}
	\begin{center}
		\begin{tabular}{l c c c c c c c c c c c c}
			\hline
			\hline
			Author &  Redshift$^a$ &  SFR Estimator  & Corrections & Type\\
			\hline
			\hline
			\citet{Chen+2009} & $z\sim0.1$  & H$_{\alpha}$/H$_{\beta}$ & S &  All\\ 
			\citet{Salim+2007} & $z\sim0.1$  & UV SED & S &  All\\ 
			\citet{Noeske+2007} & $0.2<z<1.1$  & UV+IR & S &  All\\ 
			\citet{Karim+2011} & $0.2<z<3$  & 1.4 GHz & I+S+E &  All\\ 
			\citet{Dunne+2009} & $0.45<z<2$  & 1.4 GHz & I+S+E &  All\\ 
			\citet{Kajisawa+2010} & $0.5<z<3.5$  & UV+IR & I &  All\\ 
			\citet{Whitaker+2014} & $0.5<z<3$  & UV+IR & I+S &  All\\ 
			\citet{Sobral+2014} & $z\sim2.23$  & H$_{\alpha}$  & I+S+SP &  SF\\ 
			\citet{Reddy+2012} & $2.3<z<3.7$  & UV+IR & I+S+SP &  SF\\ 
			\citet{Magdis+2010} & $z\sim3$  & FUV  & I+S+SP &  SF\\ 
			\citet{Lee+2011} & $3.3<z<4.3$  & FUV  & I+SP &  SF\\ 
			\citet{Lee+2012} & $3.9<z<5$  & FUV  & I+SP &  SF\\ 
			\citet{Gonzalez+2012} & $4<z<6$  & UV+IR & I+NE &  SF\\ 
			\citet{Salmon+2015} & $4<z<6$  &  UV SED & I+NE+E &  SF\\ 
			\citet{Bouwens+2011} & $4<z<7.2$  & FUV  &  I+S &  SF\\ 
			\citet{Duncan+2014} & $4<z<7$  & UV SED  & I+NE &  SF\\ 
			\citet{Shim+2011} & $z\sim4.4$  & H$_{\alpha}$ & I+S+SP &  SF\\ 
			\citet{Steinhardt+2014} & $z\sim5$  & UV SED  &  I+S &  SF\\ 
			\citet{Gonzalez+2010} & $z=7.2$  & UV+IR & I+NE &  SF\\ 
			This paper, Appendix D & $4<z<8$ & FUV  &  I+E+NE & SF \\
			\hline
		\end{tabular}
		\end{center}
	\label{T2}
	{\bf Notes} $^a$Indicates the redshift used in this paper.
	I=IMF; S=Star formation calibration; E=Extinction;
	NE= Nebular Emissions; SP=SPS Model
\end{table*}

\subsubsection{The Evolution of the local \gsmf}

Appendix \ref{UV_LFs} describes our inference of the \gsmf\ from $z\sim4$ to $z\sim10$. In short, 
we use several UV LFs reported in the literature together with 
stellar mass-UV luminosity relations from \citet{Duncan+2014, Song+2015,Dayal+2014}
to derive the evolution of the \gsmf\ from $z\sim4$ to $z\sim10$. We assume  
a survey area of 0.0778 deg$^2$s as in the CANDELS survey \citep[e.g.,][]{Song+2015}. 

\begin{figure}
	\vspace*{-110pt}
	\hspace*{-20pt}
	\includegraphics[height=4.7in,width=3.5in]{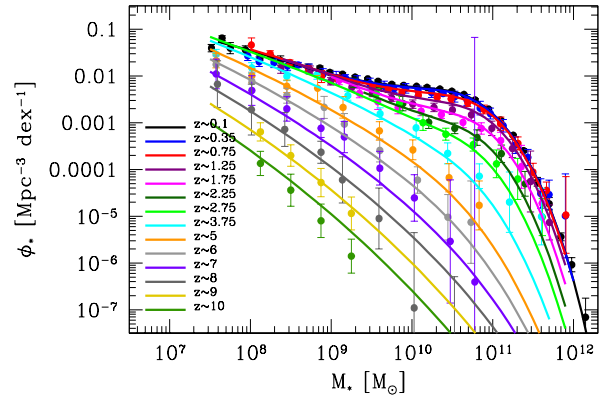}
	\vspace*{-40pt}
		\caption{
		Redshift evolution from $z\sim0.1$ to $z\sim10$ of the galaxy stellar mass function (\gsmf) derived by
		using 20 observational samples from the literature and represented with the filled circles with error bars. 
		The various $\gsmf$s have been homogenized and corrected for potential systematics that could affect 
		our results, see the text for details. Solid lines are the best fit model from a set of $3\times 10^5$ MCMC 
		trials, see Section \ref{mcmc}. 
		These fits take into account uncertainties affecting the \gsmf\ as discussed in the text. 
		Note that at lower redshifts ($z\lesssim3$) galaxies tend to pile up at $\ms\sim3\times10^{10}\msun$ due to
		the increase in the number of massive quenched galaxies at lower redshifts. 
 	}
	\label{f6}
\end{figure}

Figure \ref{f6} shows the evolution of the \gsmf\ from $z\sim0.1$ to $z\sim 10$. 
The filled circles show the mean of the observed $\gsmf$s that we use through this paper in
various redshift bins, while the errors bars represent the propagation of the individual errors from the \gsmf.
Alternatively, we also compute standard deviations from the set of \gsmf. We calculated the mean 
and the standard deviation of the observed $\gsmf$s by using the bootstrapping approach of resampling 
with replacement. We use the bootstrapping approach since it allows us to empirically derive the distribution of current 
observations of the $\gsmf$s and thus robustly infer the mean evolution of the $\gsmf$s. Methodologically, we start by choosing 
various intervals in redshift as indicated in the labels in Figure \ref{f6}. For each redshift bin, we create  $30,000$ bootstrap 
samples based on the observed distribution of all the $\gsmf$s for that redshift bin, $\phi_{g_{\rm obs}}(\ms,z)$, and
then compute the median and its corresponding standard deviation from the
distribution for a given stellar mass interval.  

A few features of the mean evolution of the observed \gsmf\ are worth mentioning at this point. 
At high redshifts the \gsmf\ is described by a Schechter function, as has been pointed out 
in previous papers \citep[see e.g.,][]{Grazian+2015, Song+2015,Duncan+2014}. At high redshifts, the faint-end slope 
becomes steeper \citep[see e.g.,][]{Song+2015,Duncan+2014}. As the galaxy population evolves, massive
galaxies tend to pile up around $\ms\sim3\times10^{10}\msun$ due to
the increasing number of massive quenched galaxies at lower redshifts \citep[see e.g.,][]{Bundy+2006,Faber+2007,Peng+2010,Pozzetti+2010,Muzzin+2013}.
These represent a second component that is well described by a
Schechter function, and thus the resulting \gsmf\ at low redshifts is better described by a double Schechter function. 

\subsection{Star Formation Rates}

In this paper, we use a compilation of 19 studies from the literature for the observed SFRs 
as a function of stellar mass at different redshifts. Table \ref{T2} lists the references that we utilize. 

\begin{figure}
\vspace*{-180pt}
\includegraphics[height=6.4in,width=5.2in]{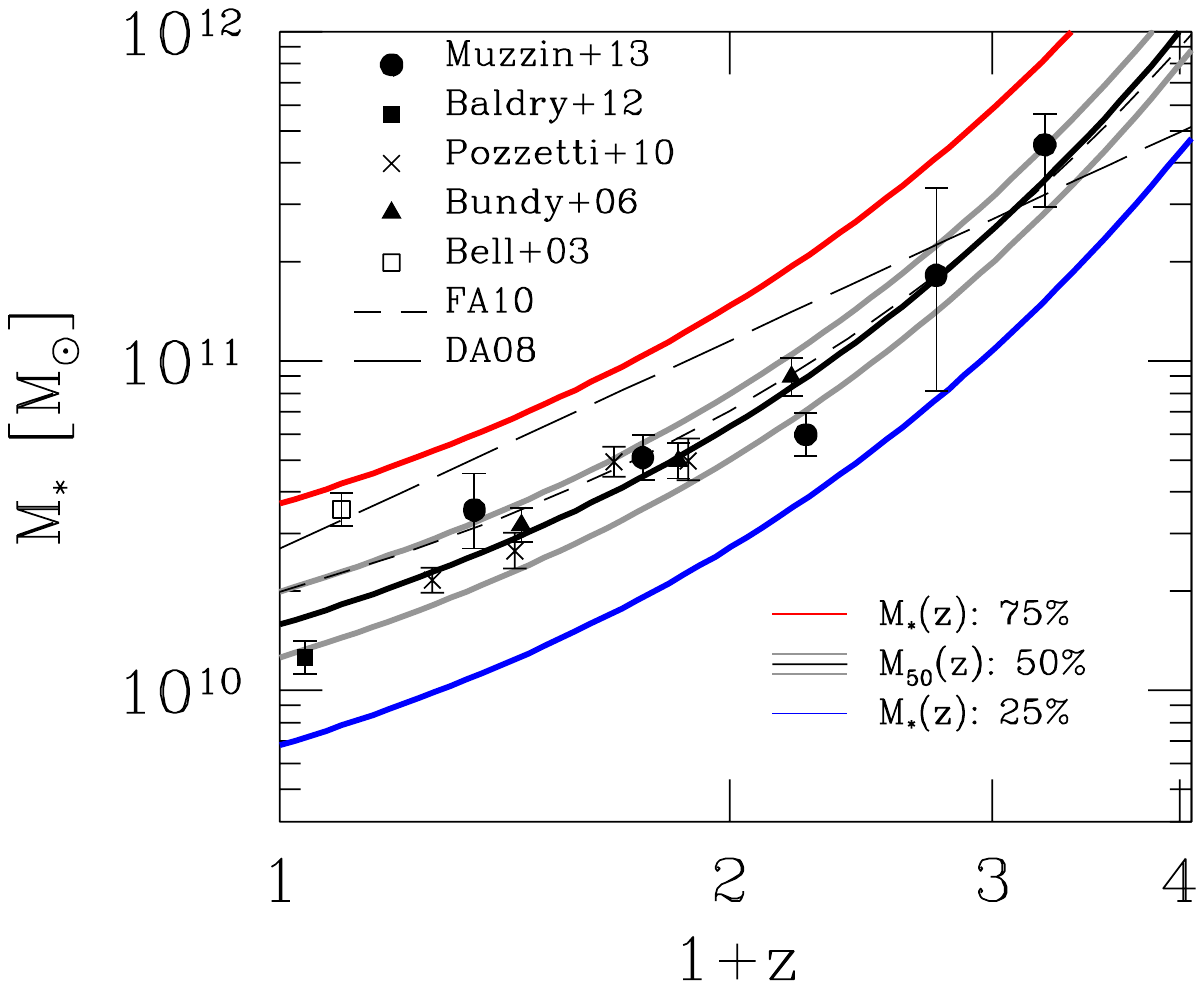}
\vspace*{-100pt}
\caption{ The stellar mass $M_{50} (z)$ at which the fractions
of blue star-forming  and red quenched galaxies are both $50\%$. 
The open square with error bars shows the transition mass for local galaxies as derived in 
\citet{Bell+2003} based on the SDSS DR2, while
the filled triangles show the transition mass derived in \citet{Bundy+2006} based on the DEEP2 survey.
 \citet{DroryAlvarez2008} based on
the FORS Deep Field survey is indicated with the long dashed line; observations from
\citet{Pozzetti+2010}  based on the COSMOS survey  are indicated with the x symbols; observations from
\citet{Baldry+2012} based on the GAMA survey are shown with a filled square; and observations from
\citet{Muzzin+2013} based on the COSMOS/ULTRAVISTA survey are shown as filled circles. 
The empirical results based on abundance matching by 
\citet{Firmani+2010a} are shown with the short dashed lines. 
The solid black line shows the relation $\log (M_{50} (z) / \msun) = 10.2 + 0.6 z$
employed in this paper, which is consistent with most of the above studies. 
The gray solid lines show the results when shifting $(M_{50} (z) / \msun) $ 
0.1 dex higher and lower.  The red (blue) curves show the stellar mass vs. $z$ where 75\% (25\%) of the galaxies are 
quenched.
 }
\label{m_trans}
\end{figure}

As for the $\gsmf$s, in order to directly compare the different SFR samples we applied some calibrations. To do so, we follow 
 \citet{Speagle+2014} who used a compilation to study star formation 
from $z\sim0$ to $z\sim6$ by correcting for different assumptions
regarding the IMF, SFR indicators, SPS models, dust extinction, emission lines and 
cosmology. The reader is referred to that paper for details on their 
calibrations. In Table \ref{T2} we indicate the specific calibrations applied to the 
data. Note that in order to constrain our model we use observations 
of the SFRs for {\it all} galaxies. Complete samples, however, for 
all galaxies are only available at $z<3$. Therefore,
here we decided to include SFRs samples from 
star-forming galaxies, especially at high $z>3$. Using only star-forming galaxies at high
redshift is not a big source of uncertainty since most of the galaxies at $z>3$ are actually 
star forming, see e.g. Figure \ref{m_trans}.  The last column of
Table \ref{T2} indicates the type of the data, namely,
if the sample is for all galaxies or for star-forming galaxies, and the redshift range.

In addition to the compiled sample for $z>3$, 
here we calculate average SFRs using again the UV LFs described in Appendix \ref{UV_LFs}.
We begin by correcting the UV rest-frame absolute magnitudes for extinction using the 
\citet{Meurer+1999} average relation 
\begin{equation}
\langle A_{\rm UV} \rangle= 4.43 + 1.99 \langle \beta \rangle,
\end{equation}
where $\langle \beta \rangle$ is the average slope of the observed UV 
continuum. We use the following relationship independent 
of redshift: $\langle \beta \rangle = - 0.11 \times (\muv + 19.5)  - 2$, which is consistent
with previous determination of the $\beta$ slope \citep[see e.g.,][]{Bouwens+2014}.
Then we calculate UV SFRs using the \citet{Kennicutt+1998} relationship
\begin{equation}
\frac{\sfr}{\msun\, {\rm yr}^{-1}}(L_{\rm UV}) = \frac{L_{\rm UV} / {\rm erg\, s^{-1} \, Hz^{-1}}}{13.9\times10^{27}}.
\end{equation}
We subtract -0.24 dex to be consistent with a \citet{Chabrier2003} IMF. Finally, we calculate the
average SFR as a function of stellar mass as
	\begin{eqnarray}
	\langle \log \sfr\ (\ms,z) \rangle =\phi_*^{-1}(\ms,z) \int P(\ms|\muv,z) \times  & &  \nonumber \\
	\log \sfr(\muv)
		 \phiuv(\muv,z)d\muv.
	\end{eqnarray}
Both the probability distribution function $P(\ms|\muv,z)$ and the function $\phi_*(\ms,z)$ are described in detail in 
Appendix \ref{UV_LFs}. We use the following intervals of integration:  $\muv\in [-17,-22.6]$ at $z=4$; 
$\muv\in [-16.4,-23]$ at $z=5$; $\muv\in [-16.75,-22.5]$ at $z=6$; 
$\muv\in [-17,-22.75]$ at $z=7$ and $\muv\in [-17.25,-22]$ at $z=8$.

\subsection{Cosmic Star Formation Rate}

We use the CSFR data compilation from \citet{MadauDickinson2014}. This data was derived from
FUV and IR rest frame luminosities by deriving empirical dust corrections to the FUV data 
in order to estimate robust CSFRs. We adjusted their data to a \citet{Chabrier2003} IMF by subtracting 
0.24 dex from their CSFRs. Finally, for $z>3$ 
we calculate the CSFR using again the UV dust-corrected LFs and SFRs described above and using  
the same integration limit as in \citet{MadauDickinson2014}. We find that our CSFR is consistent 
with the compilation derived in \citet{MadauDickinson2014} over the same redshift range.

\subsection{The Fraction of Star-Forming and Quiescent 
Galaxies}
\label{Fraction_SF_Q}

In this paper we interchangeably refer to star-forming galaxies as blue galaxies
and quiescent galaxies as red galaxies. We utilize the fraction of
blue/star-forming and red/quenched galaxies as a reference to compare with our model and thus gain
more insights on how galaxies evolve from active to passive as well as on their
structural evolution (discussed in Section \ref{Structural_Evolution}). 
For the fraction of 
quiescent galaxies $f_{\rm Q}$ we use the following relation: 
\begin{equation}
f_{\rm Q}(\ms,z) = \frac{1} {1 + (\ms / M_{50}(z))^{\alpha}},
\end{equation}
where $M_{50}$ is the transition stellar mass at which the fractions
of blue star-forming  and red quenched galaxies are both $50\%$. 
Figure \ref{m_trans} shows $M_{50}$ as a function of redshift from observations and previous constraints. 
The solid black line shows the relation $\log (M_{50} (z) / \msun) = 10.2 + 0.6 z$
that we will employ in this paper, and
the gray solid lines show the results when shifting $(M_{50} (z) / \msun)$ by
0.1 dex above and below. We will use this shift as our uncertainty in the definition for 
$\log (M_{50} (z) / \msun) $.  The red (blue) curves in the figure show the stellar mass vs. redshift 
where 75\% (25\%) of the galaxies are quenched.
Finally, we will assume that 
$\alpha = -1.3$. The transition stellar mass is such that at $z=0$ $\log (M_{50} (z) / \msun) = 10.2$
and at $z=2$ $\log (M_{50} (z) / \msun) = 11.4$.

 We note that our statistical treatment of quenched vs. star-forming galaxies is rather different from a common
approach in the literature, in which a given galaxy is considered to be quenched based on its specific star formation rate and redshift.  For example, \citet{Pandya+2017} defines transition galaxies to have sSFR between 0.6 dex (1.5$\sigma$) and 1.4 dex 
(3.5$\sigma$) below the star-forming main sequence, with fully quenched galaxies having sSFR even farther below the main sequence.  But our statistical approach does not permit this.

\section{Constraining the model}
\label{mcmc}

\begin{figure*}
	\vspace*{-270pt}
	\includegraphics[height=8.9in,width=7.3in]{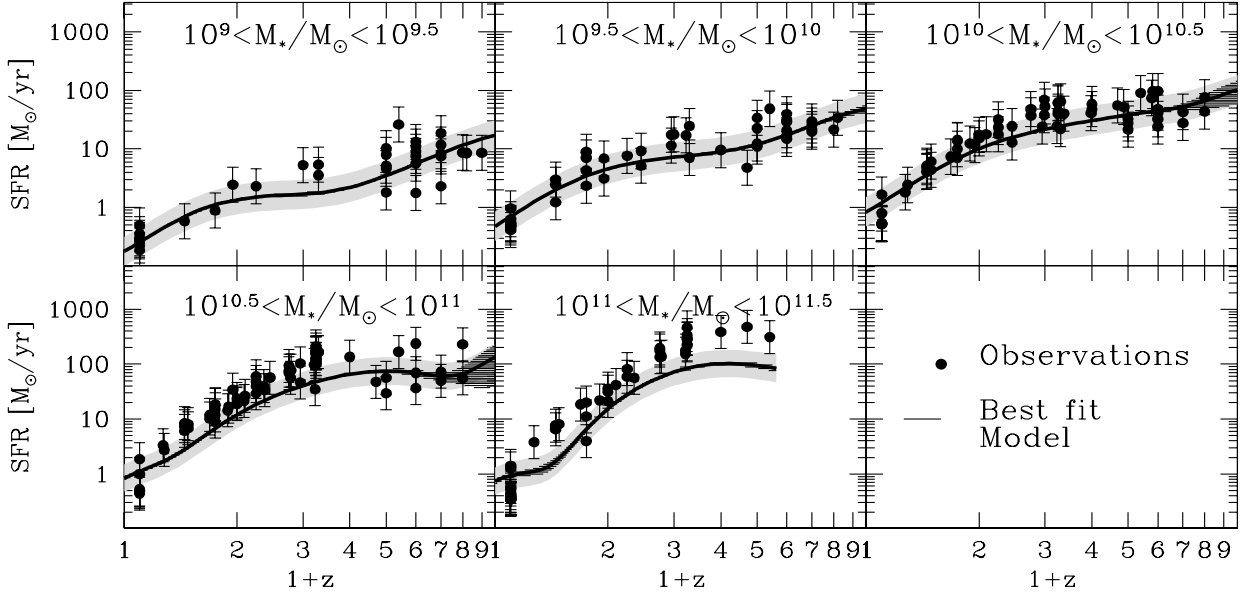}
	\vspace*{-130pt}
	\caption{Star formation rates as a function of redshift $z$ in five stellar mass bins.
	Black solid lines  shows the resulting best fit model to the SFRs implied by our approach.
	The filled circles with error bars show the observed data as described in the text, see Section \ref{mod_framework}. 
 	}
	\label{f7}
\end{figure*}

\begin{figure}
	\vspace*{-200pt}
	\includegraphics[height=7in,width=5.8in]{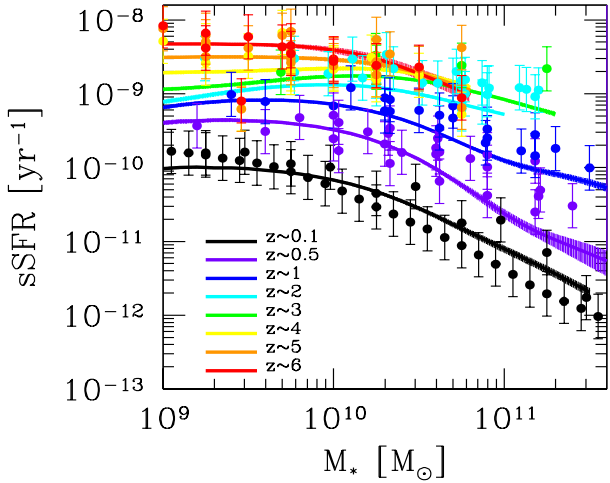}
	\vspace*{-100pt}
		\caption{
 	Specific $\sfr$s as a function of stellar mass from $z\sim0.1$ to $z\sim6$.
		The solid lines show our best fitting model while the shaded areas show the $1\sigma$
		confidence intervals using our set of MCMC trials. The filled circles 
		show the observations we utilize to constrain our model. 
 	}
	\label{sfr_all}
\end{figure}

\begin{figure}
	\vspace*{-110pt}
	\hspace*{-20pt}	
	\includegraphics[height=7in,width=5.8in]{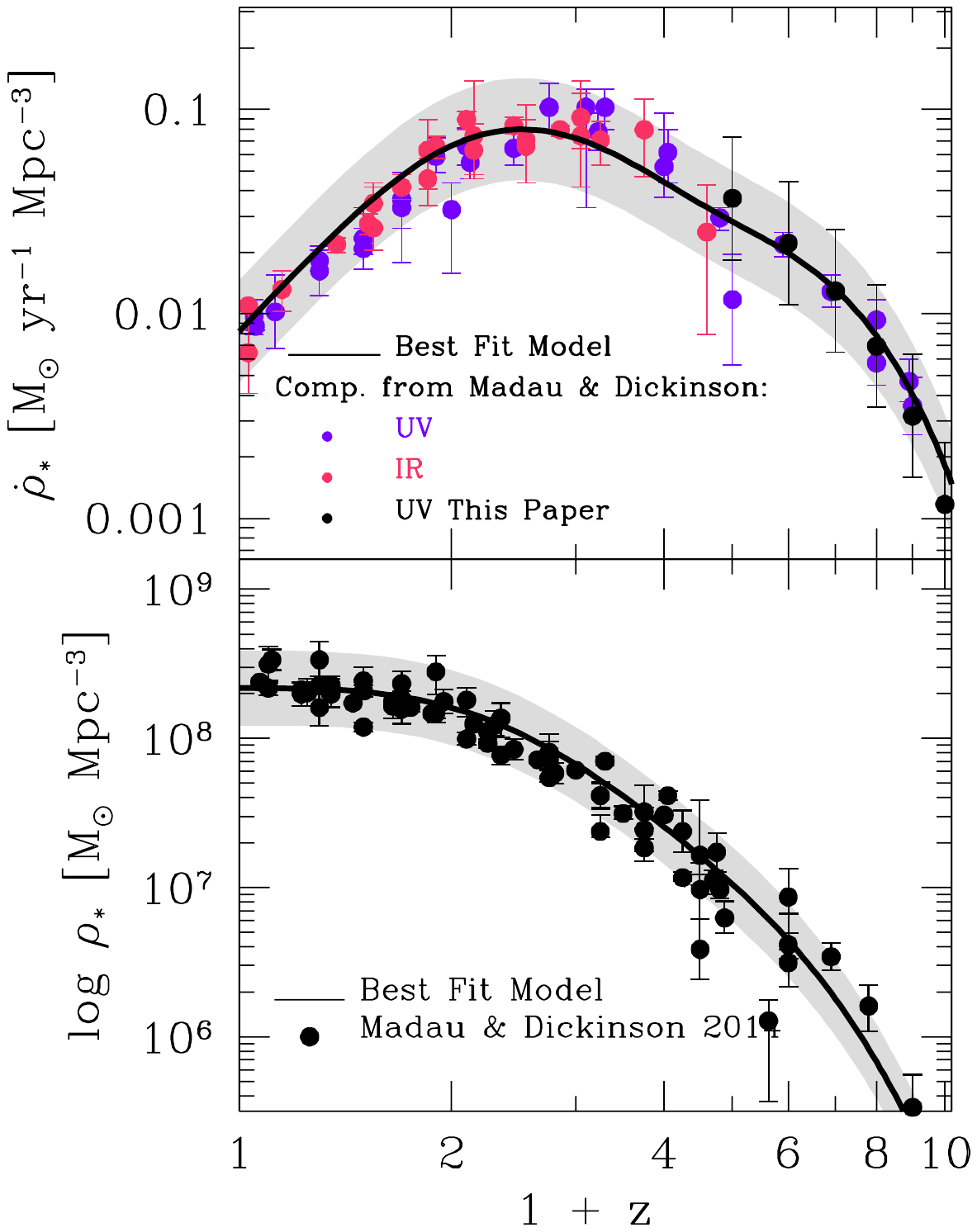}
	\vspace*{-110pt}
	\caption{{\bf Upper Panel:} Cosmic star formation rate, CSFR. The solid black line shows the resulting best fit model to the
	CSFR as described in Section \ref{csfr_model_section}. 
	Filled red and violet circles show a set of compiled observations by \citet{MadauDickinson2014} from
	FUV+IR rest frame luminosities. UV luminosities are dust-corrected.  Black solid circles
	show the results from the UV dust-corrected luminosity functions described in Appendix \ref{UV_LFs}.
	{\bf Lower Panel:} Cosmic stellar mass density.  The solid black line shows the predictions for our best fit model. 
	Filled black circles show the data points compiled in \citet{MadauDickinson2014}. All data was adjusted to the IMF
	of \citet{Chabrier2003}. In both panels, the light grey shaded 
	area shows the systematic assumed to be of $0.25$ dex. 
 }
\label{csfr_model}
\end{figure}

The galaxy population in our model is described by four properties: halo  mass \mvir, halo mass accretion rates, 
stellar mass \ms, and star formation rate SFR. In order to constrain the model
we combine several observational data sets, including the  
$\gsmf$s, the $\sfr$s and the CSFR for all galaxies. 
In this Section we describe our adopted methodology as well as the best resulting
fit parameters in our model.

In order to sample the best-fit parameters that maximize the likelihood 
function $L\propto e^{-\chi^2/2}$ we use the 
MCMC approach, described in detail in \citet{RAD13}.

We compute the total $\chi^2$ as,
\begin{equation}
	\chi^2=\chi^2_{\rm GSMF}+\chi^2_{\rm SFR} +\chi^2_{\rm CSFR} 
\end{equation}
where for the $\gsmf$s we define
\begin{equation}
	\chi^2_{\rm GSMF}=\sum_{j,i}\chi^2_{\phi_{j,i}},
\end{equation}
for the SFRs
\begin{equation}
	\chi^2_{\rm SFR}=\sum_{j,i}\chi^2_{{\rm SFR}_{j,i}},
\end{equation}
and for the CSFRs
\begin{equation}
	\chi^2_{\rm CSFR}=\sum_{i}\chi^2_{{\rm \dot{\rho}}_{i}}.
\end{equation}
In all the equations the sum over $j$ refers to different stellar mass bins while $i$ refers 
to summation over different redshifts.
The fittings are made to the data points with their error bars of each \gsmf, \sfr\ and CSFR.           

 In total our galaxy model consists of eighteen adjustable parameters. Fifteen are to model the redshift evolution
 of the \shmr, Equations (\ref{epsilonz})--(\ref{gammaz}): $\vec{p}_{\rm SHMR} = \{ \epsilon_0, \epsilon_1, \epsilon_2,
 \epsilon_3, M_{\mathcal{C}0}, M_{\mathcal{C}1}, M_{\mathcal{C}2}, \alpha, \alpha_1,\alpha_2, \delta_0, \delta_1,\delta_2,
 \gamma_0, \gamma_1 \}$; and three more to model the fraction of stellar mass growth due to in-situ star formation:  
 $\vec{p}_{\rm in\, situ} = \{M_{\rm in\, situ,0}, M_{\rm in\, situ,1}, \beta \} $.
 To sample the best fit parameters in our model we run a set of $3\times10^5$  MCMC models.  The resulting best-fit
 parameters are given in Equations (\ref{bestfit-begin}) -- (\ref{bestfit-end}).
 
 Figure \ref{f6} shows the best-fit model $\gsmf$s from $z \sim 0.1$ to $z \sim 10$ with the solid
lines as indicated by the labels.  
This figure shows the evolution of the observed $\gsmf$ based in our compiled data described in 
Section \ref{MF}.

Figure \ref{f7} shows the $\sfr$s as a function of redshift $z$ in five stellar mass bins.
The observed $\sfr$s from the literature are plotted with filled circles with error bars while the best fit
model is plotted with the solid black lines.  We also present 
our best fitting models by plotting the specific SFRs as a function of stellar mass in Figure \ref{sfr_all}.
Note that our model fits describe rather well the observations 
at all mass bins and all redshifts.

 We present the best-fit model to the CSFR in the upper Panel of Figure \ref{csfr_model}. The observed 
 CSFRs employed for constraining the model are shown with the solid circles and 
 error bars.  The lower Panel of Figure \ref{csfr_model} compares the cosmic stellar mass density 
 predicted by our model fit with the data compiled in the review by \citet{MadauDickinson2014}; the
 agreement is impressive.
 
In Appendix \ref{A1} we discuss the impact of the different assumptions employed
 in the modelling.
The best fitting parameters to our model are as follows:

\begin{equation}
	\begin{array}{l}
		\log(\epsilon(z)) = -1.758\pm0.040 +  \\ 
		\mathcal{P}(0.110\pm0.166,-0.061\pm0.029,z) 
		\times \mathcal{Q}(z)  + \\
		\mathcal{P}(-0.023\pm0.009,0,z), \\
	\end{array}
	\label{bestfit-begin}
\end{equation}
\begin{equation}
	\begin{array}{l}
		\log(M_{0}(z)) = 11.548\pm0.049  +  \\ 
		 \mathcal{P}(-1.297\pm0.225,-0.026\pm0.043,z)\times \mathcal{Q}(z),  
	\end{array}
\end{equation}
\begin{equation}
	\begin{array}{l}
		\alpha (z) = 1.975\pm0.074  +  \\ 
		 \mathcal{P}(0.714\pm0.165,0.042\pm0.017,z)\times \mathcal{Q}(z),  
	\end{array}
\end{equation}
\begin{equation}
	\begin{array}{l}
		\delta(z) =3.390\pm0.281  +  \\ 
		 \mathcal{P}(-0.472\pm0.899,-0.931\pm0.147,z)\times \mathcal{Q}(z),  
	\end{array}
\end{equation}
\begin{equation}
	\begin{array}{l}
		\gamma(z) = 0.498\pm0.044 + \mathcal{P}(-0.157\pm0.122,0,z)\times \mathcal{Q}(z), 
	\end{array}
\end{equation}
\begin{equation}
	\begin{array}{l}
		\log(M_{\rm in\, situ} (z))= 12.728\pm0.163 +  \\  
		\mathcal{P}( 2.790\pm0.163,0,z),
	\end{array}
\end{equation}
\begin{equation}
	\begin{array}{l}
		\beta(z)= 0.760\pm0.032 + \log(1+z). 
	\end{array}
	\label{bestfit-end}
\end{equation}

For our best fitting model
we find that $\chi^2= 522.8$ from a number of $N_d=488$ observational 
data points. Since our model consist of $N_p=18$ free parameters the resulting 
reduced 
$\chi^2$ is $\chi^2/{\rm d.o.f.}= 1.11$.

\section{The Galaxy-Halo Connection}

\subsection{The Stellar-to-Halo Mass Relation from $z\sim0.1$ to $z\sim10$}
\label{STHMR}

\begin{figure}
	\vspace*{-55pt}
	\hspace*{-20pt}	
	\includegraphics[height=7in,width=5.8in]{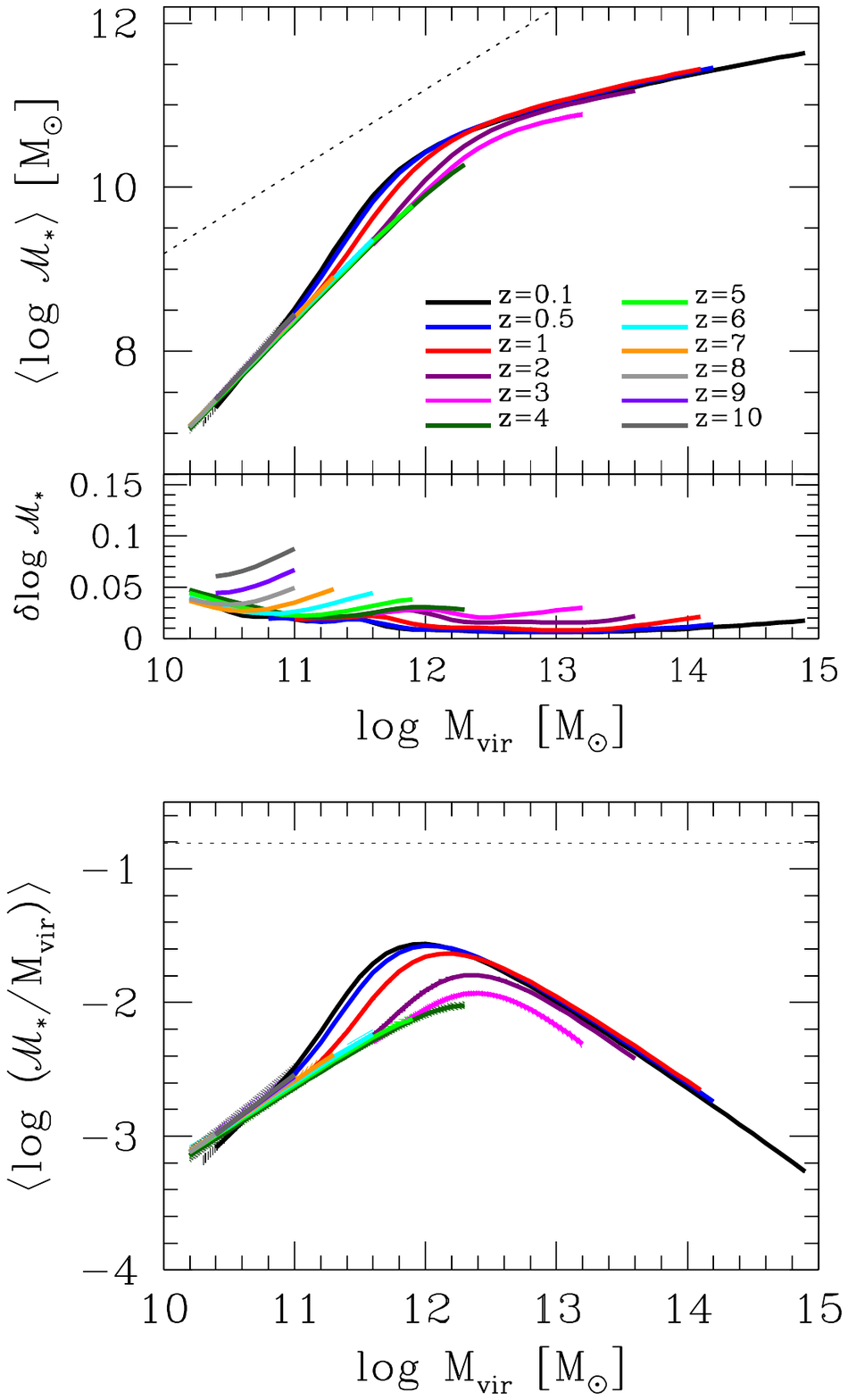}
	\vspace*{-100pt}
		\caption{
		{\bf Upper panel:} Evolution of the mean stellar-to-halo mass relation from $z = 0.1$ to $z=10$ as indicated
		in the legends. In our model
		we assume that these relations are valid both for central and satellite galaxies as explained in the text.  The relations are shown only up to 
		the largest halo mass that will be observed using the solid angles and redshift bins of the surveys from Table \ref{T1}.
		Table \ref{Tmsmh} lists the range over which our mass relations can be trust.	
		{\bf  Middle panel:}
 		1$\sigma$ confidence intervals from the $3\times10^{5}$ MCMC trials. 
		{\bf  Bottom panel:} Evolution
		of the stellar-to-halo mass ratios $\ms / \mvir$ for the same redshifts as above. The dotted lines in both panels show 
		the limits corresponding to the cosmic baryon fraction $\Omega_{\rm B} / \Omega_{\rm M} \approx 0.16$.
 	}
	\label{f8}
\end{figure}

\begin{figure*}
	\includegraphics[height=2.2in,width=7.2in]{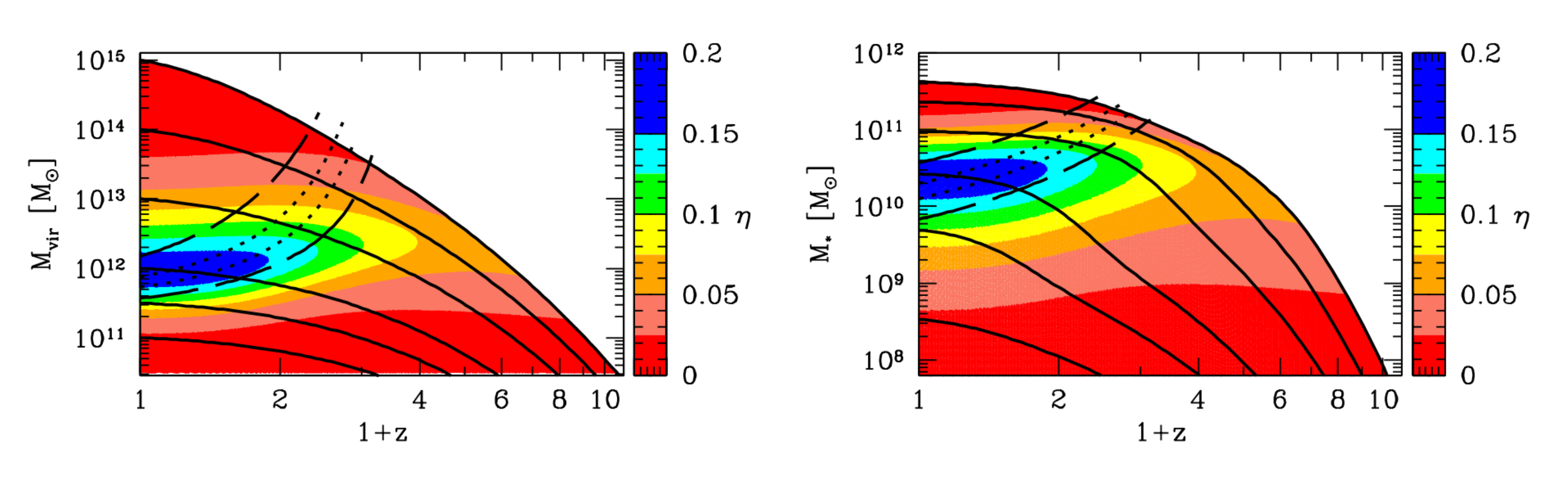}
		\caption{{\bf Left Panel:} Integral stellar conversion efficiency, defined as $\eta = f_* / f_b$, as a function
		of halo mass for progenitors at $z=0$. The black solid lines show the trajectories for progenitors with  $\mvir=10^{11},10^{11.5},10^{12},10^{13},$
		$10^{14}$ and $10^{15} $\msun. {\bf Right Panel:} Integral stellar conversion efficiency $\eta$ but now as
		a function of stellar mass for the corresponding halo progenitors. The stellar conversion efficiency
		is higher at low redshifts, $z\sim 0$, and highest for halo progenitors at $z=0.1$ between $\mvir \sim 5\times10^{11}\msun - 2\times10^{12}\msun$
		corresponding to galaxies between  $\ms \sim 10^{10}\msun-4\times10^{10}\msun$. In both panels, the dotted lines  		show $M_{50}(z)$ above which 50\% of the galaxies are 
		statistically quenched, and the upper (lower) long-dash curves show the mass vs. $z$ where 75\% (25\%) 
		of the galaxies are quenched.		 	}
	\label{sceff}
\end{figure*}

\begin{figure}
	\vspace*{-110pt}
	\hspace*{-10pt}
	\includegraphics[height=5.3in,width=4.2in]{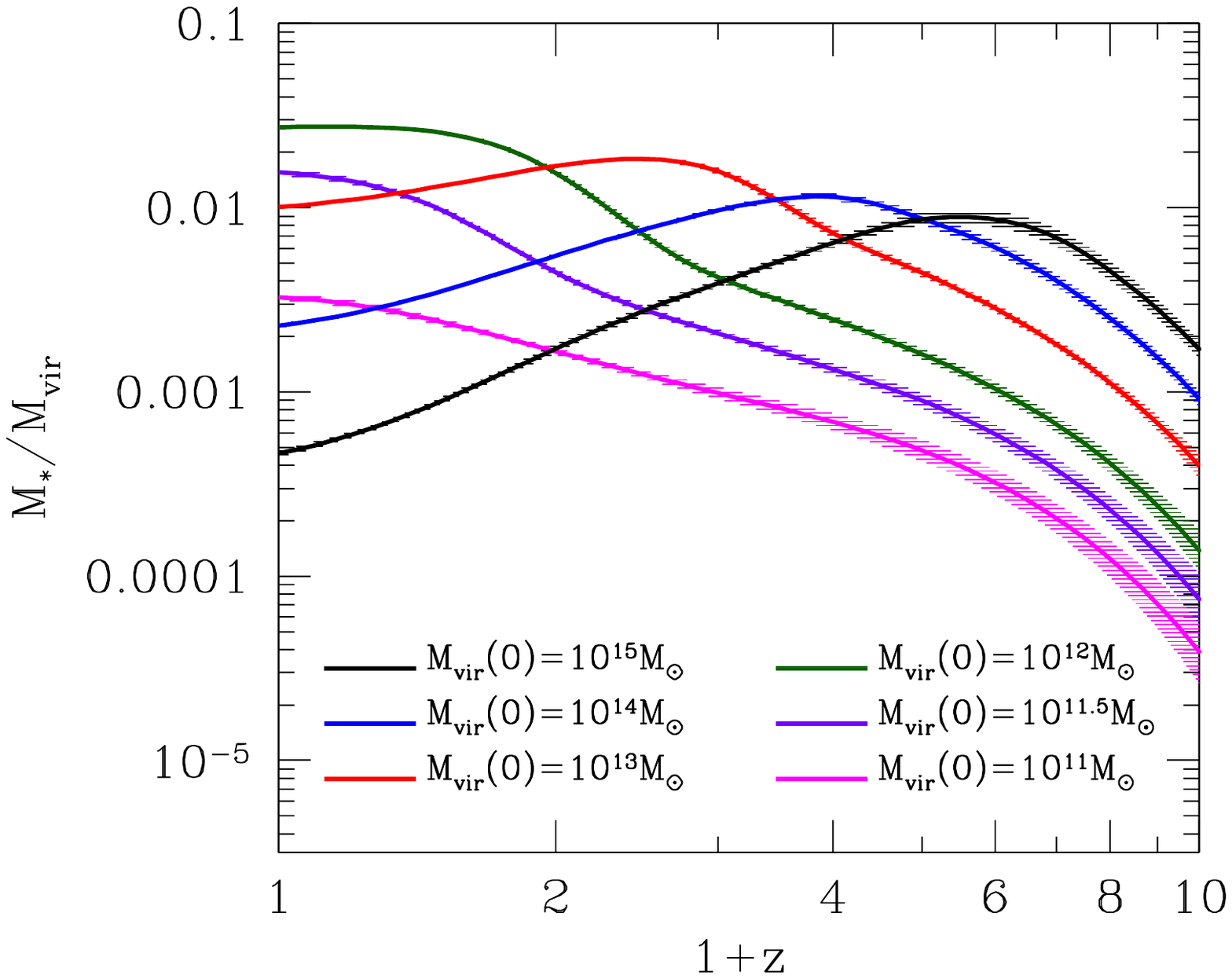}
	\vspace*{-80pt}
		\caption{
		Evolution of the stellar-to-halo mass ratio for progenitors with  $\mvir=10^{11},10^{11.5},10^{12},10^{13},$
		$10^{14}$ and $10^{15} $\msun\ at $z=0$.
 	}
	\label{ratio_msmh}
\end{figure}

The upper panel of Figure \ref{f8} shows the constrained evolution of the $\shmr$
while the lower panel shows the stellar-to-halo mass ratio  from $z\sim0.1$ to $z\sim10$. 
Recall that in the case of central
galaxies we refer to \mdm\ as the virial mass  \mvir\ of the host halo, while for satellites \mdm\ 
refers to the maximum mass \mpeak\ reached along the main progenitor assembly history. 
Consistent with previous results the $\shmr$ appears to evolve only very slowly below $z\sim1$. 
This situation is quite different between $z\sim1$ and $z\sim7$, where at a fixed halo mass
 the mean stellar mass is lower at higher redshifts. 
The middle panel of the same figure shows the 1$\sigma$ confidence intervals
 of our constrained $\shmr$s from the $3\times10^{5}$ MCMC models. Finally, we estimate at each redshift the largest
halo mass that will be observed using the solid angles and redshift bins of the surveys from Table \ref{T1}. We use these halo masses
as the upper bounds to the virial halo masses shown in Figure \ref{f8}. Finally, Table \ref{Tmsmh} lists the range over which our mass 
relations can be trust. 

\begin{table}
	\caption{Minimum and maximum halo masses over which our mass relations can be trusted.}
	\begin{center}
		\begin{tabular}{c c c}
			\hline
			\hline
			$z$ &  $\log M_{\rm vir,min}  [\msun]$ &  $\log M_{\rm vir,max}   [\msun]$\\
			\hline
			\hline
			0.1 & 10.3 & 15.0\\ 
			0.5 & 10.8 & 14.3\\ 
			1 & 11.0 & 14.1\\ 
			2 & 11.5 & 13.7\\ 
			3 & 10.6 & 13.3\\ 
			4 & 10.2 & 12.3\\ 
			5 & 10.2 & 12.0\\ 
			6 & 10.2 & 11.7\\ 
			7 & 10.2 & 11.4\\ 
			8 & 10.2 & 11.1\\ 
			9 & 10.3 & 11.1\\ 
			10 & 10.3 & 11.1\\ 
			\hline
		\end{tabular}
		\end{center}
	\label{Tmsmh}
\end{table}

The maximum of the stellar-to-halo mass ratio is around $\mvir\sim10^{12}\msun$ at $z\sim0.1$ with a value of $\sim0.03$. The maximum moves
to higher mass halos at higher redshifts up to $z\sim3$, consistent with previous studies 
 \citep[see e.g.,][]{Conroy+2009,Firmani+2010a,Behroozi+2010,Leauthaud+2012,Yang+2012,Behroozi+2013,Moster+2013,Skibba+2015}. 
 The value of the maximum of the stellar-to-halo mass ratio 
 moves to lower values with increasing redshift, decreasing by approximately a factor of $3$
between $z\sim0.1$ and $z\sim4$. At redshift $z\sim7$ the stellar-to-halo mass ratio has decreased by an 
order of magnitude. Nonetheless, given the uncertainties when deriving 
the $\gsmf$s at high redshifts $z>4$, this result should be taken with caution. 
For comparison, the dashed lines in both panels show the cosmic
baryon fraction implied by 
the \citet{Planck+2015} cosmology, $f_b = \Omega_{\rm B} / \Omega_{\rm M} \approx 0.16$. 

Next we study the integral stellar conversion efficiency, defined as $\eta = f_* / f_b$. This is shown in the left panel
of Figure \ref{sceff} for progenitors of dark matter halos with masses
between $\mvir = 10^{11}\msun$ and $\mvir = 10^{15}\msun$ at $z=0$. Dark matter halos are 
most efficient when their progenitors reached masses between $\mvir \sim 5\times10^{11}\msun- 2\times10^{12}\msun$
at $z<1$, and the stellar conversion efficiency is never larger than $\eta\sim0.2$.  Note that  \cite{Zehavi+2011} 
reached a similar conclusion when estimating the HOD model of the SDSS survey utilizing the observed galaxy clustering.
Theoretically, the characteristic halo mass of $\sim 10^{12}h^{-1}$\msun\ is expected to mark a transition above which 
the stellar conversion becomes increasingly inefficient. The reason is that at halo masses 
above $10^{12}h^{-1}$\msun\ the efficiency at which the virial shocks form and  
heat the incoming gas increases \citep[e.g.,][]{Dekel+2006}.
Additionally, in such massive galaxies the gas can be kept from cooling by the feedback from active galactic nuclei \citep[][and references 
therein]{Croton+2006,Cattaneo+2008,Henriques+2015,Somerville+2015}. Central
galaxies in massive halos are therefore expected, in a first approximation, to become passive systems roughly at the epoch when 
the halo reaches the mass of $10^{12}h^{-1}$\msun, thus the term halo mass quenching.
On the other hand, the less massive the halos, the less efficient their growth in stellar mass is expected to be due to supernova-driven gas loss in their lower gravitational potentials.

The right panel of Figure \ref{sceff} shows the stellar conversion efficiency for the 
corresponding stellar mass growth histories of the halo progenitors discussed above. 
The range of the transition stellar mass $M_{50} (z)$, defined as 
the stellar mass at which the fraction of star forming is equal to the fraction of
quenched galaxies (see Figure \ref{m_trans} and Section \ref{Structural_Evolution}), is shown by the dashed lines. 
Below these lines galaxies are more likely to be star forming. Note that the right panel of
Figure \ref{sceff} shows that  $M_{50} (z)$ roughly coincides with where $\eta$ is maximum, especially at 
low $z$. This reflects the fact that halo mass quenching is part of the physical mechanisms that  
quench galaxies in massive halos. We will come back to this point in Section \ref{halo_quenching}.

Finally, Figure \ref{ratio_msmh} shows the trajectories for the $\ms/\mvir$ ratios of progenitors of dark matter halos with masses
between $\mvir = 10^{11}\msun$ and $\mvir = 10^{15}\msun$ at $z=0$. Note that galaxies in halos above $\mvir = 10^{12}\msun$
had a maximum followed by a decline of their $\ms/\mvir$ ratio, while this ratio for galaxies in less massive halos continues
increasing today.   

\subsection{Galaxy Growth and Star-Formation Histories}
\label{growth_SFH}

\begin{figure*}
	\hspace*{-10pt}
	\includegraphics[height=2.9in,width=3.4in]{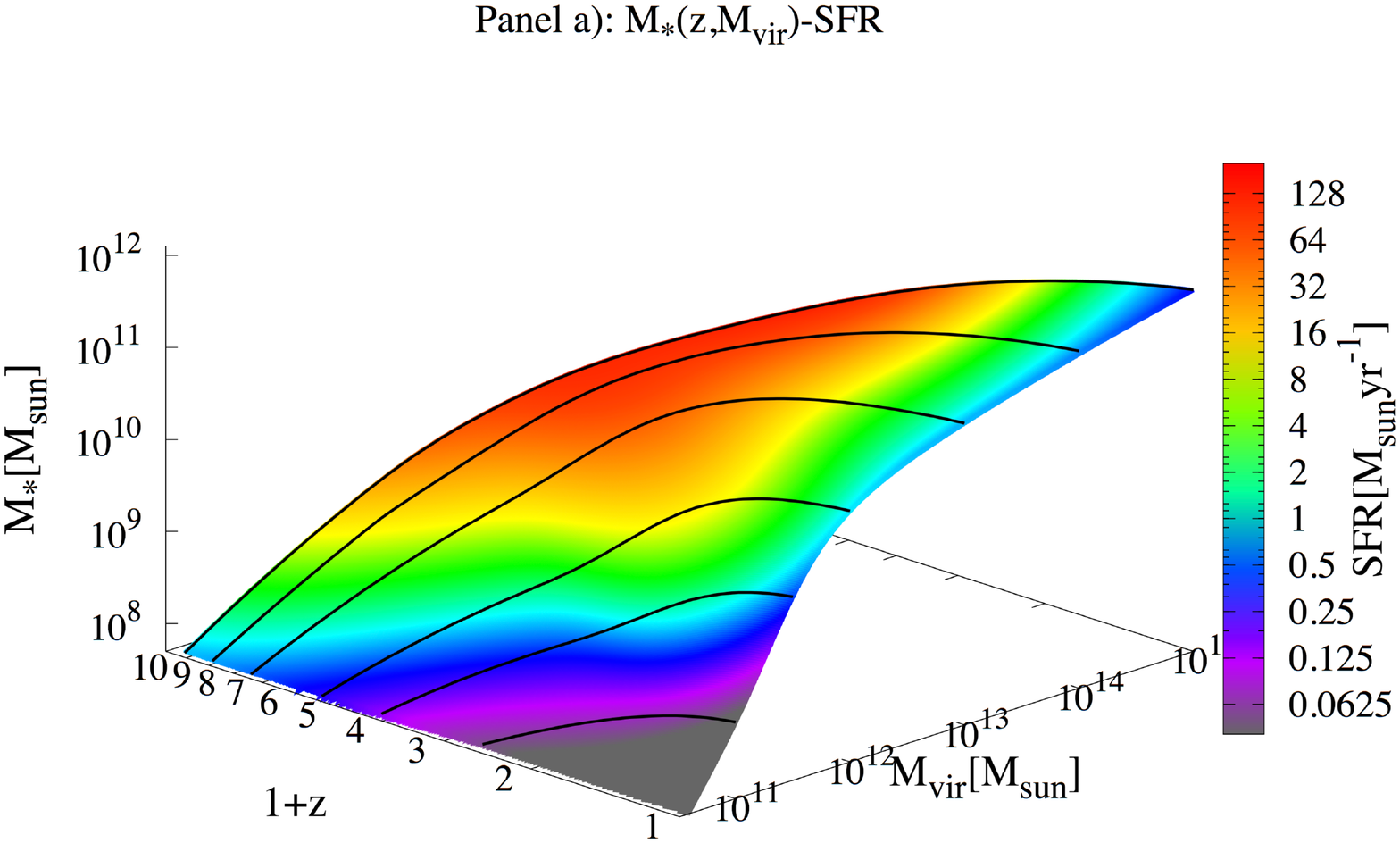}
	\hspace*{10pt}
	\includegraphics[height=2.9in,width=3.4in]{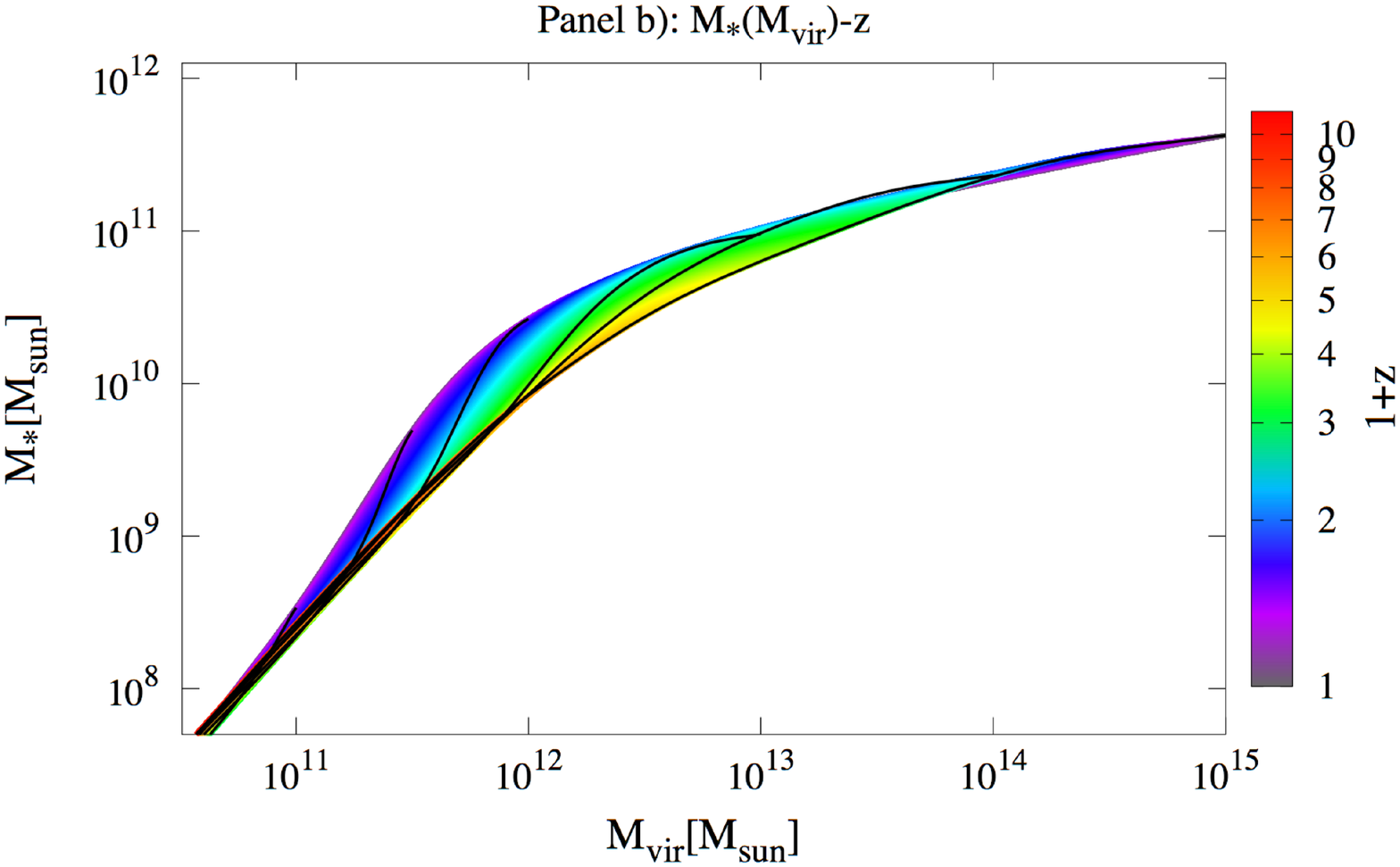}
	\hspace*{-10pt}
	\includegraphics[height=2.9in,width=3.4in]{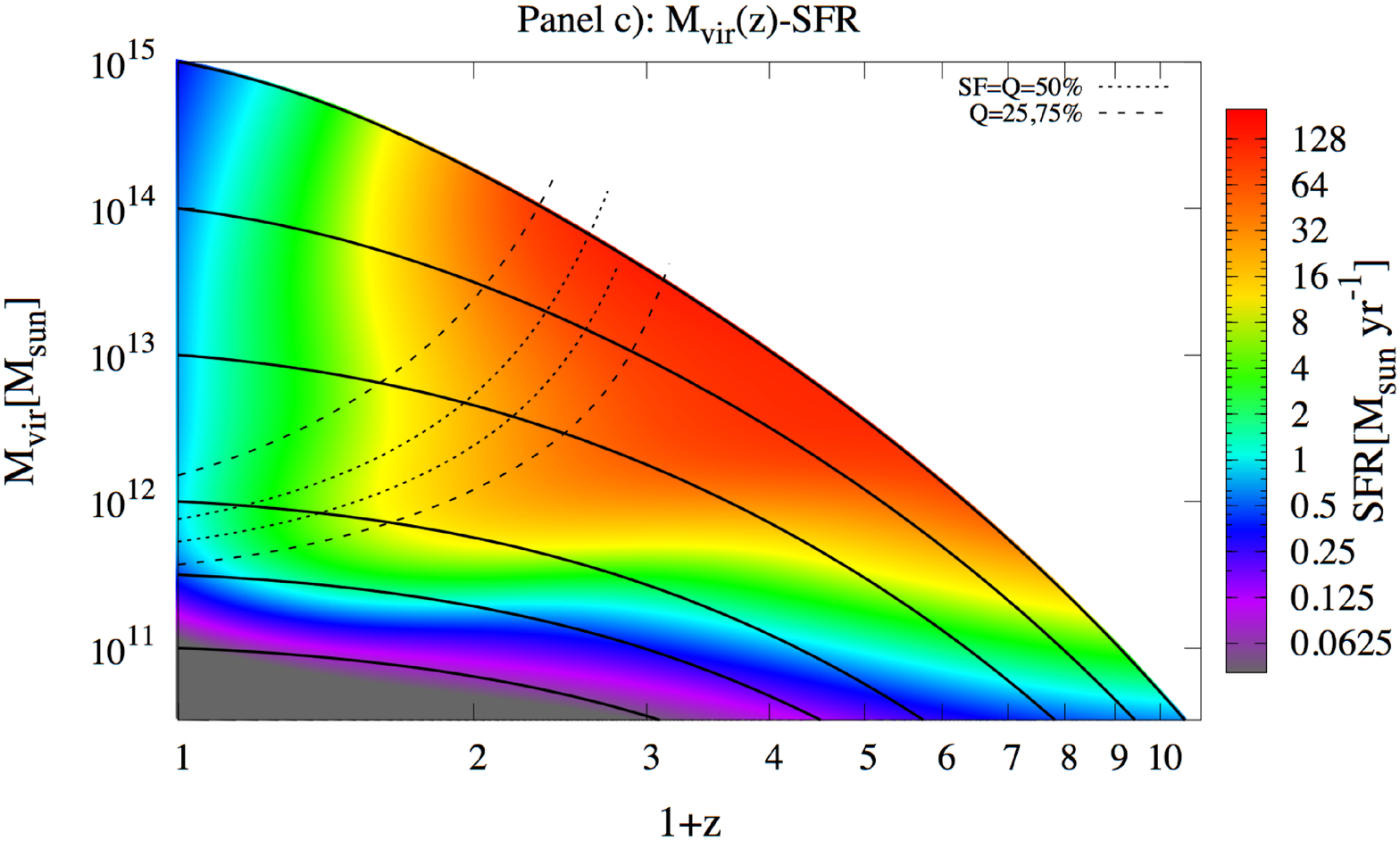}
	\hspace*{10pt}
	\includegraphics[height=2.9in,width=3.4in]{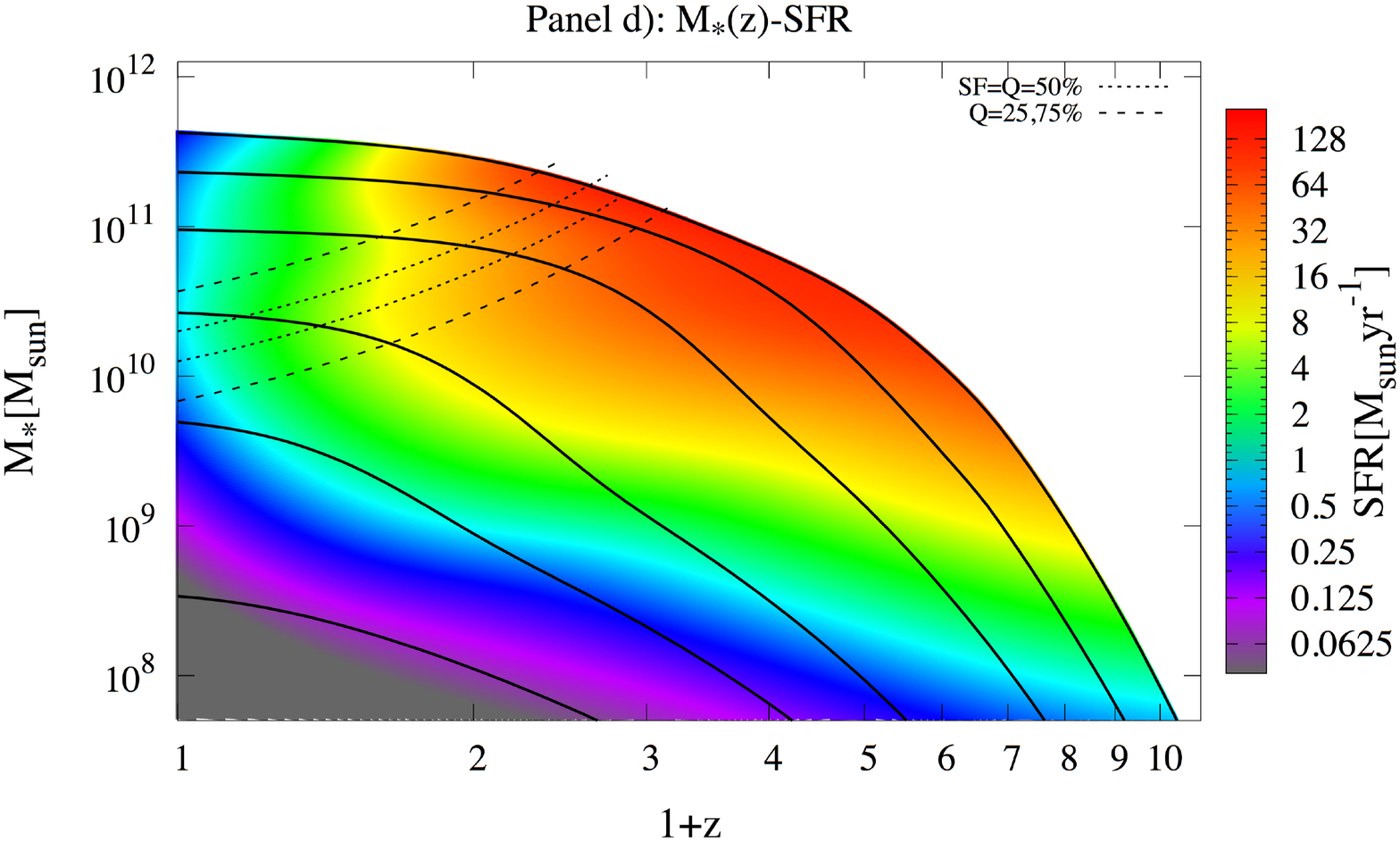}
		\caption{ Panel a): Galaxy SFRs as a function of redshift, halo mass, and stellar mass.
		The solid lines indicate the average trajectories corresponding to progenitors at $z=0$ with  
		$\mvir=10^{11},10^{11.5},10^{12},10^{13},$ $10^{14}$ and $10^{15} $\msun. The color code shows the SFRs.
		Panel b): Galaxy growth trajectories in the stellar-to-halo mass plane (this is a projection
		of Panel a) when collapsing over the redshift axis). Panel c):  Galaxy SFRs along the halo mass trajectories
		(this is a projection of the Panel a) when collapsing over the \ms\ axis). Panel d):
		Galaxy SFRs along the stellar mass trajectories 
		(this is a projection of the Panel a) when 
		collapsing over the \mvir\ axis). The dotted lines show $M_{50}(z)$ above which 50\% of the galaxies are 
		statistically quenched, and the upper (lower) long-dash curves show the mass vs. $z$ where 75\% (25\%) 
		of the galaxies are quenched.
 	}
	\label{f10}
\end{figure*}

Figure \ref{f10} shows the predicted star formation histories for progenitors of average dark matter halos at $z=0$
with masses between $\mvir = 10^{11}\msun$ and $\mvir = 10^{15}\msun$. Panel a) shows the resulting 3D 
surface for the redshift evolution of the stellar-to-halo mass relation for progenitors of dark matter halos 
at $z=0$. We color code the star formation rates as
indicated by the vertical label. For reference, the solid black lines show the average trajectories 
for progenitors with  $\mvir=10^{11},10^{11.5},10^{12},10^{13}, 10^{14}$ and $10^{15} $\msun. 
Panel b) shows the evolution of the $\shmr$s for the same progenitors while
panels c) and d) show, respectively, the projections of star formation histories as a function of halo mass and their 
corresponding stellar masses. 
Previous studies have shown related figures to the panels in Figure \ref{f10}
\citep[see e.g.,][]{Firmani+2010a,Krumholz+2012,Behroozi+2013,Yang+2013}. We note that our results are qualitatively similar
to these previous studies, updated by using more recent observational data for the
$\gsmf$s, $\sfr$s and the CSFR.  

Figure \ref{f10} shows that the star formation histories in the most massive halos, 
$\mvir\grtsim10^{13}\msun$, increased with redshift 
reaching a maximum value 
between $z\sim1-4$. On average, the most massive halos could reach SFRs as high as $\sfr\sim200$ $\msun /$yr 
in this redshift interval. 
Following this very intense period of star formation their SFR decreases, and by $z\sim0.1$ they are already 
quenched. This implies that in the most massive halos 
the stellar mass of their galaxy was already in place since $z\sim1$. 
In contrast, galaxies in low mass halos ($\mvir\sim10^{11}\msun$) on average form stars late and approximately at a constant rate, with very low rates typically below $\sfr\sim0.1$ $\msun /$yr. 
As for MW-sized halos ($\mvir\sim10^{12}\msun$), it seems that their galaxies went through various phases. The
first phase is a moderate rate of growth up to $\sim1.5$.  They reach an intense period of
star formation around $z\sim1$, with values around $\sim10$  $\msun /$yr, and then 
decline. According to our model this period of intense star formation is not associated with galaxy 
mergers as we will discuss in next subsection.  

\begin{figure*}
	\includegraphics[height=2.2in,width=7.2in]{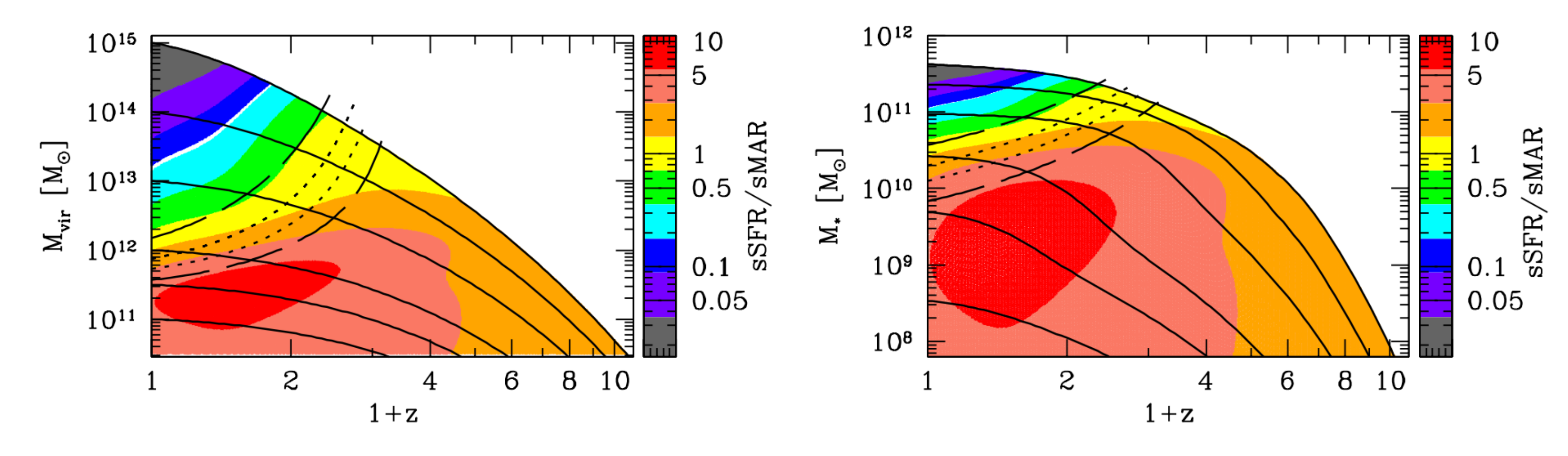}
		\caption{Halo star formation efficiencies, defined as sSFR/sMAR, as a function of halo
		mass  (left panel) and stellar mass (right panel) for halo progenitors at $z=0$. 
		The black solid lines show the trajectories for progenitors with  $\mvir=10^{11},10^{11.5},10^{12},10^{13},$
		$10^{14}$ and $10^{15} $\msun. The short dashed lines show the stellar mass
		and halo mass at which the observed fraction of star-forming galaxies is equal to the quenched fraction of
		galaxies, and the upper (lower) long-dash curves show the stellar mass vs. $z$ where 75\% (25\%) 
		of the galaxies are quenched.}
	\label{sfreff}
\end{figure*}

\begin{figure*}
	\vspace*{-240pt}
	\hspace*{-10pt}
	\includegraphics[height=9.3in,width=7.56in]{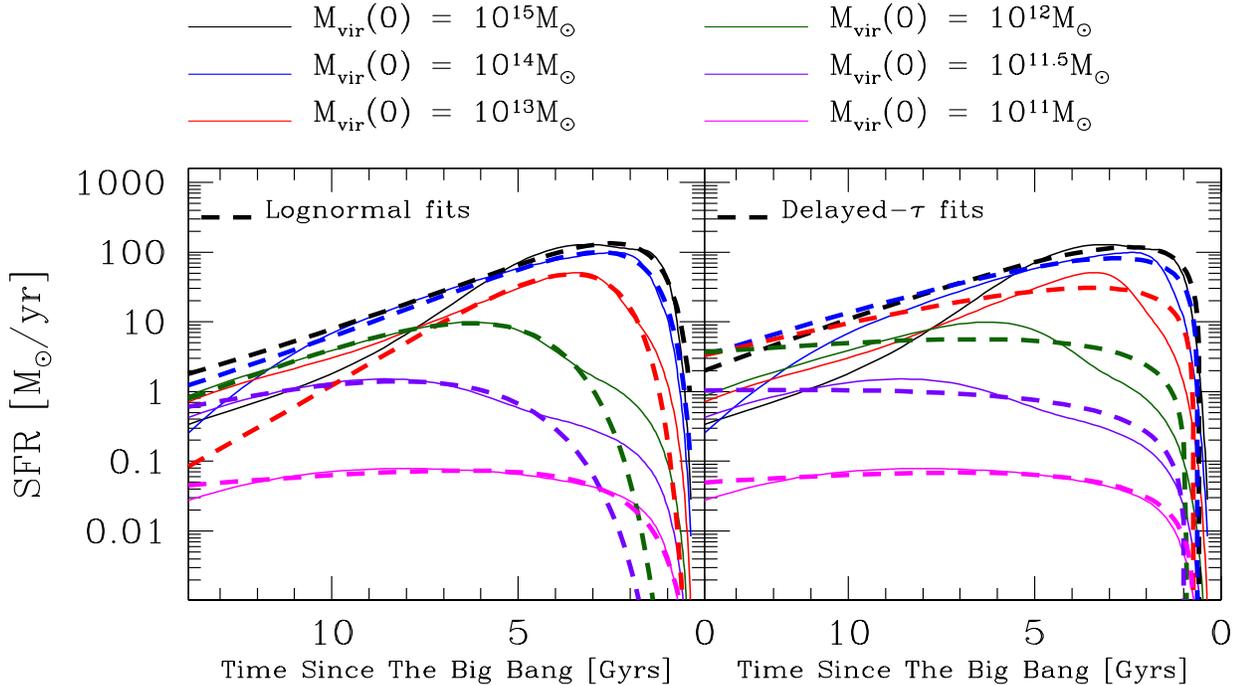}
	\vspace{-160pt}
		\caption{  Analytic fits to the star formation histories as indicated by the labels. 
		{\bf Left Panel:} Lognormal fits.
		{\bf Right Panel:} Delayed$-\tau$ fits. Note that lognormal fits describe rather well the average 
		star formation histories of galaxies in halos with masses around and below about $\mvir=10^{12}\msun$ after the first few Gyr. 
		Delayed$-\tau$ fits are poor at all but the lowest masses.       
		}
	\label{SFH_laws}
\end{figure*}

\begin{figure*}
	\includegraphics[height=2.2in,width=7.2in]{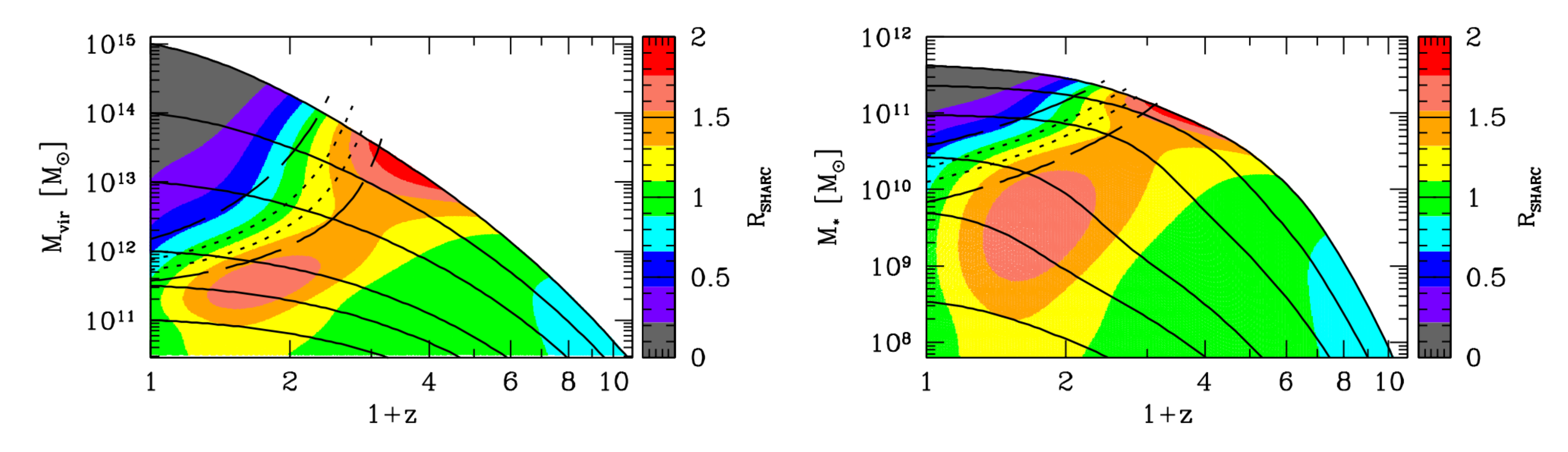}
		\caption{{\bf Left Panel:} Stellar-halo accretion rate coevolution (SHARC) assumption as a function
		of halo mass for progenitors at $z=0$. The black solid lines show the trajectories for progenitors with  $\mvir=10^{11},10^{11.5},10^{12},10^{13},$
		$10^{14}$ and $10^{15} $\msun. {\bf Right Panel:} Like the left panel, but as
		a function of stellar mass for their corresponding halo progenitors. These figures show that the SHARC assumption is a good approximation
		within a factor of $\sim2$ for star-forming galaxies, which are a majority of those below the quenching curves.
		Recall that in both panels, the dashed lines
		denote $M_{50}(z)$ below/above which 50\% of the galaxies are star-forming/quenched, and the upper (lower) long-dash curves show the stellar mass vs. $z$ where 75\% (25\%) 
		of the galaxies are quenched.
 	}
	\label{sharc}
\end{figure*}

It is interesting that panel b) shows that the spread of trajectories of the galaxies in the \shmr\ plane is very narrow, giving  
the impression that the \shmr\ is time independent.
Recall that Figure \ref{f10} shows the backwards trajectories for halos at $z=0$ 
(i.e., their progenitors) reflecting the fact that halo mass (and consequently galaxy mass) is, on average, increasing at all redshifts. 
As discussed in previous studies, a roughly time independent \shmr\ for $0\lesssim z\lesssim 4$ is consistent with observed 
evolution of the $\sfr$s, especially for star-forming main sequence galaxies \citep{Behroozi+2013c,Rodriguez-Puebla+2016a}. 

Theoretically, it is expected that dark matter halos control the growth of their host galaxies 
\citep[see e.g.][]{Bouche+2010,Dekel+2009, Dave+2012,Krumholz+2012,Dekel+2013, Dekel+2014,Mitra+2015, Feldmann+2015,FeldmannMayer2015,Rodriguez-Puebla+2016a}. 
We investigate this in the left panel of Figure \ref{sfreff} 
where we plot the ratio between the specific star formation rate (s$\sfr=\sfr/\ms$) and the specific halo mass
accretion rate (${\rm sMAR} = (d\mvir/dt) /\mvir$), i.e., s$\sfr / {\rm sMAR}$, as a function of halo mass.\footnote{Observe that the sSFR and the sMAR have units of the inverse of time. One can
interpret them as the characteristic time that it will take galaxies and halos to double their mass at a constant 
assembly rate. Therefore the ratio s$\sfr / {\rm sMAR}= t_{\rm h} / t_{\rm g}$ measures how fast galaxies are gaining stellar mass 
compared to their halos gaining total mass.} 
Hereafter, we refer to the ratio s$\sfr / {\rm sMAR}$ as 
the instantaneous halo star formation efficiency.\footnote{Do not confuse the instantaneous halo star formation efficiency with the halo 
stellar conversion efficiency $\eta = f_*/ f_b$. The former is an instantaneous quantity while the latter is an integral (cumulative)
quantity.} 
Similarly to our definition of SFRs, halo MARs were measured in time steps of 100 Myrs.
Note that halo star formation efficiencies of the order of unity imply that the assembly time for galaxies is similar to that for their dark matter halos -- 
in other words, a direct coevolution between galaxies and dark matter halos. 
In contrast, values that are in either directions much above and below unity 
imply that the galaxy stellar mass growth is disconnected from the growth of its host  dark matter halo. Recall that
this discussion is valid only for galaxies in the centers of distinct dark matter halos.  
 
The main result from Figure \ref{sfreff} is that there is not a universal halo mass (in the sense of it being time
independent) at which the halo star formation efficiency transits above and below unity, rather this transition mass
depends on redshift.  
The transition occurs around $\mvir\sim10^{12}\msun$ at $z\sim0$
and $\mvir\sim10^{13}\msun$ at $z\sim3$. 
Additionally, Figure \ref{sfreff} shows that halo star formation efficiencies of $\sim2-5$ are typical of
low mass halos. However, we find that halo star formation efficiencies as high as $\sim10$ 
are reached in halos with $\mvir\sim2\times10^{11}\msun$  between $z\sim0.1 - 1.5$. 
This implies a total disconnection between galaxies and halos. We will come back to this point below. The right panel of the same figure shows the corresponding star formation efficiencies 
as a function of stellar mass. The 50\% stellar mass at which the numbers of star-forming and quenched galaxies are equal is shown as the short dashed lines in both panels (see also Figure \ref{m_trans}). 
In Figure \ref{sfreff} we observe that galaxies are more likely to be star-forming
when their halo star formation efficiencies are above s$\sfr / {\rm sMAR}\sim1$. That is, low-mass
galaxies form stars much faster than low mass halos gain mass. In contrast, galaxies with  
star-formation efficiencies below s$\sfr / {\rm sMAR}\sim1$ are more likely to be quenched 
as a result of high mass halos growing faster compared to their host galaxies. The above behaviours are
commonly referred in the literature as downsizing in SFR and archeological or mass downsizing respectively \cite[e.g.,][ and references therein]{Conroy+2009, Firmani+2010a}.
The former implies that low-mass galaxies delayed the stellar mass assembly with respect to their halos, while 
the  latter implies that the more massive the galaxies, the earlier their mass growth was quenched while their halos continued growing. It is interesting to note
that all galaxies that are quenched today went through a phase in which they co-evolved with their host halos, i.e.,
s$\sfr / {\rm sMAR} \sim1$.

We note that the halo star formation efficiency peaks around progenitors with 
$\mvir\sim2\times10^{11}\msun$, which corresponds to galaxies with masses 
$\ms\sim (0.8 -3)\times 10^{9}\msun$. Those galaxies have very high values of s$\sfr / {\rm sMAR}\sim6-10$. Moreover, 
these galaxies spent a considerable amount of time having large values of 
s$\sfr / {\rm sMAR}$ -- of the order of few Gyrs. Then, 
the halo star formation efficiency decreases again for progenitors at $z\sim0$ 
with masses below $\mvir\sim2\times10^{11}\msun$, implying
that, at least at $z\sim0$, the halo mass $\mvir\sim2\times10^{11}\msun$ is ``special". The fact that 
in more massive halos the ratio s$\sfr / {\rm sMAR}$ decreases is not surprising, this is supported by both theoretical and observational studies which
show that they are more likely to host quenched galaxies, as we discussed above. Note, however, that the ratio s$\sfr / {\rm sMAR}$ is {\it not} always increasing as the halo mass decreases, contrary to what one might expect
by extrapolating the trends below $\mvir\sim2\times10^{11}\msun$. This may be an indication that
galaxy formation in halos with $\mvir\lesssim2\times10^{11}\msun$ is somewhat different. 

Our best fitting models can provide constraints on the functional form for the average star formation
histories (SFH) of galaxies. In observational studies, the SFHs are typically assumed to decline exponentially with 
time, but we find that our results do not support this assumption. Figure \ref{SFH_laws} shows the SFH  for progenitors with 
$\mvir=10^{11},10^{11.5},10^{12},10^{13},$ $10^{14}$ and $10^{15} $\msun. Note that the galaxy SFHs are more complex
that just a declining exponential model.  Note, however, that our results cannot rule out that the SFHs of individual galaxies
may be exponentially declining and we emphasise that in this paper we are constraining average SFHs.
Here we opted to fit the SFHs to an alternative model. In the left panel of Figure 
 \ref{SFH_laws} we present the best fitting models when SFHs are based on a lognormal function:
\begin{equation}
\sfr(t) = \frac{A_{\rm SFR}}{\sqrt{2\pi\tau^2}t}  \exp\left[{-\frac{\left(\ln t/t_0\right)^2}{2\tau^2}}\right],
\label{SHFlognorm}
\end{equation}
 where $t_0$ represents a formation
 epoch and $\tau$ is the width of the function, while the right panel 
 shows the same but when using a delayed$-\tau$ model:
\begin{equation}
\sfr(t) = A_{\rm SFR} \left(\frac{t -t_0}{\tau^2}\right)\exp\left[-\frac{t- t_0}{\tau}\right],
\label{Delayedtau}
\end{equation}
 where the parameters $t_0$ and $\tau$ have similar interpretations as above. A motivation for the 
lognormal model is that \citet{Gladders+2013, Abramson+2016,Diemer+2017} noted that the CSFR is well fitted by a lognormal model, suggesting that it could also
 describe the SFHs of galaxies. The exponentially declining $\tau$ model was introduced in pioneer works of galaxy evolution \citep[e.g.][]{Tinsley1972},
who assumed a fixed reservoir of gas that would be gradually exhausted due to in-situ star formation. Finally, the delayed-$\tau$ models 
were considered in more recent observational studies of SFR vs. \ms\ at different redshifts \citep[e.g.,][]{Noeske+2007b}.
We find the best fit parameters of the above functions by using Powell's direction set method in multi dimensions \citep{Press+1992} 
for minimization, using as constraint  the values of the SFRs of the progenitors described above. Tables \ref{T3} and  \ref{T4} list 
respectively the best fitting parameters for a lognormal and delayed$-\tau$ SFHs models.  
Our results show that the delayed$-\tau$ model describes reasonably well the SFHs of galaxies in low-mass halos, $\mvir<10^{12}$ \msun.
For halos with masses of $10^{12}$ \msun\ and larger, this model cannot capture the strong decay in SFR after the maximum seen in our 
results, especially in the most massive halos.  The lognormal model describes better the SFRs of galaxies in these halos at intermediate redshifts, but 
for the most massive halos it is unable to describe the strong SFR decay.

\begin{table}
	\caption{Best fit parameters for SFHs based on a lognormal model, see Equation (\ref{SHFlognorm}).}
	\begin{center}
		\begin{tabular}{c c c c c c}
			\hline
			\hline	
			$\mvir(z=0)$ & $A_{\rm SFR}$ [$10^{9}$\msun] & $\tau$ &  $t_0$ [Gyrs] \\
			\hline
			\hline
			 $\mvir=10^{11}\msun$ & 1.247 & 0.771 & 11.793\\
			\hline
			 $\mvir=10^{11.5}\msun$ & 12.643 & 0.404 & 9.632\\
			\hline
			 $\mvir=10^{12}\msun$ & 56.639 & 0.354 & 7.120\\
			\hline
			 $\mvir=10^{13}\msun$ & 177.245 & 0.383 & 4.101\\
			\hline
			 $\mvir=10^{14}\msun$ & 435.354 & 0.538 & 3.755\\
			\hline
			 $\mvir=10^{15}\msun$ & 577.258 & 0.584 & 3.504\\
			\hline
			\hline
		\end{tabular}
		\end{center}
	\label{T3}
\end{table}

\begin{table}
	\caption{Best fit parameters for SFHs based on a Delayed$-\tau$ model, see Equation (\ref{Delayedtau}).}
	\begin{center}
		\begin{tabular}{c c c c c c}
			\hline
			\hline	
			$\mvir(z=0)$ & $A_{\rm SFR}$ [$10^{9}$\msun] & $\tau$ [Gyrs] &  $t_0$ [Gyrs] \\
			\hline
			\hline
			 $\mvir=10^{11}\msun$ & 1.211 & 6.476 & 0.708\\
			\hline
			 $\mvir=10^{11.5}\msun$ & 29.978 & 10.452 & 0.956\\
			\hline
			 $\mvir=10^{12}\msun$ & 87.837 & 5.734 & 0.932\\
			\hline
			 $\mvir=10^{13}\msun$ & 228.725 & 2.737 & 0.723\\
			\hline
			 $\mvir=10^{14}\msun$ & 490.11 & 2.207 & 0.624\\
			\hline
			 $\mvir=10^{15}\msun$ & 604.758 & 1.891 & 0.562\\
			\hline
			\hline
		\end{tabular}
		\end{center}
	\label{T4}
\end{table}

 Finally, in Figure \ref{sharc} we color-code the ratio $R_{\rm SHARC} \equiv (\frac{\dot{M}_*}{\dot{M}_{\rm vir}})/(\frac{\partial M_*}{\partial M_{\rm vir}})$.  In \citet{Rodriguez-Puebla+2016a} 
 we showed that, on average, star-forming galaxies must have star-formation rates satisfying 
 SFR $ = \dot{M}_* / (1-R)= (\frac{\partial M_*}{\partial M_{\rm vir}})\times\dot{M}_{\rm vir} / (1-R)$, which we called stellar-halo accretion rate co-evolution (SHARC), 
 in order to be consistent with the SHMR being nearly redshift independent up to $z\sim4$. We showed that if this is true galaxy-by-galaxy for most star-forming galaxies, 
 the dispersion of the MAR in the Bolshoi-Planck simulation predicts the observed dispersion in the SFR on the Main Sequence of galaxies.  
 The SHARC assumption is equivalent to $R_{\rm SHARC} = 1$, and Figure \ref{sharc} shows that this is a good approximation ($R_{\rm SHARC} \approx 1$ to 2) 
 for star-forming galaxies (a majority of those below the quenching transition marked by the dashed lines, and a decreasing fraction of those above it).

\begin{figure*}
	\vspace*{-70pt}
	\includegraphics[height=4.6in,width=3.4in]{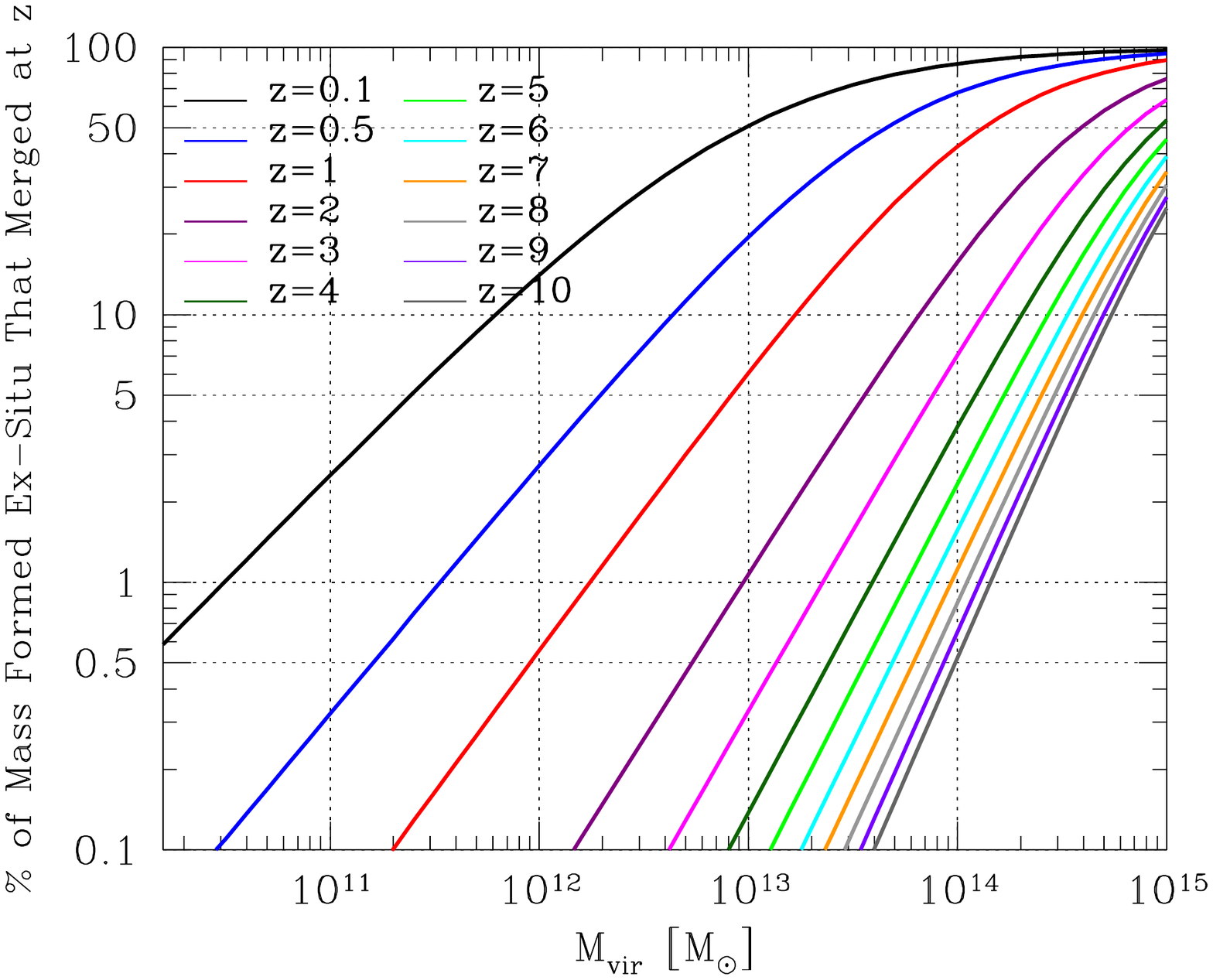}
	\includegraphics[height=4.6in,width=3.4in]{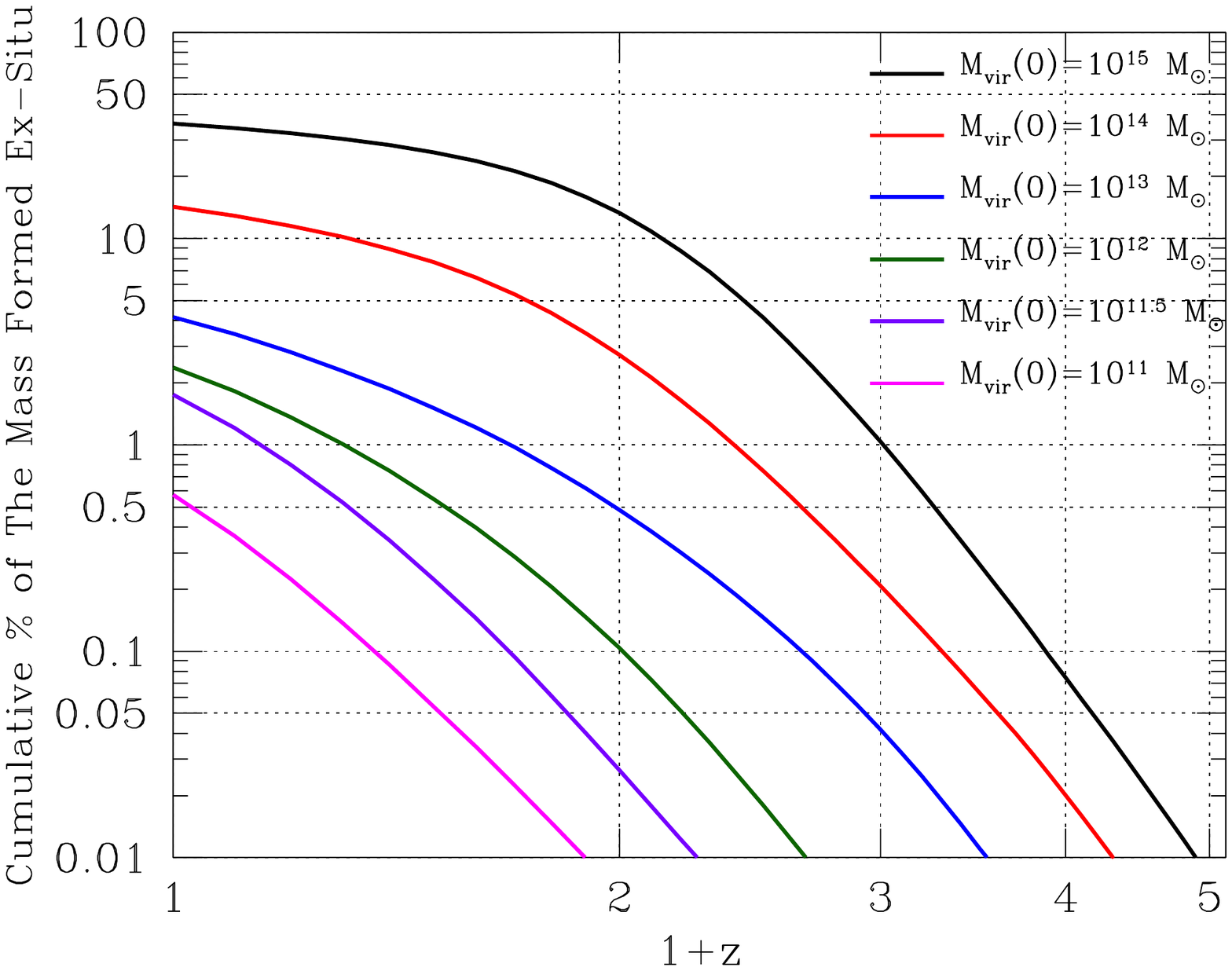}
	\vspace*{-70pt}
		\caption{{\bf Left Panel:} Instantaneous fraction of mass that formed ex-situ and was accreted by galaxy mergers as a function of the halo mass at redshift $z=0$. 
		 {\bf Right Panel:} Cumulative fraction of mass that formed ex-situ and accreted through galaxy mergers. Note that $\sim40\%$ of the final mass
		 in host galaxies of halos with $\mvir(0) = 1\times 10^{15}$ was accreted by galaxy mergers. 	 
 	}
	\label{merger_mass_fraction}
\end{figure*}

\begin{figure*}
	\vspace*{-280pt}
	\includegraphics[height=8.1in,width=6.7in]{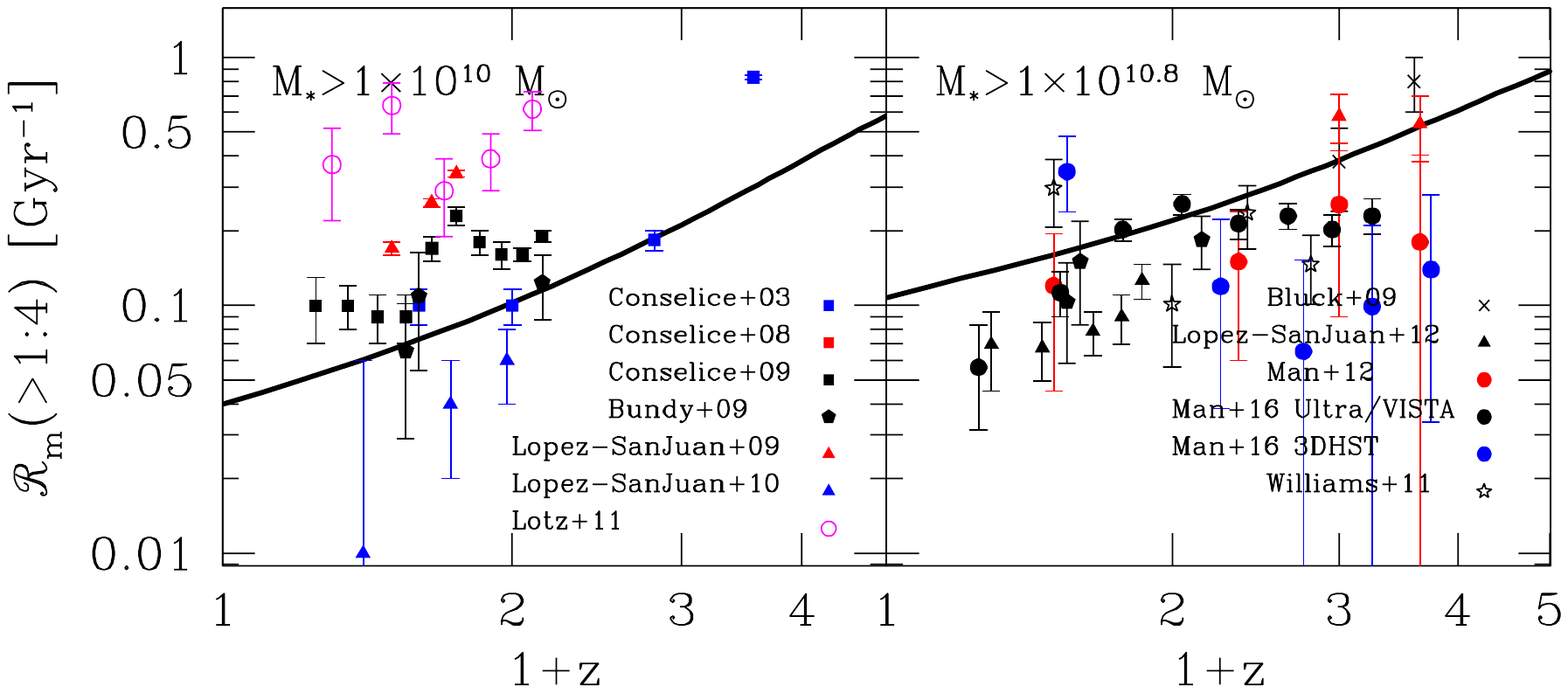}
	\vspace*{-100pt}
		\caption{{\bf Left Panel:} Galaxy major merger rate for galaxies with masses above $1\times10^{10}\ms$. Solid lines show
		the predictions based on our new \shmr\ while the different symbols show observational estimates from \citet{Conselice+2003,Conselice+2008,Conselice+2009,LPS+09}
		and \citet{LPS+10} based on galaxy asymmetries while \citet{Bundy+2009} gives the merger rate fraction from galaxy pairs. 
		{\bf Right Panel:} Similarly above but for galaxies with masses above $1\times10^{10.8}\ms$. Symbols show data from \citet{Bluck+2009} using 
		galaxy asymmetries, \citet{LPS+12,Man+2012,Man+2016}, \citet{Williams+2011} based on galaxy pairs and \citet{Lotz+2011} using the data 
		from \citet{Lotz+2008} based on the $G-M_{20}$ identification
		technique.   
		}
	\label{merger_rate}
\end{figure*}

\begin{figure}
	\vspace*{-110pt}
	\hspace*{-10pt}
	\includegraphics[height=7.2in,width=6in]{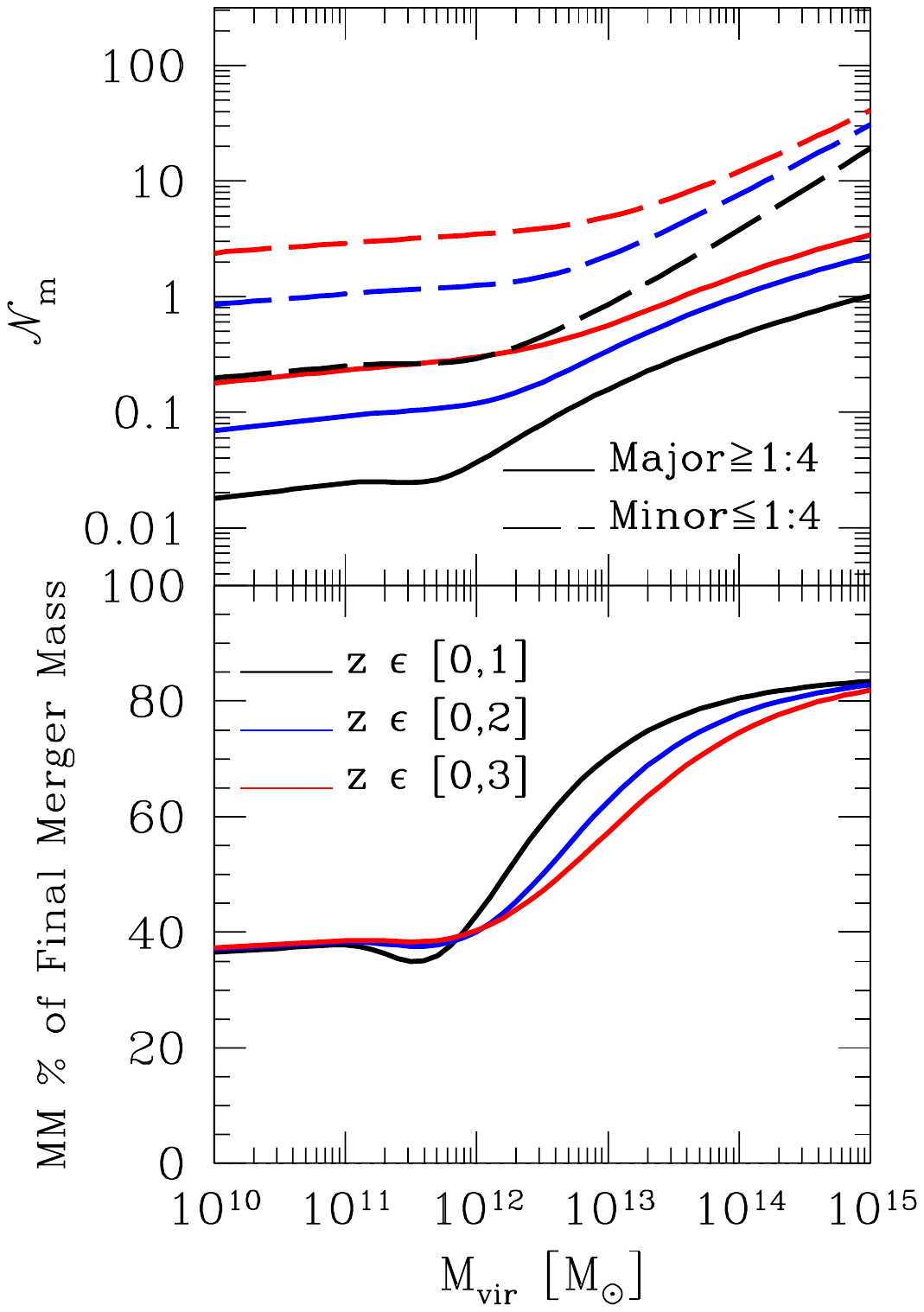}
	\vspace*{-110pt}
	\caption{ {\bf Upper Panel:} Number of major ($\mu_*\geq1/4$, solid lines) and minor ($\mu_*\leq1/4$, dashed lines) mergers as a function of halo mass 
	since $z=1$, 2 and 3 as indicated by the labels. {\bf Bottom Panel:} Fraction of final mass acquired through galaxy major mergers since  $z=1$, 2 and 3.
	While minor mergers are more frequent in massive galaxies, most of their accreted mass was acquired by major mergers.
	 }
	 \label{Nmerger}
\end{figure}

\subsection{Galaxy Mergers}
\label{Mergers}

As discussed previously, galaxies can build their masses via 
in-situ star formation and/or through the accretion of ex-situ stars from other galaxies in mergers. 
As described in Section \ref{SFH_model}, our model parameterizes the amount of stellar mass that 
galaxies formed via in-situ star formation as a function of $z$, denoted by \fmerger. The complement 
to this function, $f_{\rm ex~situ} = 1 - \fmerger$, is simply the fraction of mass in stars that were accreted via galaxy mergers. The percentage of ex-situ accreted stellar mass is presented in 
the left Panel of Figure \ref{merger_mass_fraction} as a function of halo mass and redshift. The instantaneous merger contribution to the stellar mass growth 
increases with mass at all epochs. For example,  at $z\sim0.1$ the growth of the stellar mass
in halos above $\mvir\sim2\times10^{13}\msun$ is dominated by galaxy mergers. Observe that at $z\sim0.1$ around $\sim15\%$
of the stellar mass growth in a Milky-Way sized halo is via mergers, while for smaller halos with 
$\mvir\sim10^{11}\msun$ this fraction is $\sim2.5\%$. The contribution of galaxy mergers declines 
strongly with redshift, and at $z\sim1$ halos around $\mvir\sim10^{14}\msun$ are assembling around $\sim40\%$
of their mass through mergers. 

The right Panel of Figure \ref{merger_mass_fraction} shows the average cumulative fraction of the stellar mass that was formed ex-situ. 
At $z\sim0$ the fraction of mass from galaxy mergers is $\sim36\%$, $\sim14\%$, $\sim4\%$, $\sim2.4\%$ and 
$\sim1.8\%$ for halos with $\mvir=10^{15}\msun$, $10^{14}\msun$, $10^{13}\msun$, 
$10^{12}\msun$ and  $10^{11}\msun$, respectively. Note that at $z\sim1$ the cumulative fraction of 
stellar mass that was accreted by mergers is $\sim13\%$ for the progenitor of a  $\mvir=10^{15}\msun$ halo at $z=0$,
while this is $\sim3\%$ for the progenitor of a $\mvir=10^{14}\msun$ halo. In other words, most of the mass gained
via galaxy mergers in high mass halos was from $z\leq1$. Interesting enough, below $z\sim1$ high mass halos 
have little or no star formation. Also, mergers do not appear to be responsible for the increase in the SFR of Milky-Way
sized halos that suddenly happened around $z\sim1$ as we noted in Figure \ref{sfreff}. According to Figure \ref{merger_mass_fraction} most of the mass assembly 
through mergers happened only very recently; the cumulative fraction was never higher than $\sim3\%$. Moreover, typical Milky-Way sized halos 
never experienced a major merger since $z\sim1$. Thus we conclude that mostly internal process to the galaxy are responsible for this
sudden enhanced in SFR. 

A relevant question is which type of mergers, either major or minor, are responsible for the ex-situ mass fractions presented above. 
This question has been studied in a number of previous works 
\citep{Zheng+2007b,Stewart+2009,Hopkins+2010a,Hopkins+2010b,Zavala+2012,Avila-Reese+2014,Rodriguez-Gomez+2015,Rodriguez-Gomez+2016}. 
Here, we introduce a simple model to study the impact of galaxy mergers as a function of mass and time. For this model, 
we use the results of halo merger rates as measured from $N-$body high resolution simulations convolved through the evolution of the
SHMR. Before continuing the description of our model, 
it is important to realize that not all the merged satellite's stars may necessarily end up in the 
central galaxy. 
This is because 1) some fraction of stars can be ejected from the halo if their escape velocities are
large enough and, 2) due to the disruption of satellite galaxies, i.e., those that do not merge with the central galaxy but are instead tidally
destroyed inside the halo. As a result, some central galaxies, especially the most massive ones, will be 
surrounded by a diffuse stellar structure.  In the literature this is typically referred as intra cluster light (ICL). Note that even with the most modern
instruments in telescopes, this diffuse stellar structure is very challenging to observe and typically this is not counted as part of the mass of the central
galaxy.  In our model, we use the subhalo destruction rate to define the galaxy merger rate. Measuring the disruption rate in simulations is not trivial as the measurements
will depend on the definition and on subhalo completeness. Normally, disruption is defined when subhalos
lose a significant amount of their original mass at the infall.\footnote{For example \citet{Stewart+2009}, define a merger when a subhalo
loses 90\% of its mass at accretion.} Thus there are two scenarios in which subhalos are counted as disruptions, either by being destroyed into
the halo at some radii or merged with the central galaxy of the halo. This could lead to a potential conflict when interpreting galaxy mergers 
through the subhalo destruction rate. The reason is that in cases in which the subhalo destruction happens at large radius it would lead to an overestimation 
of the galaxy merger rate since it will take some time for the host satellite to merge with the central galaxy; indeed, for some cases this time could be larger than the Hubble time.
Note, however, that we are not taking into account the ``true'' time that it would take the host satellite to merge with the central galaxy. 
Therefore, when comparing our predicted merger rate with observations one should keep in mind that these results 
could represent an upper limit to the true merger rates of galaxies. Nevertheless, as shown by \citet{Wetzel+2010} the disruption of subhalos occurs mainly 
in the inner regions of the halo where the tidal forces are strongest, 
while only a small fraction are disrupted at larger radii. 

We calculate galaxy mergers by convolving subhalo disruption rates with the evolution of the SHMRs:
	\begin{eqnarray}
		\frac{d\mathcal{R}_{\rm m}}{d\log\mu_*}(\mu_*|\mvir,z) = \int P(\mu_*|\mpeak,z)  \times  & &  \nonumber \\
		\frac{d\mathcal{R}_{\rm m}}{d\log\mu_{\rm peak}}(\mu_{\rm peak}|\mvir,z)\times d\log \mu_{\rm peak},
	\end{eqnarray}
where $d\mathcal{R}_{\rm m}/d\log\mu_{\rm peak}$ is the disruption rate per host halo per logarithmic interval in subhalo peak mass to primary halo mass
and per unit of redshift. We use the fitted relation from \citet{Behroozi+2013}. The distribution $P(\mu_*|\mpeak,z)$ is given by Equation \ref{P_ms_X}
where $\mu_*$ is the observed satellite-to-central galaxy stellar mass ratio in a halo of mass \mvir. Note that the above distribution includes 
uncertainties due to random errors from stellar masses in addition to the intrinsic scatter of the SHMR. 

Figure \ref{merger_rate} shows the galaxy major merger rates calculated using our SHMRs and compared to observations. In order to compare directly to our
model predictions, we compiled estimations of galaxy major merger from the literature based on stellar mass thresholds samples. The mass thresholds
indicated in the upper part of both panels reflect the fact that we adjusted stellar masses to a Chabrier IMF.  Observational
reports from close pairs as well as from measurements using asymmetric features in galaxies are plotted in Figure \ref{merger_rate}. 
We also adjusted galaxy merger rates
by using the cosmologically averaged observability timescales from \citet{Lotz+2011}. All samples were selected to have major
mergers defined as $\mu_*\geq 1 / 4$.  

In general, observations are consistent with our results especially for lower mass galaxies. This is encouraging given the uncertainties both 
in observations and for our model predictions. Thus, we can conclude that {\it our SHMR constrained by our semi-empirical modelling in combination with subhalo
merger rates is roughly able to account for observed galaxy merger rates.} The 
possible disagreement at large masses might just be reflecting the fact that our definition of galaxy mergers is related to the disruption rate of subhalos rather
than the ``true" central mergers. Recall that  we did not take into account the dynamical friction time for galaxies in subhalos that were destroyed at larger radius inside
of the host halo. Since this time is directly proportional to the virial circular velocity of the host halo, one expects that the  overestimation in the merger rate will be 
larger for the more massive halos. This leaves a window for
decreasing the merger rate from abundance matching but still being consistent with the data, and thus leaves some room to improve the study of mergers using our empirical model. 
 
 The number of mergers experienced by the progenitors described above from $z=0$ to $z=1$, $2$ and $3$ are presented in the upper panel of Figure \ref{Nmerger}.
 The solid lines indicate the major mergers, $\mu_*\geq 1 / 4$, while dashed lines are for the minor mergers, $\mu_*\leq 1 / 4$. The black, 
 blue and red colors are for all the mergers that happened since $z=1$, $z=2$ and $z=3$, respectively. The number of mergers is given by integrating
 the merger rate over the assembly history of the progenitor
	\begin{eqnarray}
		\mathcal{N}_{\rm m} (>\mu_*|M_{\rm vir,0},  z_0)= & &  \nonumber \\ 
		\int_{\mu_{*}}^{1} \int_{z_0}^{z} \frac{d\mathcal{R}_{\rm m}}{d\log\mu_*}(\mu_*|\mvir(z|z_0),z) dz d\log\mu_* .
	\end{eqnarray}
Figure \ref{Nmerger} shows that minor mergers happened at least an order of magnitude more frequently than major mergers. 
Galaxies in halos with present day masses $\mvir=10^{15}\msun$ experienced, on average, $\sim5$, $\sim3$, and $\sim1.5$  major mergers since $z=3$, $z=2$ and $z=1$. 
Galaxies in halos with present day masses  $\mvir=10^{14}\msun$ experienced, on average,  $\sim2$, $\sim1.5$ and $\sim0.5$ major merger since $z=3$, $z=2$ and $z=1$ while
galaxies in halos below $\mvir=10^{13}\msun$ probably 
never had a major merger since $z=1$ and $z=2$ but only one major merger since $z=3$. In contrast, most of the galaxies have suffered at least a 
minor merger since $z=2$ and at least a few since $z=3$, while central galaxies in massive halos had tens of minor mergers. 

The fact that minor mergers 
dominate over major ones is perhaps not surprising given 1) the hierarchical nature of the LCDM cosmological scenario 
and 2) the fact that the number density of low-mass galaxies is much higher than that of high-mass galaxies means that the chance of 
having minor mergers is larger. Note, however, that minor mergers do not always contribute significantly to the total stellar mass assembled by mergers.  
The bottom panel of Figure \ref{Nmerger} demonstrates this. In Milky-Way sized halos the contribution of major mergers is $\sim35\%$ of the total stellar mass acquired by mergers, while
for halos with $\mvir\geq10^{14}\msun$ the fraction asymptotes to $\sim75\%$. There is no doubt that major mergers are an important part 
of the formation history of the 
galaxies, and for that reason we will discuss further the above results in a more general context in Section \ref{Discussion_section}.

\section{The Average Structural Evolution of Galaxies} 
\label{Structural_Evolution}

The average stellar mass trajectories and star formation histories obtained in the previous sections
allow us to explore several implications that can be naturally studied in our framework. 
Our goal for this section is to study the structural evolution of the galaxies
and its relation to the properties derived above, namely, SFRs and galaxy mergers. 

\begin{figure*}
	\vspace*{-270pt}
	\includegraphics[height=8.9in,width=7.3in]{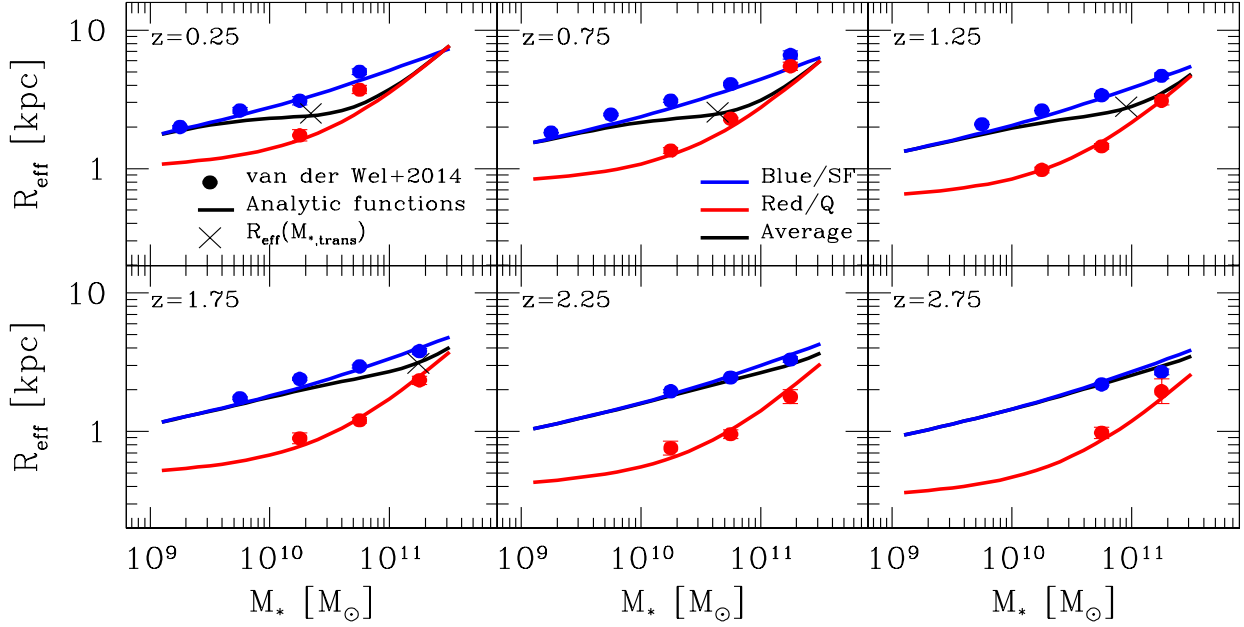}
	\vspace*{-130pt}
		\caption{ Circularized effective radius for blue star-forming galaxies and
		red quiescent galaxies for six different redshift bins. The filled circles show the 
		circularized effective radius as a function of stellar mass and redshift  from \citet{vanderWel+2014}
		based on multiwavelength photometry from the 3D-HST survey
		and HST/WFC3 imaging from CANDELS. Solid lines show
		the redshift dependence for blue and red galaxies of the local
		relation by  \citet{Mosleh+2013} based on the MPA-JHU SDSS DR7. The black solid
		lines show the average circularized effective radius as a function of stellar mass. The
		crosses show the effective radius at $M_{50}$, i.e., the stellar mass
		at which the observed star-forming fraction of galaxies is equal to the quenched fraction of galaxies. 
		Note that the effective  radius at $M_{50}$ evolves very little with redshift and is $\sim3$ kpc. 
		We utilize the plotted redshift dependences as an input to derive the average 
		galaxy's radial mass distribution as a function of stellar mass by assuming that
		blue/star-forming galaxies have a S\`ersic index $n=1$ while red/quenched galaxies
		have a S\`ersic index $n=4$ (see text for details).
 	}
	\label{reff_ms_z}
\end{figure*}

\begin{figure*}
	\vspace*{-270pt}
	\includegraphics[height=8.9in,width=7.3in]{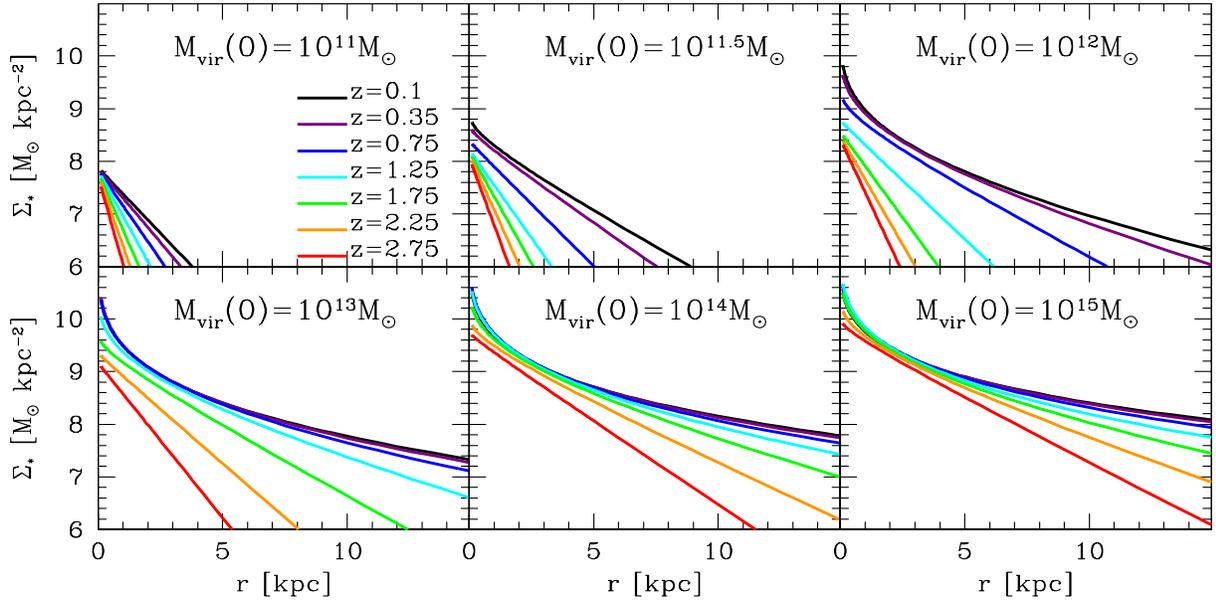}
	\vspace*{-130pt}
	\caption{Average evolution of the radial distribution of stellar mass for galaxies in
	halo progenitors with $\mvir=10^{11},10^{11.5},10^{12},10^{13},$
		$10^{14}$ and $10^{15} $\msun at $z=0$. These radial distributions can be 
		imagined as stacking all the density profiles of galaxies at a given virial mass and $z$, no matter whether galaxies 
		are spheroids or disks or a combination of both.
 	}
	\label{surface_evol}
\end{figure*}

\begin{figure*}
	\includegraphics[height=2.2in,width=7.2in]{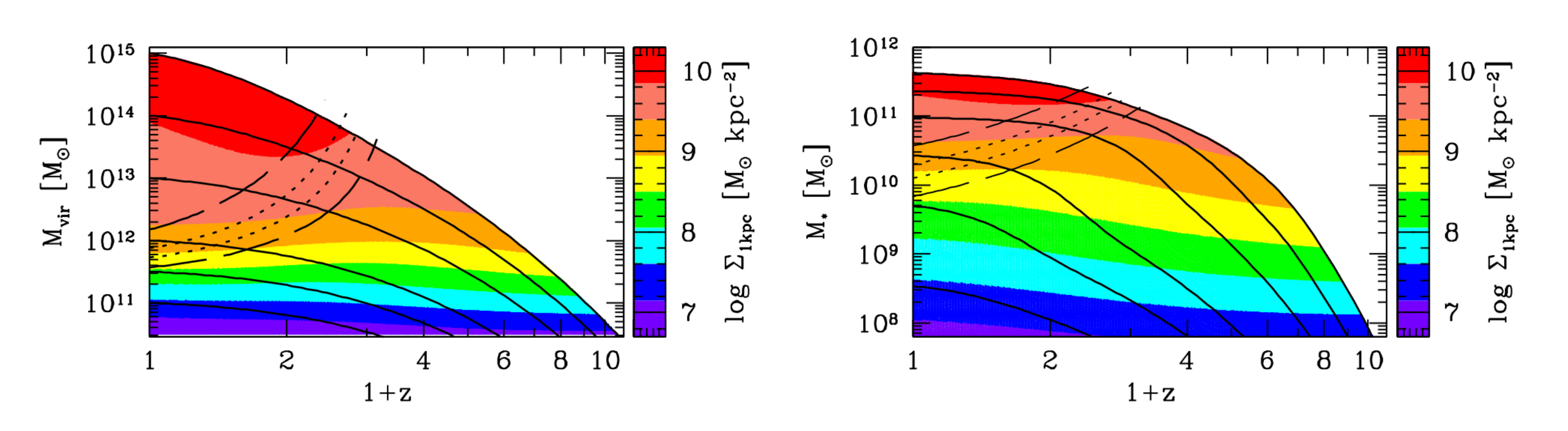}
		\caption{Integrated mass density at 1 kpc, as a function of halo
		mass  (left panel) and stellar mass (right panel) for halo progenitors. 
		The black solid lines show the trajectories for progenitors with  $\mvir=10^{11},10^{11.5},
		10^{12},10^{13},$
		$10^{14}$ and $10^{15} $\msun at $z=0$. The short-dashed curves show the stellar mass
		and halo mass at which the observed fraction of star-forming galaxies is equal to the quenched fraction of
		galaxies, and the upper (lower) long-dash curves show the stellar mass vs. $z$ where 75\% (25\%) 
		of the galaxies are quenched. }
	\label{sigma_eff}
\end{figure*}

Constraining observations that relate galaxy stellar mass to its structural 
properties are the S\`ersic index $n$ and the effective radius \reff,\footnote{The effective radius 
is defined as the 
radius that encloses half the luminosity or stellar mass of the galaxy.} both at $z=0$ and at higher redshifts 
\citep[see e.g.,][]{Bell+2012,Patel+2013,vanderWel+2012}. 
For a S\`ersic law $r^{1/n}$, 
the S\`ersic index $n$ is a parameter that controls the slope of the curvature for the radial distribution of
light/mass. Observational results based on
the SDSS have shown that when fitting the global light distribution to a S\`ersic law $r^{1/n}$ 
most of the galaxies have indices $n$ between 0.5 and 8 
\citep[see e.g.,][]{Simard+2011,Meert+2015}. However, when dividing galaxies
into two main clases -- e.g., as early- and late-types -- the radial distribution of
the light/mass is fairly well described with $n=4$ \citep{deVaucouleurs1948} and $n=1$ (exponential disc) respectively. 
In this section, we will assume for simplicity that all late-type galaxies are blue/star-forming systems 
with a S\`ersic index $n=1$ while all early-type morphologies correspond to red/quiescent galaxies
with S\`ersic index $n=4$. Hereafter, we will use these galaxy classifications interchangeably. 
While this is an oversimplification of a more complex reality, for our purpose it is accurate enough
since the relatively small fraction of galaxies that do not follow the above assumptions is not critical
for our conclusions. 

We begin by writing the explicit form of the \citet{Sersic1963} law $r^{1/n}$:
\begin{equation}
\Sigma(r) = \Sigma_0 \exp\left[-b_n\left( \left(\frac{r}{\reff}\right)^{1/n} -1 \right)\right],
\label{surface_mass}
\end{equation}
where $\Sigma_0$ is the surface mass density at the effective radius $\reff$ 
and $b_n$ has a value of $b_1 \approx 1.678$ and $b_4 \approx 7.669$ for $n = 1$ and $n = 4$
respectively. Therefore once $n$ is defined, in order to fully characterize the surface density profile 
we need to specify the two free parameters $\Sigma_0$ and \reff. 
The mass profile is given by
\begin{equation}
\ms(<r) = 2\pi \Sigma_0 \int_0^r \exp\left[-b_n\left( \left(\frac{r}{\reff}\right)^{1/n} -1 \right)\right] r dr.
\end{equation}
Notice that when $r\rightarrow\infty$ the total stellar mass \ms\ becomes a function of 
 $\Sigma_0$  and of \reff, $\ms = 2\pi \Sigma_0 F_n(\reff)$. 
The sizes of the galaxies are known to correlate with their total stellar masses. This correlation
has been observed to be shallower for late-type morphologies than for more early-type ones, 
not only at $z\sim0.1$ \citep[see e.g.][]{Shen+2003} but also at higher redshifts \citep[see e.g.][]{vanderWel+2014}. 
Technically, 
if we introduce a relation between \reff\  and \ms, $\reff = \reff(\ms,z)$, then Equation (\ref{surface_mass}) 
is completely determined and the surface mass density at the effective radius is given by  $\Sigma_0$:
$\Sigma_0  = \ms / 2\pi F_n(\reff(\ms))$. Note that the same line of reasoning can be applied at any redshift. 

As mentioned earlier, our main goal is to study the structural evolution of the galaxies and its relation to
the galaxy properties derived in the previous sections, that is, SFRs and galaxy mergers. Moreover,
we will use the average stellar mass trajectories constrained above to predict the average trajectories 
for structural evolution of galaxies. Recall that our results are based on the whole galaxy population
and there is not an explicit distinction between different populations, i.e., they only depend on mass.
The structure of galaxies at a fixed mass can be different as discussed already in the previous 
paragraph.  But we note that there is a remarkable similarity in the surface density profiles of both early- 
and late-type galaxies when their profiles are rescaled to a fixed fraction of their virial radii, at least at low redshifts \citep{Kravtsov2013,Somerville2017}.
Here we adopt a more probabilistic description by considering the contribution 
of these two populations to the average, that is, we will compute the 
stellar mass surface density by averaging over the two main populations---i.e. the early- and late-types, or equivalently quenched and star-forming galaxies---and thus obtain average 
density profiles that depend only on \ms\ and $z$. This procedure can be 
imagined as stacking all the density profiles of galaxies in a given mass bin, 
no matter whether they are spheroids or disks or a combination of both (for a recent
and similar idea see \citep{Hill+2017}). 
The average  radial distribution of galaxies with total stellar mass \ms\ is then
given by
\begin{equation}
\Sigma(r,\ms) = f_{\rm SF} (\ms)\Sigma_{\rm SF}(r,\ms)
+ f_{\rm Q}(\ms)\Sigma_{\rm Q}(r,\ms),
\end{equation}
where for simplicity we have omitted the dependence on $z$. 
The fraction of blue/star-forming and red/quenched galaxies was discussed 
in Section \ref{Fraction_SF_Q}. Additionally, 
recall that for blue/star-forming galaxies we assume $n=1$ and
for red/quenched galaxies $n=4$. 
For galaxy sizes we adopt the $\reff-\ms$
relations derived in 
\citet{Mosleh+2013} for nearby, $z\sim0.015$, blue and red galaxies.
They constructed a sample based on the MPA-JHU SDSS DR7
by selecting spectroscopic galaxies with a surface brightness limit of
$\mu_{50,r}\leq23$ mag arcsec$^{-2}$. These authors computed the half-light 
radius of galaxies by measuring directly the radius at which the flux reaches 
half of its total value. Their reported effective radii are circularized.  In order
to use a model to higher redshifts, 
we use the following redshift dependence for blue/star-forming galaxies:
\begin{equation}
R_{\rm eff,SF}(z,\ms) = R_{\rm eff,SF}(0,\ms) \left (\frac{H(z)}{H(0)}\right)^{-0.5},
\end{equation}
while for red/quenched galaxies we use
\begin{equation}
R_{\rm eff,Q}(z,\ms) = R_{\rm eff,Q}(0,\ms) \left (\frac{H(z)}{H(0)}\right)^{-0.85}.
\end{equation}
Here $R_{\rm eff,SF}(0,\ms)$ and $R_{\rm eff,Q}(0,\ms)$ are the local relations
derived in \citet{Mosleh+2013} and $H(z)$ is the redshift dependence of the Hubble parameter. 
We choose these redshift dependencies since, as
can be seen in Figure \ref{reff_ms_z}, they are consistent
with the size-mass evolution observed by \citet{vanderWel+2014}, which
constrained the redshift evolution of galaxy sizes up to redshift $z=3$ by 
using multiwavelength photometry from the 3D-HST survey \citep{Brammer+2012} 
and HST/WFC3 imaging from CANDELS \citep{Grogin+2011, Koekemoer+2011}. 
Figure \ref{reff_ms_z} compares our adopted redshift dependences with the 
circularlized half-light radius from \citet{vanderWel+2014}. More recently, \citet{Allen+2016} 
determined the  $\reff-\ms$ relation of star-forming galaxies from $z\sim1$ to $z\sim7$ by using the ZFOURGE 
survey cross-matched with CANDELS and the HST/F160W imaging. 
Their results agree with those of \citet{vanderWel+2014}, and for $z>3$ we note that their 
size--mass relation is consistent with our implied relation. 

\begin{figure*}
	\vspace*{-260pt}
	\includegraphics[height=8.1in,width=6.7in]{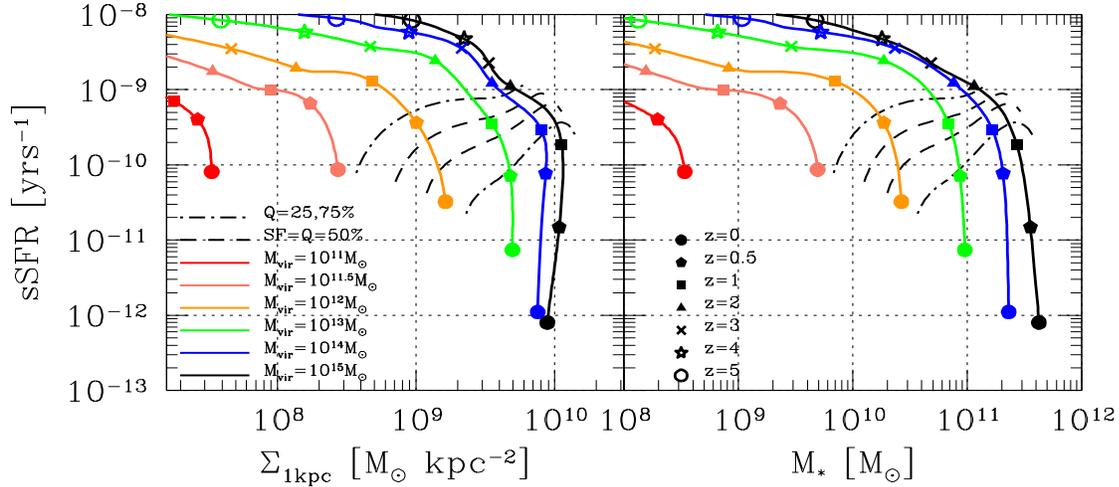}
		\vspace*{-110pt}
		\caption{{\bf Left Panel:} Trajectories for progenitors of halos with $\mvir=10^{11},10^{11.5},10^{12},10^{13},10^{14}$ and $10^{15} $\msun\ at $z=0$ 
		in the $\Sigma_1-$sSFR plane. {\bf Right Panel:} Same progenitors but in the $\ms-$sSFR plane.  The symbols
		show different redshifts as indicated by the labels.  The dashed curves show $M_{50}(z)$ below which half the galaxies are quiescent, and the upper (lower) dot-dashed curves show where 25\% (75\%) of the galaxies are quenched. 
 	}
	\label{sigma_sfr_ms}
\end{figure*}

\begin{figure*}
	\vspace*{-100pt}
	\includegraphics[height=8.1in,width=6.7in]{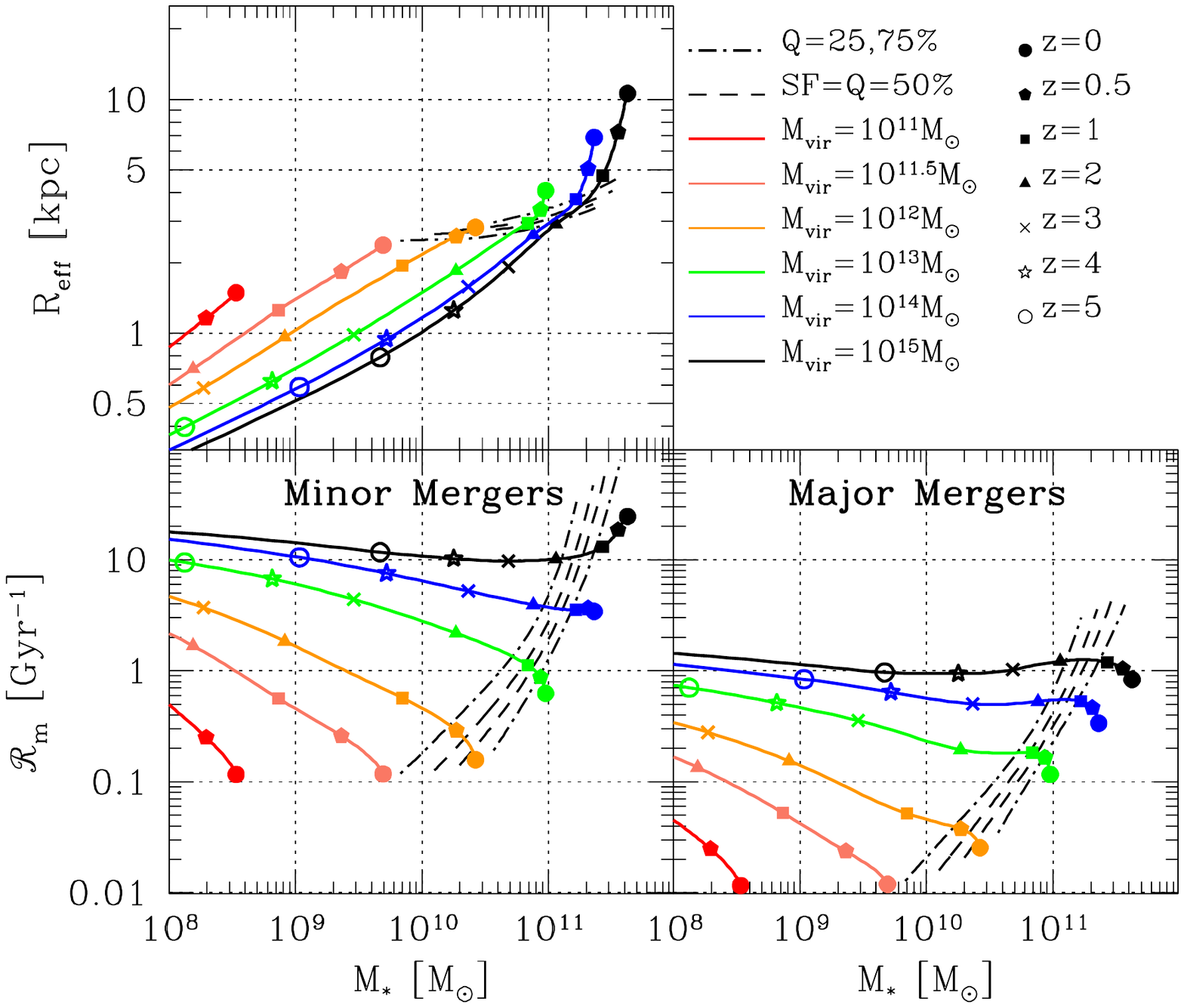}
		\vspace*{-110pt}
		\caption{ Trajectories for progenitors of halos with $\mvir=10^{11},10^{11.5},10^{12},10^{13},10^{14}$ and $10^{15} $\msun\ at $z=0$ 
		in the size-mass relation (upper left panel) and for the minor (bottom left) and major (bottom right) merger rates. As
		in Figure \ref{sigma_sfr_ms},  the symbols show different redshifts as indicated by the labels.
		The dashed lines show $M_{50}(z)$ below which most galaxies are quenched, and the upper (lower) dot-dashed 
		curves show where  25\% (75\%) of the galaxies are quenched. Progenitors of quenched galaxies
		went through two phases, initially they grew in parallel trajectories as star formers in the size-mass relation, but 
		after quenching they evolved much faster in size than in mass, resulting in a steeper relation at low $z$. 
		Presumably the high rate of minor mergers is responsible for this rapid size growth. 
 	}
	\label{reff_mr_sfr}
\end{figure*}

Our goal is to empirically study the impact of the structural properties in the 
evolution of galaxies. We will do so by inferring the average trajectories
for the density profiles of galaxies in halos of a given mass. Consider the
trajectory of a halo with final mass $M_{\rm vir,0}$ hosting a galaxy with
final total stellar mass $M_{*,0}$ at the redshift of observation $z_0$:
$ \ms(z|M_{*,0} , z_0)=\ms(z | M_{\rm vir,0}, z_0)$. Next, we use this relation
to describe the mean structural evolution of galaxies as
\begin{equation}
\Sigma(r, z|z_0) = \Sigma(r, \ms(z|M_{*,0} , z_0)).
\end{equation}

\subsection{Surface Mass Density and Size Evolution}
\label{size_evolution}

Figure \ref{surface_evol} shows the evolution of the radial stellar mass density
for galaxies in different halo progenitors at $z=0$: 
$\mvir=10^{11},10^{11.5},10^{12},10^{13},10^{14}$ and $10^{15} $\msun. 
This figure shows that, in general, galaxies form from the {\it inside out}. 
At high redshifts, most of the galaxies have S\`ersic index $n=1$, as expected, and most of the low
mass galaxies at $z\sim0$ are still $n=1$, while for more massive galaxies, the contribution of the $n=4$ 
component increases as $z$ decreases.
In the most massive halos, 
the stellar mass within $\sim4$ kpc was already in place since 
$z\sim1.25$ for  the $\mvir=10^{13}\msun$ trajectory, and since $z\sim2.5$ for the $\mvir=10^{15}\msun$ trajectory. Notice that 
Figure \ref{sfreff} shows that the star formation efficiency in a progenitor with 
$\mvir=10^{13}\msun$ at $z=0$ is s$\sfr / {\rm sMAR}\sim1$ at $z\sim1$. Recall that galaxies
with star formation efficiencies below $\sim1$ are predominately quenched systems, see 
Section \ref{growth_SFH}. Similarly, for halo progenitors with $\mvir=10^{15}\msun$ we find that 
s$\sfr / {\rm sMAR}\sim1$ at $z\sim2.5$, suggesting that
the structural evolution of galaxies is playing a role in quenching the galaxies 
\citep{Kauffmann+2003a,vanDokkum+2010,Bell+2012,Cheung+2012,Patel+2013,vanDokkum+2014,Barro+2013,Barro+2015,vanDokkum+2015}. The qualitative 
behaviour described above is consistent with previous studies of the structural evolution for progenitors of
massive galaxies at $z\sim0$ by selecting galaxies at  a constant number density
\citep[see e.g.,][]{vanDokkum+2010,vanDokkum+2013,Patel+2013}. 
 
An alternative way to study the role of the structural properties of the galaxies is  
the integrated mass density at some inner radius.  
Based on a sample from the DEEP2/AEGIS survey at $z\sim0.65$, \citet{Cheung+2012}
showed that the integrated stellar mass density within 1 kpc, $\Sigma_1$, shows a tight correlation 
with color, more than any other structural parameter, e.g., S\`ersic index, effective surface brightness, etc. 
This was later confirmed in \citet[][see also \citealp{Bell+2012}]{Fang+2013}  based on a much larger sample using the SDSS DR7. 

The left panel of Figure \ref{sigma_eff} shows  the trajectories 
for halos with  $\mvir=10^{11},10^{11.5},10^{12},10^{13}, 10^{14}$ and $10^{15} $\msun\ at $z=0$
and the color code shows
the corresponding integrated mass density at 1 kpc: $\Sigma_1 = \ms(<r_1) / \pi r_1^2$ with 
$r_1 = 1$ kpc. The right panel of Figure \ref{sigma_eff} shows the same but as a function of stellar mass. 
The short dashed lines show the empirical $M_{50}(z)$ above which most galaxies are quiescent. 
An interesting feature is revealed in this figure. Note that the values of $\Sigma_1$  are nearly constant as a function of
both halo and stellar mass. Since, on average, galaxies and their host halos 
are growing at all times, this implies that the values of $\Sigma_1$ significantly
increase with time as galaxies evolve. 
This directly implies that there is not a universal threshold value for $\Sigma_1$ at which galaxies are 
more likely to be quiescent, instead this threshold increases with $z$. 
As seen in Figure \ref{sigma_eff}, at $z\sim0$ the threshold is around
$\Sigma_1\approx10^9$ \msun kpc$^{-2}$ while at $z\sim2$ it is an order of magnitude
higher, i.e.,   $\Sigma_1\approx10^{10}$ \msun kpc$^{-2}$.
This is consistent  with Figure 2 in \citet{Barro+2015} where the value of $\Sigma_1$ above which the galaxy population
becomes dominated by quiescent galaxies also increases with redshift.

It is instructive to analyze what would be the consequences of a 
constant threshold value for $\Sigma_1$. 
Figure 4 from \citet[][see also \citealp{vanDokkum+2014}]{Fang+2013}  shows that the value of
$\Sigma_1\sim10^{9.5}$ \msun\ kpc$^2$ marks the transition above which  the 
majority of the galaxies are quiescent at $z\sim0$. Adopting 
this value, we would conclude that the halo mass transition above which the star formation
becomes more inefficient is around $\mvir\sim10^{12}\msun$ at all redshifts. As discussed earlier in this paper,
this is the expected mass above which virial shocks become more efficient. Therefore, our finding 
that the threshold value for $\Sigma_1$ (and for \mvir) increases with $z$ suggests that at high redshifts 
other mechanisms should be in play for keeping the $\sfr$ high in halos above $\mvir\sim10^{12}\msun$. 
Moreover, according to our results these mechanisms were so efficient in the centres of the galaxies 
that they had a chance to increase their central stellar densities significantly before being quenched compared to their low redshift counterparts. 
If those mechanisms are not as relevant today as they were at higher redshifts,
then one possibility is that virial shocks are the main mechanism to quench galaxies today. 
We will come back to this in Section \ref{Discussion_section}.

The left panel of 
Figure \ref{sigma_sfr_ms} shows the trajectories in the $\Sigma_1$--sSFR  
plane  for progenitors with 
$\mvir=10^{11}$, $10^{11.5}$, $10^{12}$, $10^{13}$, $10^{14}$, and $10^{15} \msun$ at $z=0$ while the right panel  shows the same
but in the \ms--sSFR plane. The different symbols
in the figure indicate the redshifts and the dashed
lines show $M_{50}(z)$ below which galaxies are quiescent. Interesting conclusions can be obtained
from this figure: {\it i }) There is a tight correlation between the sSFR and $\Sigma_1$, in other words, 
$\Sigma_1$ is an indicator of the global SFR of the galaxy. Similarly to Figure \ref{sigma_eff}, {\it ii}) 
this figure shows that once a galaxy reaches a maximum mass density at 1 kpc, $\Sigma_1$, the SFR is suppressed.  
This confirms previous studies that the evolution of the structural properties of the galaxies played a key role in the quenching
of galaxies 
\citep{Kauffmann+2003,vanDokkum+2010,Bell+2012,Cheung+2012,Patel+2013,vanDokkum+2014,Barro+2013,Barro+2015,vanDokkum+2015}.

The left panel of Figure \ref{sigma_sfr_ms} shows some hints of negative stellar mass evolution within 1 kpc: for halos $\mvir=10^{15}\msun$ we see $\Sigma_1$
evolution of $\sim-0.13$ dex from $z=0$ to $z\sim1$, while for halos $\mvir=10^{14}\msun$ we see 
$\Sigma_1$ evolution of $\sim-0.06$ dex at the same 
redshift range. \citet{vanDokkum+2014} reported similar trends based on the analysis of galaxy sizes from the SDSS, Ultra VISTA and 3D-HST 
surveys. The authors considered three different scenarios: {\it i}) central mass loss due to core-core mergers, {\it ii}) central mass loss due to stellar evolution, and 
{\it iii}) adiabatic expansion due to stellar winds \citep[also considered in][ and referred as the less likely scenario for the mass-size relation]{Damjanov+2009}. 
Unfortunately, our analysis is not detailed enough to decide for one over the others. Nonetheless, given the subtlety of this effect we suspect that mass loss due to stellar evolution (and the associated stellar winds) is the most likely explanation, although more work is needed on this.

The upper left panel in Figure \ref{reff_mr_sfr} shows the trajectories for the halo progenitors discussed above in the
size-mass relation. Note first that star-forming galaxies evolved in parallel tracks in the size-mass relation. The second thing to note is that the progenitors at $z=0$ of quenched galaxies evolved along two very 
different trajectories. When star formation was the dominant mode of evolution 
for progenitors of quenched galaxies at $z=0$, they evolved in a trajectories with a slope of $\sim0.35$. This situation  
changed dramatically when star formation was suppressed:  the slope of the  trajectory in the size-mass relation
now became $\sim 2.5$, implying that the most massive galaxies increased their size by a factor of $\sim3$ after they quenched. 
One of the most popular explanations for this upturn in the size-mass relation is dry minor mergers. Indeed, \citet{Hilz+2013}
showed that dry minor mergers of diffuse satellites embedded in dark matter halos produced slopes  $\sim 2$ consistent with
our findings. In order to investigate this further, we present the merger rate history 
for the progenitors discussed above in the bottom panels of Figure \ref{reff_mr_sfr}. The left panel plots the merger rate histories from minor mergers, $\mu_*\leq1/4$, 
while the right panel is the same but from major mergers, $\mu_*\geq1/4$. 

Similarly to Figures \ref{merger_mass_fraction} and \ref{merger_rate}, Figure \ref{reff_mr_sfr} shows that the
merger rate history is dominated by minor mergers. Moreover, in the case of the most massive halos we do observe an 
upturn in the merger rate approximately at the same epoch when the upturn in the size-mass relation happened due to
quenching. Therefore, we conclude that our semi-empirical results are not in conflict with the minor merger hypothesis. 

The next thing to note, and perhaps the most surprising, is the fact that galaxies with sizes above $R_{\rm eff}\sim 3$ kpc 
are more likely to be quenched, as can be seen by the dashed lines in the upper panel of Figure \ref{reff_mr_sfr} 
(and also in Panel f) of Figure \ref{model_comparison}). 
Recall that the dashed lines mark the $M$ vs. $z$ above/below which galaxies are likely to be quenched/star-forming.  
{\it We believe that the above result does not indicate something fundamental about how galaxies quench, but is 
just a coincidence}.  As we will discuss in Section \ref{structure_quenching},
similar trajectories in the size-mass relation have been reported in previous studies 
\citep[e.g.,][]{Patel+2013,vanDokkum+2013}. Our findings are not in conflict with those 
previous empirical results.  Figure \ref{reff_ms_z} also shows that galaxies with $M_{50}$ have, on average, effective radius of $\sim3$ kpc.  This figure shows that at all redshifts and masses, blue/star-forming galaxies have larger $R_{\rm eff}$ than red/quenched galaxies.  But note that the black curves in Figure \ref{reff_ms_z}, which represent the average effective radii at each stellar mass and redshift, coincide with the red curves for large $M_*$ at low redshifts, where the vast majority of these massive galaxies are quenched.

\section{Discussion}
\label{Discussion_section}

The semi-empirical inferences of galaxy mass and structural evolution presented here refer to the average behaviour 
of the whole galaxy population as a function of halo or stellar mass. 
While this is clearly a simplification, our results provide 
relevant clues to constrain the main processes of galaxy evolution and their dependence on galaxy mass.  
In this Section we will discuss some implications and interpretations of our results.  We start by comparing
them with some previous works.

\subsection{Comparison with Previous Studies}

In this Section we compare our \shmr\ at different redshifts to previous works. 
We divide our discussion into two main comparisons: those studies
that reported stellar mass as function of halo mass, SHMR, and those 
that have estimated the inverse of this relation, $\mvir-\ms$.
The former is typically reported in studies based on statistical approaches, namely indirect methods, as in our case, while 
the latter is more natural for studies based on direct determinations (e.g., weak gravitational lensing, galaxy clustering). 
All results were adjusted
to a Hubble parameter of $h = 0.678$ and to virial halo masses. When required, we adjusted
stellar masses to a Chabrier IMF. 

\subsubsection{Stellar-to-halo mass relationship}

\begin{figure*}
	\vspace*{-150pt}
	\includegraphics[height=8.1in,width=6.7in]{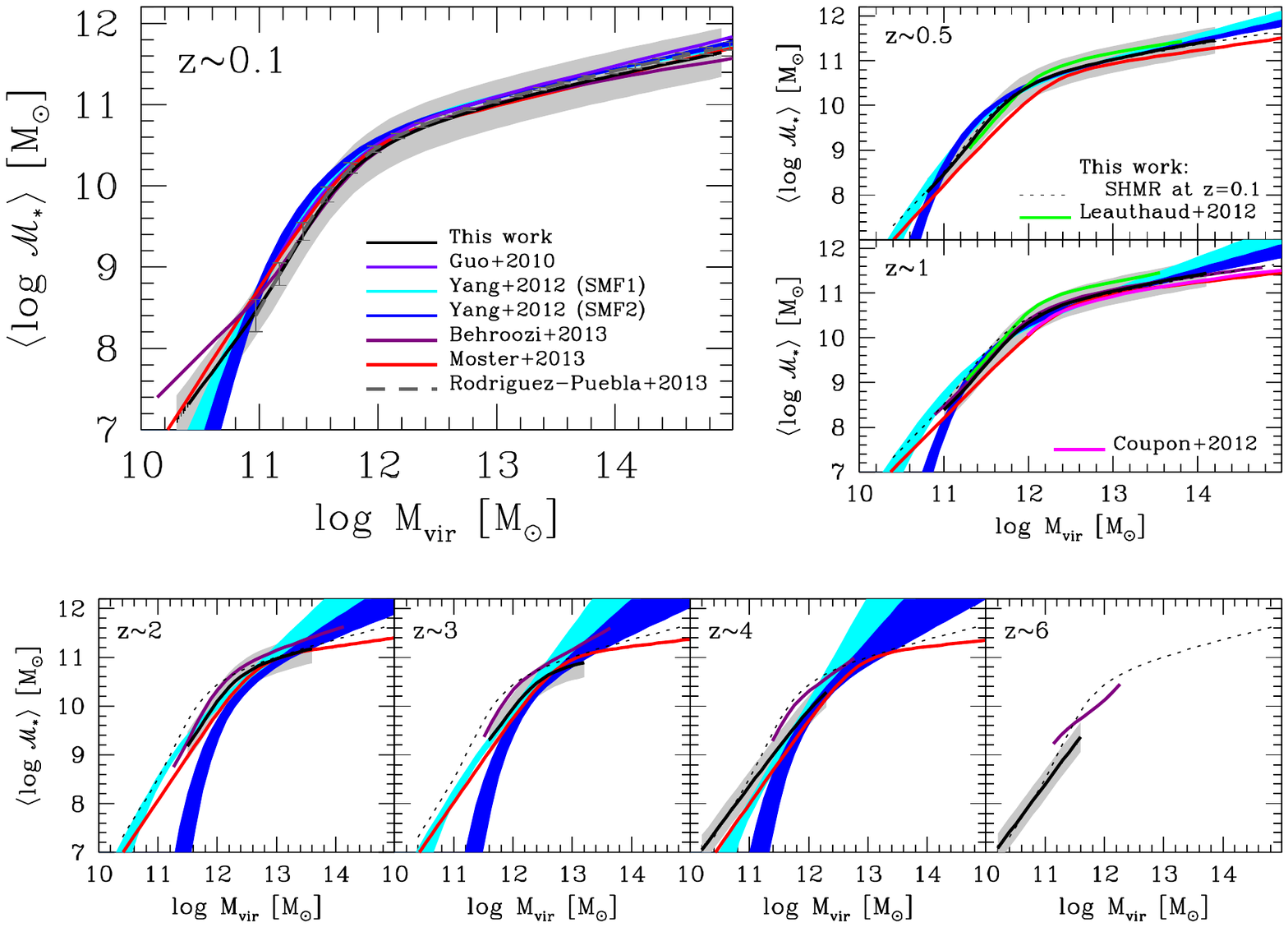}
	\vspace*{-100pt}
		\caption{ The mean logarithm of stellar mass, $\langle\log\mathcal{M_*}\rangle$, is plotted at each $\log \mvir$ and compared with previous works that reported
		galaxy stellar masses as a function of halo mass.  Our abundance matching results are shown with the solid black curve, and compared with abundance matching results from \citet{Guo+2010, Behroozi+2013};
		and \citet{Moster+2013}, shown respectively with violet, purple and red solid lines. The dark gray dashed lines show the \shmr\ at 
		$z\sim0.1$ reported in \citet{RAD13} who used the $\gsmf$s for centrals and satellite galaxies as well as the two-point correlation function to constrain  their best fit model.
		The results of \citet{Yang+2012} based on the
		evolution of the $\gsmf$, galaxy groups counts, and galaxy clustering are shown with the cyan and blue shaded regions for their SMF1 and SMF2 cases.
		Constraints from combining the \gsmf, galaxy clustering, and galaxy weak lensing from \citet{Leauthaud+2012} and \citet{Coupon+2015} are shown
		with the green and magenta lines, respectively. In all the higher redshift panels we plot with short dashed curves our  \shmr\ at $z=0.1$. 
		The gray shading shows the amplitude of the systematic errors assumed to be 0.3 dex. Note the good agreement at all redshifts between the different techniques except for SMF2, and except for SMF1 for $\mvir\grtsim10^{13}\msun$. 
 	}
	\label{msmh_comp}
\end{figure*}

We begin by comparing the mean logarithm of stellar mass $\langle\log\mathcal{M_*}\rangle$ plotted in each $\log \mvir$, (i.e., the \shmr) 
and shown for seven different redshift bins in Figure \ref{msmh_comp}. Our resulting $\shmr$s 
are shown with the solid black lines. The grey areas in all the panels show the $\sim0.3$ dex 
systematic errors. 

In Figure \ref{msmh_comp} we compare our \shmr\ with   
\citet[][constrained only at $z\sim0.1$, violet
solid line]{Guo+2010}, 
\citet[][purple solid lines]{Behroozi+2013}, and \citet[][red solid lines]{Moster+2013}. 
These authors used subhalo abundance matching to derive the $\shmr$s. 
\citet{Guo+2010} and \citet{Moster+2013} constrained the \shmr\ using only the \gsmf, while   
\citet{Behroozi+2013} included the observed specific SFRs and the CSFR. 
The dark gray dashed lines
with error bars show the \shmr\ at $z\sim0.1$ reported in \citet{RAD13} who used 
the $\gsmf$s for centrals and satellite galaxies as well as the two-point correlation function to constrain  
their best fit model, their set labeled as C. 
We also compared the \shmr\ from \citet{Yang+2012} who
used the redshift evolution of the $\gsmf$ up to $z\sim4$, 
the conditional stellar mass function at $z\sim0.1$ in various halo mass bins, and the galaxy clustering at $z\sim0.1$
in different mass bins. The cyan and blue shaded areas represent the $1\sigma$ confidence level of their results
when they used the GSMF reported in \citet{Perez-Gonzalez+2008} and from \citet{Drory+2005}, their SMF1 and SMF2
sets respectively. Additionally, we compared the \shmr\ at $z\sim0.6$ and  at $z\sim0.9$ from 
\citet{Leauthaud+2012} who combined stacked galaxy-galaxy weak lensing data and galaxy clustering at various mass 
thresholds from the COSMOS data. Finally, we compared with the \shmr\ at $z\sim0.8$ from
\citet{Coupon+2015} who combined galaxy clustering, galaxy--galaxy lensing and the stellar mass function from
observations in the Canada-France-Hawaii Telescope Lensing Survey (CFHTLenS) and VIPERS field. 

We begin our discussion by noting the broad agreement between the various
methods compared in Figure \ref{msmh_comp} for $z\lesssim1$. 
This is encouraging due to the different nature of observational constrains for each work plotted. 
Note that at $z\sim0.1$ the mass relation of dwarf galaxies (i.e., $M_{\rm vir} \lesssim 10^{11} M_\odot$) becomes slightly  
shallower \citep[but not as shallow as in][]{Behroozi+2013}, 
something that is not seen in \citet{Yang+2012,Moster+2013}; and \citet{Guo+2010}. 
The redshift evolution of our resulting $\shmr$ is more consistent with the
evolution derived in \citet{Yang+2012}, case SMF1, and \citet{Moster+2013}. When 
considering the uncertainties in the constrained relations our $\shmr$ is consistent with those
derived in  \citet{Behroozi+2013}. 

In the $z>0.1$ panels of Figure \ref{msmh_comp} we plot as a dashed curve the time-independent \shmr, i.e., the \shmr\ 
obtained at $z\sim0.1$. Note that below $z\sim2$ most
of the models as well as our mass relations are consistent with a time-independent \shmr. 
\citet{Behroozi+2013c} showed that assuming a time-independent 
\shmr\ could simply explain the cosmic star formation rate since $z=4$. In a subsequent work
\citet{Rodriguez-Puebla+2016a} extended that argument for studying the galaxies in the main sequence of galaxy 
star formation by showing that the dispersion of the 
halo mass accretion rates correctly predicts the observed dispersion of star formation rates.

\subsubsection{Halo mass-to-stellar mass relationship}

\begin{figure*}
	\vspace*{-160pt}
	\includegraphics[height=8.1in,width=6.7in]{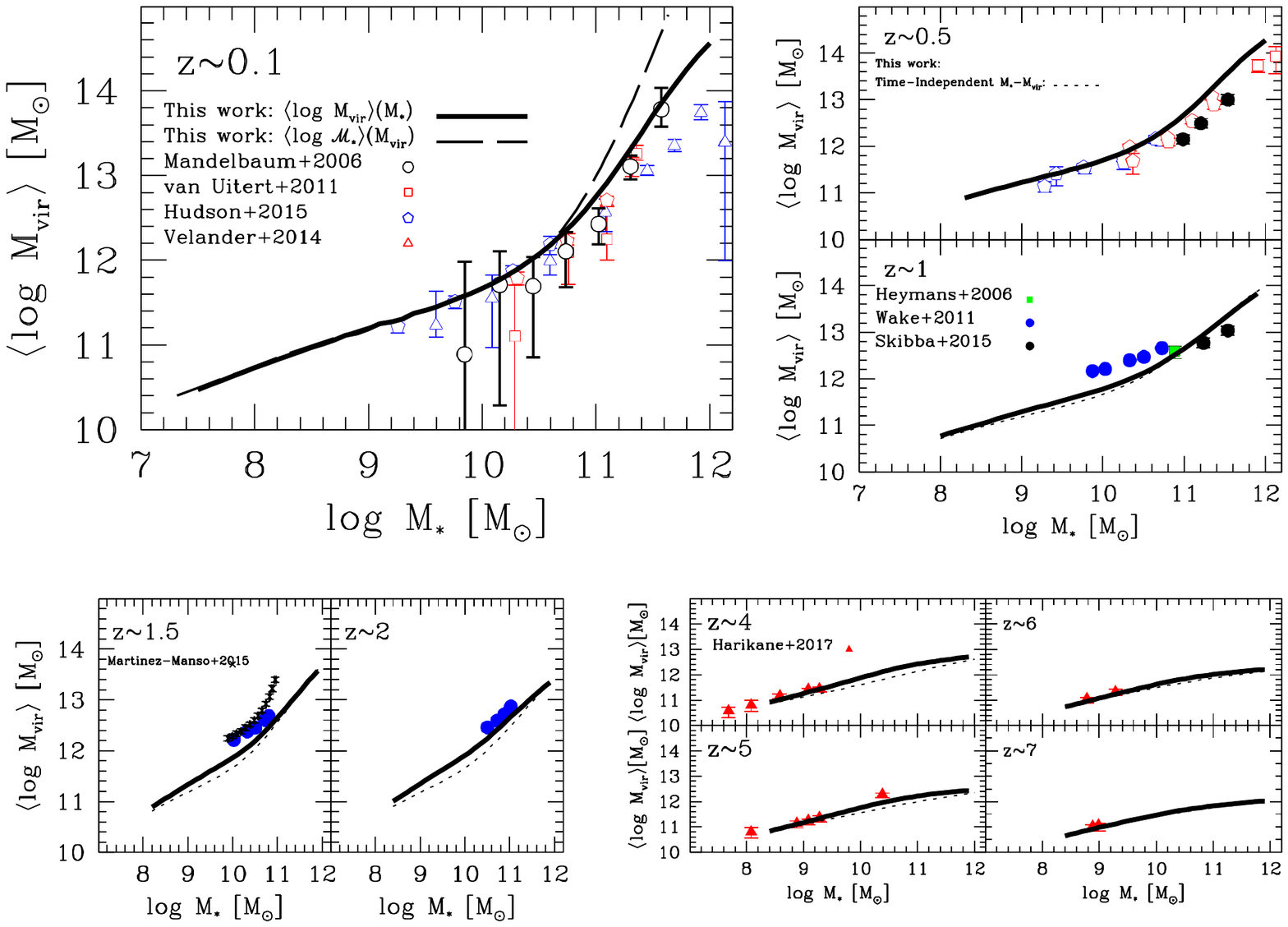}
	\vspace*{-100pt}
		\caption{ The mean logarithm of halo mass, $\langle\log\mvir\rangle$, is plotted at each $\log \ms$  and compared with previous works that reported
		halo masses as a function of galaxy stellar mass. Weak lensing studies from
		\citet{Mandelbaum+2006,vanUitert+2011,Velander+2014,Hudson+2014} and \citet{Heymans+2006}
		are shown respectively with the  black open circles, empty red squares, empty blue triangles, 
		open blue/red pentagons and green filled square. Galaxy clustering from  \citet{Wake+2011,Skibba+2015},
		and \citet{Harikane+2016} are shown with the filled blue circles, solid black circles and red solid triangles
		respectively. The dotted line shows a time-independent \ms--\mvir\ relation. The relation 
		$\langle \log \mathcal{M_*} (\mvir)\rangle$ at $z=0.1$ is shown with a long dashed line.  Table \ref{Tmhms} lists the 
		parameters for an analytic fit to  $\langle\log\mvir\rangle$  given by Equation (\ref{Invsmhms_fit}).
		}
	\label{mhms_comp}
\end{figure*}

Figure \ref{mhms_comp} shows the mean logarithm of halo mass, $\langle \log \mvir\rangle$, plotted for each $\log \ms$, i.e., we invert
the \shmr\ to obtain a \mvir--\ms\ relationship. Because the \shmr\ has scatter, inverting this relation is not as simple as just inverting the axes of the relation; we also need 
to take into account the {\it scatter around the relation}. This can be done by using Bayes' theorem by writing 
$P(\mvir|\ms,z) = P(\ms|\mvir,z) \times \phih(\mvir,z) /  \phi_g(\ms,z)$. Recall that the distribution function $P(\ms|\mvir,z)$ is assumed to be lognormal and in consequence
the distribution $P(\mvir|\ms,z) $ is expected to be different from a lognormal distribution. Using the above equation we can thus
compute $\langle\log \mvir\rangle$ as a function of $\log \ms$. In the upper left panel of Figure  \ref{mhms_comp} we compare the resulting 
$\langle \log \mvir(\ms)\rangle$ with the relation $\langle \log \mathcal{M_*} (\mvir)\rangle$ at $z=0.1$. Note that these relations are very different for $\ms\grtsim10^{10.6}\msun$. 
In general, we observe that the resulting \mvir--\ms\ relationship evolves in the direction that at a fixed stellar mass, higher stellar mass galaxies tend to have lower halo masses at higher redshifts. 

We now compare with recent determinations of the \mvir--\ms\ relationship. 
We begin by describing data obtained from galaxy weak lensing analysis. 
In Figure \ref{mhms_comp} we plot the results reported  in \citet{Mandelbaum+2006} 
from the stacked weak-lensing analysis for the SDSS DR4 at $z\sim0.1$, black open circles 
with error bars ($95\%$ of confidence intervals). \citet{Mandelbaum+2006} reported halo masses separately for
late- and early-type galaxies. Here we
estimated the average mass relation as:
$\langle\mvir\rangle(\ms)=f_l(\ms)\langle\mvir\rangle_l(\ms)+
f_e(\ms)\langle\mvir\rangle_e(\ms)$, where $f_e(\ms)$ and $f_l(\ms)$ are 
the fraction of late- and early-type galaxies in their sample. 
The corresponding values of halo masses for
late- and early-type galaxies are
 $\langle\mvir\rangle_l$ and $\langle\mvir\rangle_e$. In this case we assume that $\log\langle\mvir\rangle\sim\langle\log\mvir\rangle$
 given that the authors did not report a dispersion around the relation. 
 The empty red squares
 in the same figure show the analysis from \citet{vanUitert+2011} who
combined image data from the Red Sequence 
Cluster Survey (RCS2) and the SDSS DR7
to obtain the halo masses for late- and early-type galaxies as a function of \ms. 
Similarly to  \citet{Mandelbaum+2006} data, we also derive their mean halo masses based
on their reported fraction of early- and late-type galaxies. 
 We also include the stacked weak-lensing analysis from \citet{Velander+2014} based on the CFHTLens survey,
 empty blue triangles.
 The authors derive halo masses separately for blue and red galaxies based on
 the color-magnitude  diagram. We again derive their mean halo masses by using the reported
 fraction of blue and red galaxies. 
 Using the CFHTLenS survey \citet{Hudson+2014} also derived halo masses as a function 
 of stellar masses for blue and red galaxies separately. We showed their results with 
 the open blue and red pentagons. 
 Unfortunately, the authors do not report the fraction of blue 
 and red galaxies so we plot their mass relations separately for blue and red galaxies. 
 The green filled square in Figure \ref{mhms_comp} shows the halo mass
 derived from galaxy weak lensing at $z\sim0.8$ from \citet{Heymans+2006} by combing the 
 Chandra Deep Field South and the Hubble Space Telescope GEMS survey. 
 
 Next, we discuss halo masses obtained from galaxy clustering. \citet{Wake+2011}
used the halo occupation distribution (HOD) model  of galaxy clustering to derive halo masses
between $z=1-2$ from the NEWFIRM Medium Band Survey (NMBS), filled blue circles. 
Similarly, \citet{Skibba+2015} used the HOD model and the observed stellar mass dependent clustering of 
galaxies in the PRIMUS and DEEP2 redshift survey from $z\sim0.2$ to $z\sim1.2$ to constrain  the \mvir--\ms\ relationship, indicated
by the solid black circles. \citet{Martinez-Manso+2015} used the Deep-Field Survey
to derive the angular clustering of galaxies and obtain halo masses by modelling galaxy clustering
in the context of the HOD. Finally, \citet{Harikane+2016} estimated the angular distribution of 
of Lyman break galaxies between $z\sim4-7$ from the Hubble Legacy deep Imaging and the Subaru/Hyper Suprime-Cam data. 
Halo masses were estimated using the HOD model, filled red triangles in Figure \ref{mhms_comp}. 

\begin{table}
	\caption{Best fit parameters to the $\langle \log \mvir(\ms)\rangle$ relation given by Equation \ref{Invsmhms_fit}. Note that
	due to the low constraints at high masses for $\delta$ and $\gamma$ we kept them constant above $z>2$.}
	\begin{center}
		\begin{tabular}{c c c c c c}
			\hline
			\hline
			$z$ &  $\log M_1  [\msun]$ &  $\log M_{*,0}   [\msun]$ & $\beta$ & $\delta$ & $\gamma$\\
			\hline
			\hline
			0.10 & 12.58 & 10.90 & 0.48 & 0.29 & 1.52\\ 
			0.25 & 12.61 & 10.93 & 0.48 & 0.27 & 1.46\\ 
			0.50 & 12.68 & 10.99 & 0.48 & 0.23 & 1.39\\ 
			0.75 & 12.77 & 11.08 & 0.50 & 0.18 & 1.33\\ 
			1.00 & 12.89 & 11.19 & 0.51 & 0.12 & 1.27\\ 
			1.25 & 13.01 & 11.31 & 0.53 & 0.03 & 1.22\\ 
			1.50 & 13.15 & 11.47 & 0.54 & -0.10 & 1.17\\ 
			1.75 & 13.33 & 11.73 & 0.55 & -0.34 & 1.16\\ 
			2.00 & 13.51 & 12.14 & 0.55 & -0.44 & 0.92\\ 
			3.00 & 14.02 & 12.73 & 0.59 & -0.44 & 0.92\\ 
			4.00 & 14.97 & 14.31 & 0.60 & -0.44 & 0.92\\ 
			5.00 & 14.86 & 14.52 & 0.58 & -0.44 & 0.92\\ 
			6.00 & 17.43 & 19.69 & 0.55 & -0.44 & 0.92\\ 
			7.00 & 17.27 & 20.24 & 0.52 &  -0.44 & 0.92\\ 
			\hline
		\end{tabular}
		\end{center}
	\label{Tmhms}
\end{table}

Similarly to the determinations of the $\shmr$ compared in Figure \ref{msmh_comp},
the \mvir--\ms\ relationships described above
agree very well between each other and with our mass relations from abundance matching, specially
at $z\lesssim1$. The short dashed lines show the resulting  $\langle \log \mvir(\ms)\rangle$ when assuming a time-independent 
$\langle \log \mathcal{M_*} (\mvir)\rangle$ relation. Note that the evolution of this relation simply reflects the fact that the ratio 
$\phih(\mvir) /  \phi_g(\ms)$ is not constant with time, that is, $P(\mvir|\ms,z) = P(\ms|\mvir) \times \phih(\mvir,z) /  \phi_g(\ms,z)$
where the distribution $P(\ms|\mvir)$ is independent of time just because $\langle \log \mvir(\ms)\rangle$ relation is time-independent. 
 Do not confuse this with a $\langle \log \mvir(\ms)\rangle$ relationship that is constant in $z$, that is, {\it we are not replicating}
 the resulting $\langle \log \mvir(\ms)\rangle$ at $z=0.1$ in all the panels.  Table \ref{Tmhms} lists the best fit parameters when
$\langle \log \mvir(\ms)\rangle$ is parameterized using the function proposed in \citet{Behroozi+2010}:
\begin{eqnarray}
\langle \log \mvir(\ms)\rangle = \log M_1 + \beta \log\left(\frac{\ms}{M_{*,0}}\right) & &  \nonumber \\ 
 +\frac{\left(\frac{\ms}{M_{*,0}}\right)^{\delta}}{1+ \left(\frac{\ms}{M_{*,0}}\right)^{-\gamma}} +\frac{1}{2}.
 \label{Invsmhms_fit}
\end{eqnarray}
The accuracy of the fitting parameters on Table \ref{Tmhms} are better than $5\%$ at all redshifts.

Finally, we acknowledge that there are other direct techniques to derive halo masses
from galaxy samples. One example is to use the kinematics of satellite galaxies
as test particles of the gravitation field of the dark matter halo in which their reside 
\citep[e.g.,][]{Conroy+2007,More+2011,Wojtak+2013}. 
In the case of satellite kinematics it has been noted that this method tends to give
higher halo masses than other methods
\citep[like those described above, see e.g.,][]{More+2011, Skibba+2011, Rodriguez-Puebla+2011}.
\citet{RP+2015} argued that the difference with other techniques can be
partially explained by the relation between halo mass and the number of satellite galaxies. 
Given the challenge that would imply to show a fair comparison between our mass relations
and those from satellite kinematics, we prefer to omit that comparison in this paper.

\begin{figure*}
	\vspace*{-80pt}
	\hspace*{0pt}
	\includegraphics[height=8.1in,width=6.7in]{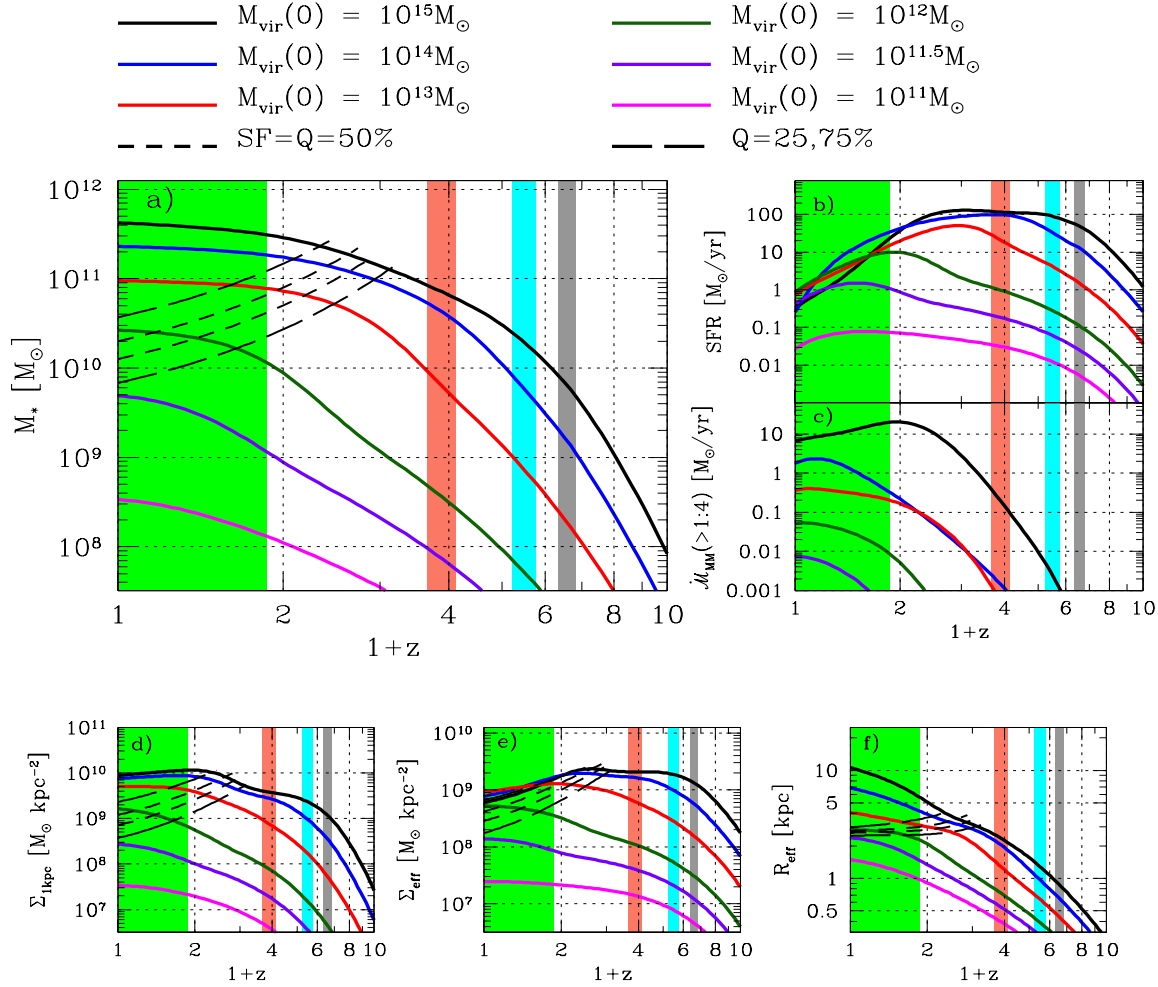}
	\vspace*{-110pt}
		\caption{Summary plot for various galaxy properties: Panel a) Stellar mass, Panel b) star formation
		history, Panel c) Stellar mass accretion from major mergers,  Panel d) Integrated  surface mass density at 
		1 kpc, Panel f) Effective surface mass density and Panel e) Effective radius. The magenta,
		violet, green, red, blue and black lines show the trajectories
		for progenitors of $\mvir=10^{11}$, $10^{11.5}$, $10^{12}$, $10^{13}$, $10^{14}$, and $10^{15} \msun$.
		The gray, cyan, light red and light green shaded areas in all the panels 
		show the epoch range at which the progenitors of
		halos of $10^{12}$, $10^{13}$, $10^{14}$, and $10^{15} \msun$ 
		reached the mass of $\mvir = 10^{11.8}-10^{12}\msun$. The short-dashed lines show $M_{50}(z)$ where half the
		galaxies are statistically quenched, while the fractions of quenched galaxies are 25$\%$ (75$\%$) at  
		the upper (lower) long-dashed lines.
		 	}
	\label{model_comparison}
\end{figure*}

\subsection{The Quenching Halo Mass Depends On Redshift}
\label{halo_quenching}

The conception of a transition halo mass above which the star formation
becomes strongly 
inefficient has been introduced by a number of theoretical studies 
\citep[e.g.,][]{WhiteFrenk1991,Birnboim+2003,Keres+2005,Dekel+2006,Cattaneo+2007}. 
These studies have shown that a stable virial shock is formed when the virial mass reaches $\mvir\sim10^{12}\msun$,
approximately independent of $z$,
through which the cosmological inflowing gas must cross, thus
resulting in the heating of the infalling gas \citep{Dekel+2006}. In the literature, this 
mechanism is typically referred as halo mass quenching. 
Indeed, this seems to apply
at $z\sim0$, where the stellar mass above which the galaxy population is observed to
be dominated by quiescent galaxies is $\ms\sim10^{10.5}\msun$,  
corresponding to a halo mass of $\mvir\sim10^{12}\msun$ (see Figure \ref{f8}).  
But as we have discussed starting with Figure \ref{m_trans}, $M_{50}$, the halo mass where 50\% of galaxies are quenched,  actually depends on redshift. 
In light of our results, here we discuss this redshift dependence.

Panel a) of Figure \ref{model_comparison} shows the average stellar mass growth history for halos 
with final masses of $\mvir=10^{11},10^{11.5},10^{12},10^{13},10^{14}$,
and $10^{15}\msun$ at $z=0$. 
The gray, cyan, light red, and green shaded areas show the redshift ranges when the progenitors
of halos with $\mvir=10^{15},  10^{14},  10^{13}$ and $10^{12} \msun$ reached the mass range 
of $\mvir=10^{11.8} - 10^{12}\msun$.  We use a halo mass range instead of a fixed mass
given that this transition mass might be not sharp \citep[e.g.,][]{Keres+2005,Dekel+2006}. 
In the same panel,
the black short-dashed lines indicate when the observed fraction of star
forming galaxies is $50\%$ while the upper and lower long-dashed lines indicate when the fraction
of quenched galaxies is $75\%$ and $25\%$.
The corresponding star formation histories for the progenitors described above are 
presented in Panel b) of the same figure. Similarly to Panel a), the shaded areas 
show  the redshift ranges when the progenitors of halos of masses $10^{12-15}M_\odot$ at $z=0$ 
reached the mass range of $\mvir=10^{11.8} - 10^{12}\msun$. 

Next, we consider the progenitors of halos at $z=0$ with masses above $\mvir>10^{12}\msun$. 
According to Figure \ref{model_comparison},
we note that, on average, the epoch at which the SFRs of those galaxies declined is
{\it rather
different} from the epoch at which their host halo reached the halo mass quenching transition. Moreover,
this depends on halo mass. This
suggests that galaxy quenching is not driven mainly by halo mass alone and that other
mechanisms are playing an important role. Note, however, that the galaxy population 
in halo progenitors at $z=0$ with masses $\mvir=10^{12}\msun$ are on the statistical 
limit of being dominated by quiescent systems. This could also suggest that the mechanisms that are
responsible for quenching are complex and that they could be actually somewhat diverse at 
different redshifts leading to the observed imprint in Figure \ref{m_trans}, namely, 
high stellar mass galaxies tend to quench earlier, which is sometimes referred as quenching downsizing. 
We will come back to this in the next subsection.

One possibility is that cold streams were able to supply enough gas to sustain the star formation
in galaxies in massive halos at high redshifts. That was suggested in a number of previous works showing that
hydrodynamical simulations predict that at high redshifts $z>1$ hot halos can be 
penetrated by cold streams 
\citep[see e.g.,][]{NagaiKravtsov2003,Keres+2005,Dekel+2006,Dekel+2009}. 
 Indeed, in Figure \ref{sfreff} we showed that the halo mass at which the
star formation efficiency sSFR/sMAR drops below unity depends on redshift 
in the direction that the higher the redshift, the larger the transition halo mass. Interestingly
enough, progenitors of Milky-Way mass halos with $\mvir=10^{12}\msun$ at $z=0$,
reached the ratio sSFR/sMAR$\sim1$ at $z\sim0$ and thus are unlikely to still be fed by cold streams, as predicted
in the cold stream theory \citep{Dekel+2006}. 

We thus conclude that 
our new $\shmr$s are consistent with the idea that at high redshifts cold streams were able
to sustain the star formation in the progenitors of the most massive galaxies 
observed today. 
Therefore, mechanisms alternative to halo mass quenching should be invoked to explain the strong 
decline in the SFR of those progenitors. 
Additionally, our new $\shmr$s suggest that  
virial shock heating might become more relevant at lower redshifts.

\subsection{The Role of the Structural Properties of the Galaxies}
\label{structure_quenching}

In Section \ref{size_evolution}, we discussed various structural properties of galaxies 
and their evolution. Our findings show that galaxies grow in central stellar mass density (within
1 kpc, $\Sigma_1$) while they are star forming, and they quench once they reach a maximum 
$\Sigma_1$. This maximum $\Sigma_1$ is higher for higher mass galaxies, and it is reached at higher redshifts, as seen 
in Figure \ref{model_comparison}. 
That is, the progenitors of massive galaxies today have
higher $\Sigma_1$ and quenched earlier than the progenitors of lower mass galaxies. 
These results strongly suggest that the quenching of 
massive galaxies is related to the central density of these galaxies, a mechanism that could be associated with 
the growth of the central supermassive black hole (SMBH) powering an AGN or quasar. 
The central density and the growth of the black hole should be 
fundamentally linked, given all the observational evidence that the masses of SMBHs are 
well correlated with the most internal properties of their host galaxies, particularly with their bulge properties \citep[see e.g.][]{Kormendy+2013}.

Our results are consistent with a scenario in which the progenitors of the most massive galaxies, $\ms\gg3\times10^{10} \msun$: 
1) have a fast growth in mass at high redshifts; 2) cold streams facilitate this growth in mass and central density, 
in spite of the hot gas halo (see the discussion in the previous subsection); 
3) SMBHs grow in their centres along with the increase in central density; 
4) the central density and overall galaxy mass growth are slowed down when a luminous AGN is switched on (quenching).  
In addition, the more massive the system, the higher the central densities (hence more massive SMBHs) must be reached in order to produce AGNs luminous enough 
to quench the galaxy \citep[e.g.,][]{Terrazas+2016}. These are just the trends seen in Figure \ref{model_comparison}. 
For lower masses, down to   $\ms\sim3\times10^{10} \msun$,  the central densities do not attain very high values, and the 
AGN is probably not luminous enough to be the main quenching mechanism.
 Instead, for these galaxies, whose halos attain the quenching
transition mass at late epochs (when cold streams are already of low efficiency), the long cooling time of the hot halo gas can be 
the main mechanism for suppressing  star formation.

In Section \ref{size_evolution}, we found that galaxies 
evolve in parallel tracks in the size-mass relation with slopes around  $d\log R_{\rm eff}/d\log\ms\sim0.35$, but once  
they quench they move in the size-mass relation with a steeper slope $d\log R_{\rm eff}/d\log\ms\sim2.5$ (which is consistent 
with the fact that minor dry mergers increase the sizes of early-types galaxies, see Figure \ref{reff_mr_sfr}). This implies that
{\it i}) galaxies are growing in very different regimes, and {\it ii}) at all times the sizes are growing, on average, ``inside-out". 
We note that this transition occurs, statistically, when galaxy's size is around $\sim3$ kpc, see panel f) in Figure \ref{model_comparison}. 
However, we do not think that this size reveals something fundamental
about their maximum density, $\Sigma_1$. 
As discussed above, the central density is associated to the SMBH growth, and hence to the luminosity of the AGN that quenches star formation. 
Thus it may be that the black hole is playing a major role in initiating or maintaining quenching, and the fact that galaxies quench when they reach, on average, an effective radius of 
 $\sim3$ kpc may just be a coincidence.

Panels d), e) and f) in Figure \ref{model_comparison} show respectively the trajectories of the halo progenitors discussed above for
$\Sigma_1$, $\Sigma_{\rm eff}$ and $R_{\rm eff}$. Recall that the black dashed lines mark $M_{50}(z)$ above/below
which  a majority of galaxies are quenched/star forming. Panels d) and f) illustrate the conclusions described above, while the trends shown
in panel e) are a direct consequence of the evolution of the size-mass relation. That is, the progenitors of
quenched galaxies evolved along two markedly different trajectories: they grew in density as star formers and
quenched when they reached a maximum of effective density, and then their density within $\reff$ decreased at low redshift
as $\reff$ increased.
Note that nothing interesting happened to the structural properties of galaxies when the progenitors of their host halos 
reached the characteristic quenching halo mass, $\mvir\sim10^{12}\msun$. 
Our findings that galaxies, on average, grew ``inside-out" and quenched when they reached a maximum density
are in excellent agreement with previous conclusions by 
\citet{Kauffmann+2003,Franx+2008,Patel+2013,vanDokkum+2014,Barro+2015}; and \citet{vanDokkum+2015}. 

Our findings, unfortunately, 
cannot be compared directly to the wet (gas inflow) compaction model  \citep{Dekel_Burket2014,Zolotov+2015,Tacchella+2016}.
This model asserts that shrinkage of the effective radius arises from central starbursts fed by dissipative processes, driven by violent disc instabilities \citep{Dekel_Burket2014}, possibly triggered by major or minor mergers \citep{Inoue+2016}, that could only happen
at higher redshifts when the gas fractions are high. In particular, this model predicts that 
in the innermost parts of the galaxies gas compaction provides the fuel to increase the central stellar mass density due 
to in-situ star formation, and when the central gas is entirely depleted the central stellar mass density should remain constant. Nonetheless,
we can compare our model indirectly with the wet compaction model by studying the $\Sigma_1$ trajectories of the progenitors
discussed earlier in this section, see panel d). We find that our results are consistent with 
the wet compaction model. Finally, as discussed in Section \ref{halo_quenching}, it is likely that cold streams can maintain the gas supply in 
massive galaxies at high redshifts. We note that cold streams and wet compaction are both phenomena that may
happen at the same time; indeed, \citet{Dekel+2009b} showed that smooth cold streams can keep a gas-rich disc 
unstable and turbulent  
at high redshifts. Moreover, the presence of an AGN/QSO is also plausible, especially in massive galaxies. 
It is not clear how these three phenomena act together, but presumably wet contraction and cold streams 
may help to feed the AGN, which in turn
depletes the gas from the galaxy more efficiently and keeps hot gas from cooling.

\subsection{The Impact of Galaxy Mergers}

In Section \ref{Mergers} we showed that growth in stellar mass is primary due to in-situ star formation. 
High mass galaxies, $\ms\sim5\times10^{11}\msun$, assembled around $\sim36\%$ of 
their final stellar mass through galaxy mergers, while galaxies in Milky-Way sized halos assembled around $2.4\%$. Additionally,
we found that minor mergers, $\mu_*\leq1/4$, are more frequent than major mergers, $\mu_*\geq1/4$, at all masses. 
For example, massive galaxies with $\ms\sim5\times10^{11}\msun$ had at least 1 major merger but $\sim20$ minor ones since $z\sim1$. 
Nonetheless, at these high masses major mergers contributed $\sim75\%$ of the total mass in mergers while at low masses this was around $\sim35\%$. 
Given the predictions presented for galaxy mergers, here we discuss their impact on the formation history of the galaxies, namely, in the context of
their star formation histories and structural properties. Panel c) of Figure \ref{model_comparison} shows the historical contribution from major mergers 
to the stellar mass accretion rate for our set of progenitors at $z=0$, described above. This figure shows that the mass growth due to major mergers peaks below $z\sim1$, 
with the redshift location of the peak of the stellar mass growth due to major mergers 
decreasing with decreasing halo mass. Today galaxies with $\ms\sim10^{11}\msun$ are on their peak of stellar mass growth due to major
mergers. 

While in this paper we focused on stellar mergers, different types of mergers
are linked to different phenomena. For example,   
hydrodynamical simulations of galaxy mergers have shown that the 
dissipational (wet) mergers result in starburst activity between $\sim100-500$ Myrs while 
(dry) dissipationless mergers, especially the major ones, will result in disturbances to the morphologies.
Of course, both types of mergers happen over the merger history of the galaxies. 
In general, we do not observe an increase of star formation triggered by major mergers, particularly at high
redshifts where the fractions of gas were high, and not even for massive galaxies.  Nevertheless, we speculate that 
the extended period of star formation in massive galaxies can be partly explained by merger-induced starbursts, since
the rise of stellar mass accreted due to major mergers in galaxies  with $\ms\sim5\times10^{11}\msun$ occurs mainly 
during the phase of high star formation, although the main contribution of mergers occurs at low redshift.  
Probably this extended period of star formation is related to gas fed by cold streams, as discussed  above.

We observe that when the progenitors of quenched galaxies start to suppress their SFRs, the 
stellar mass accretion rate became dominated by major mergers.  Galaxy major dry mergers are expected
to change morphologies \citep[e.g.][]{Toomre_Toomre1972,Barnes_Hernquist1996,Robertson+2006a,Robertson+2006b,Burkert+2008}.
This could imply that the change in morphology of massive galaxies occurred below $z\sim1$. 

Major mergers and black hole
growth are thought to be intimately correlated during the AGN phase in massive galaxies, especially at high redshift
where the gas fractions were higher. Therefore the fact that the SFR was suppressed when the contribution from
major mergers became more important is consistent with AGN feedback from the black hole
being an effective quenching mechanism. Looking to Figure \ref{model_comparison}, it is not obvious that major mergers could
trigger an AGN phase in massive galaxies. Nonetheless, our results do not reject this possibility. To see this,
we recapitule the main results from Figure \ref{Nmerger}. This Figure shows that the host galaxies in massive halos had
$\sim3$ major mergers since $z\sim3$ and had $\sim2$ since $z\sim2$. We thus conclude that our results
do not reject the possibility that AGNs were triggered by major mergers.

\section{Summary and Conclusions}
\label{summary}

This paper presents new determinations of the stellar-to-halo mass relation, \shmr, 
over a very broad redshift range, between $z\sim0$ to $z\sim10$, from abundance matching.
We use the redshift evolution of the \shmr\ together with the growth of dark matter halos
to predict the average in-situ stellar mass growth of galaxies and therefore their star formation histories,
as well as the ex-situ mass growth due to mergers. 
We used as observational constraints the redshift evolution of the galaxy stellar mass function, \gsmf s, and the 
star formation rates, SFRs. In order to get a robust determination of our $\shmr$s we used a  
large compilation of $\gsmf$s from the literature as well 
as of SFRs. We calibrated all the observations to the same basis in order to minimize
potential systematic effects that might bias our results. Specifically, all the observations were corrected to 
the IMF of \citet{Chabrier2003}, the \citet{BC03} SPS model, the \citet{Calzetti+2000} dust
attenuation model, and the cosmological parameters reported by
the Planck mission and used in the Bolshoi-Planck simulation. Corrections for surface brightness 
incompleteness were also introduced. 

We report star formation histories for a wide range of halo masses at $z=0$ and their progenitors at higher redshifts. 
We quantify both the instantaneous and the cumulative fraction of mass accreted due to mergers. 
To quantify in more detail which type dominates the accreted mass fraction we use the subhalo 
disruption rates convolved with the redshift evolution of the \shmr. 

Once we have constrained robust trajectories for galaxy progenitors, we present a study of the average evolution of the radial distribution of stellar mass as a function of mass, 
and explore the impact of the structural properties on the evolution of galaxies.
This is the first time that this has been done within the semi-empirical galaxy-halo connection approach.
To do this, we used the observed size-mass relation of local galaxies derived 
in \citet{Mosleh+2013} and determined the dependence to higher redshifts
to be consistent with the size-mass evolution observed by HST \citet{vanderWel+2014}.
Finally, we assumed that the radial distribution of stellar mass is a combination of 
S\`ersic $n=1$ and $n=4$ profiles, with fractions associated to the
population fractions of star-forming and quenched galaxies at each $z$.

In this paper we do not model the individual growth histories of galaxies,
but rather average growth histories. Specifically, we do not know ``case by case" 
how galaxies suppressed their star formation.  
Instead we adopt a more probabilistic description by inferring
which were the most likely trajectories of the galaxy progenitors as a function of mass and redshift. 

Finally, we list our main results and conclusions:
\vspace{-8pt}
\begin{itemize}
\item The stellar-to-halo mass relation (\shmr) evolves very slowly below $z\sim1$ but has a noticeable evolution
between redshift $z\sim1$ and $z\sim7$. This statement is mass dependent: the high mass end evolves more
strongly than the low mass end.  This implies that the value of the peak stellar-to-halo mass ratio
decreased approximately a factor of $\sim3$ between $z\sim0.1$ and $z\sim4$, and almost an 
order of magnitude at $z\sim10$. 

\item When comparing indirect methods for constraining the galaxy halo-mass-to-stellar-mass connection (where the halo masses are determined by abundance matching), we find a broad agreement.  This is encouraging given 
the very diverse nature of the methods that have been employed in the literature. 
Similar conclusions were drawn when comparing instead direct methods aimed at constraining the galaxy stellar-mass-to-halo-mass connection (where the halo masses are determined by gravitational lensing or clustering),

\item The star formation histories for the progenitors of today's massive elliptical galaxies in Figure \ref{f10} show 
that they reached an intense period of star formation with an average value of $\sfr\sim200$ $\msun/ {\rm yr}$ at redshifts 
between $z\sim1$ and $z\sim4$ depending on the mass; after this period their SFRs dramatically decrease.  Galaxies in Milky-Way-mass halos went through distinct phases: below $z\sim1.5$ they formed 
stars in-situ at a moderate rate of SFR, $\sim2$ $\msun/ {\rm yr}$, but then for some reason they had a period of 
intense star formation peaking around $z\sim1$ with values of SFR $\sim10$ $\msun/ {\rm yr}$, and then a smooth 
decline with modest values of the order of  SFR $\sim1$ $\msun/ {\rm yr}$ at $z\sim0$.  
At low masses, $\ms\sim3\times10^{8}\msun$, galaxies had much more constant SFRs along their histories, 
with values  of $\sfr\sim0.1$ $\msun/ {\rm yr}$.  

\item Defining the halo star formation efficiency as the ratio s$\sfr/{\rm sMAR}$, we find in Figure \ref{sfreff} that
the observed transition mass, above/below which galaxies are statistically quenched/star forming, coincides
with the mass and epoch where the ratio s$\sfr/{\rm sMAR}\sim1$. 
Perhaps surprisingly, the halo star formation efficiency peaks at $z \lesssim 1$ for halos of about $\mvir\sim2\times10^{11}\msun$, corresponding to 
galaxies with stellar masses $\ms\sim (0.8 -3)\times 10^{9}\msun$. It is not
clear why $\mvir\sim2\times10^{11}\msun$ is a ``special" mass or whether this should be expected theoretically. 
This could be a hint that galaxy formation is somewhat different below this mass. 
More work is needed in this direction. 

\item 
The typical halo mass at which the halo star formation efficiency  s$\sfr/{\rm sMAR}$ transits above and below unity
is not constant but it changes with redshift: at $z\sim0$ the transition occurs at $\mvir\sim10^{12}\msun$ 
while at $z\sim3$ it occurs at $\mvir\sim10^{13}\msun$. This result is consistent with high redshift galaxies 
in halos as massive as $\mvir\sim10^{13}\msun$ being fed by cold streams, while at lower redshifts 
the virial shocks and AGN feedback play a major role in quenching star formation in galaxies formed in
halos more massive than $\mvir\sim10^{12}\msun$.

\item Galaxy growth is primarily due to in-situ star formation. Massive galaxies with $\ms\sim5\times10^{11}\msun$
assembled around $\sim36\%$ of their final mass through galaxy mergers (ex situ) while galaxies in Milky-Way
sized halos assembled around $\sim2.4\%$. 
Minor mergers (defined as $<$1:4) are more frequent than major mergers  (defined as $>$1:4). 
Present-day galaxies of $\ms\sim5\times10^{11}$ and $\ms\sim 3\times10^{10}$ $\msun$ had on average 
$\sim 1$ and $\sim 0$ major mergers since $z=1$, but $\sim20$ and $\sim 0.5$ minor ones, respectively. 
Nonetheless, of the final accreted stellar mass, major mergers have contributed $\sim75\%$ and 35\%  in 
galaxies of these masses, respectively. Thus, major mergers played a more important role 
in massive galaxies than in lower mass galaxies.

\item On average, the radial stellar surface mass density of galaxies grows the inside out. The effective radii
increase with stellar mass as $\reff(z)\propto\ms^{0.35}(z)$ while galaxies are star forming,
but once they became statistically quenched they shift in the size-mass relation to a steeper slope 
of $\sim2.5$. We conclude that this change in slope is consistent 
with theoretical expectations of quenched galaxies predominantly evolving in size through dry minor mergers. 

\item 
The evolution of the surface stellar mass density enclosed within 1 kpc, $\Sigma_1$, 
is closely related to the evolution of the global sSFR.   
This is consistent with previous findings. 
Moreover, once galaxies reach their maximum $\Sigma_1$ value, the global SFR is suppressed. 
The larger the galaxy mass, the higher is this maximum and the higher the redshift at which it is attained:
at $z\sim0$, $\Sigma_{\rm 1, max}\sim10^9$ $\msun/{\rm kpc}^2$ for galaxies of $\ms\sim 2\times 10^9$ \msun,
while at $z\sim3$,  $\Sigma_{\rm 1, max}\sim10^{10}$ $\msun/{\rm kpc}^2$ for progenitors of present-day galaxies
of  $\ms\sim 5\times 10^{11}$ \msun.

\item Our results do not support the scenario in which galaxy major mergers cause galaxies to shrink their sizes, resulting in compact quenched spheroids at high redshifts. Instead, our results support on average a constant inside-out growth. 

\item Indirectly, we find that the size-mass evolution of quenched galaxies is consistent with the wet 
compaction model, because the trajectories in the sSFR$-\Sigma_1$ plane are very similar 
to what is observed in hydrodynamical simulations of galaxy formation. Similar conclusions were reached in 
previous studies that analyzed observations and hydro simulations of galaxy formation \citep{Barro+2015,Tacchella+2016}. 
\end{itemize}

This work presents the most updated and self-consistent determination of the \shmr\ from $z\sim 0$ to $\sim 10$ that is currently available,
as well as the inference of the average stellar mass, star formation, and merger rate histories of galaxies
as a function of mass. We have extended the semi-empirical approach to study the average structural evolution of  
galaxies and its connection to the star formation and merger rate evolution. 
The semi-empirical data presented in this paper are available at our website {\color{blue} \footnote{\url{
https://132.248.1.39/galaxy/galaxy_halo.html}}}. These data are useful
for constraining semi-analytic models and cosmological simulations of galaxy evolution,
as well as for comparing with determinations of galaxy evolution from the fossil record method applied
to large galaxy surveys or direct look-back time observations of selected galaxy populations.

\section*{Acknowledgments} 
We are grateful to Peter Behroozi, Avishai Dekel, Alejandro Gonzalez-Samaniego, Yicheng Guo, and David Koo, for useful comments and discussions and we thank to Zhu Cheng and Robert Feldman for detecting typographical errors. We thank to Robert Feldman and Rachel Somerville for suggesting table \ref{Tmsmh}. We thank to S. Ilbert, A. Muzzin, P. Santini, K. Duncan and Mimi Song for providing their
data in an electronic form. We also thank to all the people that have derived all the wonderful data (GSMF, SFR, CSFR and
galaxy sizes) utilized through this paper. We thank to Francisco Ruiz Sala for his help in our website. 
ARP has been partially supported by a UC-MEXUS Fellowship. 
JRP acknowledges support from grants HST-GO-12060.12-A-004 and HST-AR-14578.001-A. VAR acknowledges partial support from CONACyT grant (Ciencia B\'asica) 167332.  
SF acknowledges support from NSF grant AST-08-08133. We also thank to the anonymous Referee for a constructive report that helped to improve this paper.

\bibliographystyle{mn2efix.bst}
\bibliography{Bibliography}

\begin{thebibliography}{113}
\expandafter\ifx\csname natexlab\endcsname\relax\def\natexlab#1{#1}\fi

\bibitem[{{Becker}(2015)}]{Becker2015}
{Becker} M.~R., 2015, arXiv:1507.03605

\bibitem[{{Behroozi}, {Conroy} \& {Wechsler}(2010){Behroozi}, {Conroy}, \&
  {Wechsler}}]{Behroozi+2010}
{Behroozi} P.~S., {Conroy} C., {Wechsler} R.~H., 2010, \apj, 717, 379

\bibitem[{{Behroozi} \& {Silk}(2015)}]{Behroozi15}
{Behroozi} P.~S., {Silk} J., 2015, \apj, 799, 32

\bibitem[{{Behroozi}, {Wechsler} \& {Conroy}(2013{\natexlab{a}}){Behroozi},
  {Wechsler}, \& {Conroy}}]{Behroozi+2013c}
{Behroozi} P.~S., {Wechsler} R.~H., {Conroy} C., 2013{\natexlab{a}}, \apjl,
  762, L31

\bibitem[{{Behroozi}, {Wechsler} \& {Conroy}(2013{\natexlab{b}}){Behroozi},
  {Wechsler}, \& {Conroy}}]{Behroozi+2013a}
{Behroozi} P.~S., {Wechsler} R.~H., {Conroy} C., 2013{\natexlab{b}}, \apj, 770,
  57

\bibitem[{{Behroozi} {et~al}\mbox{.}(2014){Behroozi}, {Wechsler}, {Lu}, {Hahn},
  {Busha}, {Klypin}, \& {Primack}}]{Behroozi14}
{Behroozi} P.~S., {Wechsler} R.~H., {Lu} Y., {Hahn} O., {Busha} M.~T., {Klypin}
  A., {Primack} J.~R., 2014, \apj, 787, 156

\bibitem[{{Behroozi}, {Wechsler} \& {Wu}(2013){Behroozi}, {Wechsler}, \&
  {Wu}}]{Behroozi+2013d}
{Behroozi} P.~S., {Wechsler} R.~H., {Wu} H.-Y., 2013, \apj, 762, 109

\bibitem[{{Behroozi} {et~al}\mbox{.}(2013){Behroozi}, {Wechsler}, {Wu},
  {Busha}, {Klypin}, \& {Primack}}]{Behroozi+2013b}
{Behroozi} P.~S., {Wechsler} R.~H., {Wu} H.-Y., {Busha} M.~T., {Klypin} A.~A.,
  {Primack} J.~R., 2013, \apj, 763, 18

\bibitem[{{Bernardi} {et~al}\mbox{.}(2010){Bernardi}, {Shankar}, {Hyde}, {Mei},
  {Marulli}, \& {Sheth}}]{Bernardi+2010}
{Bernardi} M., {Shankar} F., {Hyde} J.~B., {Mei} S., {Marulli} F., {Sheth}
  R.~K., 2010, \mnras, 404, 2087

\bibitem[{{Blanton} {et~al}\mbox{.}(2005){Blanton}, {Lupton}, {Schlegel},
  {Strauss}, {Brinkmann}, {Fukugita}, \& {Loveday}}]{Blanton+2005}
{Blanton} M.~R., {Lupton} R.~H., {Schlegel} D.~J., {Strauss} M.~A., {Brinkmann}
  J., {Fukugita} M., {Loveday} J., 2005, \apj, 631, 208

\bibitem[{{Blumenthal} {et~al}\mbox{.}(1984){Blumenthal}, {Faber}, {Primack},
  \& {Rees}}]{BFPR}
{Blumenthal} G.~R., {Faber} S.~M., {Primack} J.~R., {Rees} M.~J., 1984, \nat,
  311, 517

\bibitem[{{Bouch{\'e}} {et~al}\mbox{.}(2010){Bouch{\'e}}, {Dekel}, {Genzel},
  {Genel}, {Cresci}, {F{\"o}rster Schreiber}, {Shapiro}, {Davies}, \&
  {Tacconi}}]{Bouche+2010}
{Bouch{\'e}} N. {et~al.}, 2010, \apj, 718, 1001

\bibitem[{{Boylan-Kolchin} {et~al}\mbox{.}(2009){Boylan-Kolchin}, {Springel},
  {White}, {Jenkins}, \& {Lemson}}]{MillenniumII}
{Boylan-Kolchin} M., {Springel} V., {White} S.~D.~M., {Jenkins} A., {Lemson}
  G., 2009, \mnras, 398, 1150

\bibitem[{{Brennan} {et~al}\mbox{.}(2015){Brennan}, {Pandya}, {Somerville},
  {Barro}, {Taylor}, {Wuyts}, {Bell}, {Dekel}, {Ferguson}, {McIntosh},
  {Papovich}, \& {Primack}}]{Brennan+2015}
{Brennan} R. {et~al.}, 2015, \mnras, 451, 2933

\bibitem[{{Bryan} \& {Norman}(1998)}]{BryanNorman}
{Bryan} G.~L., {Norman} M.~L., 1998, \apj, 495, 80

\bibitem[{{Bullock} {et~al}\mbox{.}(2001){Bullock}, {Dekel}, {Kolatt},
  {Kravtsov}, {Klypin}, {Porciani}, \& {Primack}}]{Bullock+2001}
{Bullock} J.~S., {Dekel} A., {Kolatt} T.~S., {Kravtsov} A.~V., {Klypin} A.~A.,
  {Porciani} C., {Primack} J.~R., 2001, \apj, 555, 240

\bibitem[{{Chabrier}(2003)}]{Chabrier2003}
{Chabrier} G., 2003, \pasp, 115, 763

\bibitem[{{Conroy} \& {Wechsler}(2009)}]{Conroy+2009}
{Conroy} C., {Wechsler} R.~H., 2009, \apj, 696, 620

\bibitem[{{Conroy}, {Wechsler} \& {Kravtsov}(2006){Conroy}, {Wechsler}, \&
  {Kravtsov}}]{Conroy+2006}
{Conroy} C., {Wechsler} R.~H., {Kravtsov} A.~V., 2006, \apj, 647, 201

\bibitem[{{Cuesta} {et~al}\mbox{.}(2008){Cuesta}, {Prada}, {Klypin}, \&
  {Moles}}]{Cuesta+2008}
{Cuesta} A.~J., {Prada} F., {Klypin} A., {Moles} M., 2008, \mnras, 389, 385

\bibitem[{{Daddi} {et~al}\mbox{.}(2007){Daddi}, {Dickinson}, {Morrison},
  {Chary}, {Cimatti}, {Elbaz}, {Frayer}, {Renzini}, {Pope}, {Alexander},
  {Bauer}, {Giavalisco}, {Huynh}, {Kurk}, \& {Mignoli}}]{Daddi+2007}
{Daddi} E. {et~al.}, 2007, \apj, 670, 156

\bibitem[{{Dav{\'e}}, {Finlator} \& {Oppenheimer}(2011){Dav{\'e}}, {Finlator},
  \& {Oppenheimer}}]{Dave+2011}
{Dav{\'e}} R., {Finlator} K., {Oppenheimer} B.~D., 2011, \mnras, 416, 1354

\bibitem[{{Dav{\'e}}, {Finlator} \& {Oppenheimer}(2012){Dav{\'e}}, {Finlator},
  \& {Oppenheimer}}]{Dave+2012}
{Dav{\'e}} R., {Finlator} K., {Oppenheimer} B.~D., 2012, \mnras, 421, 98

\bibitem[{{Dekel} {et~al}\mbox{.}(2009){Dekel}, {Birnboim}, {Engel},
  {Freundlich}, {Goerdt}, {Mumcuoglu}, {Neistein}, {Pichon}, {Teyssier}, \&
  {Zinger}}]{Dekel+2009}
{Dekel} A. {et~al.}, 2009, \nat, 457, 451

\bibitem[{{Dekel} \& {Mandelker}(2014)}]{DekelMandelker2014}
{Dekel} A., {Mandelker} N., 2014, \mnras, 444, 2071

\bibitem[{{Dekel} {et~al}\mbox{.}(2013){Dekel}, {Zolotov}, {Tweed}, {Cacciato},
  {Ceverino}, \& {Primack}}]{Dekel+2013}
{Dekel} A., {Zolotov} A., {Tweed} D., {Cacciato} M., {Ceverino} D., {Primack}
  J.~R., 2013, \mnras, 435, 999

\bibitem[{{Diemand}, {Kuhlen} \& {Madau}(2007){Diemand}, {Kuhlen}, \&
  {Madau}}]{Diemand+2007}
{Diemand} J., {Kuhlen} M., {Madau} P., 2007, \apj, 667, 859

\bibitem[{{Diemer}, {More} \& {Kravtsov}(2013){Diemer}, {More}, \&
  {Kravtsov}}]{Diemer+2013}
{Diemer} B., {More} S., {Kravtsov} A.~V., 2013, \apj, 766, 25

\bibitem[{{Elbaz} {et~al}\mbox{.}(2007){Elbaz}, {Daddi}, {Le Borgne},
  {Dickinson}, {Alexander}, {Chary}, {Starck}, {Brandt}, {Kitzbichler},
  {MacDonald}, {Nonino}, {Popesso}, {Stern}, \& {Vanzella}}]{Elbaz+2007}
{Elbaz} D. {et~al.}, 2007, \aap, 468, 33

\bibitem[{{Faber}(1984)}]{Faber84}
{Faber} S.~M., 1984, in Large-Scale Structure of the Universe, {Setti} G., {Van
  Hove} L., eds., p. 187

\bibitem[{{Faucher-Gigu{\`e}re}, {Kere{\v s}} \&
  {Ma}(2011){Faucher-Gigu{\`e}re}, {Kere{\v s}}, \& {Ma}}]{Faucher+2011}
{Faucher-Gigu{\`e}re} C.-A., {Kere{\v s}} D., {Ma} C.-P., 2011, \mnras, 417,
  2982

\bibitem[{{Feldmann}(2015)}]{Feldmann2015}
{Feldmann} R., 2015, \mnras, 449, 3274

\bibitem[{{Finkelstein} {et~al}\mbox{.}(2015){Finkelstein}, {Song}, {Behroozi},
  {Somerville}, {Papovich}, {Milosavljevic}, {Dekel}, {Narayanan}, {Ashby},
  {Cooray}, {Fazio}, {Ferguson}, {Koekemoer}, {Salmon}, \&
  {Willner}}]{Finkelstein15}
{Finkelstein} S.~L. {et~al.}, 2015, arXiv:1504.00005

\bibitem[{{Firmani} \& {Avila-Reese}(2010)}]{Firmani+2010a}
{Firmani} C., {Avila-Reese} V., 2010, \apj, 723, 755

\bibitem[{{Fontanot} {et~al}\mbox{.}(2009){Fontanot}, {De Lucia}, {Monaco},
  {Somerville}, \& {Santini}}]{Fontanot+2009}
{Fontanot} F., {De Lucia} G., {Monaco} P., {Somerville} R.~S., {Santini} P.,
  2009, \mnras, 397, 1776

\bibitem[{{Forbes} {et~al}\mbox{.}(2014){Forbes}, {Krumholz}, {Burkert}, \&
  {Dekel}}]{Forbes+2014}
{Forbes} J.~C., {Krumholz} M.~R., {Burkert} A., {Dekel} A., 2014, \mnras, 443,
  168

\bibitem[{{Gao}, {Springel} \& {White}(2005){Gao}, {Springel}, \&
  {White}}]{Gao+2005}
{Gao} L., {Springel} V., {White} S.~D.~M., 2005, \mnras, 363, L66

\bibitem[{{Genzel} {et~al}\mbox{.}(2015){Genzel}, {Tacconi}, {Lutz},
  {Saintonge}, {Berta}, {Magnelli}, {Combes}, {Garc{\'{\i}}a-Burillo}, {Neri},
  {Bolatto}, {Contini}, {Lilly}, {Boissier}, {Boone}, {Bouch{\'e}}, {Bournaud},
  {Burkert}, {Carollo}, {Colina}, {Cooper}, {Cox}, {Feruglio}, {F{\"o}rster
  Schreiber}, {Freundlich}, {Gracia-Carpio}, {Juneau}, {Kovac}, {Lippa},
  {Naab}, {Salome}, {Renzini}, {Sternberg}, {Walter}, {Weiner}, {Weiss}, \&
  {Wuyts}}]{Genzel+2015}
{Genzel} R. {et~al.}, 2015, \apj, 800, 20

\bibitem[{{Gonz{\'a}lez-Samaniego}
  {et~al}\mbox{.}(2014){Gonz{\'a}lez-Samaniego}, {Col{\'{\i}}n}, {Avila-Reese},
  {Rodr{\'{\i}}guez-Puebla}, \& {Valenzuela}}]{GS+2014}
{Gonz{\'a}lez-Samaniego} A., {Col{\'{\i}}n} P., {Avila-Reese} V.,
  {Rodr{\'{\i}}guez-Puebla} A., {Valenzuela} O., 2014, \apj, 785, 58

\bibitem[{{Gottloeber} \& {Klypin}(2008)}]{Gottloeber+2008}
{Gottloeber} S., {Klypin} A., 2008, ArXiv e-prints

\bibitem[{{Hearin}, {Behroozi} \& {van den Bosch}(2015){Hearin}, {Behroozi}, \&
  {van den Bosch}}]{HearinBehroozivdB}
{Hearin} A.~P., {Behroozi} P.~S., {van den Bosch} F.~C., 2015, ArXiv e-prints

\bibitem[{{Hearin} \& {Watson}(2013)}]{HearinWatson2013}
{Hearin} A.~P., {Watson} D.~F., 2013, \mnras, 435, 1313

\bibitem[{{Hearin}, {Watson} \& {van den Bosch}(2014){Hearin}, {Watson}, \&
  {van den Bosch}}]{HearinWatsonvdB+2014}
{Hearin} A.~P., {Watson} D.~F., {van den Bosch} F.~C., 2014, ArXiv e-prints

\bibitem[{{Henry} {et~al}\mbox{.}(2013){Henry}, {Scarlata}, {Dom{\'{\i}}nguez},
  {Malkan}, {Martin}, {Siana}, {Atek}, {Bedregal}, {Colbert}, {Rafelski},
  {Ross}, {Teplitz}, {Bunker}, {Dressler}, {Hathi}, {Masters}, {McCarthy}, \&
  {Straughn}}]{Henry+2013}
{Henry} A. {et~al.}, 2013, \apjl, 776, L27

\bibitem[{{Huang} \& {Kauffmann}(2014)}]{Huang+2014}
{Huang} M.-L., {Kauffmann} G., 2014, \mnras, 443, 1329

\bibitem[{{Huang} \& {Kauffmann}(2015)}]{Huang+2015}
{Huang} M.-L., {Kauffmann} G., 2015, \mnras, 450, 1375

\bibitem[{{Hudson} {et~al}\mbox{.}(2013){Hudson}, {Gillis}, {Coupon},
  {Hildebrandt}, {Erben}, {Heymans}, {Hoekstra}, {Kitching}, {Mellier},
  {Miller}, {Van Waerbeke}, {Bonnett}, {Fu}, {Kuijken}, {Rowe}, {Schrabback},
  {Semboloni}, {van Uitert}, \& {Velander}}]{Hudson+2014}
{Hudson} M.~J. {et~al.}, 2013, ArXiv e-prints

\bibitem[{{Ilbert} {et~al}\mbox{.}(2015){Ilbert}, {Arnouts}, {Le Floc'h},
  {Aussel}, {Bethermin}, {Capak}, {Hsieh}, {Kajisawa}, {Karim}, {Le F{\`e}vre},
  {Lee}, {Lilly}, {McCracken}, {Michel-Dansac}, {Moutard}, {Renzini},
  {Salvato}, {Sanders}, {Scoville}, {Sheth}, {Silverman}, {Smol{\v c}i{\'c}},
  {Taniguchi}, \& {Tresse}}]{Ilbert+2015}
{Ilbert} O. {et~al.}, 2015, \aap, 579, A2

\bibitem[{{Karim} {et~al}\mbox{.}(2011){Karim}, {Schinnerer},
  {Mart{\'{\i}}nez-Sansigre}, {Sargent}, {van der Wel}, {Rix}, {Ilbert},
  {Smol{\v c}i{\'c}}, {Carilli}, {Pannella}, {Koekemoer}, {Bell}, \&
  {Salvato}}]{Karim+2011}
{Karim} A. {et~al.}, 2011, \apj, 730, 61

\bibitem[{{Kauffmann} {et~al}\mbox{.}(2013){Kauffmann}, {Li}, {Zhang}, \&
  {Weinmann}}]{Kauffmann+2013}
{Kauffmann} G., {Li} C., {Zhang} W., {Weinmann} S., 2013, \mnras, 430, 1447

\bibitem[{{Kelson}(2014)}]{Kelson2014}
{Kelson} D.~D., 2014, ArXiv e-prints

\bibitem[{{Klypin} {et~al}\mbox{.}(2014){Klypin}, {Yepes}, {Gottlober},
  {Prada}, \& {Hess}}]{Klypin+2014}
{Klypin} A., {Yepes} G., {Gottlober} S., {Prada} F., {Hess} S., 2014, ArXiv
  e-prints

\bibitem[{{Klypin}, {Trujillo-Gomez} \& {Primack}(2011){Klypin},
  {Trujillo-Gomez}, \& {Primack}}]{Klypin+2011}
{Klypin} A.~A., {Trujillo-Gomez} S., {Primack} J., 2011, \apj, 740, 102

\bibitem[{{Kravtsov} {et~al}\mbox{.}(2004){Kravtsov}, {Berlind}, {Wechsler},
  {Klypin}, {Gottl{\"o}ber}, {Allgood}, \& {Primack}}]{Kravtsov+2004}
{Kravtsov} A.~V., {Berlind} A.~A., {Wechsler} R.~H., {Klypin} A.~A.,
  {Gottl{\"o}ber} S., {Allgood} B., {Primack} J.~R., 2004, \apj, 609, 35

\bibitem[{{Kravtsov}, {Klypin} \& {Khokhlov}(1997){Kravtsov}, {Klypin}, \&
  {Khokhlov}}]{Kravtsov+1997}
{Kravtsov} A.~V., {Klypin} A.~A., {Khokhlov} A.~M., 1997, \apjs, 111, 73

\bibitem[{{Krumholz} \& {Dekel}(2012)}]{Krumholz+2012}
{Krumholz} M.~R., {Dekel} A., 2012, \apj, 753, 16

\bibitem[{{Leauthaud} {et~al}\mbox{.}(2012){Leauthaud}, {Tinker}, {Bundy},
  {Behroozi}, {Massey}, {Rhodes}, {George}, {Kneib}, {Benson}, {Wechsler},
  {Busha}, {Capak}, {Cort{\^e}s}, {Ilbert}, {Koekemoer}, {Le F{\`e}vre},
  {Lilly}, {McCracken}, {Salvato}, {Schrabback}, {Scoville}, {Smith}, \&
  {Taylor}}]{Leauthaud+2012}
{Leauthaud} A. {et~al.}, 2012, \apj, 744, 159

\bibitem[{{Lee} {et~al}\mbox{.}(2012){Lee}, {Ferguson}, {Wiklind}, {Dahlen},
  {Dickinson}, {Giavalisco}, {Grogin}, {Papovich}, {Messias}, {Guo}, \&
  {Lin}}]{Lee+2012}
{Lee} K.-S. {et~al.}, 2012, \apj, 752, 66

\bibitem[{{Lilly} {et~al}\mbox{.}(2013){Lilly}, {Carollo}, {Pipino}, {Renzini},
  \& {Peng}}]{Lilly+2013}
{Lilly} S.~J., {Carollo} C.~M., {Pipino} A., {Renzini} A., {Peng} Y., 2013,
  \apj, 772, 119

\bibitem[{{Madau} \& {Dickinson}(2014)}]{Madau+2014}
{Madau} P., {Dickinson} M., 2014, \araa, 52, 415

\bibitem[{{Maiolino} {et~al}\mbox{.}(2008){Maiolino}, {Nagao}, {Grazian},
  {Cocchia}, {Marconi}, {Mannucci}, {Cimatti}, {Pipino}, {Ballero}, {Calura},
  {Chiappini}, {Fontana}, {Granato}, {Matteucci}, {Pastorini}, {Pentericci},
  {Risaliti}, {Salvati}, \& {Silva}}]{Maiolino+2008}
{Maiolino} R. {et~al.}, 2008, \aap, 488, 463

\bibitem[{{Mannucci} {et~al}\mbox{.}(2010){Mannucci}, {Cresci}, {Maiolino},
  {Marconi}, \& {Gnerucci}}]{Mannucci+2010}
{Mannucci} F., {Cresci} G., {Maiolino} R., {Marconi} A., {Gnerucci} A., 2010,
  \mnras, 408, 2115

\bibitem[{{Marchesini} {et~al}\mbox{.}(2009){Marchesini}, {van Dokkum},
  {F{\"o}rster Schreiber}, {Franx}, {Labb{\'e}}, \& {Wuyts}}]{Marchesini+2009}
{Marchesini} D., {van Dokkum} P.~G., {F{\"o}rster Schreiber} N.~M., {Franx} M.,
  {Labb{\'e}} I., {Wuyts} S., 2009, \apj, 701, 1765

\bibitem[{{Mitra}, {Dav{\'e}} \& {Finlator}(2015){Mitra}, {Dav{\'e}}, \&
  {Finlator}}]{Mitra+2015}
{Mitra} S., {Dav{\'e}} R., {Finlator} K., 2015, \mnras, 452, 1184

\bibitem[{{Mo}, {van den Bosch} \& {White}(2010){Mo}, {van den Bosch}, \&
  {White}}]{MvdBW}
{Mo} H., {van den Bosch} F.~C., {White} S., 2010, {Galaxy Formation and
  Evolution}. Cambridge, UK: Cambridge University Press, 2010

\bibitem[{{More}, {Diemer} \& {Kravtsov}(2015){More}, {Diemer}, \&
  {Kravtsov}}]{More+2015}
{More} S., {Diemer} B., {Kravtsov} A., 2015, ArXiv e-prints

\bibitem[{{More} {et~al}\mbox{.}(2011){More}, {van den Bosch}, {Cacciato},
  {Skibba}, {Mo}, \& {Yang}}]{More+2011}
{More} S., {van den Bosch} F.~C., {Cacciato} M., {Skibba} R., {Mo} H.~J.,
  {Yang} X., 2011, \mnras, 410, 210

\bibitem[{{Mortlock} {et~al}\mbox{.}(2011){Mortlock}, {Conselice}, {Bluck},
  {Bauer}, {Gr{\"u}tzbauch}, {Buitrago}, \& {Ownsworth}}]{Mortlock+2011}
{Mortlock} A., {Conselice} C.~J., {Bluck} A.~F.~L., {Bauer} A.~E.,
  {Gr{\"u}tzbauch} R., {Buitrago} F., {Ownsworth} J., 2011, \mnras, 413, 2845

\bibitem[{{Moster}, {Naab} \& {White}(2013){Moster}, {Naab}, \&
  {White}}]{Moster+2013}
{Moster} B.~P., {Naab} T., {White} S.~D.~M., 2013, \mnras, 428, 3121

\bibitem[{{Moustakas} {et~al}\mbox{.}(2013){Moustakas}, {Coil}, {Aird},
  {Blanton}, {Cool}, {Eisenstein}, {Mendez}, {Wong}, {Zhu}, \&
  {Arnouts}}]{Moustakas+2013}
{Moustakas} J. {et~al.}, 2013, \apj, 767, 50

\bibitem[{{Navarro}, {Frenk} \& {White}(1997){Navarro}, {Frenk}, \&
  {White}}]{NFW}
{Navarro} J.~F., {Frenk} C.~S., {White} S.~D.~M., 1997, \apj, 490, 493

\bibitem[{{Noeske} {et~al}\mbox{.}(2007){Noeske}, {Weiner}, {Faber},
  {Papovich}, {Koo}, {Somerville}, {Bundy}, {Conselice}, {Newman},
  {Schiminovich}, {Le Floc'h}, {Coil}, {Rieke}, {Lotz}, {Primack}, {Barmby},
  {Cooper}, {Davis}, {Ellis}, {Fazio}, {Guhathakurta}, {Huang}, {Kassin},
  {Martin}, {Phillips}, {Rich}, {Small}, {Willmer}, \& {Wilson}}]{Noeske+2007}
{Noeske} K.~G. {et~al.}, 2007, \apjl, 660, L43

\bibitem[{{P{\'e}rez-Gonz{\'a}lez}
  {et~al}\mbox{.}(2008){P{\'e}rez-Gonz{\'a}lez}, {Rieke}, {Villar}, {Barro},
  {Blaylock}, {Egami}, {Gallego}, {Gil de Paz}, {Pascual}, {Zamorano}, \&
  {Donley}}]{PG+2008}
{P{\'e}rez-Gonz{\'a}lez} P.~G. {et~al.}, 2008, \apj, 675, 234

\bibitem[{{Planck Collaboration} {et~al}\mbox{.}(2014){Planck Collaboration},
  {Ade}, {Aghanim}, {Armitage-Caplan}, {Arnaud}, {Ashdown}, {Atrio-Barandela},
  {Aumont}, {Baccigalupi}, {Banday}, \& et~al.}]{Planck13}
{Planck Collaboration} {et~al.}, 2014, \aap, 571, A16

\bibitem[{{Planck Collaboration} {et~al}\mbox{.}(2015){Planck Collaboration},
  {Ade}, {Aghanim}, {Arnaud}, {Ashdown}, {Aumont}, {Baccigalupi}, {Banday},
  {Barreiro}, {Bartlett}, \& et~al.}]{Planck15}
{Planck Collaboration} {et~al.}, 2015, ArXiv e-prints

\bibitem[{{Porter} {et~al}\mbox{.}(2014){Porter}, {Somerville}, {Primack}, \&
  {Johansson}}]{Porter+2014}
{Porter} L.~A., {Somerville} R.~S., {Primack} J.~R., {Johansson} P.~H., 2014,
  \mnras, 444, 942

\bibitem[{{Prada} {et~al}\mbox{.}(2006){Prada}, {Klypin}, {Simonneau},
  {Betancort-Rijo}, {Patiri}, {Gottl{\"o}ber}, \& {Sanchez-Conde}}]{Prada+2006}
{Prada} F., {Klypin} A.~A., {Simonneau} E., {Betancort-Rijo} J., {Patiri} S.,
  {Gottl{\"o}ber} S., {Sanchez-Conde} M.~A., 2006, \apj, 645, 1001

\bibitem[{Primack(1984)}]{Primack84}
Primack J.~R., 1984, Proc. Int. Sch. Phys. Fermi, 92, 140

\bibitem[{{Reddick} {et~al}\mbox{.}(2013){Reddick}, {Wechsler}, {Tinker}, \&
  {Behroozi}}]{Reddick+2013}
{Reddick} R.~M., {Wechsler} R.~H., {Tinker} J.~L., {Behroozi} P.~S., 2013,
  \apj, 771, 30

\bibitem[{{Reddy} {et~al}\mbox{.}(2012){Reddy}, {Dickinson}, {Elbaz},
  {Morrison}, {Giavalisco}, {Ivison}, {Papovich}, {Scott}, {Buat},
  {Burgarella}, {Charmandaris}, {Daddi}, {Magdis}, {Murphy}, {Altieri},
  {Aussel}, {Dannerbauer}, {Dasyra}, {Hwang}, {Kartaltepe}, {Leiton},
  {Magnelli}, \& {Popesso}}]{Reddy+2012}
{Reddy} N. {et~al.}, 2012, \apj, 744, 154

\bibitem[{{Rodr{\'{\i}}guez-Puebla}, {Avila-Reese} \&
  {Drory}(2013){Rodr{\'{\i}}guez-Puebla}, {Avila-Reese}, \& {Drory}}]{RAD13}
{Rodr{\'{\i}}guez-Puebla} A., {Avila-Reese} V., {Drory} N., 2013, \apj, 767, 92

\bibitem[{{Rodr{\'{\i}}guez-Puebla}
  {et~al}\mbox{.}(2015){Rodr{\'{\i}}guez-Puebla}, {Avila-Reese}, {Yang},
  {Foucaud}, {Drory}, \& {Jing}}]{RP+2015}
{Rodr{\'{\i}}guez-Puebla} A., {Avila-Reese} V., {Yang} X., {Foucaud} S.,
  {Drory} N., {Jing} Y.~P., 2015, \apj, 799, 130

\bibitem[{{Rodr{\'{\i}}guez-Puebla}, {Drory} \&
  {Avila-Reese}(2012){Rodr{\'{\i}}guez-Puebla}, {Drory}, \&
  {Avila-Reese}}]{RDA12}
{Rodr{\'{\i}}guez-Puebla} A., {Drory} N., {Avila-Reese} V., 2012, \apj, 756, 2

\bibitem[{{Salim} {et~al}\mbox{.}(2007){Salim}, {Rich}, {Charlot},
  {Brinchmann}, {Johnson}, {Schiminovich}, {Seibert}, {Mallery}, {Heckman},
  {Forster}, {Friedman}, {Martin}, {Morrissey}, {Neff}, {Small}, {Wyder},
  {Bianchi}, {Donas}, {Lee}, {Madore}, {Milliard}, {Szalay}, {Welsh}, \&
  {Yi}}]{Salim+2007}
{Salim} S. {et~al.}, 2007, \apjs, 173, 267

\bibitem[{{S{\'a}nchez} {et~al}\mbox{.}(2013){S{\'a}nchez}, {Rosales-Ortega},
  {Jungwiert}, {Iglesias-P{\'a}ramo}, {V{\'{\i}}lchez}, {Marino}, {Walcher},
  {Husemann}, {Mast}, {Monreal-Ibero}, {Cid Fernandes}, {P{\'e}rez},
  {Gonz{\'a}lez Delgado}, {Garc{\'{\i}}a-Benito}, {Galbany}, {van de Ven},
  {Jahnke}, {Flores}, {Bland-Hawthorn}, {L{\'o}pez-S{\'a}nchez}, {Stanishev},
  {Miralles-Caballero}, {D{\'{\i}}az}, {S{\'a}nchez-Blazquez}, {Moll{\'a}},
  {Gallazzi}, {Papaderos}, {Gomes}, {Gruel}, {P{\'e}rez}, {Ruiz-Lara},
  {Florido}, {de Lorenzo-C{\'a}ceres}, {Mendez-Abreu}, {Kehrig}, {Roth},
  {Ziegler}, {Alves}, {Wisotzki}, {Kupko}, {Quirrenbach}, {Bomans}, \& {Califa
  Collaboration}}]{Sanchez+2013}
{S{\'a}nchez} S.~F. {et~al.}, 2013, \aap, 554, A58

\bibitem[{{Sanders} {et~al}\mbox{.}(2015){Sanders}, {Shapley}, {Kriek},
  {Reddy}, {Freeman}, {Coil}, {Siana}, {Mobasher}, {Shivaei}, {Price}, \& {de
  Groot}}]{Sanders+2015}
{Sanders} R.~L. {et~al.}, 2015, \apj, 799, 138

\bibitem[{{Santini} {et~al}\mbox{.}(2009){Santini}, {Fontana}, {Grazian},
  {Salimbeni}, {Fiore}, {Fontanot}, {Boutsia}, {Castellano}, {Cristiani}, {de
  Santis}, {Gallozzi}, {Giallongo}, {Menci}, {Nonino}, {Paris}, {Pentericci},
  \& {Vanzella}}]{Santini+2009}
{Santini} P. {et~al.}, 2009, \aap, 504, 751

\bibitem[{{Sargent} {et~al}\mbox{.}(2014){Sargent}, {Daddi}, {B{\'e}thermin},
  {Aussel}, {Magdis}, {Hwang}, {Juneau}, {Elbaz}, \& {da Cunha}}]{Sargent+2014}
{Sargent} M.~T. {et~al.}, 2014, \apj, 793, 19

\bibitem[{{Schreiber} {et~al}\mbox{.}(2015){Schreiber}, {Pannella}, {Elbaz},
  {B{\'e}thermin}, {Inami}, {Dickinson}, {Magnelli}, {Wang}, {Aussel}, {Daddi},
  {Juneau}, {Shu}, {Sargent}, {Buat}, {Faber}, {Ferguson}, {Giavalisco},
  {Koekemoer}, {Magdis}, {Morrison}, {Papovich}, {Santini}, \&
  {Scott}}]{Schreiber+2015}
{Schreiber} C. {et~al.}, 2015, \aap, 575, A74

\bibitem[{{Shankar} {et~al}\mbox{.}(2014){Shankar}, {Guo}, {Bouillot},
  {Rettura}, {Meert}, {Buchan}, {Kravtsov}, {Bernardi}, {Sheth}, {Vikram},
  {Marchesini}, {Behroozi}, {Zheng}, {Maraston}, {Ascaso}, {Lemaux}, {Capozzi},
  {Huertas-Company}, {Gal}, {Lubin}, {Conselice}, {Carollo}, \&
  {Cattaneo}}]{Shankar+2014}
{Shankar} F. {et~al.}, 2014, \apjl, 797, L27

\bibitem[{{Speagle} {et~al}\mbox{.}(2014){Speagle}, {Steinhardt}, {Capak}, \&
  {Silverman}}]{Speagle+2014}
{Speagle} J.~S., {Steinhardt} C.~L., {Capak} P.~L., {Silverman} J.~D., 2014,
  \apjs, 214, 15

\bibitem[{{Springel} {et~al}\mbox{.}(2005){Springel}, {White}, {Jenkins},
  {Frenk}, {Yoshida}, {Gao}, {Navarro}, {Thacker}, {Croton}, {Helly},
  {Peacock}, {Cole}, {Thomas}, {Couchman}, {Evrard}, {Colberg}, \&
  {Pearce}}]{Millennium}
{Springel} V. {et~al.}, 2005, \nat, 435, 629

\bibitem[{{Stark} {et~al}\mbox{.}(2013){Stark}, {Schenker}, {Ellis},
  {Robertson}, {McLure}, \& {Dunlop}}]{Stark+2013}
{Stark} D.~P., {Schenker} M.~A., {Ellis} R., {Robertson} B., {McLure} R.,
  {Dunlop} J., 2013, \apj, 763, 129

\bibitem[{{Steidel} {et~al}\mbox{.}(2014){Steidel}, {Rudie}, {Strom},
  {Pettini}, {Reddy}, {Shapley}, {Trainor}, {Erb}, {Turner}, {Konidaris},
  {Kulas}, {Mace}, {Matthews}, \& {McLean}}]{Steidel+2014}
{Steidel} C.~C. {et~al.}, 2014, \apj, 795, 165

\bibitem[{{Taghizadeh-Popp} {et~al}\mbox{.}(2015){Taghizadeh-Popp}, {Fall},
  {White}, \& {Szalay}}]{Popp15}
{Taghizadeh-Popp} M., {Fall} S.~M., {White} R.~L., {Szalay} A.~S., 2015, \apj,
  801, 14

\bibitem[{{Tinker} {et~al}\mbox{.}(2013){Tinker}, {Leauthaud}, {Bundy},
  {George}, {Behroozi}, {Massey}, {Rhodes}, \& {Wechsler}}]{Tinker+2013}
{Tinker} J.~L., {Leauthaud} A., {Bundy} K., {George} M.~R., {Behroozi} P.,
  {Massey} R., {Rhodes} J., {Wechsler} R.~H., 2013, \apj, 778, 93

\bibitem[{{Tremonti} {et~al}\mbox{.}(2004){Tremonti}, {Heckman}, {Kauffmann},
  {Brinchmann}, {Charlot}, {White}, {Seibert}, {Peng}, {Schlegel}, {Uomoto},
  {Fukugita}, \& {Brinkmann}}]{Tremonti+2004}
{Tremonti} C.~A. {et~al.}, 2004, \apj, 613, 898

\bibitem[{{Troncoso} {et~al}\mbox{.}(2014){Troncoso}, {Maiolino}, {Sommariva},
  {Cresci}, {Mannucci}, {Marconi}, {Meneghetti}, {Grazian}, {Cimatti},
  {Fontana}, {Nagao}, \& {Pentericci}}]{Troncoso+2014}
{Troncoso} P. {et~al.}, 2014, \aap, 563, A58

\bibitem[{{Vale} \& {Ostriker}(2004)}]{ValeOstriker04}
{Vale} A., {Ostriker} J.~P., 2004, \mnras, 353, 189

\bibitem[{{van den Bosch}, {Tormen} \& {Giocoli}(2005){van den Bosch},
  {Tormen}, \& {Giocoli}}]{vandenBosch+2005}
{van den Bosch} F.~C., {Tormen} G., {Giocoli} C., 2005, \mnras, 359, 1029

\bibitem[{{Watson} \& {Conroy}(2013)}]{Watson+2013}
{Watson} D.~F., {Conroy} C., 2013, ArXiv e-prints

\bibitem[{{Wechsler} {et~al}\mbox{.}(2002){Wechsler}, {Bullock}, {Primack},
  {Kravtsov}, \& {Dekel}}]{Wechsler+2002}
{Wechsler} R.~H., {Bullock} J.~S., {Primack} J.~R., {Kravtsov} A.~V., {Dekel}
  A., 2002, \apj, 568, 52

\bibitem[{{Wechsler} {et~al}\mbox{.}(2006){Wechsler}, {Zentner}, {Bullock},
  {Kravtsov}, \& {Allgood}}]{Wechsler+2006}
{Wechsler} R.~H., {Zentner} A.~R., {Bullock} J.~S., {Kravtsov} A.~V., {Allgood}
  B., 2006, \apj, 652, 71

\bibitem[{{Wetzel} \& {Nagai}(2014)}]{WetzelNagai2015}
{Wetzel} A.~R., {Nagai} D., 2014, ArXiv e-prints

\bibitem[{{Wetzel} {et~al}\mbox{.}(2013){Wetzel}, {Tinker}, {Conroy}, \& {van
  den Bosch}}]{Wetzel+2013}
{Wetzel} A.~R., {Tinker} J.~L., {Conroy} C., {van den Bosch} F.~C., 2013,
  \mnras, 432, 336

\bibitem[{{Whitaker} {et~al}\mbox{.}(2014){Whitaker}, {Franx}, {Leja}, {van
  Dokkum}, {Henry}, {Skelton}, {Fumagalli}, {Momcheva}, {Brammer}, {Labb{\'e}},
  {Nelson}, \& {Rigby}}]{Whitaker+2014}
{Whitaker} K.~E. {et~al.}, 2014, \apj, 795, 104

\bibitem[{{Wuyts} {et~al}\mbox{.}(2014){Wuyts}, {Kurk}, {F{\"o}rster
  Schreiber}, {Genzel}, {Wisnioski}, {Bandara}, {Wuyts}, {Beifiori}, {Bender},
  {Brammer}, {Burkert}, {Buschkamp}, {Carollo}, {Chan}, {Davies}, {Eisenhauer},
  {Fossati}, {Kulkarni}, {Lang}, {Lilly}, {Lutz}, {Mancini}, {Mendel},
  {Momcheva}, {Naab}, {Nelson}, {Renzini}, {Rosario}, {Saglia}, {Seitz},
  {Sharples}, {Sternberg}, {Tacchella}, {Tacconi}, {van Dokkum}, \&
  {Wilman}}]{Wuyts+2014}
{Wuyts} E. {et~al.}, 2014, \apjl, 789, L40

\bibitem[{{Yang}, {Mo} \& {van den Bosch}(2009){Yang}, {Mo}, \& {van den
  Bosch}}]{Yang+2009b}
{Yang} X., {Mo} H.~J., {van den Bosch} F.~C., 2009, \apj, 695, 900

\bibitem[{{Yang} {et~al}\mbox{.}(2007){Yang}, {Mo}, {van den Bosch},
  {Pasquali}, {Li}, \& {Barden}}]{Yang+2007}
{Yang} X., {Mo} H.~J., {van den Bosch} F.~C., {Pasquali} A., {Li} C., {Barden}
  M., 2007, \apj, 671, 153

\bibitem[{{Yang} {et~al}\mbox{.}(2012){Yang}, {Mo}, {van den Bosch}, {Zhang},
  \& {Han}}]{Yang+2012}
{Yang} X., {Mo} H.~J., {van den Bosch} F.~C., {Zhang} Y., {Han} J., 2012, \apj,
  752, 41

\bibitem[{{Zahid} {et~al}\mbox{.}(2012){Zahid}, {Dima}, {Kewley}, {Erb}, \&
  {Dav{\'e}}}]{Zahid+2012}
{Zahid} H.~J., {Dima} G.~I., {Kewley} L.~J., {Erb} D.~K., {Dav{\'e}} R., 2012,
  \apj, 757, 54

\bibitem[{{Zahid} {et~al}\mbox{.}(2014){Zahid}, {Dima}, {Kudritzki}, {Kewley},
  {Geller}, {Hwang}, {Silverman}, \& {Kashino}}]{Zahid+2014}
{Zahid} H.~J., {Dima} G.~I., {Kudritzki} R.-P., {Kewley} L.~J., {Geller} M.~J.,
  {Hwang} H.~S., {Silverman} J.~D., {Kashino} D., 2014, \apj, 791, 130

\bibitem[{{Zolotov} {et~al}\mbox{.}(2015){Zolotov}, {Dekel}, {Mandelker},
  {Tweed}, {Inoue}, {DeGraf}, {Ceverino}, {Primack}, {Barro}, \&
  {Faber}}]{Zolotov+2015}
{Zolotov} A. {et~al.}, 2015, \mnras, 450, 2327

\end{thebibliography}


\begin{thebibliography}{257}
\expandafter\ifx\csname natexlab\endcsname\relax\def\natexlab#1{#1}\fi

\bibitem[{{Abramson} {et~al}\mbox{.}(2016){Abramson}, {Gladders}, {Dressler},
  {Oemler}, {Poggianti}, \& {Vulcani}}]{Abramson+2016}
{Abramson} L.~E., {Gladders} M.~D., {Dressler} A., {Oemler}, Jr. A.,
  {Poggianti} B., {Vulcani} B., 2016, \apj, 832, 7

\bibitem[{{Allen} {et~al}\mbox{.}(2016){Allen}, {Kacprzak}, {Glazebrook},
  {Labbe}, {Tran}, {Spitler}, {Cowley}, {Nanayakkara}, {Papovich}, {Quadri},
  {Straatman}, {Tilvi}, \& {van Dokkum}}]{Allen+2016}
{Allen} R.~J. {et~al.}, 2016, ArXiv e-prints

\bibitem[{{Avila-Reese}, {Zavala} \& {Lacerna}(2014){Avila-Reese}, {Zavala}, \&
  {Lacerna}}]{Avila-Reese+2014}
{Avila-Reese} V., {Zavala} J., {Lacerna} I., 2014, \mnras, 441, 417

\bibitem[{{Baldry} {et~al}\mbox{.}(2012){Baldry}, {Driver}, {Loveday},
  {Taylor}, {Kelvin}, {Liske}, {Norberg}, {Robotham}, {Brough}, {Hopkins},
  {Bamford}, {Peacock}, {Bland-Hawthorn}, {Conselice}, {Croom}, {Jones},
  {Parkinson}, {Popescu}, {Prescott}, {Sharp}, \& {Tuffs}}]{Baldry+2012}
{Baldry} I.~K. {et~al.}, 2012, \mnras, 421, 621

\bibitem[{{Barnes} \& {Hernquist}(1996)}]{Barnes_Hernquist1996}
{Barnes} J.~E., {Hernquist} L., 1996, \apj, 471, 115

\bibitem[{{Barro} {et~al}\mbox{.}(2015){Barro}, {Faber}, {Koo}, {Dekel},
  {Fang}, {Trump}, {Perez-Gonzalez}, {Pacifici}, {Primack}, {Somerville},
  {Yan}, {Guo}, {Liu}, {Ceverino}, {Kocevski}, \& {McGrath}}]{Barro+2015}
{Barro} G. {et~al.}, 2015, ArXiv e-prints

\bibitem[{{Barro} {et~al}\mbox{.}(2013){Barro}, {Faber},
  {P{\'e}rez-Gonz{\'a}lez}, {Koo}, {Williams}, {Kocevski}, {Trump}, {Mozena},
  {McGrath}, {van der Wel}, {Wuyts}, {Bell}, {Croton}, {Ceverino}, {Dekel},
  {Ashby}, {Cheung}, {Ferguson}, {Fontana}, {Fang}, {Giavalisco}, {Grogin},
  {Guo}, {Hathi}, {Hopkins}, {Huang}, {Koekemoer}, {Kartaltepe}, {Lee},
  {Newman}, {Porter}, {Primack}, {Ryan}, {Rosario}, {Somerville}, {Salvato}, \&
  {Hsu}}]{Barro+2013}
{Barro} G. {et~al.}, 2013, \apj, 765, 104

\bibitem[{{Bastian}, {Covey} \& {Meyer}(2010){Bastian}, {Covey}, \&
  {Meyer}}]{Bastian+2010}
{Bastian} N., {Covey} K.~R., {Meyer} M.~R., 2010, \araa, 48, 339

\bibitem[{{Behroozi}, {Conroy} \& {Wechsler}(2010){Behroozi}, {Conroy}, \&
  {Wechsler}}]{Behroozi+2010}
{Behroozi} P.~S., {Conroy} C., {Wechsler} R.~H., 2010, \apj, 717, 379

\bibitem[{{Behroozi}, {Wechsler} \& {Conroy}(2013{\natexlab{a}}){Behroozi},
  {Wechsler}, \& {Conroy}}]{Behroozi+2013c}
{Behroozi} P.~S., {Wechsler} R.~H., {Conroy} C., 2013{\natexlab{a}}, \apjl,
  762, L31

\bibitem[{{Behroozi}, {Wechsler} \& {Conroy}(2013{\natexlab{b}}){Behroozi},
  {Wechsler}, \& {Conroy}}]{Behroozi+2013}
{Behroozi} P.~S., {Wechsler} R.~H., {Conroy} C., 2013{\natexlab{b}}, \apj, 770,
  57

\bibitem[{{Behroozi} {et~al}\mbox{.}(2013){Behroozi}, {Wechsler}, {Wu},
  {Busha}, {Klypin}, \& {Primack}}]{Behroozi+2013b}
{Behroozi} P.~S., {Wechsler} R.~H., {Wu} H.-Y., {Busha} M.~T., {Klypin} A.~A.,
  {Primack} J.~R., 2013, \apj, 763, 18

\bibitem[{{Bell} \& {de Jong}(2001)}]{BellJong2001}
{Bell} E.~F., {de Jong} R.~S., 2001, \apj, 550, 212

\bibitem[{{Bell} {et~al}\mbox{.}(2003){Bell}, {McIntosh}, {Katz}, \&
  {Weinberg}}]{Bell+2003}
{Bell} E.~F., {McIntosh} D.~H., {Katz} N., {Weinberg} M.~D., 2003, \apjs, 149,
  289

\bibitem[{{Bell} {et~al}\mbox{.}(2012){Bell}, {van der Wel}, {Papovich},
  {Kocevski}, {Lotz}, {McIntosh}, {Kartaltepe}, {Faber}, {Ferguson},
  {Koekemoer}, {Grogin}, {Wuyts}, {Cheung}, {Conselice}, {Dekel}, {Dunlop},
  {Giavalisco}, {Herrington}, {Koo}, {McGrath}, {de Mello}, {Rix}, {Robaina},
  \& {Williams}}]{Bell+2012}
{Bell} E.~F. {et~al.}, 2012, \apj, 753, 167

\bibitem[{{Bell} {et~al}\mbox{.}(2007){Bell}, {Zheng}, {Papovich}, {Borch},
  {Wolf}, \& {Meisenheimer}}]{Bell+2007}
{Bell} E.~F., {Zheng} X.~Z., {Papovich} C., {Borch} A., {Wolf} C.,
  {Meisenheimer} K., 2007, \apj, 663, 834

\bibitem[{{Berlind} \& {Weinberg}(2002)}]{Berlind+2002}
{Berlind} A.~A., {Weinberg} D.~H., 2002, \apj, 575, 587

\bibitem[{{Bernardi} {et~al}\mbox{.}(2016){Bernardi}, {Meert}, {Sheth},
  {Huertas-Company}, {Maraston}, {Shankar}, \& {Vikram}}]{Bernardi+2016}
{Bernardi} M., {Meert} A., {Sheth} R.~K., {Huertas-Company} M., {Maraston} C.,
  {Shankar} F., {Vikram} V., 2016, \mnras, 455, 4122

\bibitem[{{Bernardi} {et~al}\mbox{.}(2013){Bernardi}, {Meert}, {Sheth},
  {Vikram}, {Huertas-Company}, {Mei}, \& {Shankar}}]{Bernardi+2013}
{Bernardi} M., {Meert} A., {Sheth} R.~K., {Vikram} V., {Huertas-Company} M.,
  {Mei} S., {Shankar} F., 2013, \mnras, 436, 697

\bibitem[{{Bernardi} {et~al}\mbox{.}(2010){Bernardi}, {Shankar}, {Hyde}, {Mei},
  {Marulli}, \& {Sheth}}]{Bernardi+2010}
{Bernardi} M., {Shankar} F., {Hyde} J.~B., {Mei} S., {Marulli} F., {Sheth}
  R.~K., 2010, \mnras, 404, 2087

\bibitem[{{Birnboim} \& {Dekel}(2003)}]{Birnboim+2003}
{Birnboim} Y., {Dekel} A., 2003, \mnras, 345, 349

\bibitem[{{Blanton} {et~al}\mbox{.}(2011){Blanton}, {Kazin}, {Muna}, {Weaver},
  \& {Price-Whelan}}]{Blanton+2011}
{Blanton} M.~R., {Kazin} E., {Muna} D., {Weaver} B.~A., {Price-Whelan} A.,
  2011, \aj, 142, 31

\bibitem[{{Blanton} {et~al}\mbox{.}(2005{\natexlab{a}}){Blanton}, {Lupton},
  {Schlegel}, {Strauss}, {Brinkmann}, {Fukugita}, \& {Loveday}}]{Blanton2005}
{Blanton} M.~R., {Lupton} R.~H., {Schlegel} D.~J., {Strauss} M.~A., {Brinkmann}
  J., {Fukugita} M., {Loveday} J., 2005{\natexlab{a}}, \apj, 631, 208

\bibitem[{{Blanton} {et~al}\mbox{.}(2005{\natexlab{b}}){Blanton}, {Schlegel},
  {Strauss}, {Brinkmann}, {Finkbeiner}, {Fukugita}, {Gunn}, {Hogg},
  {Ivezi{\'c}}, {Knapp}, {Lupton}, {Munn}, {Schneider}, {Tegmark}, \&
  {Zehavi}}]{Blanton+2005}
{Blanton} M.~R. {et~al.}, 2005{\natexlab{b}}, \aj, 129, 2562

\bibitem[{{Bluck} {et~al}\mbox{.}(2009){Bluck}, {Conselice}, {Bouwens},
  {Daddi}, {Dickinson}, {Papovich}, \& {Yan}}]{Bluck+2009}
{Bluck} A.~F.~L., {Conselice} C.~J., {Bouwens} R.~J., {Daddi} E., {Dickinson}
  M., {Papovich} C., {Yan} H., 2009, \mnras, 394, L51

\bibitem[{{Bouch{\'e}} {et~al}\mbox{.}(2010){Bouch{\'e}}, {Dekel}, {Genzel},
  {Genel}, {Cresci}, {F{\"o}rster Schreiber}, {Shapiro}, {Davies}, \&
  {Tacconi}}]{Bouche+2010}
{Bouch{\'e}} N. {et~al.}, 2010, \apj, 718, 1001

\bibitem[{{Bouwens} {et~al}\mbox{.}(2007){Bouwens}, {Illingworth}, {Franx}, \&
  {Ford}}]{Bouwens+2007}
{Bouwens} R.~J., {Illingworth} G.~D., {Franx} M., {Ford} H., 2007, \apj, 670,
  928

\bibitem[{{Bouwens} {et~al}\mbox{.}(2011){Bouwens}, {Illingworth}, {Oesch},
  {Labb{\'e}}, {Trenti}, {van Dokkum}, {Franx}, {Stiavelli}, {Carollo},
  {Magee}, \& {Gonzalez}}]{Bouwens+2011}
{Bouwens} R.~J. {et~al.}, 2011, \apj, 737, 90

\bibitem[{{Bouwens} {et~al}\mbox{.}(2014){Bouwens}, {Illingworth}, {Oesch},
  {Labb{\'e}}, {van Dokkum}, {Trenti}, {Franx}, {Smit}, {Gonzalez}, \&
  {Magee}}]{Bouwens+2014}
{Bouwens} R.~J. {et~al.}, 2014, \apj, 793, 115

\bibitem[{{Bouwens} {et~al}\mbox{.}(2015){Bouwens}, {Illingworth}, {Oesch},
  {Trenti}, {Labb{\'e}}, {Bradley}, {Carollo}, {van Dokkum}, {Gonzalez},
  {Holwerda}, {Franx}, {Spitler}, {Smit}, \& {Magee}}]{Bouwens+2015}
{Bouwens} R.~J. {et~al.}, 2015, \apj, 803, 34

\bibitem[{{Bouwens} {et~al}\mbox{.}(2016){Bouwens}, {Oesch}, {Labb{\'e}},
  {Illingworth}, {Fazio}, {Coe}, {Holwerda}, {Smit}, {Stefanon}, {van Dokkum},
  {Trenti}, {Ashby}, {Huang}, {Spitler}, {Straatman}, {Bradley}, \&
  {Magee}}]{Bouwens+2016}
{Bouwens} R.~J. {et~al.}, 2016, \apj, 830, 67

\bibitem[{{Bowler} {et~al}\mbox{.}(2014){Bowler}, {Dunlop}, {McLure}, {Rogers},
  {McCracken}, {Milvang-Jensen}, {Furusawa}, {Fynbo}, {Taniguchi}, {Afonso},
  {Bremer}, \& {Le F{\`e}vre}}]{Bowler+2014}
{Bowler} R.~A.~A. {et~al.}, 2014, \mnras, 440, 2810

\bibitem[{{Brammer} {et~al}\mbox{.}(2012){Brammer}, {van Dokkum}, {Franx},
  {Fumagalli}, {Patel}, {Rix}, {Skelton}, {Kriek}, {Nelson}, {Schmidt},
  {Bezanson}, {da Cunha}, {Erb}, {Fan}, {F{\"o}rster Schreiber}, {Illingworth},
  {Labb{\'e}}, {Leja}, {Lundgren}, {Magee}, {Marchesini}, {McCarthy},
  {Momcheva}, {Muzzin}, {Quadri}, {Steidel}, {Tal}, {Wake}, {Whitaker}, \&
  {Williams}}]{Brammer+2012}
{Brammer} G.~B. {et~al.}, 2012, \apjs, 200, 13

\bibitem[{{Bruzual}(2007)}]{BC07}
{Bruzual} G., 2007, in Astronomical Society of the Pacific Conference Series,
  Vol. 374, From Stars to Galaxies: Building the Pieces to Build Up the
  Universe, {Vallenari} A., {Tantalo} R., {Portinari} L., {Moretti} A., eds.,
  p. 303

\bibitem[{{Bruzual} \& {Charlot}(2003)}]{BC03}
{Bruzual} G., {Charlot} S., 2003, \mnras, 344, 1000

\bibitem[{{Bundy} {et~al}\mbox{.}(2006){Bundy}, {Ellis}, {Conselice}, {Taylor},
  {Cooper}, {Willmer}, {Weiner}, {Coil}, {Noeske}, \&
  {Eisenhardt}}]{Bundy+2006}
{Bundy} K. {et~al.}, 2006, \apj, 651, 120

\bibitem[{{Bundy} {et~al}\mbox{.}(2009){Bundy}, {Fukugita}, {Ellis}, {Targett},
  {Belli}, \& {Kodama}}]{Bundy+2009}
{Bundy} K., {Fukugita} M., {Ellis} R.~S., {Targett} T.~A., {Belli} S., {Kodama}
  T., 2009, \apj, 697, 1369

\bibitem[{{Burkert} {et~al}\mbox{.}(2008){Burkert}, {Naab}, {Johansson}, \&
  {Jesseit}}]{Burkert+2008}
{Burkert} A., {Naab} T., {Johansson} P.~H., {Jesseit} R., 2008, \apj, 685, 897

\bibitem[{{Calzetti} {et~al}\mbox{.}(2000){Calzetti}, {Armus}, {Bohlin},
  {Kinney}, {Koornneef}, \& {Storchi-Bergmann}}]{Calzetti+2000}
{Calzetti} D., {Armus} L., {Bohlin} R.~C., {Kinney} A.~L., {Koornneef} J.,
  {Storchi-Bergmann} T., 2000, \apj, 533, 682

\bibitem[{{Casey}, {Narayanan} \& {Cooray}(2014){Casey}, {Narayanan}, \&
  {Cooray}}]{Casey+2014}
{Casey} C.~M., {Narayanan} D., {Cooray} A., 2014, \physrep, 541, 45

\bibitem[{{Cattaneo} {et~al}\mbox{.}(2007){Cattaneo}, {Blaizot}, {Weinberg},
  {Kere{\v s}}, {Colombi}, {Dav{\'e}}, {Devriendt}, {Guiderdoni}, \&
  {Katz}}]{Cattaneo+2007}
{Cattaneo} A. {et~al.}, 2007, \mnras, 377, 63

\bibitem[{{Cattaneo} {et~al}\mbox{.}(2008){Cattaneo}, {Dekel}, {Faber}, \&
  {Guiderdoni}}]{Cattaneo+2008}
{Cattaneo} A., {Dekel} A., {Faber} S.~M., {Guiderdoni} B., 2008, \mnras, 389,
  567

\bibitem[{{Chabrier}(2003)}]{Chabrier2003}
{Chabrier} G., 2003, \pasp, 115, 763

\bibitem[{{Charlot} \& {Fall}(2000)}]{Charlot_Fall2000}
{Charlot} S., {Fall} S.~M., 2000, \apj, 539, 718

\bibitem[{{Chen} {et~al}\mbox{.}(2009){Chen}, {Wild}, {Kauffmann}, {Blaizot},
  {Davis}, {Noeske}, {Wang}, \& {Willmer}}]{Chen+2009}
{Chen} Y.-M., {Wild} V., {Kauffmann} G., {Blaizot} J., {Davis} M., {Noeske} K.,
  {Wang} J.-M., {Willmer} C., 2009, \mnras, 393, 406

\bibitem[{{Cheung} {et~al}\mbox{.}(2012){Cheung}, {Faber}, {Koo}, {Dutton},
  {Simard}, {McGrath}, {Huang}, {Bell}, {Dekel}, {Fang}, {Salim}, {Barro},
  {Bundy}, {Coil}, {Cooper}, {Conselice}, {Davis}, {Dom{\'{\i}}nguez},
  {Kassin}, {Kocevski}, {Koekemoer}, {Lin}, {Lotz}, {Newman}, {Phillips},
  {Rosario}, {Weiner}, \& {Willmer}}]{Cheung+2012}
{Cheung} E. {et~al.}, 2012, \apj, 760, 131

\bibitem[{{Conroy}(2013)}]{Conroy2013}
{Conroy} C., 2013, \araa, 51, 393

\bibitem[{{Conroy} {et~al}\mbox{.}(2013){Conroy}, {Dutton}, {Graves}, {Mendel},
  \& {van Dokkum}}]{Conroy+2013b}
{Conroy} C., {Dutton} A.~A., {Graves} G.~J., {Mendel} J.~T., {van Dokkum}
  P.~G., 2013, \apjl, 776, L26

\bibitem[{{Conroy}, {Gunn} \& {White}(2009){Conroy}, {Gunn}, \&
  {White}}]{Conroy+2009a}
{Conroy} C., {Gunn} J.~E., {White} M., 2009, \apj, 699, 486

\bibitem[{{Conroy} {et~al}\mbox{.}(2007){Conroy}, {Prada}, {Newman}, {Croton},
  {Coil}, {Conselice}, {Cooper}, {Davis}, {Faber}, {Gerke}, {Guhathakurta},
  {Klypin}, {Koo}, \& {Yan}}]{Conroy+2007}
{Conroy} C. {et~al.}, 2007, \apj, 654, 153

\bibitem[{{Conroy} \& {Wechsler}(2009)}]{Conroy+2009}
{Conroy} C., {Wechsler} R.~H., 2009, \apj, 696, 620

\bibitem[{{Conroy}, {Wechsler} \& {Kravtsov}(2006){Conroy}, {Wechsler}, \&
  {Kravtsov}}]{Conroy+2006}
{Conroy} C., {Wechsler} R.~H., {Kravtsov} A.~V., 2006, \apj, 647, 201

\bibitem[{{Conselice} {et~al}\mbox{.}(2003){Conselice}, {Bershady},
  {Dickinson}, \& {Papovich}}]{Conselice+2003}
{Conselice} C.~J., {Bershady} M.~A., {Dickinson} M., {Papovich} C., 2003, \aj,
  126, 1183

\bibitem[{{Conselice}, {Rajgor} \& {Myers}(2008){Conselice}, {Rajgor}, \&
  {Myers}}]{Conselice+2008}
{Conselice} C.~J., {Rajgor} S., {Myers} R., 2008, \mnras, 386, 909

\bibitem[{{Conselice}, {Yang} \& {Bluck}(2009){Conselice}, {Yang}, \&
  {Bluck}}]{Conselice+2009}
{Conselice} C.~J., {Yang} C., {Bluck} A.~F.~L., 2009, \mnras, 394, 1956

\bibitem[{{Contreras} {et~al}\mbox{.}(2017){Contreras}, {Zehavi}, {Baugh},
  {Padilla}, \& {Norberg}}]{Contreras+2017}
{Contreras} S., {Zehavi} I., {Baugh} C.~M., {Padilla} N., {Norberg} P., 2017,
  \mnras, 465, 2833

\bibitem[{{Cooray}(2006)}]{Cooray+2006}
{Cooray} A., 2006, \mnras, 365, 842

\bibitem[{{Cooray} \& {Sheth}(2002)}]{Cooray+2002}
{Cooray} A., {Sheth} R., 2002, \physrep, 372, 1

\bibitem[{{Correa} {et~al}\mbox{.}(2015){Correa}, {Wyithe}, {Schaye}, \&
  {Duffy}}]{Correa+2015}
{Correa} C.~A., {Wyithe} J.~S.~B., {Schaye} J., {Duffy} A.~R., 2015, \mnras,
  452, 1217

\bibitem[{{Coupon} {et~al}\mbox{.}(2015){Coupon}, {Arnouts}, {van Waerbeke},
  {Moutard}, {Ilbert}, {van Uitert}, {Erben}, {Garilli}, {Guzzo}, {Heymans},
  {Hildebrandt}, {Hoekstra}, {Kilbinger}, {Kitching}, {Mellier}, {Miller},
  {Scodeggio}, {Bonnett}, {Branchini}, {Davidzon}, {De Lucia}, {Fritz}, {Fu},
  {Hudelot}, {Hudson}, {Kuijken}, {Leauthaud}, {Le F{\`e}vre}, {McCracken},
  {Moscardini}, {Rowe}, {Schrabback}, {Semboloni}, \& {Velander}}]{Coupon+2015}
{Coupon} J. {et~al.}, 2015, \mnras, 449, 1352

\bibitem[{{Croton} {et~al}\mbox{.}(2006){Croton}, {Springel}, {White}, {De
  Lucia}, {Frenk}, {Gao}, {Jenkins}, {Kauffmann}, {Navarro}, \&
  {Yoshida}}]{Croton+2006}
{Croton} D.~J. {et~al.}, 2006, \mnras, 365, 11

\bibitem[{{Damjanov} {et~al}\mbox{.}(2009){Damjanov}, {McCarthy}, {Abraham},
  {Glazebrook}, {Yan}, {Mentuch}, {Le Borgne}, {Savaglio}, {Crampton},
  {Murowinski}, {Juneau}, {Carlberg}, {J{\o}rgensen}, {Roth}, {Chen}, \&
  {Marzke}}]{Damjanov+2009}
{Damjanov} I. {et~al.}, 2009, \apj, 695, 101

\bibitem[{{Dav{\'e}}, {Finlator} \& {Oppenheimer}(2012){Dav{\'e}}, {Finlator},
  \& {Oppenheimer}}]{Dave+2012}
{Dav{\'e}} R., {Finlator} K., {Oppenheimer} B.~D., 2012, \mnras, 421, 98

\bibitem[{{Dayal} {et~al}\mbox{.}(2014){Dayal}, {Ferrara}, {Dunlop}, \&
  {Pacucci}}]{Dayal+2014}
{Dayal} P., {Ferrara} A., {Dunlop} J.~S., {Pacucci} F., 2014, \mnras, 445, 2545

\bibitem[{{de Vaucouleurs}(1948)}]{deVaucouleurs1948}
{de Vaucouleurs} G., 1948, Annales d'Astrophysique, 11, 247

\bibitem[{{Dekel} \& {Birnboim}(2006)}]{Dekel+2006}
{Dekel} A., {Birnboim} Y., 2006, \mnras, 368, 2

\bibitem[{{Dekel} {et~al}\mbox{.}(2009){Dekel}, {Birnboim}, {Engel},
  {Freundlich}, {Goerdt}, {Mumcuoglu}, {Neistein}, {Pichon}, {Teyssier}, \&
  {Zinger}}]{Dekel+2009}
{Dekel} A. {et~al.}, 2009, \nat, 457, 451

\bibitem[{{Dekel} \& {Burkert}(2014)}]{Dekel_Burket2014}
{Dekel} A., {Burkert} A., 2014, \mnras, 438, 1870

\bibitem[{{Dekel} \& {Mandelker}(2014)}]{Dekel+2014}
{Dekel} A., {Mandelker} N., 2014, \mnras, 444, 2071

\bibitem[{{Dekel}, {Sari} \& {Ceverino}(2009){Dekel}, {Sari}, \&
  {Ceverino}}]{Dekel+2009b}
{Dekel} A., {Sari} R., {Ceverino} D., 2009, \apj, 703, 785

\bibitem[{{Dekel} {et~al}\mbox{.}(2013){Dekel}, {Zolotov}, {Tweed}, {Cacciato},
  {Ceverino}, \& {Primack}}]{Dekel+2013}
{Dekel} A., {Zolotov} A., {Tweed} D., {Cacciato} M., {Ceverino} D., {Primack}
  J.~R., 2013, \mnras, 435, 999

\bibitem[{{Diemer} {et~al}\mbox{.}(2017){Diemer}, {Sparre}, {Abramson}, \&
  {Torrey}}]{Diemer+2017}
{Diemer} B., {Sparre} M., {Abramson} L.~E., {Torrey} P., 2017, \apj, 839, 26

\bibitem[{{Drory} \& {Alvarez}(2008)}]{DroryAlvarez2008}
{Drory} N., {Alvarez} M., 2008, \apj, 680, 41

\bibitem[{{Drory} {et~al}\mbox{.}(2009){Drory}, {Bundy}, {Leauthaud},
  {Scoville}, {Capak}, {Ilbert}, {Kartaltepe}, {Kneib}, {McCracken}, {Salvato},
  {Sanders}, {Thompson}, \& {Willott}}]{Drory+2009}
{Drory} N. {et~al.}, 2009, \apj, 707, 1595

\bibitem[{{Drory} {et~al}\mbox{.}(2005){Drory}, {Salvato}, {Gabasch}, {Bender},
  {Hopp}, {Feulner}, \& {Pannella}}]{Drory+2005}
{Drory} N., {Salvato} M., {Gabasch} A., {Bender} R., {Hopp} U., {Feulner} G.,
  {Pannella} M., 2005, \apjl, 619, L131

\bibitem[{{Duncan} {et~al}\mbox{.}(2014){Duncan}, {Conselice}, {Mortlock},
  {Hartley}, {Guo}, {Ferguson}, {Dav{\'e}}, {Lu}, {Ownsworth}, {Ashby},
  {Dekel}, {Dickinson}, {Faber}, {Giavalisco}, {Grogin}, {Kocevski},
  {Koekemoer}, {Somerville}, \& {White}}]{Duncan+2014}
{Duncan} K. {et~al.}, 2014, \mnras, 444, 2960

\bibitem[{{Dunne} {et~al}\mbox{.}(2009){Dunne}, {Ivison}, {Maddox},
  {Cirasuolo}, {Mortier}, {Foucaud}, {Ibar}, {Almaini}, {Simpson}, \&
  {McLure}}]{Dunne+2009}
{Dunne} L. {et~al.}, 2009, \mnras, 394, 3

\bibitem[{{Faber} {et~al}\mbox{.}(2007){Faber}, {Willmer}, {Wolf}, {Koo},
  {Weiner}, {Newman}, {Im}, {Coil}, {Conroy}, {Cooper}, {Davis}, {Finkbeiner},
  {Gerke}, {Gebhardt}, {Groth}, {Guhathakurta}, {Harker}, {Kaiser}, {Kassin},
  {Kleinheinrich}, {Konidaris}, {Kron}, {Lin}, {Luppino}, {Madgwick},
  {Meisenheimer}, {Noeske}, {Phillips}, {Sarajedini}, {Schiavon}, {Simard},
  {Szalay}, {Vogt}, \& {Yan}}]{Faber+2007}
{Faber} S.~M. {et~al.}, 2007, \apj, 665, 265

\bibitem[{{Fang} {et~al}\mbox{.}(2013){Fang}, {Faber}, {Koo}, \&
  {Dekel}}]{Fang+2013}
{Fang} J.~J., {Faber} S.~M., {Koo} D.~C., {Dekel} A., 2013, \apj, 776, 63

\bibitem[{{Feldmann}(2015)}]{Feldmann+2015}
{Feldmann} R., 2015, \mnras, 449, 3274

\bibitem[{{Feldmann} \& {Mayer}(2015)}]{FeldmannMayer2015}
{Feldmann} R., {Mayer} L., 2015, \mnras, 446, 1939

\bibitem[{{Finkelstein} {et~al}\mbox{.}(2015){Finkelstein}, {Ryan}, {Papovich},
  {Dickinson}, {Song}, {Somerville}, {Ferguson}, {Salmon}, {Giavalisco},
  {Koekemoer}, {Ashby}, {Behroozi}, {Castellano}, {Dunlop}, {Faber}, {Fazio},
  {Fontana}, {Grogin}, {Hathi}, {Jaacks}, {Kocevski}, {Livermore}, {McLure},
  {Merlin}, {Mobasher}, {Newman}, {Rafelski}, {Tilvi}, \&
  {Willner}}]{Finkelstein+2015}
{Finkelstein} S.~L. {et~al.}, 2015, \apj, 810, 71

\bibitem[{{Fioc} \& {Rocca-Volmerange}(1997)}]{Fioc_RoccaVolmerange1997}
{Fioc} M., {Rocca-Volmerange} B., 1997, \aap, 326, 950

\bibitem[{{Firmani} \& {Avila-Reese}(2010)}]{Firmani+2010a}
{Firmani} C., {Avila-Reese} V., 2010, \apj, 723, 755

\bibitem[{{Franx} {et~al}\mbox{.}(2008){Franx}, {van Dokkum}, {F{\"o}rster
  Schreiber}, {Wuyts}, {Labb{\'e}}, \& {Toft}}]{Franx+2008}
{Franx} M., {van Dokkum} P.~G., {F{\"o}rster Schreiber} N.~M., {Wuyts} S.,
  {Labb{\'e}} I., {Toft} S., 2008, \apj, 688, 770

\bibitem[{{Gladders} {et~al}\mbox{.}(2013){Gladders}, {Oemler}, {Dressler},
  {Poggianti}, {Vulcani}, \& {Abramson}}]{Gladders+2013}
{Gladders} M.~D., {Oemler} A., {Dressler} A., {Poggianti} B., {Vulcani} B.,
  {Abramson} L., 2013, \apj, 770, 64

\bibitem[{{Gonz{\'a}lez} {et~al}\mbox{.}(2012){Gonz{\'a}lez}, {Bouwens},
  {Labb{\'e}}, {Illingworth}, {Oesch}, {Franx}, \& {Magee}}]{Gonzalez+2012}
{Gonz{\'a}lez} V., {Bouwens} R.~J., {Labb{\'e}} I., {Illingworth} G., {Oesch}
  P., {Franx} M., {Magee} D., 2012, \apj, 755, 148

\bibitem[{{Gonz{\'a}lez} {et~al}\mbox{.}(2011){Gonz{\'a}lez}, {Labb{\'e}},
  {Bouwens}, {Illingworth}, {Franx}, \& {Kriek}}]{Gonzalez+2011}
{Gonz{\'a}lez} V., {Labb{\'e}} I., {Bouwens} R.~J., {Illingworth} G., {Franx}
  M., {Kriek} M., 2011, \apjl, 735, L34

\bibitem[{{Gonz{\'a}lez} {et~al}\mbox{.}(2010){Gonz{\'a}lez}, {Labb{\'e}},
  {Bouwens}, {Illingworth}, {Franx}, {Kriek}, \& {Brammer}}]{Gonzalez+2010}
{Gonz{\'a}lez} V., {Labb{\'e}} I., {Bouwens} R.~J., {Illingworth} G., {Franx}
  M., {Kriek} M., {Brammer} G.~B., 2010, \apj, 713, 115

\bibitem[{{Grazian} {et~al}\mbox{.}(2015){Grazian}, {Fontana}, {Santini},
  {Dunlop}, {Ferguson}, {Castellano}, {Amorin}, {Ashby}, {Barro}, {Behroozi},
  {Boutsia}, {Caputi}, {Chary}, {Dekel}, {Dickinson}, {Faber}, {Fazio},
  {Finkelstein}, {Galametz}, {Giallongo}, {Giavalisco}, {Grogin}, {Guo},
  {Kocevski}, {Koekemoer}, {Koo}, {Lee}, {Lu}, {Merlin}, {Mobasher}, {Nonino},
  {Papovich}, {Paris}, {Pentericci}, {Reddy}, {Renzini}, {Salmon}, {Salvato},
  {Sommariva}, {Song}, \& {Vanzella}}]{Grazian+2015}
{Grazian} A. {et~al.}, 2015, \aap, 575, A96

\bibitem[{{Grogin} {et~al}\mbox{.}(2011){Grogin}, {Kocevski}, {Faber},
  {Ferguson}, {Koekemoer}, {Riess}, {Acquaviva}, {Alexander}, {Almaini},
  {Ashby}, {Barden}, {Bell}, {Bournaud}, {Brown}, {Caputi}, {Casertano},
  {Cassata}, {Castellano}, {Challis}, {Chary}, {Cheung}, {Cirasuolo},
  {Conselice}, {Roshan Cooray}, {Croton}, {Daddi}, {Dahlen}, {Dav{\'e}}, {de
  Mello}, {Dekel}, {Dickinson}, {Dolch}, {Donley}, {Dunlop}, {Dutton}, {Elbaz},
  {Fazio}, {Filippenko}, {Finkelstein}, {Fontana}, {Gardner}, {Garnavich},
  {Gawiser}, {Giavalisco}, {Grazian}, {Guo}, {Hathi}, {H{\"a}ussler},
  {Hopkins}, {Huang}, {Huang}, {Jha}, {Kartaltepe}, {Kirshner}, {Koo}, {Lai},
  {Lee}, {Li}, {Lotz}, {Lucas}, {Madau}, {McCarthy}, {McGrath}, {McIntosh},
  {McLure}, {Mobasher}, {Moustakas}, {Mozena}, {Nandra}, {Newman}, {Niemi},
  {Noeske}, {Papovich}, {Pentericci}, {Pope}, {Primack}, {Rajan},
  {Ravindranath}, {Reddy}, {Renzini}, {Rix}, {Robaina}, {Rodney}, {Rosario},
  {Rosati}, {Salimbeni}, {Scarlata}, {Siana}, {Simard}, {Smidt}, {Somerville},
  {Spinrad}, {Straughn}, {Strolger}, {Telford}, {Teplitz}, {Trump}, {van der
  Wel}, {Villforth}, {Wechsler}, {Weiner}, {Wiklind}, {Wild}, {Wilson},
  {Wuyts}, {Yan}, \& {Yun}}]{Grogin+2011}
{Grogin} N.~A. {et~al.}, 2011, \apjs, 197, 35

\bibitem[{{Guo} {et~al}\mbox{.}(2016){Guo}, {Zheng}, {Behroozi}, {Zehavi},
  {Chuang}, {Comparat}, {Favole}, {Gottloeber}, {Klypin}, {Prada},
  {Rodr{\'{\i}}guez-Torres}, {Weinberg}, \& {Yepes}}]{Guo+2016}
{Guo} H. {et~al.}, 2016, \mnras, 459, 3040

\bibitem[{{Guo} {et~al}\mbox{.}(2010){Guo}, {White}, {Li}, \&
  {Boylan-Kolchin}}]{Guo+2010}
{Guo} Q., {White} S., {Li} C., {Boylan-Kolchin} M., 2010, \mnras, 404, 1111

\bibitem[{{Harikane} {et~al}\mbox{.}(2016){Harikane}, {Ouchi}, {Ono}, {More},
  {Saito}, {Lin}, {Coupon}, {Shimasaku}, {Shibuya}, {Price}, {Lin}, {Hsieh},
  {Ishigaki}, {Komiyama}, {Silverman}, {Takata}, {Tamazawa}, \&
  {Toshikawa}}]{Harikane+2016}
{Harikane} Y. {et~al.}, 2016, \apj, 821, 123

\bibitem[{{Hearin} {et~al}\mbox{.}(2013{\natexlab{a}}){Hearin}, {Watson},
  {Becker}, {Reyes}, {Berlind}, \& {Zentner}}]{Hearin+2013b}
{Hearin} A.~P., {Watson} D.~F., {Becker} M.~R., {Reyes} R., {Berlind} A.~A.,
  {Zentner} A.~R., 2013{\natexlab{a}}, ArXiv e-prints

\bibitem[{{Hearin} {et~al}\mbox{.}(2013{\natexlab{b}}){Hearin}, {Zentner},
  {Berlind}, \& {Newman}}]{Hearin+2013}
{Hearin} A.~P., {Zentner} A.~R., {Berlind} A.~A., {Newman} J.~A.,
  2013{\natexlab{b}}, \mnras, 433, 659

\bibitem[{{Henriques} {et~al}\mbox{.}(2015){Henriques}, {White}, {Thomas},
  {Angulo}, {Guo}, {Lemson}, {Springel}, \& {Overzier}}]{Henriques+2015}
{Henriques} B.~M.~B., {White} S.~D.~M., {Thomas} P.~A., {Angulo} R., {Guo} Q.,
  {Lemson} G., {Springel} V., {Overzier} R., 2015, \mnras, 451, 2663

\bibitem[{{Heymans} {et~al}\mbox{.}(2006){Heymans}, {Bell}, {Rix}, {Barden},
  {Borch}, {Caldwell}, {McIntosh}, {Meisenheimer}, {Peng}, {Wolf}, {Beckwith},
  {H{\"a}u{\ss}ler}, {Jahnke}, {Jogee}, {S{\'a}nchez}, {Somerville}, \&
  {Wisotzki}}]{Heymans+2006}
{Heymans} C. {et~al.}, 2006, \mnras, 371, L60

\bibitem[{{Hill} {et~al}\mbox{.}(2017){Hill}, {Muzzin}, {Franx}, {Clauwens},
  {Schreiber}, {Marchesini}, {Stefanon}, {Labbe}, {Brammer}, {Caputi}, {Fynbo},
  {Milvang-Jensen}, {Skelton}, {van Dokkum}, \& {Whitaker}}]{Hill+2017}
{Hill} A.~R. {et~al.}, 2017, ArXiv e-prints

\bibitem[{{Hilz}, {Naab} \& {Ostriker}(2013){Hilz}, {Naab}, \&
  {Ostriker}}]{Hilz+2013}
{Hilz} M., {Naab} T., {Ostriker} J.~P., 2013, \mnras, 429, 2924

\bibitem[{{Hogg}(1999)}]{Hogg1999}
{Hogg} D.~W., 1999, ArXiv Astrophysics e-prints

\bibitem[{{Hopkins} {et~al}\mbox{.}(2010{\natexlab{a}}){Hopkins}, {Bundy},
  {Croton}, {Hernquist}, {Keres}, {Khochfar}, {Stewart}, {Wetzel}, \&
  {Younger}}]{Hopkins+2010a}
{Hopkins} P.~F. {et~al.}, 2010{\natexlab{a}}, \apj, 715, 202

\bibitem[{{Hopkins} {et~al}\mbox{.}(2010{\natexlab{b}}){Hopkins}, {Croton},
  {Bundy}, {Khochfar}, {van den Bosch}, {Somerville}, {Wetzel}, {Keres},
  {Hernquist}, {Stewart}, {Younger}, {Genel}, \& {Ma}}]{Hopkins+2010b}
{Hopkins} P.~F. {et~al.}, 2010{\natexlab{b}}, \apj, 724, 915

\bibitem[{{Huang} {et~al}\mbox{.}(2017){Huang}, {Fall}, {Ferguson}, {van der
  Wel}, {Grogin}, {Koekemoer}, {Lee}, {P{\'e}rez-Gonz{\'a}lez}, \&
  {Wuyts}}]{Huang+2017}
{Huang} K.-H. {et~al.}, 2017, \apj, 838, 6

\bibitem[{{Hudson} {et~al}\mbox{.}(2013){Hudson}, {Gillis}, {Coupon},
  {Hildebrandt}, {Erben}, {Heymans}, {Hoekstra}, {Kitching}, {Mellier},
  {Miller}, {Van Waerbeke}, {Bonnett}, {Fu}, {Kuijken}, {Rowe}, {Schrabback},
  {Semboloni}, {van Uitert}, \& {Velander}}]{Hudson+2014}
{Hudson} M.~J. {et~al.}, 2013, ArXiv e-prints

\bibitem[{{Ibarra-Medel} {et~al}\mbox{.}(2016){Ibarra-Medel}, {S{\'a}nchez},
  {Avila-Reese}, {Hern{\'a}ndez-Toledo}, {Gonz{\'a}lez}, {Drory}, {Bundy},
  {Bizyaev}, {Cano-D{\'{\i}}az}, {Malanushenko}, {Pan}, {Roman-Lopes}, \&
  {Thomas}}]{Ibarra-Medel+2016}
{Ibarra-Medel} H.~J. {et~al.}, 2016, \mnras, 463, 2799

\bibitem[{{Ilbert} {et~al}\mbox{.}(2013){Ilbert}, {McCracken}, {Le F{\`e}vre},
  {Capak}, {Dunlop}, {Karim}, {Renzini}, {Caputi}, {Boissier}, {Arnouts},
  {Aussel}, {Comparat}, {Guo}, {Hudelot}, {Kartaltepe}, {Kneib}, {Krogager},
  {Le Floc'h}, {Lilly}, {Mellier}, {Milvang-Jensen}, {Moutard}, {Onodera},
  {Richard}, {Salvato}, {Sanders}, {Scoville}, {Silverman}, {Taniguchi},
  {Tasca}, {Thomas}, {Toft}, {Tresse}, {Vergani}, {Wolk}, \&
  {Zirm}}]{Ilbert+2013}
{Ilbert} O. {et~al.}, 2013, \aap, 556, A55

\bibitem[{{Inoue} {et~al}\mbox{.}(2016){Inoue}, {Dekel}, {Mandelker},
  {Ceverino}, {Bournaud}, \& {Primack}}]{Inoue+2016}
{Inoue} S., {Dekel} A., {Mandelker} N., {Ceverino} D., {Bournaud} F., {Primack}
  J., 2016, \mnras, 456, 2052

\bibitem[{{Jing}, {Mo} \& {B{\"o}rner}(1998){Jing}, {Mo}, \&
  {B{\"o}rner}}]{Jing+1998}
{Jing} Y.~P., {Mo} H.~J., {B{\"o}rner} G., 1998, \apj, 494, 1

\bibitem[{{Kajisawa} {et~al}\mbox{.}(2010){Kajisawa}, {Ichikawa}, {Yamada},
  {Uchimoto}, {Yoshikawa}, {Akiyama}, \& {Onodera}}]{Kajisawa+2010}
{Kajisawa} M., {Ichikawa} T., {Yamada} T., {Uchimoto} Y.~K., {Yoshikawa} T.,
  {Akiyama} M., {Onodera} M., 2010, \apj, 723, 129

\bibitem[{{Karim} {et~al}\mbox{.}(2011){Karim}, {Schinnerer},
  {Mart{\'{\i}}nez-Sansigre}, {Sargent}, {van der Wel}, {Rix}, {Ilbert},
  {Smol{\v c}i{\'c}}, {Carilli}, {Pannella}, {Koekemoer}, {Bell}, \&
  {Salvato}}]{Karim+2011}
{Karim} A. {et~al.}, 2011, \apj, 730, 61

\bibitem[{{Kauffmann} {et~al}\mbox{.}(2003{\natexlab{a}}){Kauffmann},
  {Heckman}, {White}, {Charlot}, {Tremonti}, {Brinchmann}, {Bruzual}, {Peng},
  {Seibert}, {Bernardi}, {Blanton}, {Brinkmann}, {Castander}, {Cs{\'a}bai},
  {Fukugita}, {Ivezic}, {Munn}, {Nichol}, {Padmanabhan}, {Thakar}, {Weinberg},
  \& {York}}]{Kauffmann+2003}
{Kauffmann} G. {et~al.}, 2003{\natexlab{a}}, \mnras, 341, 33

\bibitem[{{Kauffmann} {et~al}\mbox{.}(2003{\natexlab{b}}){Kauffmann},
  {Heckman}, {White}, {Charlot}, {Tremonti}, {Peng}, {Seibert}, {Brinkmann},
  {Nichol}, {SubbaRao}, \& {York}}]{Kauffmann+2003a}
{Kauffmann} G. {et~al.}, 2003{\natexlab{b}}, \mnras, 341, 54

\bibitem[{{Kennicutt} \& {Evans}(2012)}]{Kennicutt+2012}
{Kennicutt} R.~C., {Evans} N.~J., 2012, \araa, 50, 531

\bibitem[{{Kennicutt}(1998)}]{Kennicutt+1998}
{Kennicutt}, Jr. R.~C., 1998, \apj, 498, 541

\bibitem[{{Kere{\v s}} {et~al}\mbox{.}(2005){Kere{\v s}}, {Katz}, {Weinberg},
  \& {Dav{\'e}}}]{Keres+2005}
{Kere{\v s}} D., {Katz} N., {Weinberg} D.~H., {Dav{\'e}} R., 2005, \mnras, 363,
  2

\bibitem[{{Klypin}, {Trujillo-Gomez} \& {Primack}(2011){Klypin},
  {Trujillo-Gomez}, \& {Primack}}]{Klypin+2011}
{Klypin} A.~A., {Trujillo-Gomez} S., {Primack} J., 2011, \apj, 740, 102

\bibitem[{{Koekemoer} {et~al}\mbox{.}(2011){Koekemoer}, {Faber}, {Ferguson},
  {Grogin}, {Kocevski}, {Koo}, {Lai}, {Lotz}, {Lucas}, {McGrath}, {Ogaz},
  {Rajan}, {Riess}, {Rodney}, {Strolger}, {Casertano}, {Castellano}, {Dahlen},
  {Dickinson}, {Dolch}, {Fontana}, {Giavalisco}, {Grazian}, {Guo}, {Hathi},
  {Huang}, {van der Wel}, {Yan}, {Acquaviva}, {Alexander}, {Almaini}, {Ashby},
  {Barden}, {Bell}, {Bournaud}, {Brown}, {Caputi}, {Cassata}, {Challis},
  {Chary}, {Cheung}, {Cirasuolo}, {Conselice}, {Roshan Cooray}, {Croton},
  {Daddi}, {Dav{\'e}}, {de Mello}, {de Ravel}, {Dekel}, {Donley}, {Dunlop},
  {Dutton}, {Elbaz}, {Fazio}, {Filippenko}, {Finkelstein}, {Frazer}, {Gardner},
  {Garnavich}, {Gawiser}, {Gruetzbauch}, {Hartley}, {H{\"a}ussler},
  {Herrington}, {Hopkins}, {Huang}, {Jha}, {Johnson}, {Kartaltepe},
  {Khostovan}, {Kirshner}, {Lani}, {Lee}, {Li}, {Madau}, {McCarthy},
  {McIntosh}, {McLure}, {McPartland}, {Mobasher}, {Moreira}, {Mortlock},
  {Moustakas}, {Mozena}, {Nandra}, {Newman}, {Nielsen}, {Niemi}, {Noeske},
  {Papovich}, {Pentericci}, {Pope}, {Primack}, {Ravindranath}, {Reddy},
  {Renzini}, {Rix}, {Robaina}, {Rosario}, {Rosati}, {Salimbeni}, {Scarlata},
  {Siana}, {Simard}, {Smidt}, {Snyder}, {Somerville}, {Spinrad}, {Straughn},
  {Telford}, {Teplitz}, {Trump}, {Vargas}, {Villforth}, {Wagner}, {Wandro},
  {Wechsler}, {Weiner}, {Wiklind}, {Wild}, {Wilson}, {Wuyts}, \&
  {Yun}}]{Koekemoer+2011}
{Koekemoer} A.~M. {et~al.}, 2011, \apjs, 197, 36

\bibitem[{{Kormendy} \& {Ho}(2013)}]{Kormendy+2013}
{Kormendy} J., {Ho} L.~C., 2013, \araa, 51, 511

\bibitem[{{Kravtsov}(2013)}]{Kravtsov2013}
{Kravtsov} A.~V., 2013, \apjl, 764, L31

\bibitem[{{Kravtsov} {et~al}\mbox{.}(2004){Kravtsov}, {Berlind}, {Wechsler},
  {Klypin}, {Gottl{\"o}ber}, {Allgood}, \& {Primack}}]{Kravtsov+2004}
{Kravtsov} A.~V., {Berlind} A.~A., {Wechsler} R.~H., {Klypin} A.~A.,
  {Gottl{\"o}ber} S., {Allgood} B., {Primack} J.~R., 2004, \apj, 609, 35

\bibitem[{{Kroupa}(2001)}]{Kroupa2001}
{Kroupa} P., 2001, \mnras, 322, 231

\bibitem[{{Krumholz} \& {Dekel}(2012)}]{Krumholz+2012}
{Krumholz} M.~R., {Dekel} A., 2012, \apj, 753, 16

\bibitem[{{Lahav} {et~al}\mbox{.}(1991){Lahav}, {Lilje}, {Primack}, \&
  {Rees}}]{Lahav+1991}
{Lahav} O., {Lilje} P.~B., {Primack} J.~R., {Rees} M.~J., 1991, \mnras, 251,
  128

\bibitem[{{Leauthaud} {et~al}\mbox{.}(2011){Leauthaud}, {Tinker}, {Behroozi},
  {Busha}, \& {Wechsler}}]{Leauthaud+2011}
{Leauthaud} A., {Tinker} J., {Behroozi} P.~S., {Busha} M.~T., {Wechsler} R.~H.,
  2011, \apj, 738, 45

\bibitem[{{Leauthaud} {et~al}\mbox{.}(2012){Leauthaud}, {Tinker}, {Bundy},
  {Behroozi}, {Massey}, {Rhodes}, {George}, {Kneib}, {Benson}, {Wechsler},
  {Busha}, {Capak}, {Cort{\^e}s}, {Ilbert}, {Koekemoer}, {Le F{\`e}vre},
  {Lilly}, {McCracken}, {Salvato}, {Schrabback}, {Scoville}, {Smith}, \&
  {Taylor}}]{Leauthaud+2012}
{Leauthaud} A. {et~al.}, 2012, \apj, 744, 159

\bibitem[{{Lee} {et~al}\mbox{.}(2011){Lee}, {Dey}, {Reddy}, {Brown},
  {Gonzalez}, {Jannuzi}, {Cooper}, {Fan}, {Bian}, {Glikman}, {Stern},
  {Brodwin}, \& {Cooray}}]{Lee+2011}
{Lee} K.-S. {et~al.}, 2011, \apj, 733, 99

\bibitem[{{Lee} {et~al}\mbox{.}(2012){Lee}, {Ferguson}, {Wiklind}, {Dahlen},
  {Dickinson}, {Giavalisco}, {Grogin}, {Papovich}, {Messias}, {Guo}, \&
  {Lin}}]{Lee+2012}
{Lee} K.-S. {et~al.}, 2012, \apj, 752, 66

\bibitem[{{Leja} {et~al}\mbox{.}(2015){Leja}, {van Dokkum}, {Franx}, \&
  {Whitaker}}]{Leja+2015}
{Leja} J., {van Dokkum} P.~G., {Franx} M., {Whitaker} K.~E., 2015, \apj, 798,
  115

\bibitem[{{Li} \& {White}(2009)}]{Li+2009}
{Li} C., {White} S.~D.~M., 2009, \mnras, 398, 2177

\bibitem[{{Li} {et~al}\mbox{.}(2016){Li}, {Zhang}, {Yang}, {Wang}, {Tweed},
  {Liu}, {Yang}, {Shi}, {Lu}, {Luo}, \& {Wei}}]{Li+2016}
{Li} S.-J. {et~al.}, 2016, Research in Astronomy and Astrophysics, 16, 130

\bibitem[{{L{\'o}pez-Sanjuan} {et~al}\mbox{.}(2009){L{\'o}pez-Sanjuan},
  {Balcells}, {P{\'e}rez-Gonz{\'a}lez}, {Barro}, {Garc{\'{\i}}a-Dab{\'o}},
  {Gallego}, \& {Zamorano}}]{LPS+09}
{L{\'o}pez-Sanjuan} C., {Balcells} M., {P{\'e}rez-Gonz{\'a}lez} P.~G., {Barro}
  G., {Garc{\'{\i}}a-Dab{\'o}} C.~E., {Gallego} J., {Zamorano} J., 2009, \aap,
  501, 505

\bibitem[{{L{\'o}pez-Sanjuan} {et~al}\mbox{.}(2010){L{\'o}pez-Sanjuan},
  {Balcells}, {P{\'e}rez-Gonz{\'a}lez}, {Barro}, {Garc{\'{\i}}a-Dab{\'o}},
  {Gallego}, \& {Zamorano}}]{LPS+10}
{L{\'o}pez-Sanjuan} C., {Balcells} M., {P{\'e}rez-Gonz{\'a}lez} P.~G., {Barro}
  G., {Garc{\'{\i}}a-Dab{\'o}} C.~E., {Gallego} J., {Zamorano} J., 2010, \apj,
  710, 1170

\bibitem[{{L{\'o}pez-Sanjuan} {et~al}\mbox{.}(2012){L{\'o}pez-Sanjuan}, {Le
  F{\`e}vre}, {Ilbert}, {Tasca}, {Bridge}, {Cucciati}, {Kampczyk}, {Pozzetti},
  {Xu}, {Carollo}, {Contini}, {Kneib}, {Lilly}, {Mainieri}, {Renzini},
  {Sanders}, {Scodeggio}, {Scoville}, {Taniguchi}, {Zamorani}, {Aussel},
  {Bardelli}, {Bolzonella}, {Bongiorno}, {Capak}, {Caputi}, {de la Torre}, {de
  Ravel}, {Franzetti}, {Garilli}, {Iovino}, {Knobel}, {Kova{\v c}},
  {Lamareille}, {Le Borgne}, {Le Brun}, {Le Floc'h}, {Maier}, {McCracken},
  {Mignoli}, {Pell{\'o}}, {Peng}, {P{\'e}rez-Montero}, {Presotto},
  {Ricciardelli}, {Salvato}, {Silverman}, {Tanaka}, {Tresse}, {Vergani},
  {Zucca}, {Barnes}, {Bordoloi}, {Cappi}, {Cimatti}, {Coppa}, {Koekemoer},
  {Liu}, {Moresco}, {Nair}, {Oesch}, {Schawinski}, \& {Welikala}}]{LPS+12}
{L{\'o}pez-Sanjuan} C. {et~al.}, 2012, \aap, 548, A7

\bibitem[{{Lotz} {et~al}\mbox{.}(2008){Lotz}, {Davis}, {Faber}, {Guhathakurta},
  {Gwyn}, {Huang}, {Koo}, {Le Floc'h}, {Lin}, {Newman}, {Noeske}, {Papovich},
  {Willmer}, {Coil}, {Conselice}, {Cooper}, {Hopkins}, {Metevier}, {Primack},
  {Rieke}, \& {Weiner}}]{Lotz+2008}
{Lotz} J.~M. {et~al.}, 2008, \apj, 672, 177

\bibitem[{{Lotz} {et~al}\mbox{.}(2011){Lotz}, {Jonsson}, {Cox}, {Croton},
  {Primack}, {Somerville}, \& {Stewart}}]{Lotz+2011}
{Lotz} J.~M., {Jonsson} P., {Cox} T.~J., {Croton} D., {Primack} J.~R.,
  {Somerville} R.~S., {Stewart} K., 2011, \apj, 742, 103

\bibitem[{{Madau} \& {Dickinson}(2014)}]{MadauDickinson2014}
{Madau} P., {Dickinson} M., 2014, \araa, 52, 415

\bibitem[{{Magdis} {et~al}\mbox{.}(2010){Magdis}, {Rigopoulou}, {Huang}, \&
  {Fazio}}]{Magdis+2010}
{Magdis} G.~E., {Rigopoulou} D., {Huang} J.-S., {Fazio} G.~G., 2010, \mnras,
  401, 1521

\bibitem[{{Man} {et~al}\mbox{.}(2012){Man}, {Toft}, {Zirm}, {Wuyts}, \& {van
  der Wel}}]{Man+2012}
{Man} A.~W.~S., {Toft} S., {Zirm} A.~W., {Wuyts} S., {van der Wel} A., 2012,
  \apj, 744, 85

\bibitem[{{Man}, {Zirm} \& {Toft}(2016){Man}, {Zirm}, \& {Toft}}]{Man+2016}
{Man} A.~W.~S., {Zirm} A.~W., {Toft} S., 2016, \apj, 830, 89

\bibitem[{{Mandelbaum} {et~al}\mbox{.}(2006){Mandelbaum}, {Seljak},
  {Kauffmann}, {Hirata}, \& {Brinkmann}}]{Mandelbaum+2006}
{Mandelbaum} R., {Seljak} U., {Kauffmann} G., {Hirata} C.~M., {Brinkmann} J.,
  2006, \mnras, 368, 715

\bibitem[{{Maraston}(2005)}]{Maraston+2005}
{Maraston} C., 2005, \mnras, 362, 799

\bibitem[{{Maraston} {et~al}\mbox{.}(2006){Maraston}, {Daddi}, {Renzini},
  {Cimatti}, {Dickinson}, {Papovich}, {Pasquali}, \& {Pirzkal}}]{Maraston+2006}
{Maraston} C., {Daddi} E., {Renzini} A., {Cimatti} A., {Dickinson} M.,
  {Papovich} C., {Pasquali} A., {Pirzkal} N., 2006, \apj, 652, 85

\bibitem[{{Marchesini} {et~al}\mbox{.}(2009){Marchesini}, {van Dokkum},
  {F{\"o}rster Schreiber}, {Franx}, {Labb{\'e}}, \& {Wuyts}}]{Marchesini+2009}
{Marchesini} D., {van Dokkum} P.~G., {F{\"o}rster Schreiber} N.~M., {Franx} M.,
  {Labb{\'e}} I., {Wuyts} S., 2009, \apj, 701, 1765

\bibitem[{{Mart{\'{\i}}nez-Delgado}
  {et~al}\mbox{.}(2016){Mart{\'{\i}}nez-Delgado}, {L{\"a}sker}, {Sharina},
  {Toloba}, {Fliri}, {Beaton}, {Valls-Gabaud}, {Karachentsev}, {Chonis},
  {Grebel}, {Forbes}, {Romanowsky}, {Gallego-Laborda}, {Teuwen},
  {G{\'o}mez-Flechoso}, {Wang}, {Guhathakurta}, {Kaisin}, \&
  {Ho}}]{Martinez-Delgado+2016}
{Mart{\'{\i}}nez-Delgado} D. {et~al.}, 2016, \aj, 151, 96

\bibitem[{{Martinez-Manso} {et~al}\mbox{.}(2015){Martinez-Manso}, {Gonzalez},
  {Ashby}, {Stanford}, {Brodwin}, {Holder}, \& {Stern}}]{Martinez-Manso+2015}
{Martinez-Manso} J., {Gonzalez} A.~H., {Ashby} M.~L.~N., {Stanford} S.~A.,
  {Brodwin} M., {Holder} G.~P., {Stern} D., 2015, \mnras, 446, 169

\bibitem[{{Masaki}, {Lin} \& {Yoshida}(2013){Masaki}, {Lin}, \&
  {Yoshida}}]{Masaki+2013}
{Masaki} S., {Lin} Y.-T., {Yoshida} N., 2013, \mnras, 436, 2286

\bibitem[{{Matthee} {et~al}\mbox{.}(2017){Matthee}, {Schaye}, {Crain},
  {Schaller}, {Bower}, \& {Theuns}}]{Matthee+2017}
{Matthee} J., {Schaye} J., {Crain} R.~A., {Schaller} M., {Bower} R., {Theuns}
  T., 2017, \mnras, 465, 2381

\bibitem[{{McLure} {et~al}\mbox{.}(2013){McLure}, {Dunlop}, {Bowler},
  {Curtis-Lake}, {Schenker}, {Ellis}, {Robertson}, {Koekemoer}, {Rogers},
  {Ono}, {Ouchi}, {Charlot}, {Wild}, {Stark}, {Furlanetto}, {Cirasuolo}, \&
  {Targett}}]{McLure+2013}
{McLure} R.~J. {et~al.}, 2013, \mnras, 432, 2696

\bibitem[{{Meert}, {Vikram} \& {Bernardi}(2015){Meert}, {Vikram}, \&
  {Bernardi}}]{Meert+2015}
{Meert} A., {Vikram} V., {Bernardi} M., 2015, \mnras, 446, 3943

\bibitem[{{Mendel} {et~al}\mbox{.}(2014){Mendel}, {Simard}, {Palmer},
  {Ellison}, \& {Patton}}]{Mendel+2014}
{Mendel} J.~T., {Simard} L., {Palmer} M., {Ellison} S.~L., {Patton} D.~R.,
  2014, \apjs, 210, 3

\bibitem[{{Meurer}, {Heckman} \& {Calzetti}(1999){Meurer}, {Heckman}, \&
  {Calzetti}}]{Meurer+1999}
{Meurer} G.~R., {Heckman} T.~M., {Calzetti} D., 1999, \apj, 521, 64

\bibitem[{{Micic}, {Martinovi{\'c}} \& {Sinha}(2016){Micic}, {Martinovi{\'c}},
  \& {Sinha}}]{Micic+2016}
{Micic} M., {Martinovi{\'c}} N., {Sinha} M., 2016, \mnras, 461, 3322

\bibitem[{{Mitra}, {Dav{\'e}} \& {Finlator}(2015){Mitra}, {Dav{\'e}}, \&
  {Finlator}}]{Mitra+2015}
{Mitra} S., {Dav{\'e}} R., {Finlator} K., 2015, \mnras, 452, 1184

\bibitem[{{Mo}, {van den Bosch} \& {White}(2010){Mo}, {van den Bosch}, \&
  {White}}]{Mo+2010}
{Mo} H., {van den Bosch} F.~C., {White} S., 2010, {Galaxy Formation and
  Evolution}

\bibitem[{{More} {et~al}\mbox{.}(2011){More}, {van den Bosch}, {Cacciato},
  {Skibba}, {Mo}, \& {Yang}}]{More+2011}
{More} S., {van den Bosch} F.~C., {Cacciato} M., {Skibba} R., {Mo} H.~J.,
  {Yang} X., 2011, \mnras, 410, 210

\bibitem[{{Mortlock} {et~al}\mbox{.}(2011){Mortlock}, {Conselice}, {Bluck},
  {Bauer}, {Gr{\"u}tzbauch}, {Buitrago}, \& {Ownsworth}}]{Mortlock+2011}
{Mortlock} A., {Conselice} C.~J., {Bluck} A.~F.~L., {Bauer} A.~E.,
  {Gr{\"u}tzbauch} R., {Buitrago} F., {Ownsworth} J., 2011, \mnras, 413, 2845

\bibitem[{{Mosleh}, {Williams} \& {Franx}(2013){Mosleh}, {Williams}, \&
  {Franx}}]{Mosleh+2013}
{Mosleh} M., {Williams} R.~J., {Franx} M., 2013, \apj, 777, 117

\bibitem[{{Moster}, {Naab} \& {White}(2013){Moster}, {Naab}, \&
  {White}}]{Moster+2013}
{Moster} B.~P., {Naab} T., {White} S.~D.~M., 2013, \mnras, 428, 3121

\bibitem[{{Moster} {et~al}\mbox{.}(2010){Moster}, {Somerville}, {Maulbetsch},
  {van den Bosch}, {Macci{\`o}}, {Naab}, \& {Oser}}]{Moster+2010}
{Moster} B.~P., {Somerville} R.~S., {Maulbetsch} C., {van den Bosch} F.~C.,
  {Macci{\`o}} A.~V., {Naab} T., {Oser} L., 2010, \apj, 710, 903

\bibitem[{{Moustakas} {et~al}\mbox{.}(2013){Moustakas}, {Coil}, {Aird},
  {Blanton}, {Cool}, {Eisenstein}, {Mendez}, {Wong}, {Zhu}, \&
  {Arnouts}}]{Moustakas+2013}
{Moustakas} J. {et~al.}, 2013, \apj, 767, 50

\bibitem[{{Muzzin} {et~al}\mbox{.}(2013){Muzzin}, {Marchesini}, {Stefanon},
  {Franx}, {McCracken}, {Milvang-Jensen}, {Dunlop}, {Fynbo}, {Brammer},
  {Labb{\'e}}, \& {van Dokkum}}]{Muzzin+2013}
{Muzzin} A. {et~al.}, 2013, \apj, 777, 18

\bibitem[{{Muzzin} {et~al}\mbox{.}(2009){Muzzin}, {Marchesini}, {van Dokkum},
  {Labb{\'e}}, {Kriek}, \& {Franx}}]{Muzzin+2009}
{Muzzin} A., {Marchesini} D., {van Dokkum} P.~G., {Labb{\'e}} I., {Kriek} M.,
  {Franx} M., 2009, \apj, 701, 1839

\bibitem[{{Nagai} \& {Kravtsov}(2003)}]{NagaiKravtsov2003}
{Nagai} D., {Kravtsov} A.~V., 2003, \apj, 587, 514

\bibitem[{{Neistein} {et~al}\mbox{.}(2011){Neistein}, {Li}, {Khochfar},
  {Weinmann}, {Shankar}, \& {Boylan-Kolchin}}]{Neistein+2011}
{Neistein} E., {Li} C., {Khochfar} S., {Weinmann} S.~M., {Shankar} F.,
  {Boylan-Kolchin} M., 2011, \mnras, 416, 1486

\bibitem[{{Noeske} {et~al}\mbox{.}(2007{\natexlab{a}}){Noeske}, {Faber},
  {Weiner}, {Koo}, {Primack}, {Dekel}, {Papovich}, {Conselice}, {Le Floc'h},
  {Rieke}, {Coil}, {Lotz}, {Somerville}, \& {Bundy}}]{Noeske+2007b}
{Noeske} K.~G. {et~al.}, 2007{\natexlab{a}}, \apjl, 660, L47

\bibitem[{{Noeske} {et~al}\mbox{.}(2007{\natexlab{b}}){Noeske}, {Weiner},
  {Faber}, {Papovich}, {Koo}, {Somerville}, {Bundy}, {Conselice}, {Newman},
  {Schiminovich}, {Le Floc'h}, {Coil}, {Rieke}, {Lotz}, {Primack}, {Barmby},
  {Cooper}, {Davis}, {Ellis}, {Fazio}, {Guhathakurta}, {Huang}, {Kassin},
  {Martin}, {Phillips}, {Rich}, {Small}, {Willmer}, \& {Wilson}}]{Noeske+2007}
{Noeske} K.~G. {et~al.}, 2007{\natexlab{b}}, \apjl, 660, L43

\bibitem[{{Oesch} {et~al}\mbox{.}(2012){Oesch}, {Bouwens}, {Illingworth},
  {Gonzalez}, {Trenti}, {van Dokkum}, {Franx}, {Labb{\'e}}, {Carollo}, \&
  {Magee}}]{Oesch+2012}
{Oesch} P.~A. {et~al.}, 2012, \apj, 759, 135

\bibitem[{{Pandya} {et~al}\mbox{.}(2016){Pandya}, {Brennan}, {Somerville},
  {Choi}, {Barro}, {Wuyts}, {Taylor}, {Behroozi}, {Kirkpatrick}, {Faber},
  {Primack}, {Koo}, {McIntosh}, {Kocevski}, {Bell}, {Dekel}, {Fang},
  {Ferguson}, {Grogin}, {Koekemoer}, {Lu}, {Mantha}, {Mobasher}, {Newman},
  {Pacifici}, {Papovich}, {van der Wel}, \& {Yesuf}}]{Pandya+2017}
{Pandya} V. {et~al.}, 2016, ArXiv e-prints

\bibitem[{{Papastergis} {et~al}\mbox{.}(2012){Papastergis}, {Cattaneo},
  {Huang}, {Giovanelli}, \& {Haynes}}]{Papastergis+2012}
{Papastergis} E., {Cattaneo} A., {Huang} S., {Giovanelli} R., {Haynes} M.~P.,
  2012, \apj, 759, 138

\bibitem[{{Papovich} {et~al}\mbox{.}(2011){Papovich}, {Finkelstein},
  {Ferguson}, {Lotz}, \& {Giavalisco}}]{Papovich+2011}
{Papovich} C., {Finkelstein} S.~L., {Ferguson} H.~C., {Lotz} J.~M.,
  {Giavalisco} M., 2011, \mnras, 412, 1123

\bibitem[{{Patel} {et~al}\mbox{.}(2013){Patel}, {van Dokkum}, {Franx},
  {Quadri}, {Muzzin}, {Marchesini}, {Williams}, {Holden}, \&
  {Stefanon}}]{Patel+2013}
{Patel} S.~G. {et~al.}, 2013, \apj, 766, 15

\bibitem[{{Peng} {et~al}\mbox{.}(2010){Peng}, {Lilly}, {Kova{\v c}},
  {Bolzonella}, {Pozzetti}, {Renzini}, {Zamorani}, {Ilbert}, {Knobel},
  {Iovino}, {Maier}, {Cucciati}, {Tasca}, {Carollo}, {Silverman}, {Kampczyk},
  {de Ravel}, {Sanders}, {Scoville}, {Contini}, {Mainieri}, {Scodeggio},
  {Kneib}, {Le F{\`e}vre}, {Bardelli}, {Bongiorno}, {Caputi}, {Coppa}, {de la
  Torre}, {Franzetti}, {Garilli}, {Lamareille}, {Le Borgne}, {Le Brun},
  {Mignoli}, {Perez Montero}, {Pello}, {Ricciardelli}, {Tanaka}, {Tresse},
  {Vergani}, {Welikala}, {Zucca}, {Oesch}, {Abbas}, {Barnes}, {Bordoloi},
  {Bottini}, {Cappi}, {Cassata}, {Cimatti}, {Fumana}, {Hasinger}, {Koekemoer},
  {Leauthaud}, {Maccagni}, {Marinoni}, {McCracken}, {Memeo}, {Meneux}, {Nair},
  {Porciani}, {Presotto}, \& {Scaramella}}]{Peng+2010}
{Peng} Y.-j. {et~al.}, 2010, \apj, 721, 193

\bibitem[{{P{\'e}rez} {et~al}\mbox{.}(2013){P{\'e}rez}, {Cid Fernandes},
  {Gonz{\'a}lez Delgado}, {Garc{\'{\i}}a-Benito}, {S{\'a}nchez}, {Husemann},
  {Mast}, {Rod{\'o}n}, {Kupko}, {Backsmann}, {de Amorim}, {van de Ven},
  {Walcher}, {Wisotzki}, {Cortijo-Ferrero}, \& {CALIFA
  Collaboration}}]{Perez+2013}
{P{\'e}rez} E. {et~al.}, 2013, \apjl, 764, L1

\bibitem[{{P{\'e}rez-Gonz{\'a}lez}
  {et~al}\mbox{.}(2008){P{\'e}rez-Gonz{\'a}lez}, {Rieke}, {Villar}, {Barro},
  {Blaylock}, {Egami}, {Gallego}, {Gil de Paz}, {Pascual}, {Zamorano}, \&
  {Donley}}]{Perez-Gonzalez+2008}
{P{\'e}rez-Gonz{\'a}lez} P.~G. {et~al.}, 2008, \apj, 675, 234

\bibitem[{{Planck Collaboration} {et~al}\mbox{.}(2016){Planck Collaboration},
  {Ade}, {Aghanim}, {Arnaud}, {Ashdown}, {Aumont}, {Baccigalupi}, {Banday},
  {Barreiro}, {Bartlett}, \& et~al.}]{Planck+2015}
{Planck Collaboration} {et~al.}, 2016, \aap, 594, A13

\bibitem[{{Pozzetti} {et~al}\mbox{.}(2010){Pozzetti}, {Bolzonella}, {Zucca},
  {Zamorani}, {Lilly}, {Renzini}, {Moresco}, {Mignoli}, {Cassata}, {Tasca},
  {Lamareille}, {Maier}, {Meneux}, {Halliday}, {Oesch}, {Vergani}, {Caputi},
  {Kova{\v c}}, {Cimatti}, {Cucciati}, {Iovino}, {Peng}, {Carollo}, {Contini},
  {Kneib}, {Le F{\'e}vre}, {Mainieri}, {Scodeggio}, {Bardelli}, {Bongiorno},
  {Coppa}, {de la Torre}, {de Ravel}, {Franzetti}, {Garilli}, {Kampczyk},
  {Knobel}, {Le Borgne}, {Le Brun}, {Pell{\`o}}, {Perez Montero},
  {Ricciardelli}, {Silverman}, {Tanaka}, {Tresse}, {Abbas}, {Bottini}, {Cappi},
  {Guzzo}, {Koekemoer}, {Leauthaud}, {Maccagni}, {Marinoni}, {McCracken},
  {Memeo}, {Porciani}, {Scaramella}, {Scarlata}, \& {Scoville}}]{Pozzetti+2010}
{Pozzetti} L. {et~al.}, 2010, \aap, 523, A13

\bibitem[{{Press} \& {Schechter}(1974)}]{PressSchechter1974}
{Press} W.~H., {Schechter} P., 1974, \apj, 187, 425

\bibitem[{{Press} {et~al}\mbox{.}(1992){Press}, {Teukolsky}, {Vetterling}, \&
  {Flannery}}]{Press+1992}
{Press} W.~H., {Teukolsky} S.~A., {Vetterling} W.~T., {Flannery} B.~P., 1992,
  {Numerical recipes in FORTRAN. The art of scientific computing}

\bibitem[{{Reddick} {et~al}\mbox{.}(2013){Reddick}, {Wechsler}, {Tinker}, \&
  {Behroozi}}]{Reddick+2013}
{Reddick} R.~M., {Wechsler} R.~H., {Tinker} J.~L., {Behroozi} P.~S., 2013,
  \apj, 771, 30

\bibitem[{{Reddy} {et~al}\mbox{.}(2012){Reddy}, {Pettini}, {Steidel},
  {Shapley}, {Erb}, \& {Law}}]{Reddy+2012}
{Reddy} N.~A., {Pettini} M., {Steidel} C.~C., {Shapley} A.~E., {Erb} D.~K.,
  {Law} D.~R., 2012, \apj, 754, 25

\bibitem[{{Robertson} {et~al}\mbox{.}(2006{\natexlab{a}}){Robertson},
  {Bullock}, {Cox}, {Di Matteo}, {Hernquist}, {Springel}, \&
  {Yoshida}}]{Robertson+2006b}
{Robertson} B., {Bullock} J.~S., {Cox} T.~J., {Di Matteo} T., {Hernquist} L.,
  {Springel} V., {Yoshida} N., 2006{\natexlab{a}}, \apj, 645, 986

\bibitem[{{Robertson} {et~al}\mbox{.}(2006{\natexlab{b}}){Robertson}, {Cox},
  {Hernquist}, {Franx}, {Hopkins}, {Martini}, \& {Springel}}]{Robertson+2006a}
{Robertson} B., {Cox} T.~J., {Hernquist} L., {Franx} M., {Hopkins} P.~F.,
  {Martini} P., {Springel} V., 2006{\natexlab{b}}, \apj, 641, 21

\bibitem[{{Rodriguez-Gomez} {et~al}\mbox{.}(2015){Rodriguez-Gomez}, {Genel},
  {Vogelsberger}, {Sijacki}, {Pillepich}, {Sales}, {Torrey}, {Snyder},
  {Nelson}, {Springel}, {Ma}, \& {Hernquist}}]{Rodriguez-Gomez+2015}
{Rodriguez-Gomez} V. {et~al.}, 2015, \mnras, 449, 49

\bibitem[{{Rodriguez-Gomez} {et~al}\mbox{.}(2016){Rodriguez-Gomez},
  {Pillepich}, {Sales}, {Genel}, {Vogelsberger}, {Zhu}, {Wellons}, {Nelson},
  {Torrey}, {Springel}, {Ma}, \& {Hernquist}}]{Rodriguez-Gomez+2016}
{Rodriguez-Gomez} V. {et~al.}, 2016, \mnras, 458, 2371

\bibitem[{{Rodr{\'{\i}}guez-Puebla}, {Avila-Reese} \&
  {Drory}(2013){Rodr{\'{\i}}guez-Puebla}, {Avila-Reese}, \& {Drory}}]{RAD13}
{Rodr{\'{\i}}guez-Puebla} A., {Avila-Reese} V., {Drory} N., 2013, \apj, 767, 92

\bibitem[{{Rodr{\'{\i}}guez-Puebla}
  {et~al}\mbox{.}(2011){Rodr{\'{\i}}guez-Puebla}, {Avila-Reese}, {Firmani}, \&
  {Col{\'{\i}}n}}]{Rodriguez-Puebla+2011}
{Rodr{\'{\i}}guez-Puebla} A., {Avila-Reese} V., {Firmani} C., {Col{\'{\i}}n}
  P., 2011, \rmxaa, 47, 235

\bibitem[{{Rodr{\'{\i}}guez-Puebla}
  {et~al}\mbox{.}(2015){Rodr{\'{\i}}guez-Puebla}, {Avila-Reese}, {Yang},
  {Foucaud}, {Drory}, \& {Jing}}]{RP+2015}
{Rodr{\'{\i}}guez-Puebla} A., {Avila-Reese} V., {Yang} X., {Foucaud} S.,
  {Drory} N., {Jing} Y.~P., 2015, \apj, 799, 130

\bibitem[{{Rodr{\'{\i}}guez-Puebla}
  {et~al}\mbox{.}(2016{\natexlab{a}}){Rodr{\'{\i}}guez-Puebla}, {Behroozi},
  {Primack}, {Klypin}, {Lee}, \& {Hellinger}}]{Rodriguez-Puebla+2016}
{Rodr{\'{\i}}guez-Puebla} A., {Behroozi} P., {Primack} J., {Klypin} A., {Lee}
  C., {Hellinger} D., 2016{\natexlab{a}}, \mnras, 462, 893

\bibitem[{{Rodr{\'{\i}}guez-Puebla}, {Drory} \&
  {Avila-Reese}(2012){Rodr{\'{\i}}guez-Puebla}, {Drory}, \&
  {Avila-Reese}}]{RDA12}
{Rodr{\'{\i}}guez-Puebla} A., {Drory} N., {Avila-Reese} V., 2012, \apj, 756, 2

\bibitem[{{Rodr{\'{\i}}guez-Puebla}
  {et~al}\mbox{.}(2016{\natexlab{b}}){Rodr{\'{\i}}guez-Puebla}, {Primack},
  {Behroozi}, \& {Faber}}]{Rodriguez-Puebla+2016a}
{Rodr{\'{\i}}guez-Puebla} A., {Primack} J.~R., {Behroozi} P., {Faber} S.~M.,
  2016{\natexlab{b}}, \mnras, 455, 2592

\bibitem[{{Salim} {et~al}\mbox{.}(2007){Salim}, {Rich}, {Charlot},
  {Brinchmann}, {Johnson}, {Schiminovich}, {Seibert}, {Mallery}, {Heckman},
  {Forster}, {Friedman}, {Martin}, {Morrissey}, {Neff}, {Small}, {Wyder},
  {Bianchi}, {Donas}, {Lee}, {Madore}, {Milliard}, {Szalay}, {Welsh}, \&
  {Yi}}]{Salim+2007}
{Salim} S. {et~al.}, 2007, \apjs, 173, 267

\bibitem[{{Salmon} {et~al}\mbox{.}(2015){Salmon}, {Papovich}, {Finkelstein},
  {Tilvi}, {Finlator}, {Behroozi}, {Dahlen}, {Dav{\'e}}, {Dekel}, {Dickinson},
  {Ferguson}, {Giavalisco}, {Long}, {Lu}, {Mobasher}, {Reddy}, {Somerville}, \&
  {Wechsler}}]{Salmon+2015}
{Salmon} B. {et~al.}, 2015, \apj, 799, 183

\bibitem[{{Salpeter}(1955)}]{Salpeter1955}
{Salpeter} E.~E., 1955, \apj, 121, 161

\bibitem[{{Santini} {et~al}\mbox{.}(2012){Santini}, {Fontana}, {Grazian},
  {Salimbeni}, {Fontanot}, {Paris}, {Boutsia}, {Castellano}, {Fiore},
  {Gallozzi}, {Giallongo}, {Koekemoer}, {Menci}, {Pentericci}, \&
  {Somerville}}]{Santini+2012}
{Santini} P. {et~al.}, 2012, \aap, 538, A33

\bibitem[{{Schenker} {et~al}\mbox{.}(2013){Schenker}, {Robertson}, {Ellis},
  {Ono}, {McLure}, {Dunlop}, {Koekemoer}, {Bowler}, {Ouchi}, {Curtis-Lake},
  {Rogers}, {Schneider}, {Charlot}, {Stark}, {Furlanetto}, \&
  {Cirasuolo}}]{Schenker+2013}
{Schenker} M.~A. {et~al.}, 2013, \apj, 768, 196

\bibitem[{{Scoccimarro} {et~al}\mbox{.}(2001){Scoccimarro}, {Sheth}, {Hui}, \&
  {Jain}}]{Scoccimarro+2001}
{Scoccimarro} R., {Sheth} R.~K., {Hui} L., {Jain} B., 2001, \apj, 546, 20

\bibitem[{{Seljak}(2000)}]{Seljak+2000}
{Seljak} U., 2000, \mnras, 318, 203

\bibitem[{{S{\'e}rsic}(1963)}]{Sersic1963}
{S{\'e}rsic} J.~L., 1963, Boletin de la Asociacion Argentina de Astronomia La
  Plata Argentina, 6, 41

\bibitem[{{Shankar} {et~al}\mbox{.}(2014){Shankar}, {Guo}, {Bouillot},
  {Rettura}, {Meert}, {Buchan}, {Kravtsov}, {Bernardi}, {Sheth}, {Vikram},
  {Marchesini}, {Behroozi}, {Zheng}, {Maraston}, {Ascaso}, {Lemaux}, {Capozzi},
  {Huertas-Company}, {Gal}, {Lubin}, {Conselice}, {Carollo}, \&
  {Cattaneo}}]{Shankar+2014}
{Shankar} F. {et~al.}, 2014, \apjl, 797, L27

\bibitem[{{Shankar} {et~al}\mbox{.}(2006){Shankar}, {Lapi}, {Salucci}, {De
  Zotti}, \& {Danese}}]{Shankar+2006}
{Shankar} F., {Lapi} A., {Salucci} P., {De Zotti} G., {Danese} L., 2006, \apj,
  643, 14

\bibitem[{{Shen} {et~al}\mbox{.}(2003){Shen}, {Mo}, {White}, {Blanton},
  {Kauffmann}, {Voges}, {Brinkmann}, \& {Csabai}}]{Shen+2003}
{Shen} S., {Mo} H.~J., {White} S.~D.~M., {Blanton} M.~R., {Kauffmann} G.,
  {Voges} W., {Brinkmann} J., {Csabai} I., 2003, \mnras, 343, 978

\bibitem[{{Shim} {et~al}\mbox{.}(2011){Shim}, {Chary}, {Dickinson}, {Lin},
  {Spinrad}, {Stern}, \& {Yan}}]{Shim+2011}
{Shim} H., {Chary} R.-R., {Dickinson} M., {Lin} L., {Spinrad} H., {Stern} D.,
  {Yan} C.-H., 2011, \apj, 738, 69

\bibitem[{{Simard} {et~al}\mbox{.}(2011){Simard}, {Mendel}, {Patton},
  {Ellison}, \& {McConnachie}}]{Simard+2011}
{Simard} L., {Mendel} J.~T., {Patton} D.~R., {Ellison} S.~L., {McConnachie}
  A.~W., 2011, \apjs, 196, 11

\bibitem[{{Skibba} {et~al}\mbox{.}(2015){Skibba}, {Coil}, {Mendez}, {Blanton},
  {Bray}, {Cool}, {Eisenstein}, {Guo}, {Miyaji}, {Moustakas}, \&
  {Zhu}}]{Skibba+2015}
{Skibba} R.~A. {et~al.}, 2015, \apj, 807, 152

\bibitem[{{Skibba} {et~al}\mbox{.}(2011){Skibba}, {van den Bosch}, {Yang},
  {More}, {Mo}, \& {Fontanot}}]{Skibba+2011}
{Skibba} R.~A., {van den Bosch} F.~C., {Yang} X., {More} S., {Mo} H.,
  {Fontanot} F., 2011, \mnras, 410, 417

\bibitem[{{Smith}(2012)}]{Smith2012}
{Smith} R.~E., 2012, \mnras, 426, 531

\bibitem[{{Sobral} {et~al}\mbox{.}(2014){Sobral}, {Best}, {Smail}, {Mobasher},
  {Stott}, \& {Nisbet}}]{Sobral+2014}
{Sobral} D., {Best} P.~N., {Smail} I., {Mobasher} B., {Stott} J., {Nisbet} D.,
  2014, \mnras, 437, 3516

\bibitem[{{Somerville} {et~al}\mbox{.}(2017){Somerville}, {Behroozi}, {Pandya},
  {Dekel}, {Faber}, {Fontana}, {Huang}, {Koekemoer}, {Koo},
  {P{\'e}rez-Gonz{\'a}lez}, {Primack}, {Santini}, {Taylor}, \& {van der
  Wel}}]{Somerville2017}
{Somerville} R.~S. {et~al.}, 2017, ArXiv e-prints

\bibitem[{{Somerville} \& {Dav{\'e}}(2015)}]{Somerville+2015}
{Somerville} R.~S., {Dav{\'e}} R., 2015, \araa, 53, 51

\bibitem[{{Song} {et~al}\mbox{.}(2015){Song}, {Finkelstein}, {Ashby},
  {Grazian}, {Lu}, {Papovich}, {Salmon}, {Somerville}, {Dickinson}, {Duncan},
  {Faber}, {Fazio}, {Ferguson}, {Fontana}, {Guo}, {Hathi}, {Lee}, {Merlin}, \&
  {Willner}}]{Song+2015}
{Song} M. {et~al.}, 2015, ArXiv e-prints

\bibitem[{{Speagle} {et~al}\mbox{.}(2014){Speagle}, {Steinhardt}, {Capak}, \&
  {Silverman}}]{Speagle+2014}
{Speagle} J.~S., {Steinhardt} C.~L., {Capak} P.~L., {Silverman} J.~D., 2014,
  \apjs, 214, 15

\bibitem[{{Stark} {et~al}\mbox{.}(2009){Stark}, {Ellis}, {Bunker}, {Bundy},
  {Targett}, {Benson}, \& {Lacy}}]{Stark+2009}
{Stark} D.~P., {Ellis} R.~S., {Bunker} A., {Bundy} K., {Targett} T., {Benson}
  A., {Lacy} M., 2009, \apj, 697, 1493

\bibitem[{{Steinhardt} {et~al}\mbox{.}(2014){Steinhardt}, {Speagle}, {Capak},
  {Silverman}, {Carollo}, {Dunlop}, {Hashimoto}, {Hsieh}, {Ilbert}, {Le Fevre},
  {Le Floc'h}, {Lee}, {Lin}, {Lin}, {Masters}, {McCracken}, {Nagao}, {Petric},
  {Salvato}, {Sanders}, {Scoville}, {Sheth}, {Strauss}, \&
  {Taniguchi}}]{Steinhardt+2014}
{Steinhardt} C.~L. {et~al.}, 2014, \apjl, 791, L25

\bibitem[{{Stewart} {et~al}\mbox{.}(2009){Stewart}, {Bullock}, {Barton}, \&
  {Wechsler}}]{Stewart+2009}
{Stewart} K.~R., {Bullock} J.~S., {Barton} E.~J., {Wechsler} R.~H., 2009, \apj,
  702, 1005

\bibitem[{{Tacchella} {et~al}\mbox{.}(2015){Tacchella}, {Carollo}, {Renzini},
  {Schreiber}, {Lang}, {Wuyts}, {Cresci}, {Dekel}, {Genzel}, {Lilly},
  {Mancini}, {Newman}, {Onodera}, {Shapley}, {Tacconi}, {Woo}, \&
  {Zamorani}}]{Tacchella+2015}
{Tacchella} S. {et~al.}, 2015, Science, 348, 314

\bibitem[{{Tacchella} {et~al}\mbox{.}(2016){Tacchella}, {Dekel}, {Carollo},
  {Ceverino}, {DeGraf}, {Lapiner}, {Mandelker}, \& {Primack
  Joel}}]{Tacchella+2016}
{Tacchella} S., {Dekel} A., {Carollo} C.~M., {Ceverino} D., {DeGraf} C.,
  {Lapiner} S., {Mandelker} N., {Primack Joel} R., 2016, \mnras, 457, 2790

\bibitem[{{Terrazas} {et~al}\mbox{.}(2016){Terrazas}, {Bell}, {Henriques},
  {White}, {Cattaneo}, \& {Woo}}]{Terrazas+2016}
{Terrazas} B.~A., {Bell} E.~F., {Henriques} B.~M.~B., {White} S.~D.~M.,
  {Cattaneo} A., {Woo} J., 2016, \apjl, 830, L12

\bibitem[{{Tinker} {et~al}\mbox{.}(2008){Tinker}, {Kravtsov}, {Klypin},
  {Abazajian}, {Warren}, {Yepes}, {Gottl{\"o}ber}, \& {Holz}}]{Tinker+2008}
{Tinker} J., {Kravtsov} A.~V., {Klypin} A., {Abazajian} K., {Warren} M.,
  {Yepes} G., {Gottl{\"o}ber} S., {Holz} D.~E., 2008, \apj, 688, 709

\bibitem[{{Tinker} {et~al}\mbox{.}(2016){Tinker}, {Brownstein}, {Guo},
  {Leauthaud}, {Maraston}, {Masters}, {Montero-Dorta}, {Thomas}, {Tojeiro},
  {Weiner}, {Zehavi}, \& {Olmstead}}]{Tinker+2016}
{Tinker} J.~L. {et~al.}, 2016, ArXiv e-prints

\bibitem[{{Tinker} {et~al}\mbox{.}(2013){Tinker}, {Leauthaud}, {Bundy},
  {George}, {Behroozi}, {Massey}, {Rhodes}, \& {Wechsler}}]{Tinker+2013}
{Tinker} J.~L., {Leauthaud} A., {Bundy} K., {George} M.~R., {Behroozi} P.,
  {Massey} R., {Rhodes} J., {Wechsler} R.~H., 2013, \apj, 778, 93

\bibitem[{{Tinsley}(1972)}]{Tinsley1972}
{Tinsley} B.~M., 1972, \aap, 20, 383

\bibitem[{{Tomczak} {et~al}\mbox{.}(2014){Tomczak}, {Quadri}, {Tran},
  {Labb{\'e}}, {Straatman}, {Papovich}, {Glazebrook}, {Allen}, {Brammer},
  {Kacprzak}, {Kawinwanichakij}, {Kelson}, {McCarthy}, {Mehrtens}, {Monson},
  {Persson}, {Spitler}, {Tilvi}, \& {van Dokkum}}]{Tomczak+2014}
{Tomczak} A.~R. {et~al.}, 2014, \apj, 783, 85

\bibitem[{{Tonry} {et~al}\mbox{.}(2000){Tonry}, {Blakeslee}, {Ajhar}, \&
  {Dressler}}]{Tonry+2000}
{Tonry} J.~L., {Blakeslee} J.~P., {Ajhar} E.~A., {Dressler} A., 2000, \apj,
  530, 625

\bibitem[{{Toomre} \& {Toomre}(1972)}]{Toomre_Toomre1972}
{Toomre} A., {Toomre} J., 1972, \apj, 178, 623

\bibitem[{{Vale} \& {Ostriker}(2004)}]{ValeOstriker2004}
{Vale} A., {Ostriker} J.~P., 2004, \mnras, 353, 189

\bibitem[{{van der Burg}, {Hildebrandt} \& {Erben}(2010){van der Burg},
  {Hildebrandt}, \& {Erben}}]{vanderBurg+2010}
{van der Burg} R.~F.~J., {Hildebrandt} H., {Erben} T., 2010, \aap, 523, A74

\bibitem[{{van der Wel} {et~al}\mbox{.}(2012){van der Wel}, {Bell},
  {H{\"a}ussler}, {McGrath}, {Chang}, {Guo}, {McIntosh}, {Rix}, {Barden},
  {Cheung}, {Faber}, {Ferguson}, {Galametz}, {Grogin}, {Hartley}, {Kartaltepe},
  {Kocevski}, {Koekemoer}, {Lotz}, {Mozena}, {Peth}, \&
  {Peng}}]{vanderWel+2012}
{van der Wel} A. {et~al.}, 2012, \apjs, 203, 24

\bibitem[{{van der Wel} {et~al}\mbox{.}(2014){van der Wel}, {Franx}, {van
  Dokkum}, {Skelton}, {Momcheva}, {Whitaker}, {Brammer}, {Bell}, {Rix},
  {Wuyts}, {Ferguson}, {Holden}, {Barro}, {Koekemoer}, {Chang}, {McGrath},
  {H{\"a}ussler}, {Dekel}, {Behroozi}, {Fumagalli}, {Leja}, {Lundgren},
  {Maseda}, {Nelson}, {Wake}, {Patel}, {Labb{\'e}}, {Faber}, {Grogin}, \&
  {Kocevski}}]{vanderWel+2014}
{van der Wel} A. {et~al.}, 2014, \apj, 788, 28

\bibitem[{{van Dokkum} {et~al}\mbox{.}(2015{\natexlab{a}}){van Dokkum},
  {Abraham}, {Merritt}, {Zhang}, {Geha}, \& {Conroy}}]{vanDokkum+2015c}
{van Dokkum} P.~G., {Abraham} R., {Merritt} A., {Zhang} J., {Geha} M., {Conroy}
  C., 2015{\natexlab{a}}, \apjl, 798, L45

\bibitem[{{van Dokkum} {et~al}\mbox{.}(2014){van Dokkum}, {Bezanson}, {van der
  Wel}, {Nelson}, {Momcheva}, {Skelton}, {Whitaker}, {Brammer}, {Conroy},
  {F{\"o}rster Schreiber}, {Fumagalli}, {Kriek}, {Labb{\'e}}, {Leja},
  {Marchesini}, {Muzzin}, {Oesch}, \& {Wuyts}}]{vanDokkum+2014}
{van Dokkum} P.~G. {et~al.}, 2014, \apj, 791, 45

\bibitem[{{van Dokkum} {et~al}\mbox{.}(2013){van Dokkum}, {Leja}, {Nelson},
  {Patel}, {Skelton}, {Momcheva}, {Brammer}, {Whitaker}, {Lundgren},
  {Fumagalli}, {Conroy}, {F{\"o}rster Schreiber}, {Franx}, {Kriek},
  {Labb{\'e}}, {Marchesini}, {Rix}, {van der Wel}, \& {Wuyts}}]{vanDokkum+2013}
{van Dokkum} P.~G. {et~al.}, 2013, \apjl, 771, L35

\bibitem[{{van Dokkum} {et~al}\mbox{.}(2015{\natexlab{b}}){van Dokkum},
  {Nelson}, {Franx}, {Oesch}, {Momcheva}, {Brammer}, {F{\"o}rster Schreiber},
  {Skelton}, {Whitaker}, {van der Wel}, {Bezanson}, {Fumagalli}, {Illingworth},
  {Kriek}, {Leja}, \& {Wuyts}}]{vanDokkum+2015}
{van Dokkum} P.~G. {et~al.}, 2015{\natexlab{b}}, \apj, 813, 23

\bibitem[{{van Dokkum} {et~al}\mbox{.}(2010){van Dokkum}, {Whitaker},
  {Brammer}, {Franx}, {Kriek}, {Labb{\'e}}, {Marchesini}, {Quadri}, {Bezanson},
  {Illingworth}, {Muzzin}, {Rudnick}, {Tal}, \& {Wake}}]{vanDokkum+2010}
{van Dokkum} P.~G. {et~al.}, 2010, \apj, 709, 1018

\bibitem[{{van Uitert} {et~al}\mbox{.}(2011){van Uitert}, {Hoekstra},
  {Velander}, {Gilbank}, {Gladders}, \& {Yee}}]{vanUitert+2011}
{van Uitert} E., {Hoekstra} H., {Velander} M., {Gilbank} D.~G., {Gladders}
  M.~D., {Yee} H.~K.~C., 2011, \aap, 534, A14

\bibitem[{{Velander} {et~al}\mbox{.}(2014){Velander}, {van Uitert}, {Hoekstra},
  {Coupon}, {Erben}, {Heymans}, {Hildebrandt}, {Kitching}, {Mellier}, {Miller},
  {Van Waerbeke}, {Bonnett}, {Fu}, {Giodini}, {Hudson}, {Kuijken}, {Rowe},
  {Schrabback}, \& {Semboloni}}]{Velander+2014}
{Velander} M. {et~al.}, 2014, \mnras, 437, 2111

\bibitem[{{Wake} {et~al}\mbox{.}(2011){Wake}, {Whitaker}, {Labb{\'e}}, {van
  Dokkum}, {Franx}, {Quadri}, {Brammer}, {Kriek}, {Lundgren}, {Marchesini}, \&
  {Muzzin}}]{Wake+2011}
{Wake} D.~A. {et~al.}, 2011, \apj, 728, 46

\bibitem[{{Wetzel}, {Tinker} \& {Conroy}(2012){Wetzel}, {Tinker}, \&
  {Conroy}}]{Wetzel+2012}
{Wetzel} A.~R., {Tinker} J.~L., {Conroy} C., 2012, \mnras, 424, 232

\bibitem[{{Wetzel} \& {White}(2010)}]{Wetzel+2010}
{Wetzel} A.~R., {White} M., 2010, \mnras, 403, 1072

\bibitem[{{Whitaker} {et~al}\mbox{.}(2014){Whitaker}, {Franx}, {Leja}, {van
  Dokkum}, {Henry}, {Skelton}, {Fumagalli}, {Momcheva}, {Brammer}, {Labb{\'e}},
  {Nelson}, \& {Rigby}}]{Whitaker+2014}
{Whitaker} K.~E. {et~al.}, 2014, \apj, 795, 104

\bibitem[{{White} \& {Frenk}(1991)}]{WhiteFrenk1991}
{White} S.~D.~M., {Frenk} C.~S., 1991, \apj, 379, 52

\bibitem[{{Williams}, {Quadri} \& {Franx}(2011){Williams}, {Quadri}, \&
  {Franx}}]{Williams+2011}
{Williams} R.~J., {Quadri} R.~F., {Franx} M., 2011, \apjl, 738, L25

\bibitem[{{Willott} {et~al}\mbox{.}(2013){Willott}, {McLure}, {Hibon},
  {Bielby}, {McCracken}, {Kneib}, {Ilbert}, {Bonfield}, {Bruce}, \&
  {Jarvis}}]{Willott+2013}
{Willott} C.~J. {et~al.}, 2013, \aj, 145, 4

\bibitem[{{Wojtak} \& {Mamon}(2013)}]{Wojtak+2013}
{Wojtak} R., {Mamon} G.~A., 2013, \mnras, 428, 2407

\bibitem[{{Yagi} {et~al}\mbox{.}(2016){Yagi}, {Koda}, {Komiyama}, \&
  {Yamanoi}}]{Yagi+2016}
{Yagi} M., {Koda} J., {Komiyama} Y., {Yamanoi} H., 2016, \apjs, 225, 11

\bibitem[{{Yang}, {Mo} \& {van den Bosch}(2003){Yang}, {Mo}, \& {van den
  Bosch}}]{Yang+2003}
{Yang} X., {Mo} H.~J., {van den Bosch} F.~C., 2003, \mnras, 339, 1057

\bibitem[{{Yang}, {Mo} \& {van den Bosch}(2008){Yang}, {Mo}, \& {van den
  Bosch}}]{Yang+2008}
{Yang} X., {Mo} H.~J., {van den Bosch} F.~C., 2008, \apj, 676, 248

\bibitem[{{Yang}, {Mo} \& {van den Bosch}(2009{\natexlab{a}}){Yang}, {Mo}, \&
  {van den Bosch}}]{Yang+2009b}
{Yang} X., {Mo} H.~J., {van den Bosch} F.~C., 2009{\natexlab{a}}, \apj, 695,
  900

\bibitem[{{Yang}, {Mo} \& {van den Bosch}(2009{\natexlab{b}}){Yang}, {Mo}, \&
  {van den Bosch}}]{Yang+2009a}
{Yang} X., {Mo} H.~J., {van den Bosch} F.~C., 2009{\natexlab{b}}, \apj, 693,
  830

\bibitem[{{Yang} {et~al}\mbox{.}(2013){Yang}, {Mo}, {van den Bosch}, {Bonaca},
  {Li}, {Lu}, {Lu}, \& {Lu}}]{Yang+2013}
{Yang} X., {Mo} H.~J., {van den Bosch} F.~C., {Bonaca} A., {Li} S., {Lu} Y.,
  {Lu} Y., {Lu} Z., 2013, \apj, 770, 115

\bibitem[{{Yang} {et~al}\mbox{.}(2012){Yang}, {Mo}, {van den Bosch}, {Zhang},
  \& {Han}}]{Yang+2012}
{Yang} X., {Mo} H.~J., {van den Bosch} F.~C., {Zhang} Y., {Han} J., 2012, \apj,
  752, 41

\bibitem[{{Zavala} {et~al}\mbox{.}(2012){Zavala}, {Avila-Reese}, {Firmani}, \&
  {Boylan-Kolchin}}]{Zavala+2012}
{Zavala} J., {Avila-Reese} V., {Firmani} C., {Boylan-Kolchin} M., 2012, \mnras,
  427, 1503

\bibitem[{{Zehavi} {et~al}\mbox{.}(2011){Zehavi}, {Zheng}, {Weinberg},
  {Blanton}, {Bahcall}, {Berlind}, {Brinkmann}, {Frieman}, {Gunn}, {Lupton},
  {Nichol}, {Percival}, {Schneider}, {Skibba}, {Strauss}, {Tegmark}, \&
  {York}}]{Zehavi+2011}
{Zehavi} I. {et~al.}, 2011, \apj, 736, 59

\bibitem[{{Zehavi} {et~al}\mbox{.}(2005){Zehavi}, {Zheng}, {Weinberg},
  {Frieman}, {Berlind}, {Blanton}, {Scoccimarro}, {Sheth}, {Strauss}, {Kayo},
  {Suto}, {Fukugita}, {Nakamura}, {Bahcall}, {Brinkmann}, {Gunn}, {Hennessy},
  {Ivezi{\'c}}, {Knapp}, {Loveday}, {Meiksin}, {Schlegel}, {Schneider},
  {Szapudi}, {Tegmark}, {Vogeley}, {York}, \& {SDSS
  Collaboration}}]{Zehavi+2005}
{Zehavi} I. {et~al.}, 2005, \apj, 630, 1

\bibitem[{{Zheng} {et~al}\mbox{.}(2005){Zheng}, {Berlind}, {Weinberg},
  {Benson}, {Baugh}, {Cole}, {Dav{\'e}}, {Frenk}, {Katz}, \&
  {Lacey}}]{Zheng+2005}
{Zheng} Z. {et~al.}, 2005, \apj, 633, 791

\bibitem[{{Zheng}, {Coil} \& {Zehavi}(2007){Zheng}, {Coil}, \&
  {Zehavi}}]{Zheng+2007b}
{Zheng} Z., {Coil} A.~L., {Zehavi} I., 2007, \apj, 667, 760

\bibitem[{{Zolotov} {et~al}\mbox{.}(2015){Zolotov}, {Dekel}, {Mandelker},
  {Tweed}, {Inoue}, {DeGraf}, {Ceverino}, {Primack}, {Barro}, \&
  {Faber}}]{Zolotov+2015}
{Zolotov} A. {et~al.}, 2015, \mnras, 450, 2327

\end{thebibliography}

 \appendix

\section{Assumptions in the modelling of the CSFR}
\label{A1}

\begin{figure}
\vspace*{-250pt}
\hspace*{-20pt}
\includegraphics[height=6.7in,width=5.3in]{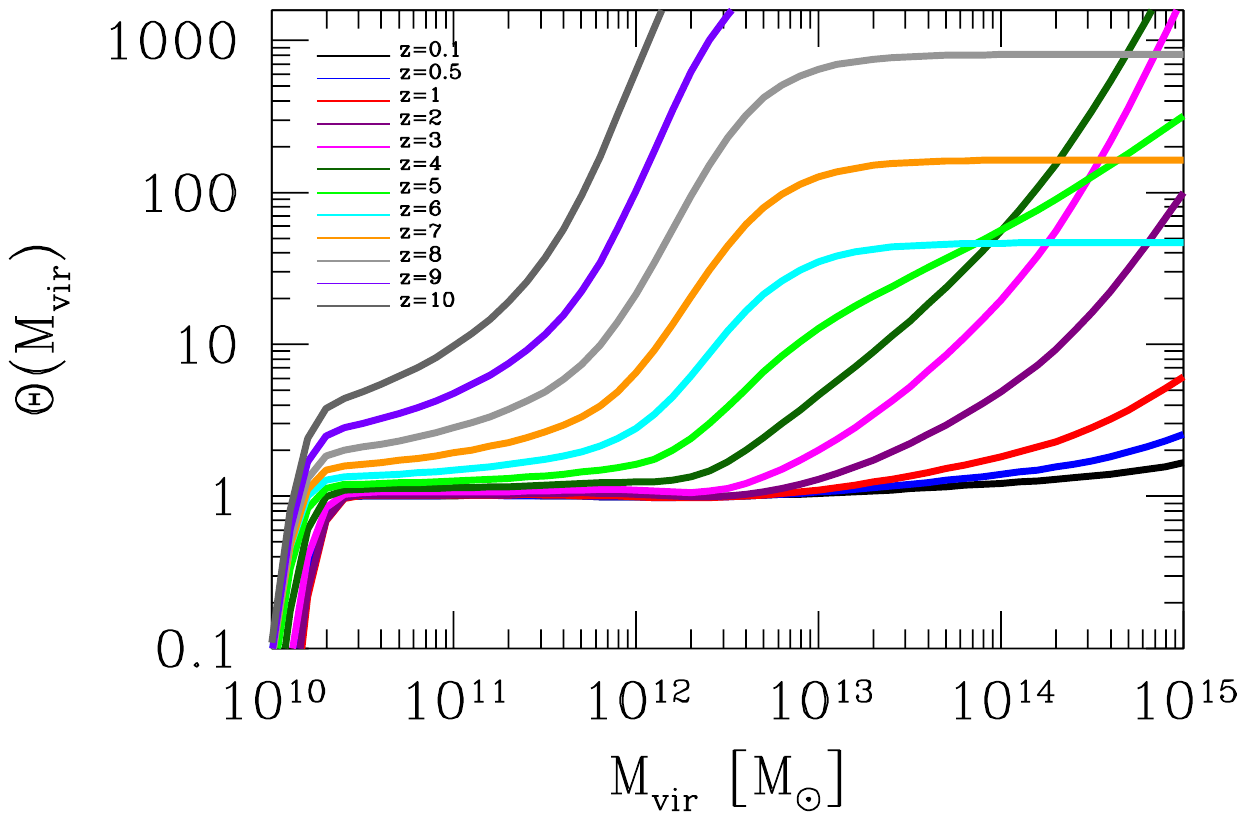}
\vspace*{-100pt}
\caption{Redshift evolution of the function $\Theta(\mdm)$
which maps the intrinsic CSFR to the observed one.  Note that
$\Theta(\mdm)=1$ implies that the observed CSFR is identical
to the intrinsic CSFR. 
}
\label{theta_model}
\end{figure}

\begin{figure}
\vspace*{-230pt}
\includegraphics[height=6.7in,width=5.5in]{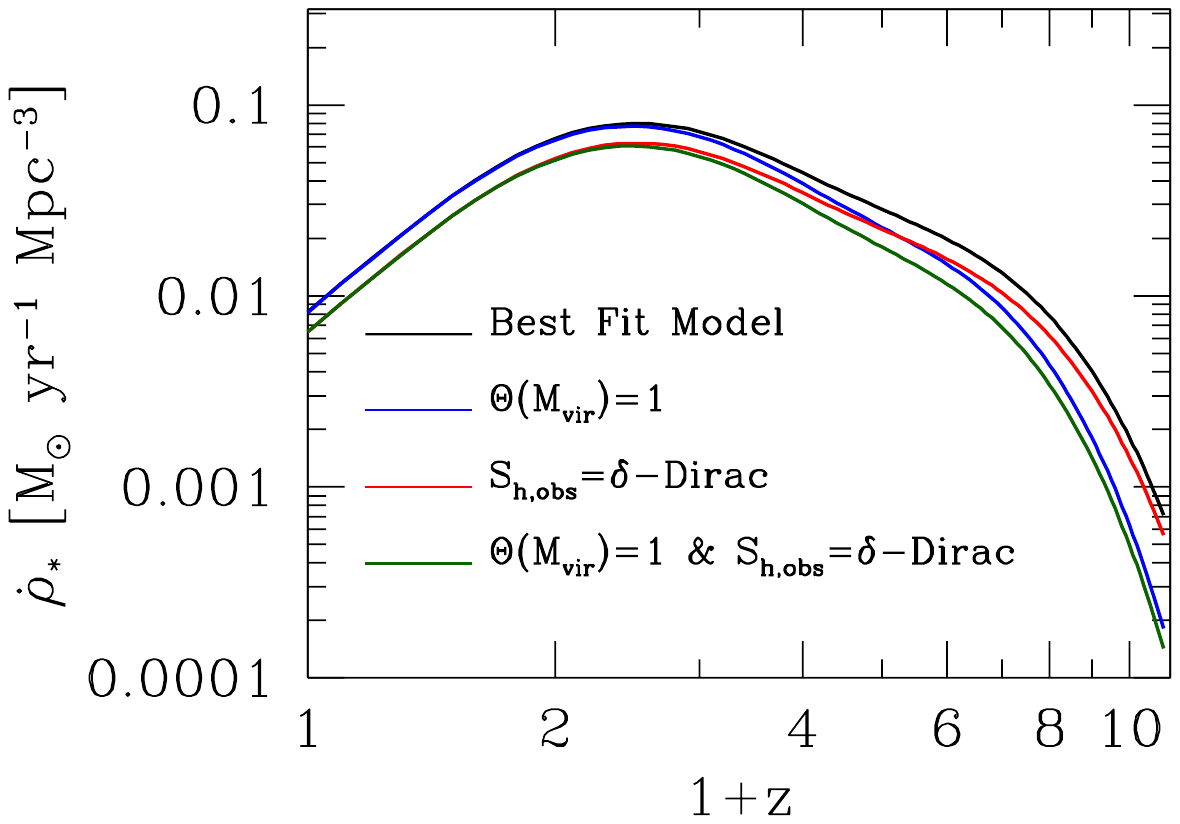}
\vspace*{-100pt}
\caption{The cosmic star formation rate, CSFR. The solid black line shows the resulting best fit model to the
CSFR as described in Section \ref{csfr_model_section}. The blue solid lines shows when ignoring 
random errors in stellar mass determinations while the red solid shows when assuming the distribution 
of star-formation rate as a Dirac$-\delta$ distribution function. The green solid line shows when both effects
are totally ignore. 
}
\label{CSFR_impact}
\end{figure}

In this Section we present the redshift dependence of the function 
$\Theta(\mdm)$ described in Section \ref{mod_framework}. We 
show this in Figure \ref{theta_model} for our best fitting model. 
Recall that this function depends strongly on the ratio $\phigobs / \phig$,
which is the reason for such a strong dependence on halo mass and redshift. Note that this 
function maps the intrinsic CSFR (the one inferred in the absence of random errors) 
to the observed CSFR. Figure \ref{CSFR_impact} shows the CSFR for our best fitting model,
in which $\Theta(\mdm)\neq0$. When ignoring random errors, $\Theta(\mdm)=1$. 
Clearly, ignoring random errors will result in an extra source of uncertainty in the modelling
of the CSFR. Note that this is more important at high redshifts where the CSFR could be underestimated   
by an order of magnitude. 

Figure \ref{CSFR_impact} also shows the impact of assuming the distribution of star-formation rate
as a Dirac$-\delta$ function. The error in the modelling of the CSFR is around $\sim30\%$.

\section{Dark matter halos}
\label{App_DMH}

In this Section we present the theoretical ingredients to fully characterize the model 
described in Section \ref{mod_framework}, namely,
the halo mass function and halo mass assembly. 

\subsection{Halo Mass Functions}

\begin{figure}
	\vspace*{-130pt}
	\hspace*{-20pt}
	\includegraphics[height=5.3in,width=4in]{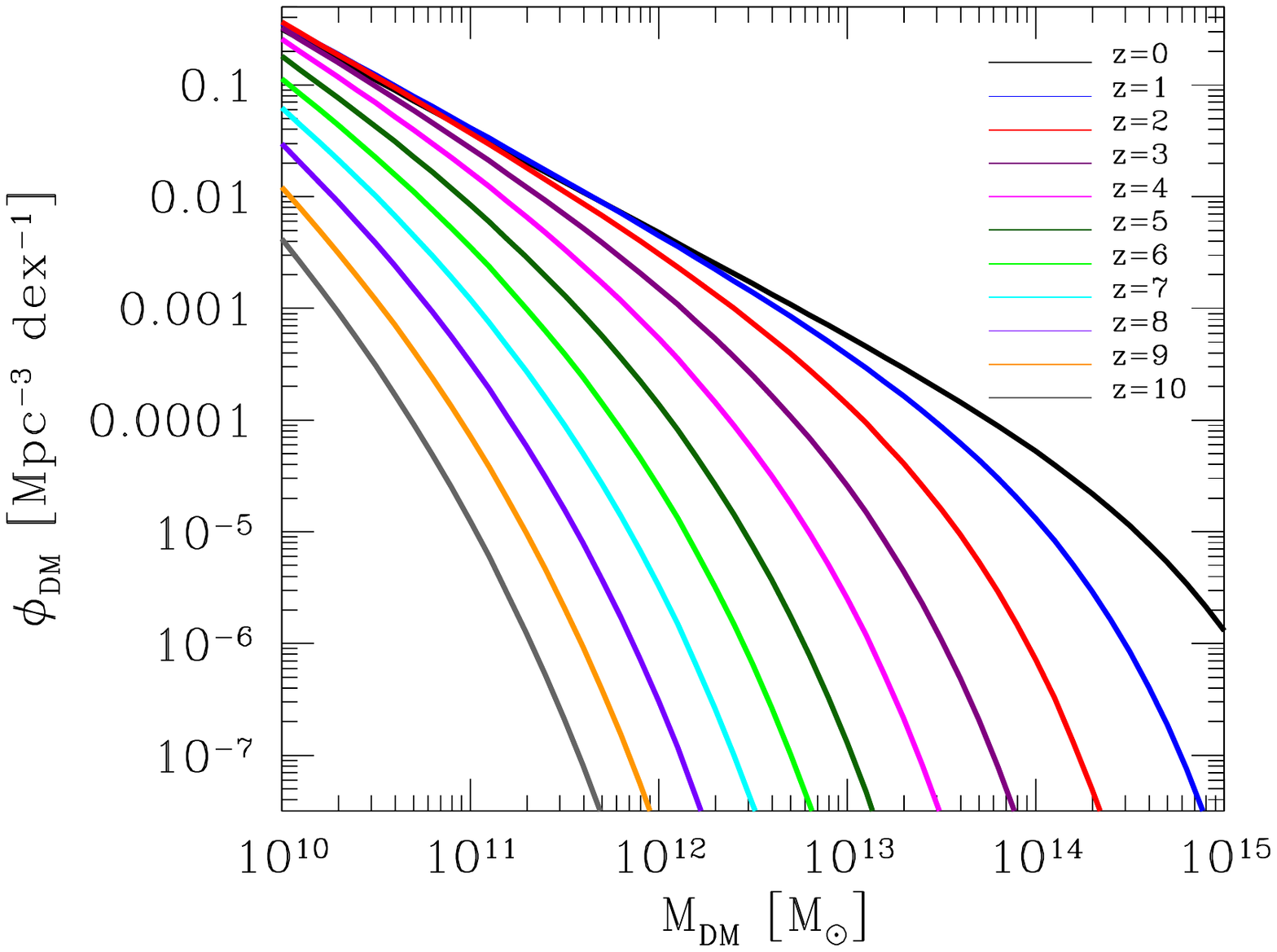}
	\vspace*{-60pt}
		\caption{
		 Total number density of halos and subhalos, $\phiDM (\mdm) d\mdm$, 
		 from $z = 0$ to $z = 10$.  \mdm\ should be interpreted as the virial mass, 
		 \mvir, for distinct halos and \mpeak\ for subhalos.. For central halos we are 
		using the \citet{Tinker+2008} model
		with the parameters updated in \citet{Rodriguez-Puebla+2016} based on 
		large Bolshoi-Planck and MultiDark-Planck cosmological simulations using the 
		cosmological parameters from the Planck
		mission. For subhalos we use the maximum mass reached along the main progenitor
		assembly, denoted as \mpeak. 
 	}
	\label{f1}
\end{figure}

The abundance of dark matter halos has been studied in a great detail in a number of previous
studies since the pioneer work in \citet{PressSchechter1974}.

In this paper we will denote the comoving number density of halos within the 
mass range $\mvir$ and $\mvir+d\mvir$ as $\phih$(\mvir) d\mvir. Theoretically this is given by
	\begin{equation}
		\phih(\mvir) d\mvir= f ( \sigma ) \frac {\rhom}{M_{\rm vir}^2} 
		\left|\frac{d\ln\sigma^{-1}}{d\ln \mvir}\right| d\mvir,
		\label{hmf}
	\end{equation}
where $\rhom$ is the mean matter density and $\sigma$ 
is the amplitude of the perturbations. Following \citet{Klypin+2011}, we find in 
\citet{Rodriguez-Puebla+2016} that to a high accuracy $\sigma$ it is given
by 	
	\begin{equation}
		\sigma(\mvir,a) = \frac{17.111y^{0.405} D(a)}{1+1.306y^{0.22}+6.218y^{0.317}},
		\label{sigma_mvir_fit}
	\end{equation}
with $y =  \mvir / 10^{12}h^{-1}\msun$. The above equation is only valid for the cosmology 
studied in this paper. The linear growth-rate factor, denoted by $D(a)$, is given 
by the expression
	\begin{equation}
		D(a) = \frac{g(a)}{g(1)},
	\end{equation}
where to a good approximation $g(a)$ is given by \citep{Lahav+1991}:
	\begin{equation}
		g(a) =\frac{ \frac{5}{2}\Omega_{\rm m}(a)a}{\Omega_{\rm m}(a) - \Omega_{\Lambda}(a) + [1 + \frac{1}{2} \Omega_{\rm m}(a) ] / [1 + \frac{1}{70}  \Omega_{\Lambda}(a) ] }.
	\end{equation}
The function  $f(\sigma)$ is given by the 
parametrization in \citet{Tinker+2008}:
	\begin{equation}
		f(\sigma) = A \left[ \left(\frac{\sigma}{b}\right)^{-a} + 1 \right] e^{-c / \sigma^2}.
	\label{Tinker}
	\end{equation}
In this paper we use the updated values for the parameters $A, a,b$ and $c$ reported 
in \citet{Rodriguez-Puebla+2016} and given by:
	\begin{equation}
		A = 0.144  - 0.011 z + 0.003 z^2,
	\end{equation}
	\begin{equation}
		a = 1.351  - 0.068 z + 0.006 z^2,
	\end{equation}
	\begin{equation}
		b = 3.113  - 0.077 z - 0.013 z^2,
	\end{equation}
	\begin{equation}
		c = 1.187  - 0.009 z.
	\end{equation}

For the comoving number density of subhalos within the 
mass range $\log\mpeak$ and $\log\mpeak+d\log\mpeak$ we use the fitting model 
proposed in \citet{Behroozi+2013} with the updated parameters in \citet{Rodriguez-Puebla+2016}
	\begin{eqnarray}
		\phisub(\mpeak) d\log\mpeak = C_{\rm sub}(z) G(\mvir,z) \times  & &  \nonumber \\
		\phih(\mvir) d\log\mvir,
	\end{eqnarray}
where 
	\begin{eqnarray}
		\log C_{\rm sub}(z) = -0.0863 + 0.0087 a - 0.0113 a^2 -   & &  \nonumber \\
		0.0039 a^3 + 0.0004 a^4,
	\label{f16fita}
	\end{eqnarray}
and 
	\begin{equation}
	 	 G(\mvir,z) = X^{0.0724}  \exp ( - X^{0.2206}),
		\label{f16fitb}
	\end{equation}
where $X = \mvir / M_{\rm cut}(z)$. The fitting function for $ M_{\rm cut}(z)$ is
given by
	\begin{eqnarray}
		\log(M_{\rm cut}(z)) = 11.9046 - 0.6364 z + 0.02069 z^2 +  & &  \nonumber \\
		0.0220 z^3 - 0.0012 z^4 .
		\label{f16fitc}
	\end{eqnarray}

Figure \ref{f1} shows the predicted redshift evolution of  
$\phiDM (\mdm)$ using the equations described in this section from $z=0$ to $z=10$. 

\subsection{Halo Mass Assembly}

\begin{figure}
	\vspace*{-190pt}
	\hspace*{-20pt}
	\includegraphics[height=6.4in,width=5.2in]{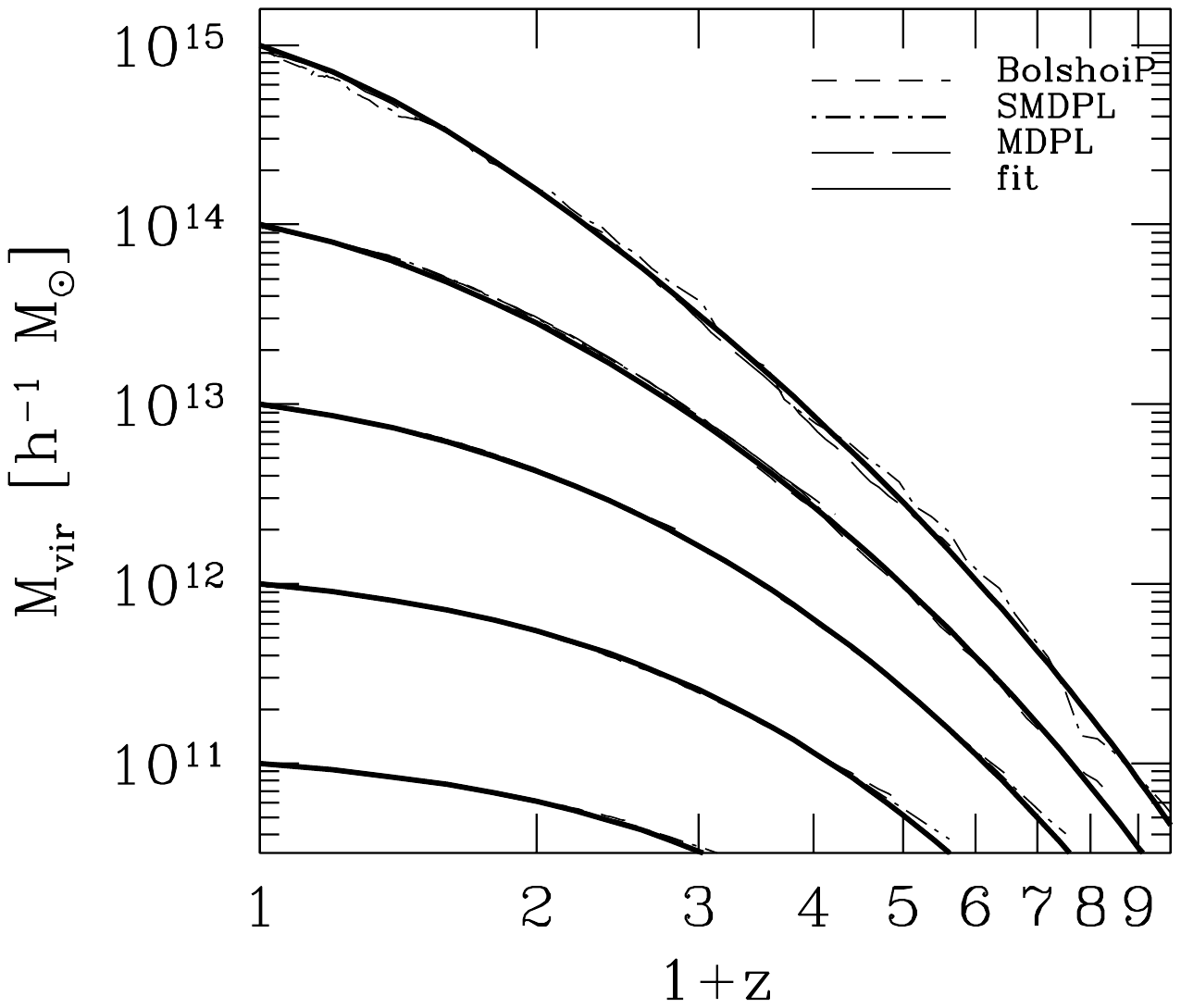}
	\vspace*{-80pt}
		\caption{Median halo mass growth for progenitors $z = 0$ with
		masses of $\mvir=10^{11},10^{12},10^{13},$
		$10^{14}$ and $10^{15} h^{-1}$\msun, solid lines. Fits to simulations are shown
		with the dotted lines. 
 	}
	\label{f2}
\end{figure}

The rate at which dark matter halos grow will determine the rate at which the cosmological 
baryonic inflow material reaches the interstellar medium of a galaxy. Eventually, when necessary conditions are satisfied,
some of this cosmological baryonic material will be transformed into stars. As described in Section \ref{SFH_model}, 
we use the growth of dark matter halos to predict the star formation 
histories of galaxies without modelling how the cold gas in the galaxy is converted into stars. 

Figure \ref{f2} compares the medians of the halo mass growth for progenitors at $z = 0$ with 
masses of $\mvir=10^{11},10^{12},10^{13}, 10^{14}$ and $10^{15}h^{-1}$\msun, 
for the \bpl\ (dashed solid line) \smdpl\ (dot-dash line) and \mdpl\ (long dashed line) simulations with the
fitting functions reported in \citet{Rodriguez-Puebla+2016} (solid line). 
In order to characterize the growth of dark matter halos \citet{Rodriguez-Puebla+2016} used 
the fitting function from \citet{Behroozi+2013}
	\begin{equation}
		\mvir (M_{\rm vir,0}, z) = M_{\rm 13} (z) 10 ^{f(M_{\rm vir,0}, z)}
		\label{AMH}
	\end{equation}
where 
	\begin{equation}
		M_{\rm 13} (z) = 10^{13.6}h^{-1}\msun(1+z)^{2.755}(1+\frac{z}{2})^{-6.351}\exp{(-0.413 z)}
	\end{equation}
	
	\begin{equation}
		f(\mvir,z) = \log\left(\frac{M_{\rm vir,0}}{M_{\rm 13} (0)}\right)\frac{g(M_{\rm vir,0},1)}{g(M_{\rm vir,0},a)}
	\end{equation}

	\begin{equation}
		g(M_{\rm vir,0},a) = 1 + \exp[-3.676 (a-a_0(M_{\rm vir,0}))]
	\end{equation}
	
	\begin{equation}
		a_0(M_{\rm vir,0}) = 0.592 - \log\left[\left(\frac{10^{15.7}h^{-1}\msun}{M_{\rm vir,0}}\right)^{0.113}+1\right].
	\end{equation}
We can generalize the above function to characterize the halo mass growth of any progenitor at any redshift $z_0$ 
by moving the origin from $z$ to $z \rightarrow z - z_0$, \citep[see e.g., ][]{Correa+2015}. 
The above change leaves Equation \ref{AMH} as:
	\begin{equation}
		\mvir (M_{\rm vir,0}, z - z_0) = M_{\rm 13} (z - z_0) 10 ^{f(M_{\rm vir,0}, z - z_0)}.
		\label{AMH}
	\end{equation}
This is the Equation that we will use for the growth of halos. Note that halo mass accretion rates 
can be derived by taking the derivative of Equation \ref{AMH} with respect to the time. 

\section{The surface brightness correction at redshifts $\sim0.1$}
\label{MFs_SDSS}

\begin{figure}
	\vspace*{-200pt}
	\hspace*{-20pt}
	\includegraphics[height=6.7in,width=5.5in]{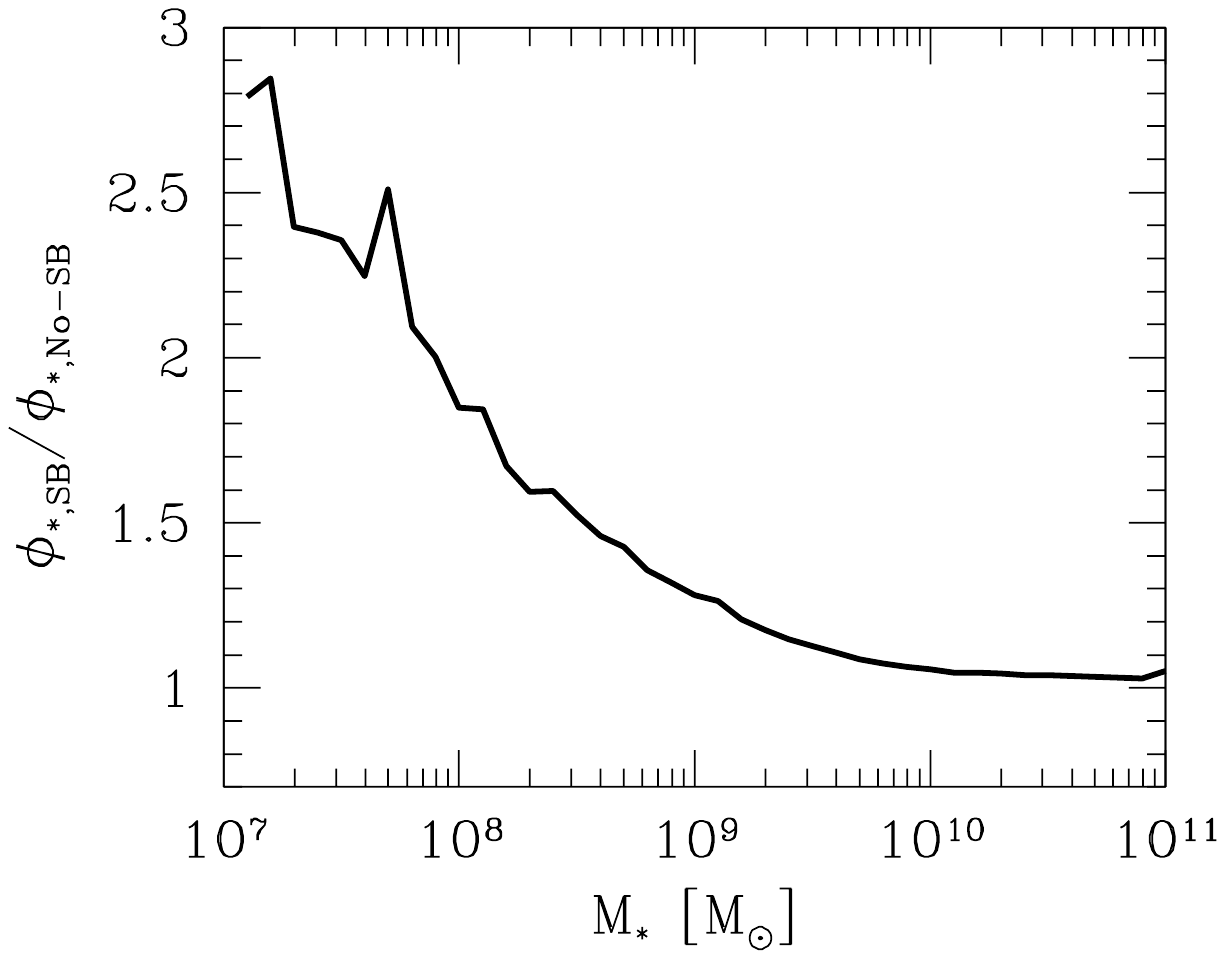}
	\vspace*{-110pt}
		\caption{Impact of
		surface brightness corrections at $z\sim 0.1$. 
		This figure shows the ratio between the \gsmf\ corrected
		for surface brightness incompleteness to the one where
		this correction was ignored.  This correction could be up
		to a factor of $\sim2-3$ at the lowest masses. This correction
		becomes more important for galaxies below $\ms\sim10^{8.5}\msun$ while
		at large masses it is unimportant and the ratio asymptotes to $\sim1$, as expected. 
 	}
	\label{f3}
\end{figure}

In this paper we are interested in constraining the galaxy stellar mass to halo mass 
relation over a wide dynamical mass range, i.e., from dwarf galaxies to giant ellipticals that
are on the centres of big clusters. 
For that reason we impose the following conditions for our GSMF at $z\sim0.1$:  
(1) We will require that it should be complete to the lowest masses and 
(2) sample the largest volume possible 
(in order to avoid sample variance and Poisson variance, \citealp[see e.g.,][]{Smith2012}). 
To do this, we use the 
\gsmf\ derived in Rodriguez-Puebla et al. (in prep.) that has been corrected for the fraction of missing 
galaxies due to surface brightness limits in the SDSS DR7. In short, 
in order to satisfy the above requirements, Rodriguez-Puebla et al. (in prep.)
uses two samples from the SDSS. 
Our first sample consist of a small volume ($0.0033<z<0.05$) 
carefully constructed to study very low mass/luminosity 
galaxies \citep{Blanton2005}. We follow closely the 
methodology described in \citet{Blanton2005} for the correction 
due to surface brightness incompleteness as a function of
\ms. Our second 
galaxy sample consists of the main galaxy sample of the SDSS DR7
in the redshift range $0.01 < z < 0.2$. This volume is large enough to
study intermediate to high mass galaxies without introducing large errors due to 
sample variance and Poission variance. 
Finally, the resulting mass function is 
the combination of the \gsmf\ for low mass galaxies from $\ms\sim10^7\msun$ to 
$\ms\sim10^{9}\msun$ and for the main galaxy sample from $\ms\sim10^{9}\msun$ to
$\ms\sim10^{12.2}\msun$. 

Figure \ref{f3} shows the comparison between the resulting \gsmf\ described above, i.e.,
corrected by surface brightness incompleteness and when ignoring this correction. This
correction could be up to a factor of $\sim2-3$ at lowest masses, as can be seen in Figure \ref{f3}. 

\section{The \gsmf\ at redshifts $> 4$ from the UV LFs}
 \label{UV_LFs} 
  
\begin{figure*}
	\vspace*{-270pt}
	\includegraphics[height=8.9in,width=7.3in]{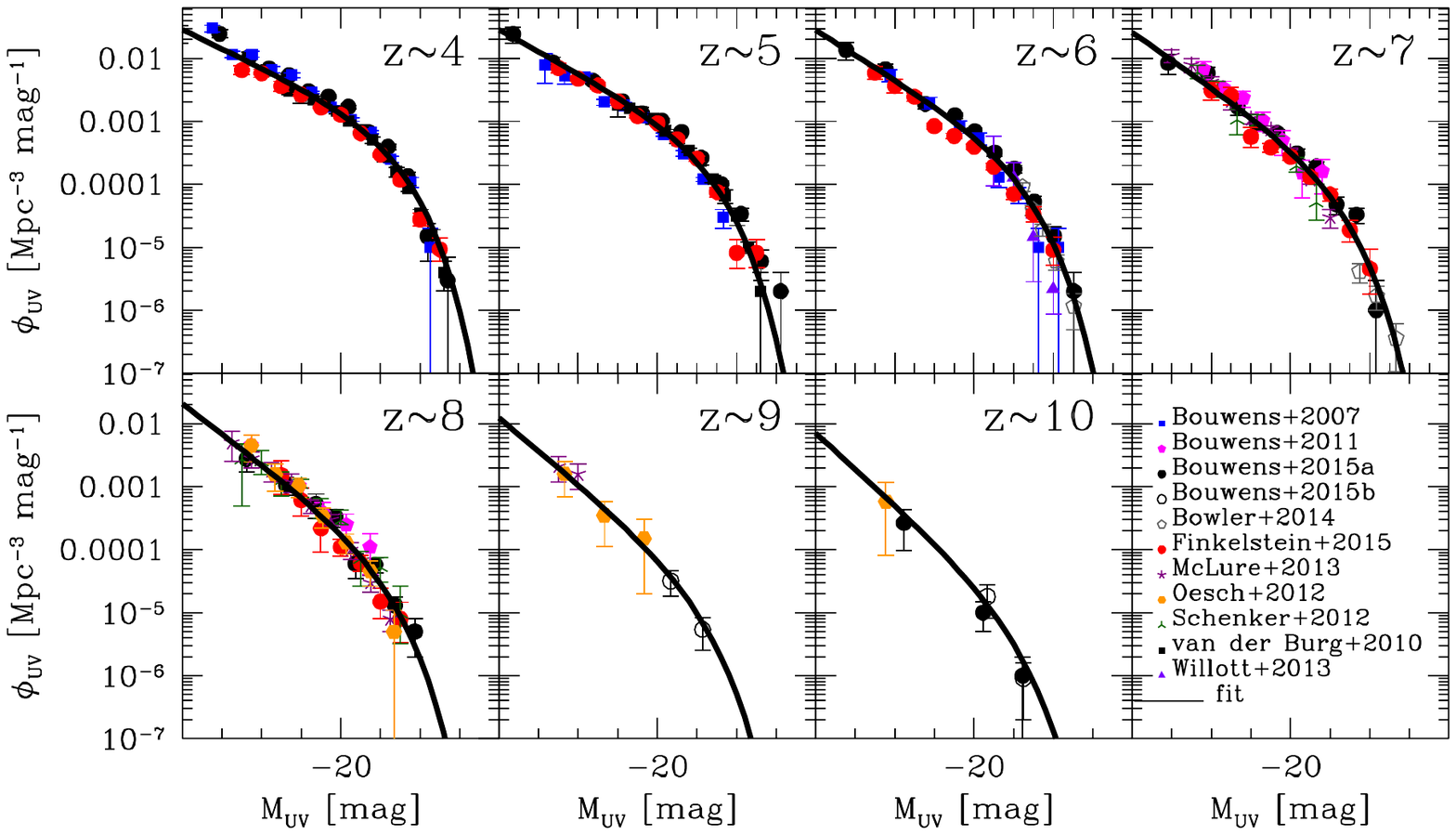}
	\vspace*{-130pt}
		\caption{
		Evolution of the rest-frame UV luminosity function from $z\sim4$ to $z\sim10$. 
		The different symbols show several luminosity functions from the literature as indicated
		in the legends. The black solid lines show the best fit Schechter function 
		from a set of $2\times10^{5}$ MCMC models. Observational data are from \citet{Bouwens+2007,Bouwens+2011,Bouwens+2015,Bouwens+2016,Bowler+2014,Finkelstein+2015,McLure+2013,Oesch+2012,Schenker+2013,vanderBurg+2010} and \citet{Willott+2013}.
 	}
	\label{f4}
\end{figure*}

\begin{figure*}
	\vspace*{-370pt}
	\includegraphics[height=8.9in,width=7.3in]{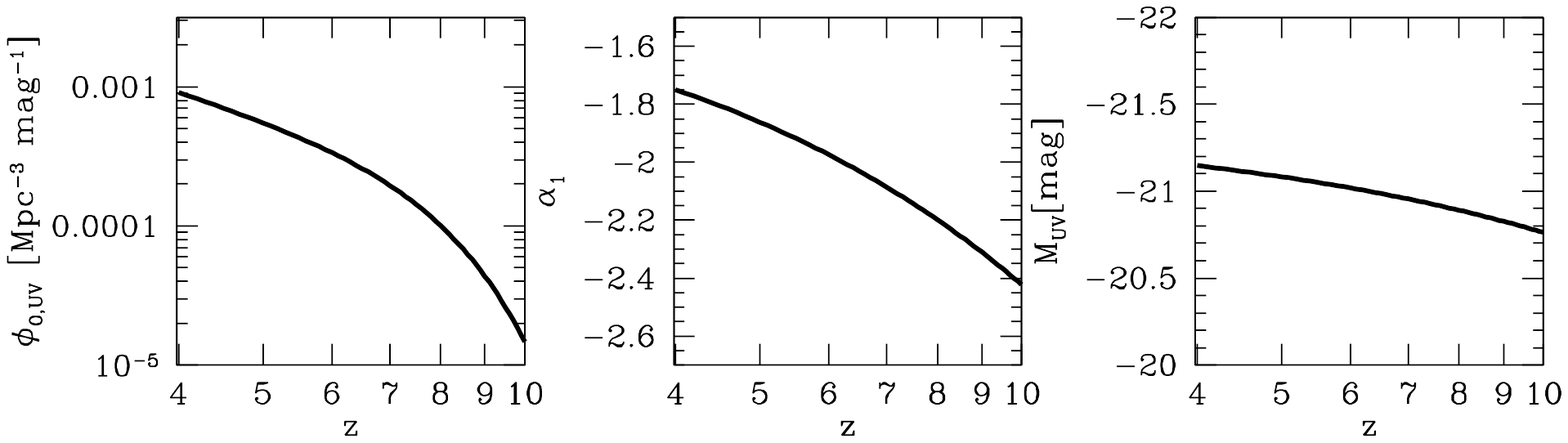}
	\vspace*{-130pt}
		\caption{
		Redshift dependence of the best fit Schechter function parameters $\phi_{\rm UV}$, $\alpha$ and
		$M_{\rm UV} ^*$, see Equations \ref{UV_Eq1}-\ref{UV_Eq3}. Note the slow evolution of  $M_{\rm UV} ^*$ with redshift, while
		there is a strong evolution in $\phi_{\rm UV}$ and a moderate evolution with $\alpha$. 
 	}
	\label{f5}
\end{figure*}

\begin{figure*}
	\vspace*{-270pt}
	\includegraphics[height=8.9in,width=7.3in]{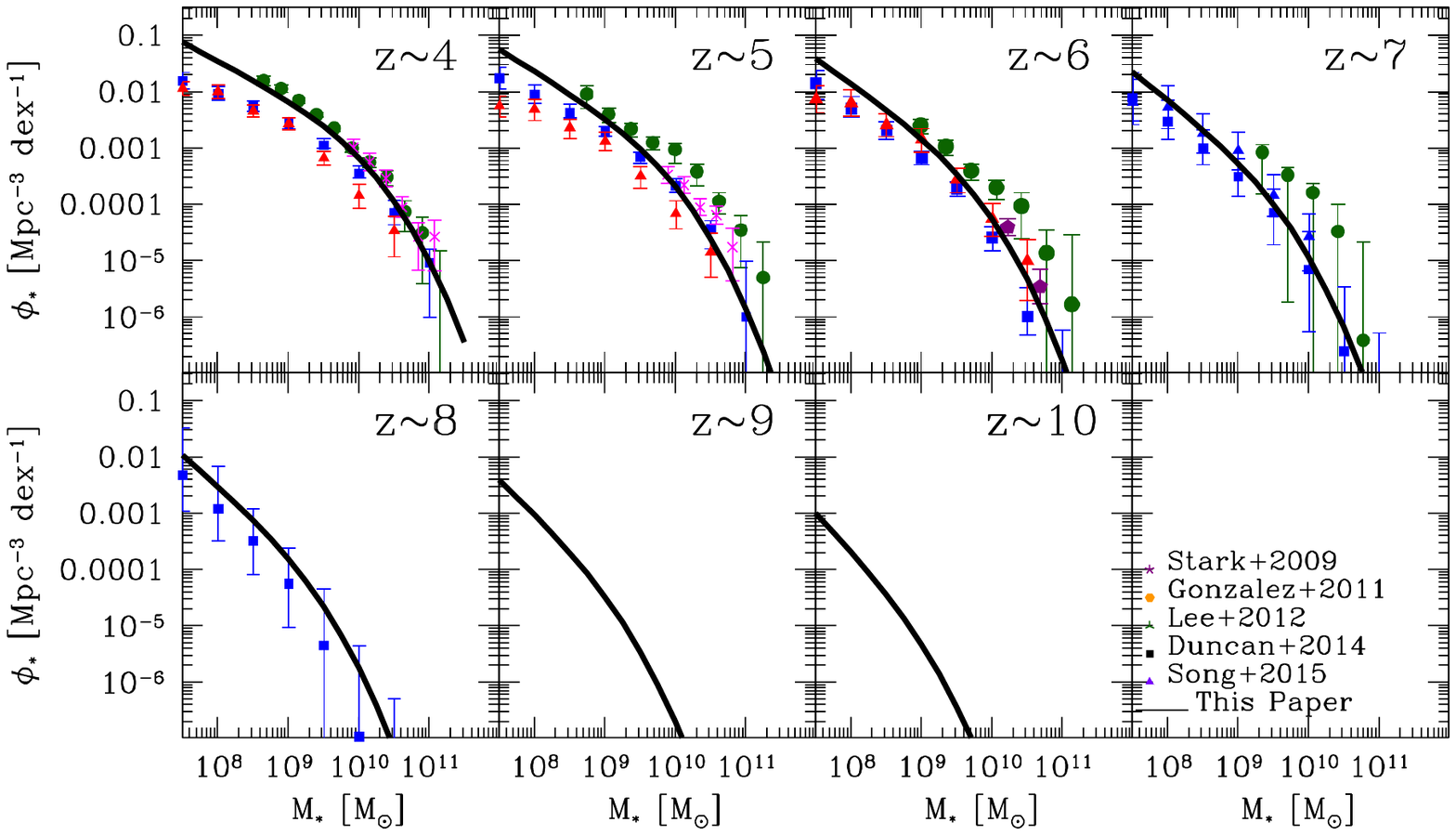}
	\vspace*{-130pt}
		\caption{
		Redshift evolution of the galaxy stellar mass function from $z\sim4$ to $z\sim10$ implied by the 
		rest-frame UV luminosity function and mass-to-light ratios by including nebular emissions, see the 
		text for details, filled circles with error bars. The different symbols show several galaxy stellar mass functions 
		functions from the literature as indicated in the legends.  Our galaxy stellar mass function is consistent with 
		studies. }
	\label{f4a}
\end{figure*}

In recent years the discovery of galaxies at very high redshifts through the Lyman
break technique has permitted study of the evolution of the ultraviolet luminosity functions (UV LFs)
out to $z\sim10$. Figure \ref{f4} shows the redshift evolution of the rest-frame UV LFs
from several measurements in the literature \citep[see also,][]{MadauDickinson2014}. 
Moreover, recent studies have combined UV with near-infrared observations to
derive individual stellar masses \citep[see e.g.,][]{Duncan+2014,Song+2015}. 
In this subsection we will (1) characterize   
the observed redshift evolution of the UV LF based on the data plotted in Figure \ref{f4}, and then 
(2) use this with the observed stellar mass-UV luminosity relations
to derive derive the 
\gsmf\ from $z\sim4$ to $z\sim10$. 

We begin by describing the evolution of the UV LF.
The solid lines in Figure \ref{f4} shows the best-fitting parameters to a Schechter function
that evolves with redshift 
\begin{equation}
	\phiuv(\muv,z) d\muv =  \phinormuv(z) x_{\rm UV}^{1+\alpha(z)}e^{-x_{\rm UV}(z)}  d\muv,
\end{equation}
where $\log\xuv(z) = 0.4 (\mcuv(z) - \muv)$. In the above model $\phinormuv(z)$ is the normalization,
$\alpha(z)$ the slope at the faint end and $\mcuv(z)$ the characteristic luminosity which breaks
the luminosity function from a power law to an exponential decay. Note that all these parameters
should depend on redshift $z$.

In order to derive the best-fitting parameters described above
we combine all the UV LFs plotted in Figure \ref{f4} from $z\sim4$ to $z\sim10$. We sample
the best-fit parameters that maximize the likelihood function by using
the Markov Chain Monte Carlo (MCMC) method as described in \citet{RAD13}, see also Section \ref{mcmc}. We run a set of
$2\times10^{5}$ MCMC models to sample our parameter space. 
The resulting redshift evolution for 
the normalization  \phinormuv\ is:
\begin{eqnarray}
	 \log(\phinormuv) = \log\left(\frac{\ln 10}{2.5}\right) + (-3.038\pm0.032)  +& &  \nonumber \\
	(-0.235 \pm0.029)z_4 + (0.016\pm0.009)z_4^2+ &  \nonumber \\
	 + (-0.005 \pm 0.001)z_4^3, 
	 \label{UV_Eq1}
\end{eqnarray}
for the faint end slope $\alpha$ is:
\begin{eqnarray}
	\alpha = (-1.748\pm0.020) +(-0.114\pm0.017) z_4, 
	\label{UV_Eq2}
\end{eqnarray}
and for the characteristic luminosity \mcuv\ we find:
\begin{eqnarray}
	\mcuv = (-21.143\pm0.048) + (-0.057 \pm 0.042)z_4.
	\label{UV_Eq3}
\end{eqnarray}
In the above equations we define $z_4 = z -4$. Figure \ref{f5} shows the redshift dependence of the best fit 
Schechter function parameters $\phi_{\rm UV}$, $\alpha$ and
$M_{\rm UV} ^*$. Note the slow evolution of  $M_{\rm UV} ^*$ with redshift while
there is a strong evolution in $\phi_{\rm UV}$ and a moderate evolution with $\alpha$. 

The next step in our program is to derive the \gsmf\ by using the observed stellar mass-UV luminosity relations.
Based on deep near-infrared observations of the GOODS South field (part of the CANDELS) 
and on optical data, \citet{Duncan+2014} studied the stellar mass-UV luminosity relations from
$z\sim4$ to $z\sim7$. The authors estimated stellar masses for every galaxy in the sample by fitting
the observed spectral energy distribution with stellar synthesis population models by including 
nebular lines and continuum emissions. The authors found that the 
mass-to-UV light ratios are well described by a simple power law. While
these authors have found shallower slopes compared to previous studies, their results seems 
to be consistent with new determinations \citep[see e.g., ][]{Song+2015}. Based on the best fitting
values reported in Table 3 in \citet{Duncan+2014} for the slopes and zero points restricted to the brightest galaxies ($\muv<-19.5$) 
of the stellar mass-UV luminosity relations, we find the following redshift evolution. 
\begin{equation}
	\log\left(\mathcal{M}_*(\muv) / \msun\right) = -0.19 z + 9.82 - 0.45\times (\muv + 19).
\end{equation}
We will assume that this relation is valid up to $z=10$. We have checked that the parameters
at $z\sim8$ are consistent with those in \citet{Song+2015}. This parameterization is also
very similar to the one reported in \citet{Dayal+2014} based on a simple galaxy formation model to reproduce the evolution of 
the luminosity function and valid up to $z\sim12$. Nevertheless, we caution that this
 this relation could give wrong stellar masses above $z\sim9$. 

Finally, we estimate the redshift evolution of the \gsmf\ from $z\sim4$ to $z\sim10$ by using the
redshift evolution of the UV LFs and the mass-to-light ratios described above. Specifically,
the \gsmf\ can be obtained as
\begin{equation}
	\phigal(\ms,z) d\log\ms = d\log\ms \int P(\ms|\muv,z) d\muv, 
\end{equation}
where
	\begin{eqnarray}
	P(\ms|\muv,z) = \frac{1}{\sqrt{2\pi\sigma_{\rm UV}^2}} \times  & &  \nonumber \\
	\exp\left[-\frac{1}{2\sigma_{\rm UV}^2}\log^2\left(\frac{\ms}{\mathcal{M}_*(\muv,z)}\right)\right],
	\end{eqnarray}
with $\sigma_{\rm UV} = 0.4$ dex and independent of redshift according to \citet{Duncan+2014,Song+2015}.

Figure \ref{f4a} shows the evolution  of the \gsmf\  implied by the rest-frame UV LF and the mass-to-light ratios as described
above. We compare with some previous determinations in the literature as indicated by the labels
and listed in Table \ref{T1}. We find a good agreement with previous determinations. Finally, we compute the error bars
in our \gsmf\ by adding in quadratures errors from the fits to the UV-LF (we used the MCMC chain to do so), poissonian 
and cosmic variance. To estimate poissonian error bars we assume a survey area similar to CANDELS. We use the 
$\Lambda$CDM power spectrum and the CANDELS survey area to estimate error bars from cosmic
variance.

\end{document}